\pdfoutput=1

\documentclass[11pt,twoside,a4paper,cmspaper,final,collab]{cms-tdr}
\def\svnVersion{382756}\def\cmsCernNoTag{CERN-EP/2016-160}\def\cmsCernDate{\today}\def\cmsMessage{Published in the Journal of Instrumentation as \href{http://dx.doi.org/10.1088/1748-0221/12/01/P01020}{\doi{10.1088/1748-0221/12/01/P01020.}}}

\begin{document}\cmsNoteHeader{TRG-12-001}

\hyphenation{had-ron-i-za-tion}
\hyphenation{cal-or-i-me-ter}
\hyphenation{de-vices}
\RCS$Revision: 381922 $
\RCS$HeadURL: svn+ssh://svn.cern.ch/reps/tdr2/papers/TRG-12-001/trunk/TRG-12-001.tex $
\RCS$Id: TRG-12-001.tex 381922 2017-01-20 20:22:36Z wittich $
\title{The CMS trigger system}

\providecommand{\alpT}{\ensuremath{\alpha_\mathrm{T}}\xspace}
\providecommand{\ST}{\ensuremath{S_\mathrm{T}}\xspace}
\renewcommand{\RNINE}{\ensuremath{\mathrm{R}_\mathrm{9}}\xspace}

\date{\today}

\abstract{This paper describes the CMS trigger system and its
  performance during Run 1 of the LHC.  The trigger system consists of
  two levels designed to select events of potential physics interest
  from a GHz (MHz) interaction rate of proton-proton (heavy ion)
  collisions. The first level of the trigger is implemented in
  hardware, and selects events containing detector signals consistent
  with an electron, photon, muon, $\tau$ lepton, jet, or missing
  transverse energy. A programmable menu of up to 128 object-based
  algorithms is used to select events for subsequent processing.  The
  trigger thresholds are adjusted to the LHC instantaneous luminosity
  during data taking in order to restrict the output rate to 100\unit{kHz},
  the upper limit imposed by the CMS readout electronics. The second
  level, implemented in software, further refines the purity of the
  output stream, selecting an average rate of 400\unit{Hz} for offline event
  storage. The objectives, strategy and performance of the trigger
  system during the LHC Run 1 are described.}

\hypersetup{%
pdfauthor={CMS Collaboration},%
pdftitle={The CMS trigger system},%
pdfsubject={CMS},%
pdfkeywords={CMS, physics, software, computing, trigger}}

\maketitle

\tableofcontents
\newpage

\section{Introduction}\label{chap:intro}
\label{sec:intro}
The Compact Muon Solenoid (CMS)~\cite{Chatrchyan:2008zzk} is a
multipurpose detector designed for the precision measurement of
leptons, photons, and jets, among other physics objects, in
proton-proton as well as heavy ion collisions at the CERN
LHC~\cite{LHC}. The LHC is designed to collide protons at a
center-of-mass energy of 14\TeV and a luminosity of $10^{34}\percms$.
At design luminosity, the pp interaction rate exceeds
1\unit{GHz}.  Only a small fraction of these collisions contain events
of interest to the CMS physics program, and only a small fraction of
those can be stored for later offline analysis. It is the job of the
trigger system to select the interesting events for offline storage
from the bulk of the inelastic collision events.

To select events of potential physics interest~\cite{DAQ-TDR}, the
CMS trigger utilizes two levels while, for comparison, ATLAS uses a
three-tiered system~\cite{ATLAS-trig}. The first level (L1) of the CMS
trigger is implemented in custom hardware, and selects events
containing candidate objects, \eg, ionization deposits consistent with
a muon, or energy clusters consistent with an electron, photon, $\tau$ lepton,
missing transverse energy (\MET), or jet. Collisions with possibly
large momentum transfer can be selected by, \eg, using the scalar sum
of the jet transverse momenta (\HT).

The final event selection is based on a programmable menu where, by
means of up to 128 algorithms utilizing those candidate objects,
events are passed to the second level
(high-level trigger, HLT). The
thresholds of the first level are adjusted during data taking in
response to the value of the LHC instantaneous luminosity so as to
restrict the output rate to 100\unit{kHz}~\cite{DAQ-TDR}, the upper
limit imposed by the CMS readout electronics. The HLT, implemented in
software, further refines the purity of the physics objects, and
selects an average rate of 400\unit{Hz} for offline storage.  The
overall output rate of the L1 trigger and HLT can be adjusted by
prescaling the number of events that pass the selection criteria of
specific algorithms.  In addition to collecting collision data, the
trigger and data acquisition systems record information for the
monitoring of the detector.

\begin{figure}[btp]
  \centering
  \includegraphics[width=\textwidth]{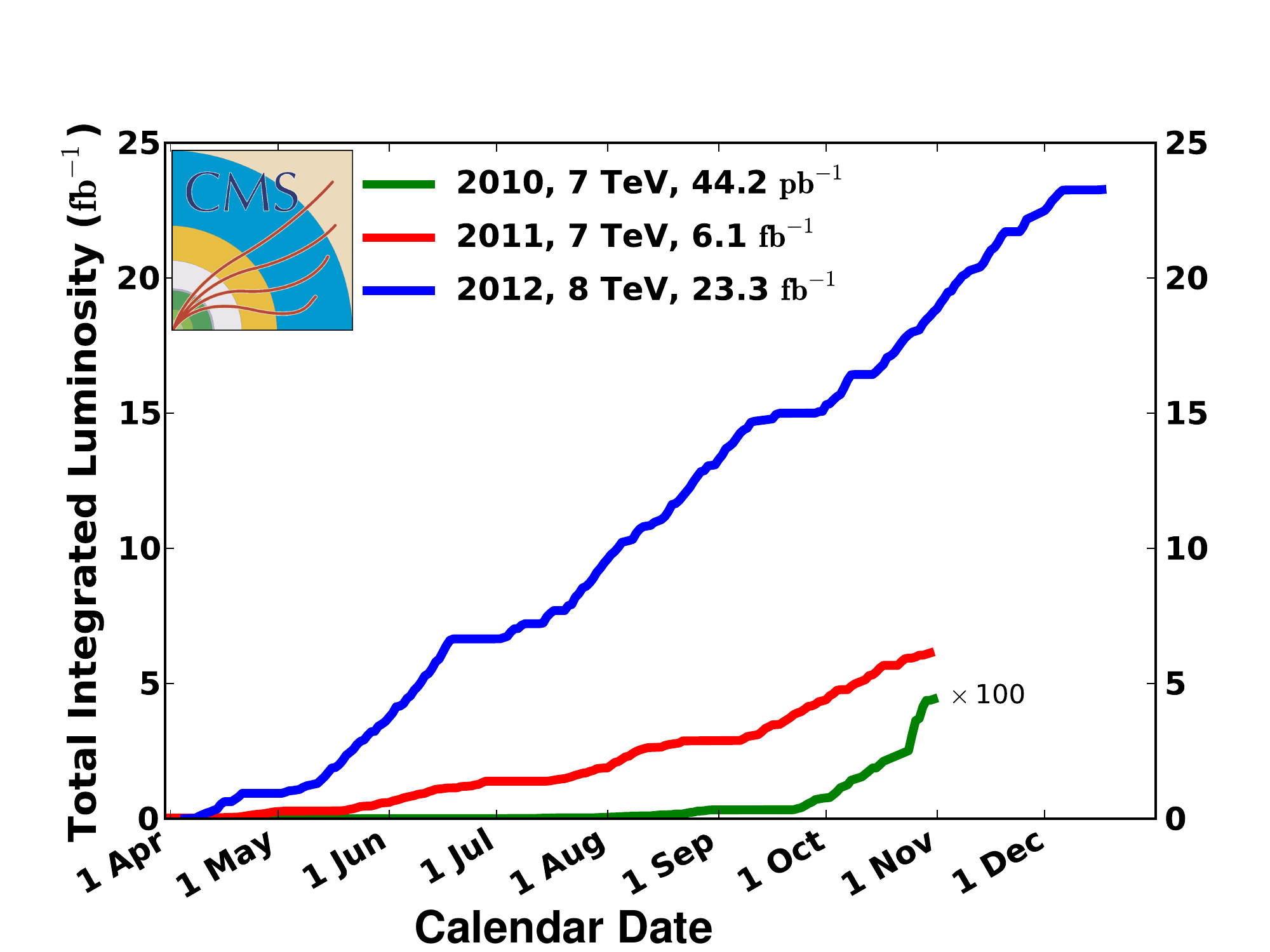}
  \includegraphics[width=\textwidth]{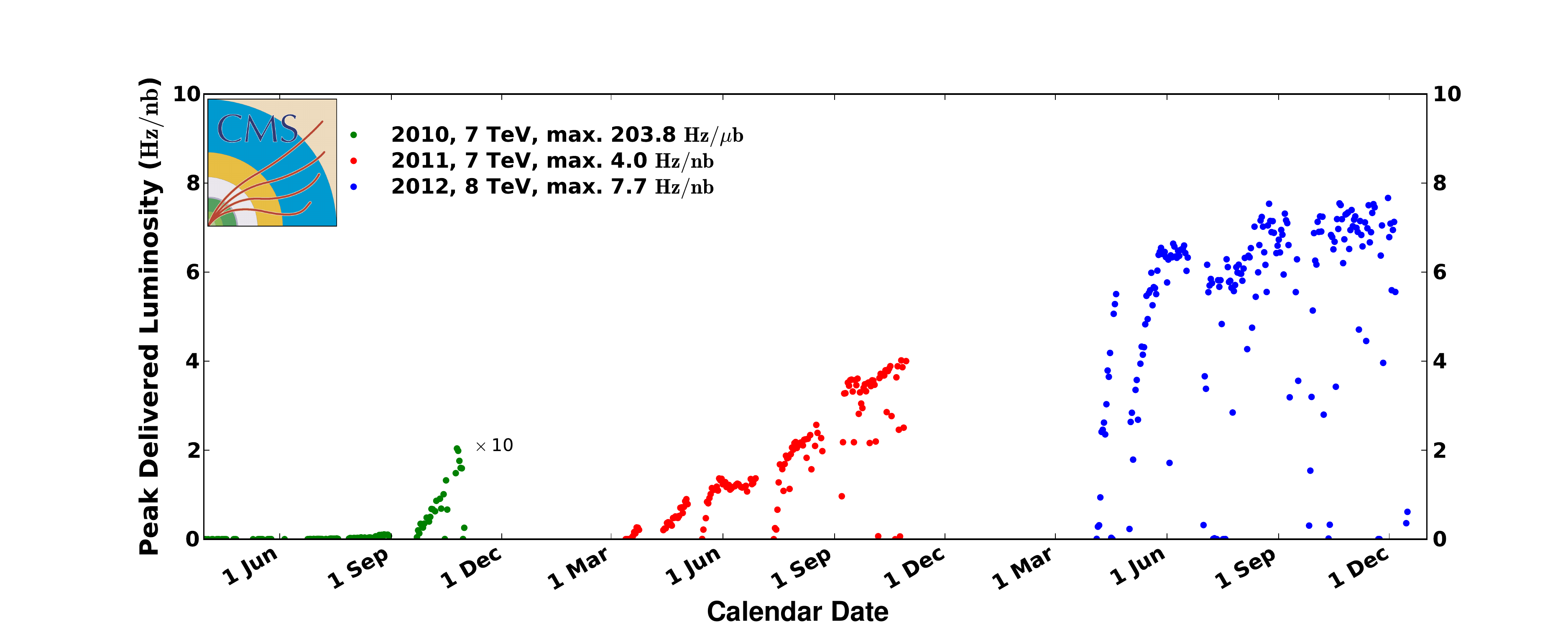}
  \caption{Integrated (top) and peak (bottom) proton-proton
    luminosities as a function of time for calendar years
    2010--2012. The 2010 integrated (instantaneous) luminosity is
    multiplied by a factor of 100 (10). In the lower plot,
    $1\unit{Hz/nb}$ corresponds to $10^{33}\percms$.}
  \label{fig:lumi2012}
\end{figure}

After commissioning periods at 0.9 and 2.36\TeV in 2009, the first
long running periods were at a center-of-mass energy of 7\TeV in 2010
and 2011, and 8\TeV in 2012.  These proton-proton data, together with
the first ion running periods (PbPb at 2.76\TeV, and
pPb at 5.02\TeV), are referred to collectively as Run~1. During this
period, the CMS trigger system selected interesting pp physics events at
maximum instantaneous luminosities of $2.1\times 10^{32}\percms$
(2010), $4\times 10^{33}\percms$ (2011), and
$7.7\times 10^{33}\percms$ (2012), corresponding to 0.2, 4, and
$7.7\,\mathrm{Hz\,nb}^{-1}$.  Figure~\ref{fig:lumi2012} shows the pp
integrated and peak luminosities as a function of time for calendar
years 2010, 2011 and 2012. While the  nominal bunch crossing (BX)
frequency is 40\unit{MHz}, corresponding to 25\unit{ns} between individual bunch
collisions, the bunch spacing during regular running was never less
than 50\unit{ns} through Run~1. The highest number of
collisions per BX (known as ``pileup'') averaged over a data run in
2011 and 2012 was 16.15 and 34.55, respectively, while the pileup
averages over the year were 9 (21) in 2011 (2012).

The trigger system is also used during heavy ion running. The
conditions for PbPb collisions are significantly different from those
in the pp case. The instantaneous luminosity delivered by the LHC in
the 2010 (2011) PbPb running period was $3\times 10^{25}$
($5\times 10^{26}$)\percms, resulting in maximum interaction rates of
250\unit{Hz} (4\unit{kHz}), much lower than in pp running, with a negligible
pileup probability and an inter-bunch spacing of 500\unit{ns}
(200\unit{ns}). During the pPb run in 2013, an instantaneous luminosity of
$10^{29}\percms$ was achieved, corresponding to an interaction
rate of 200\unit{kHz}, again with a very low pileup probability. Due to
the large data size in these events, the readout rate of the
detector is limited to 3\unit{kHz} in heavy ion collisions.

This document is organized as follows.
Section~\ref{sec:cmstrigger_intro} describes the CMS trigger system
(L1 and HLT) in detail. Section~\ref{sec:objid} gives an overview of
the methods, algorithms, and logic used to identify physics signatures
of interest in LHC collisions, and to select events accordingly. The
physics performance achieved with the CMS trigger system is outlined
in Section~\ref{sec:hpa} based on  examples of several physics
analyses. In Section~\ref{sec:triggermenus}, details of the L1 and HLT
menus are given, together with the objectives and strategies to assemble
those menus. The operation and evolution of the trigger system during
the first years of the LHC running is described in
Section~\ref{sec:operations}. A summary is given in
Section~\ref{sec:summary}.

\subsection{The CMS detector}
\label{sec:cmsdetector}

The central feature of the CMS apparatus is a superconducting solenoid, of
6\unit{m} internal diameter, providing a magnetic field of
3.8\unit{T}. Within the superconducting solenoid volume are a silicon
pixel and strip tracker, a lead tungstate crystal electromagnetic
calorimeter (ECAL), and a brass/scintillator hadron calorimeter (HCAL). Muons are
measured in gas-ionization detectors embedded in the steel return
yoke. Extensive forward calorimetry complements the coverage provided
by the barrel and endcap detectors. The missing transverse momentum
vector is defined as the projection on the plane perpendicular to the
beams of the negative vector sum of the momenta of all reconstructed
particles in an event. Its magnitude is referred to as \ETmiss. The
transverse momentum vector is defined as the projection on the plane
perpendicular to the beams of the negative vector sum of the momenta
of all reconstructed particles in an event. Its magnitude is referred
to as \ET. A more detailed description of the CMS detector, together
with a definition of the coordinate system used and the relevant
kinematic variables, can be found in Ref.~\cite{Chatrchyan:2008zzk}.

\section{The trigger system}
\label{sec:cmstrigger_intro}
The trigger system is comprised of an L1 hardware trigger and an HLT
array of commercially available computers running high-level physics
algorithms. In this section we describe the design of the combined
L1--HLT system.

\subsection{The L1 trigger overview}
\label{sec:l1overview}

The L1 trigger is a hardware system with a fixed latency. Within
4\mus of a collision, the system must decide if an
event should be tentatively accepted or rejected using
information from the calorimeter and muon detectors.

\begin{figure}[tbp]
  \centering
    \includegraphics[width=0.8\textwidth]{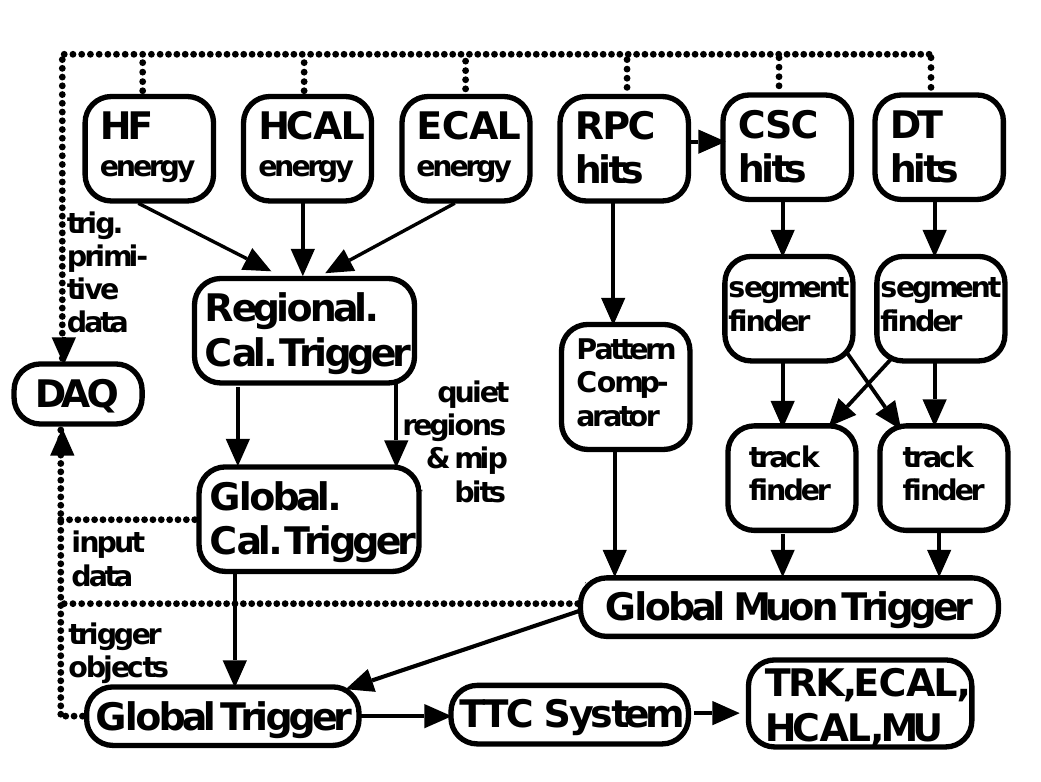}
    \caption{Overview of the CMS L1 trigger system. Data from the
      forward (HF) and barrel (HCAL) hadronic calorimeters, and from
      the electromagnetic calorimeter (ECAL), are processed first
      regionally (RCT) and then globally (GCT). Energy deposits (hits)
      from the resistive-plate chambers (RPC), cathode strip chambers
      (CSC), and drift tubes (DT) are processed either via a pattern
      comparator or via a system of segment- and track-finders and
      sent onwards to a global muon trigger (GMT). The
      information from the GCT and GMT is combined in a global trigger
      (GT), which makes the final trigger decision. This decision is
      sent to the tracker (TRK), ECAL, HCAL or muon systems (MU) via
      the trigger, timing and control (TTC) system. The data
      acquisition system (DAQ) reads data from various subsystems for
      offline storage. MIP stands for minimum-ionizing particle.}
    \label{fig:L1-over}
\end{figure}

A schematic of the L1 trigger is shown in
Fig.~\ref{fig:L1-over}. The trigger primitives (TP) from
electromagnetic and hadron calorimeters (ECAL and HCAL) and from the
muon detectors (drift tubes (DT), cathode strip chambers (CSC) and
resistive-plate chambers (RPC)) are processed in several steps before
the combined event information is evaluated in the global trigger (GT)
and a decision is made whether to accept the event or not.

The L1 calorimeter trigger comprises two stages, a regional
calorimeter trigger (RCT) and a global calorimeter trigger (GCT). The
RCT receives the transverse energies and quality flags from over
8000 ECAL and HCAL towers (Sec.~\ref{sec:ecaltpg} and
\ref{sec:hcaltpg}), giving trigger coverage over $\abs{\eta}<5$.
The RCT processes this information in parallel and sends as output
$\Pe/\Pgg$ candidates and regional \ET sums based on $4{\times}4$
towers~\cite{Trig-TDR}.
The GCT sorts the $\Pe/\Pgg$ candidates further, finds
jets (classified as central, forward, and tau) using the \ET sums, and
calculates global quantities such as \MET. It sends as output four
$\Pe/\Pgg$ candidates each of two types, isolated and
nonisolated, four each of central, tau, and forward jets, and several
global quantities.

Each of the three muon detector systems in CMS participates in the L1
muon trigger to ensure good coverage and redundancy. For the DT and
CSC systems ($\abs{\eta}<1.2$ and $\abs{\eta}>0.9$, respectively), the
front-end trigger electronics identifies track segments from the hit
information registered in multiple detector planes of a single
measurement station. These segments are collected and then transmitted
via optical fibers to regional track finders in the electronics
service cavern, which then applies pattern recognition algorithms that
identifies muon candidates and measure their momenta from the amount
they bend in the magnetic field of the flux-return yoke of the
solenoid. Information is shared between
the DT track finder (DTTF) and CSC track finder (CSCTF) for efficient
coverage in the region of overlap between the two systems at
$\abs{\eta}\approx1$. The hits from the RPCs ($\abs{\eta}<1.6$) are directly
sent from the front-end electronics to pattern comparator trigger
(PACT) logic boards that identify muon candidates. The three regional
track finders sort the identified muon candidates and transmit to the
global muon trigger (GMT) up to 4 (CSCTF, DTTF) or 8 (RPC) candidates
every bunch crossing. Each candidate is assigned a \PT and quality
code as well as an ($\eta$,$\phi$) position in the muon system (with a
granularity of ${\approx}0.05$). The GMT then merges muon candidates found
by more than one system to eliminate a single candidate passing
multiple-muon triggers (with several options on how to select \PT
between the candidates). The GMT also performs a further quality
assignment so that, at the final trigger stage, candidates can be
discarded if their quality is low and they are reconstructed only by
one muon track finder.

The GT is the final step of the CMS L1 trigger system and implements
a  menu of triggers, a set of selection requirements
applied to the final list of objects (\ie,  electrons/photons,
muons, jets, or $\tau$ leptons), required by the algorithms of the HLT
algorithms to meet the physics data-taking objectives. This menu
includes trigger criteria ranging from simple single-object selections
with \ET above a preset threshold to selections requiring coincidences
of several objects with topological conditions among them. A maximum
of 128 separate selections can be implemented in a menu.

\subsection{The L1 calorimeter trigger system}
The following is the description of the reconstruction of ECAL and HCAL energy deposits used in the L1 trigger chain followed by describing the RCT and GCT processing steps operating on these trigger primitives.

\subsubsection{The ECAL trigger primitives}
\label{sec:ecaltpg}
The ECAL trigger primitives are computed from a barrel (EB) and two
endcaps (EE), comprising 75\,848 lead tungstate (PbWO$_4$) scintillating
crystals equipped with avalanche photodiode (APD) or vacuum
phototriode (VPT) light detectors in the EB and EE, respectively. A
preshower detector (ES), based on silicon sensors, is placed in front
of the endcap crystals to aid particle identification. The ECAL is
highly segmented, is radiation tolerant and has a compact and hermetic
structure, covering the pseudorapidity range of $\abs{\eta} < 3.0$. Its
target resolution is 0.5\% for high-energy electrons/photons. It
provides excellent identification and energy measurements of electrons
and photons, which are crucial to searches for many new physics
signatures.
In the EB, five strips of five crystals (along the azimuthal direction) are
combined into trigger towers (TTs) forming a $5{\times}5$ array
of crystals. The
transverse energy detected by the crystals in a single TT is summed
into a TP by the front-end electronics and sent to off-detector
trigger concentrator cards (TCC) via optical fibers. In the EE,
trigger primitive computation is completed in the TCCs, which must
perform a mapping between the collected pseudo-strips trigger data
from the different supercrystals and the associated trigger towers.

\paragraph{Mitigation of crystal transparency changes at the trigger level}
\label{sec:trans}

Under irradiation, the ECAL crystals lose some of their transparency,
part of which is recovered when the radiation exposure stops (\eg,
between LHC fills). The effect of this is that the response of the
ECAL varies with time. This variation is accounted for by the use of a
laser system that frequently monitors the transparency
of each crystal~\cite{ecallaser} and allows for offline corrections to the
measured energies to be made~\cite{Chatrchyan:2013dga}.
In 2011, the levels of radiation in ECAL were quite small, and no
corrections to the response were made at L1. From 2012 onwards, where
the response losses were larger, particularly in the EE, corrections
to the TT energies were calculated
and applied on a weekly basis in order to maintain  high trigger
efficiency and low trigger thresholds.

\subsubsection{HCAL trigger primitives}
\label{sec:hcaltpg}
The HCAL TPs are computed out of the digital samples of the detector
pulses by the trigger primitive generator (TPG).  In the barrel, one
trigger primitive corresponds to one HCAL readout, whereas raw data
from the two depth-segmented detector readout elements are summed in
the endcap hadron calorimeter.  For the forward hadron calorimeter
(HF), up to 12 readouts are summed to form one trigger primitive. One
of the most important tasks of the TPG is to assign a precise bunch
crossing to detector pulses, which span over several clock
periods. The bunch crossing assignment uses a digital filtering
technique applied to the energy samples, followed by a peak finder
algorithm.  The amplitude filters are realized using a sliding sum of
2 consecutive samples. A single sample is used for HF where the
signals are faster.  The peak finder selects those samples of the
filtered pulse that are larger than the two nearest neighbors. The
amplitudes of the peak and peak+1 time slices are used as an estimator
of the pulse energy. The position of the peak-filtered sample in the data
pipeline flow determines the timing. The transverse energy
of each HCAL trigger tower is calculated on a 10-bit linear scale. In
case of overflow, the \ET is set to the scale maximum. Before
transmission to the RCT, the 10-bit trigger tower \ET is converted to
a programmable 8-bit compressed nonlinear scale in order to minimize
the trigger data flux to the regional trigger. This data compression
leads to a degradation in the trigger energy resolution of less than
5\%. The energy in\GeV is obtained from the ADC count by converting
the ADC count into fC, subtracting the pedestal and correcting for the
gain of each individual channel. Finally, a correction factor is
applied to compensate for the fraction of signal charge not captured
in the two time-slice sum.
\subsubsection{Regional calorimeter trigger system}

The CMS L1 electron/photon ($\Pe/\Pgg$), $\tau$ lepton, jet,
$\HT$ (where $\HT = \sum \pt^{\text{jets}}$ is the scalar sum of the \pt of all jets with
 $\pt>10$\GeV and $\abs{\eta}<3$),
and missing \ET
trigger decisions are based on input from the L1 regional calorimeter trigger
(RCT)~\cite{Trig-TDR,klab2007,klab2008,klab2009}. Eighteen crates of custom RCT
electronics process data for the barrel, endcap,
and forward calorimeters, with a separate crate for LHC clock distribution.

Twenty-four bits comprising two 8-bit calorimeter energies, either two ECAL fine-grain (FG)
bits or two HCAL minimum ionizing particle (MIP) bits,
an LHC bunch crossing bit, and 5 bits of error detection code, are sent from the ECAL, HCAL,
and HF calorimeter back-end electronics to the nearby RCT racks on 1.2\unit{Gbaud} copper links.
This is done using one of the four 24-bit channels of the Vitesse 7216-1 serial transceiver
chip on the calorimeter output and the RCT input, for 8 channels of calorimeter data per chip.
The RCT V7216-1 chips are mounted on receiver mezzanine cards located on each of 7
receiver cards (RC) and the single-jet summary cards (JSC) for all 18
RCT crates.

The RCT design includes five high-speed custom GaAs
application-specific integrated circuits (ASICs), which were designed
and manufactured by Vitesse Semiconductor: a phase ASIC, an adder
ASIC, a boundary
scan ASIC, a sort ASIC, and an electron isolation
ASIC~\cite{smith2000}.

The RC has eight receiver mezzanine cards for the HCAL and ECAL data,
four per subsystem. On the mezzanine, the V7216-1 converts the serial
data to 120\unit{MHz} TTL parallel data.  Eight phase ASICs on the RC align
and synchronize the data received on four channels of parallel data
from the Vitesse 7216-1, check for data transmission errors, and
convert 120\unit{MHz} TTL to 160\unit{MHz} emitter-coupled logic (ECL) parallel
data. Lookup tables (LUTs) convert 17 bits of input (8 bits from ECAL,
HCAL and the FG bit) for two separate paths. They rescale the incoming
ECAL energies, and set quality bits for the $\Pe/\Pgg$ path (a
tower-level logical OR of the ECAL FG bits and a limit on fractional
energy in the HCAL), and rescale and sum HCAL and ECAL for the
regional sums path.  On the RC, the boundary scan ASIC aligns the
$\Pe/\Pgg$ tower energy data with data shared on cables between
RCT crates adjacent in $\eta$ and $\phi$, and makes copies so that
each of 7 electron isolation cards (EIC) receives 28 central and 32
adjacent towers via the custom 160\unit{MHz} backplane.  The HCAL+ECAL
summed towers are added together to form $4{\times}4$ trigger tower sums
by three adder ASICs, which sum up eight 11-bit energies in 25\unit{ns},
while providing bits for overflows. The tower sums are then  sent to the JSC via the
backplane for further processing. A logical OR of the MIP bits over
the same $4{\times}4$ trigger tower regions is sent to the JSC.

The EIC receives the 32 central tower and 28 neighboring trigger tower
data from the RCs via the backplane. The electron isolation algorithm
is implemented in the electron isolation ASIC, which can handle four
7-bit electromagnetic energies, a veto bit, and nearest neighbor
energies every 6.25\unit{ns}. It finds up to four electron candidates
in two $4{\times}4$ trigger tower regions, two isolated and two
non-isolated. These candidates are then transmitted via the backplane
to the JSC for further processing. In this way the $\Pe/\Pgg$
algorithm is seamless across the entire calorimeter.

\begin{figure}[tbph]
\centering
\includegraphics[width=4in]{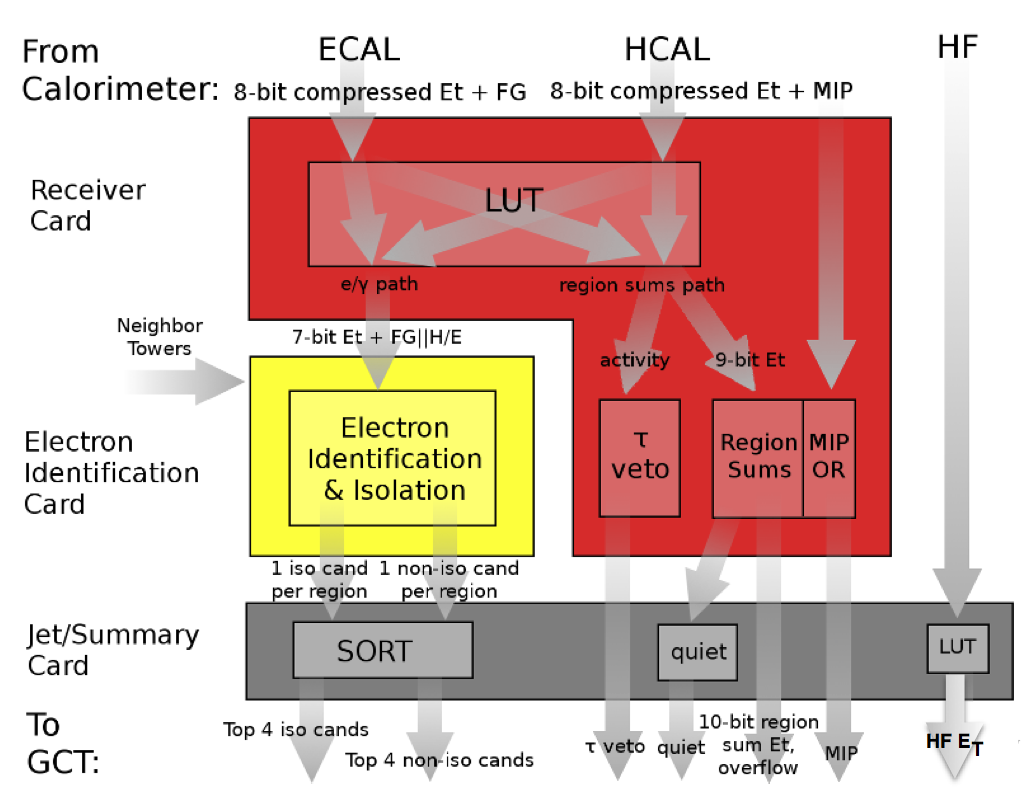}
\caption{Block diagram of the regional calorimeter trigger (RCT)
  system showing the data flow through the different cards in a RCT
  crate. At the top is the input from the calorimeters; at the bottom
  is the data transmitted to the global calorimeter trigger
  (GCT). Data exchanged on the backplane is shown as arrows between
  cards. Data from neighboring towers come via the backplane, but may
  come over cables from adjoining crates. }
\label{fig:rct}
\end{figure}

The JSC receives 28 $\Pe/\Pgg$ candidates, 14 sums, and has a
single mezzanine card to receive eight HF TPs and quality bits. The JSC
rescales the HF data using a lookup table and delays the data so
that it is in time with the 14 regional \ET   sums when they are sent
to the GCT for the jet finding and calculation of global quantities
such as $\HT$ and missing \ET.  In addition, for muon isolation, a
quiet bit is set for each region and forwarded with the MIP bits on
the same cables as the electron candidates.  The 28 electron
candidates (14 isolated and non-isolated) are sorted in \ET in two
stages of sort ASICs on the JSC, and the top four of each type are
transmitted to the GCT for further sorting. A block diagram of this
dataflow is shown in Fig.~\ref{fig:rct}.

Finally, a master clock crate (MCC) and cards are located in one of
the ten RCT racks to provide clock and control signal
distribution. Input to the system is provided by the CMS trigger
timing and control (TTC) system.  This provides the LHC clock, bunch
crossing zero (BC0), and other CMS synchronization signals via an
optical fiber from a TTC VME interface board which can internally generate or
receive these signals from either a local trigger controller board
(LTC) or from the CMS GT.

The MCC includes a clock input card (CIC) with an LHC TTC receiver
mezzanine (TTCrm) to receive the TTC clocks and signals via the fiber
and set the global alignment of the signals. The CIC feeds fan-out
cards, a clock fan-out card midlevel (CFCm) and a clock fan-out card
to crates (CFCc) to align and distribute the signals to the individual
crates via low-skew cable. Adjustable delays on these two cards allow
fine-tuning of the signals to the individual crates.

\subsubsection{Global calorimeter trigger system}
\label{sec:gct}

The GCT is the last stage of the L1 calorimeter trigger chain.  A detailed description of the GCT design, implementation and commissioning is provided in several conference papers~\cite{stettler:2006,iles:2006,foudas:2007,iles:2008,tapper:2008,brooke:2009} that describe the changes in design since the CMS trigger technical design report~\cite{Trig-TDR}.

The trigger objects computed by the GCT from data supplied by the RCT
are listed below and described in subsequent paragraphs:
\begin{itemize}
\item four isolated and four non-isolated electrons/photons of highest transverse energy;
\item four central, four forward, and four tau jets of highest transverse energy;
\item total transverse energy ($S_\mathrm{T}$),
\begin{equation*}
S_\mathrm{T} \equiv \sum \et,
\end{equation*}
calculated as the scalar sum of the \ET of all calorimeter deposits;
$H_\mathrm{T}$ (see Section 1); and ($\ETmiss$);
\item missing jet transverse energy; summing of feature bits and transverse energies in the HF calorimeter.
\end{itemize}

The electron/photon sort operation must determine the four highest transverse energy objects from 72 candidates supplied by the RCT,
for both isolated and non-isolated electrons/photons.

To sort the jets, the GCT must first perform jet finding and calibrate the clustered jet energies. The jets are created from the 396 regional transverse energy sums supplied by the RCT. These are the sum of contributions from both the hadron and electromagnetic calorimeters. This is a substantial extension of the GCT capability beyond that specified in Ref.~\cite{Trig-TDR}. The jet finding and subsequent sort is challenging because of the large data volume and the need to share or duplicate data between processing regions to perform cluster finding. The latter can require data flows of a similar magnitude to the incoming data volume depending on the clustering method used. The clusters, defined as the sum of $3{\times}3$ regions, are located using a new method~\cite{iles:2006} that requires substantially less data sharing than the previously proposed sliding window method~\cite{oldGCT}. Jets are subdivided into central, forward, and tau jets based on the RCT tau veto bits and the jet
pseudorapidity.

The GCT must also calculate some additional quantities. The total transverse energy is the sum of all regional transverse energies.  The total missing transverse energy \MET is calculated by splitting the regional transverse energy values into their $x$ and $y$ components and summing the components in quadrature. The resulting vector, after a rotation of $180^\circ$, provides the magnitude and angle of the missing energy.  The jet transverse energy \HT and missing jet transverse energy  are the corresponding sums over all clustered jets found.

Finally two quantities are calculated for the forward calorimeters. The transverse energy is summed for the two rings of regions
closest to the beam pipe in both positive and negative  pseudorapidities. The number of regions in the same rings with the
fine-grain bit is also counted.

In addition to these tasks the GCT acts as a readout device for both
itself and the RCT by storing information until receipt of an L1 accept
(L1A) and then sending the information to the DAQ.

The GCT input data volume and processing requirements did not allow all data to be concentrated in one processing unit. Thus, many large field programmable gate arrays (FPGA) across multiple discrete electronics cards are necessary to reduce the data volume in stages. The cards must be connected
together to allow data sharing and to eventually concentrate the data into a single location for the sort algorithms.

The latency allowed is 24 bunch crossings for jets and 15 bunch crossings for electrons/photons. Using many layers of high-speed
serial links to transport the large data volumes between FPGAs was not possible since these typically require several clock cycles
to serialize/deserialize the data and thus they have to be used sparingly to keep the latency low. The final architecture uses
high-speed optical links (1.6\unit{Gb/sec}) to transmit the data and then concentrates the data in the main processing FPGAs, followed by standard FPGA I/O to connect to downstream FPGAs.

\begin{figure}[tbph]
\centering
\includegraphics[width=0.6\textwidth]{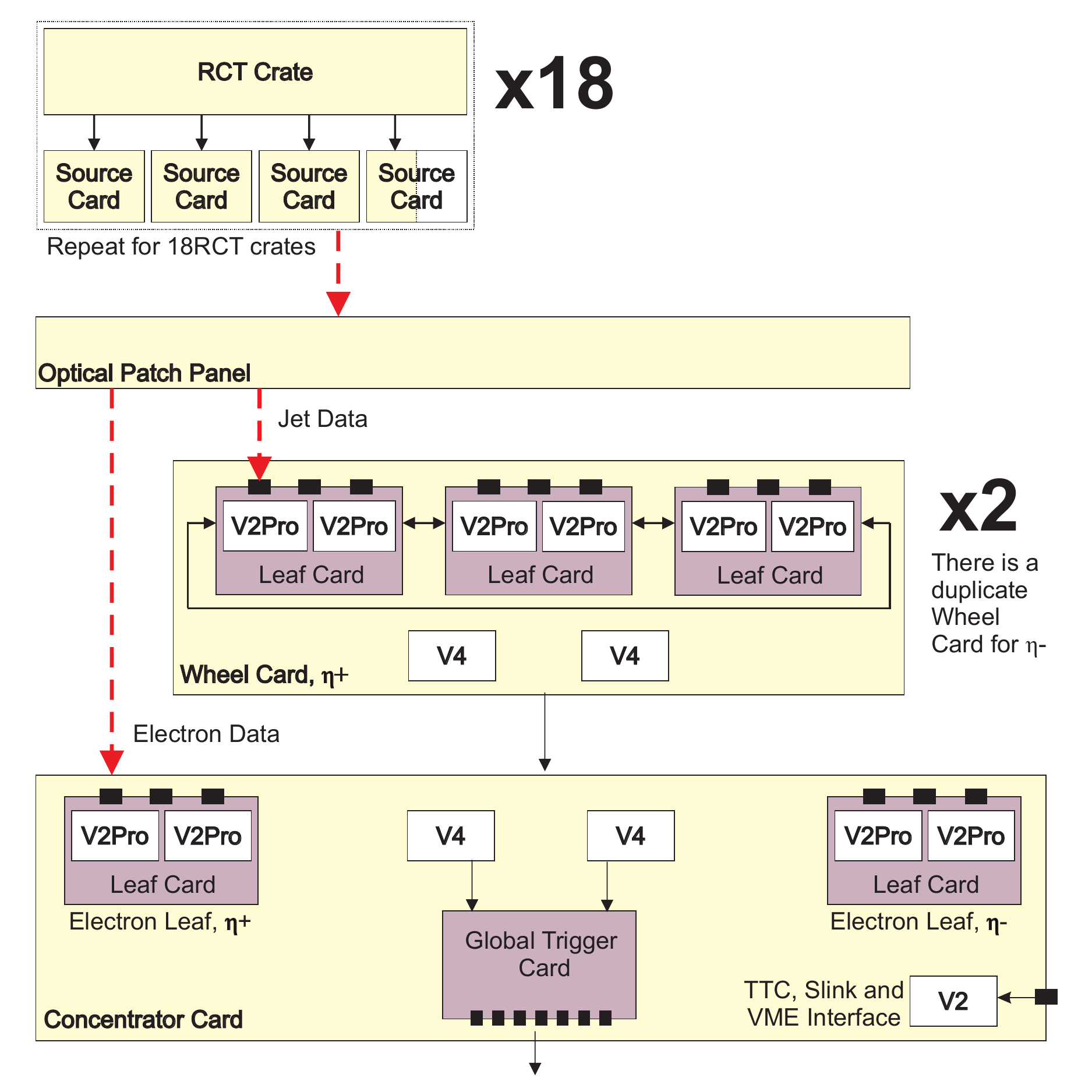}
\caption{A schematic of the global calorimeter trigger (GCT) system,
  showing the data flow through the various component cards.}
\label{fig:gct}
\end{figure}

Figure~\ref{fig:gct} shows a diagram of the GCT system data flow. The input to the GCT is 18 RCT crates.
The 63 source cards retransmit the data on optical high-speed serial links (shown by dashed arrows).
For each RCT crate, the electron data are transmitted on 3  fibers and the jet data on 10 fibers.
There are two main trigger data paths: electron and jet.

The jet data are sent to leaf cards (configured for jet finding)
mounted on the wheel cards. The leaf cards are connected in a circle
to search for clustered jets in one half of the CMS calorimeter
(either in the positive or the negative $\eta$). The wheel card
collects the results from three leaf cards, sorts the clustered jets,
and forwards the data to the concentrator card. A more detailed description of each component is given below.

\begin{itemize}

\item Source card. The 6 differential ECL cables per RCT crate are fed into source cards, each receiving up to two RCT cables and transmitting the data over four fiber links. This has several advantages: it allows the source cards to be electrically isolated from the main GCT system, the different data within the RCT cables to be rearranged, a large amount of information to be concentrated so that it can be delivered to the processing FPGAs on leaf cards, and data to be duplicated.

\item Leaf card. The leaf card is the main processing block in the GCT design. The most difficult task in the GCT is the jet finding. This is made simpler by concentrating the data in as few FPGAs as possible. Consequently, each leaf card has two Xilinx Virtex II Pro FPGAs each with 16 multi-gigabit transceivers that are used to bring the raw data in. Three Agilent 12 channel receivers provide the opto-electronic interface. The large standard I/O capacity is used to transmit the data to the wheel card.

\item Wheel card. There are two wheel cards, one for each half of the detector. They act as carriers for three leaf cards and further concentrate the data. They sum the energy values, sort the 54  clustered jets by transverse energy into the three types (forward,  central, tau). The wheel cards then forward the information to the  concentrator card via high-speed Samtec low-voltage differential signal (LVDS) cables.

\item Concentrator card. The concentrator card performs similar
  actions to that of the wheel card after which it transmits the
  resulting trigger objects to the GT and stores the information in a
  pipeline until receipt of an L1A signal. The concentrator card also
  carries two leaf cards that process the electron data. These leaf
  cards record the incoming RCT data in a pipeline memory until
  receipt of an L1A signal and perform a fast sort on the incoming data. The interface to the GT is via a mezzanine card which transmits data over 16 fiber links running at 3 Gb/s.
\end{itemize}

The CMS L1 calorimeter trigger chain does not use information from
other L1 subsystems, \ie, the L1 muon trigger, which is
described in the next section. L1 calorimeter and muon information is
combined to a final L1 trigger decision in the GT~(Sec.~\ref{sec:global_trigger_desc}).

\subsection{The L1 muon trigger system}
\label{sec:l1muon}
All three CMS muon detectors contribute to the L1 trigger decision.
Details on how the flow of information from the DTs, CSCs, and RPCs is
processed to build full muon tracks within each system, and how tracks
are combined together by the GMT to provide final muon trigger candidates,
are given below.

\subsubsection{Muon local trigger segments}

Whereas RPC trigger tracks are built by the pattern comparator trigger
(PACT) using information coming from detector hits directly, local
trigger track segments (primitives) are formed within DT and CSC
detectors prior to the transmission to the respective track finders.

In the case of the DTs, local trigger (DTLT) track segments are
reconstructed by electronics installed on the detector. Each of the
250 DTs is equipped with a mini-crate hosting readout and trigger
electronics and implemented with  custom ASIC~\cite{DTbti,DTTraco}
and
programmable ASIC~\cite{DTTSM} devices.  Up to two DTLT per BX in the
transverse plane can be generated by one chamber; DTLT information
includes the radial position, the bending angle, and information about
the reconstruction quality (\ie, the number of DT layers used to build
a track segment).  Additionally, hits along the longitudinal direction are
calculated; in this case only a position is calculated as the track is
assumed to be pointing to the vertex.  The DTLT electronics is capable
of highly efficient (94\%) BX identification~\cite{DTtestbeam,Chatrchyan:2008zzk}, which is a challenging
task given that single hits are collected with up to ${\approx}400\unit{ns}$ drift
time.  A fine grained synchronization of the DTLT clock to the LHC
beams is needed to ensure proper BX identification~\cite{DTtestbeamFineSync, DTcraftSync}.

The DTLT segments are received by the trigger sector collector (TSC) system,
installed on the balconies surrounding the detector and implemented using
flash-based FPGAs~\cite{DTtsc}.
The TSC consists of 60 modules, each receiving local trigger data from one
DT sector (the four or five detectors within the same muon barrel slice, called wheel,
and covering $30^\circ$ in azimuthal angle): trigger segments are synchronized and
transmitted over 6\unit{Gb/s} optical links per sector, to the underground
counting room, where optical receiver modules perform deserialization and
deliver data to the DT track finder (DTTF) system.

For the CSCs, local charged-track (LCT) segments, constructed
separately from the cathode (CLCT) and anode (ALCT) hits of a detector,
are correlated in the trigger motherboard (TMB) when both segments
exist within a detector.  A CLCT provides information on the azimuthal
position of a track segment, while an ALCT provides information on the
radial distance of a segment from the beam line, as well as precise
timing information. A maximum of two LCTs can be sent from each
detector per bunch crossing. The segments from nine detectors are collected
by a muon port card (MPC) residing in the same VME crate as the
TMBs. The MPC accepts up to 18 LCTs and sorts them down to the best
three before transmission over an optical fiber to the CSC
track finder (CSCTF). There are 60 MPCs, one in each peripheral crate.

More detailed description of the DT and CSC local trigger segment reconstruction and
performance in LHC collisions is given in Ref.~\cite{Chatrchyan:2013sba}.

\subsubsection{Drift tube track finder}

The DTTF processes the DTLT information in order to reconstruct muon
track candidates measured in several concentric rings of detectors,
called stations, and assigns a transverse momentum value to the track
candidates~\cite{DTdttf}.  First, the position and bending of each DTLT is
used to compute, via a LUT, that expected position at the outer
stations (in case of the fourth station layer, the extrapolation is
done inward towards the third one). The position of actual DTLTs is
compared to the expected one and accepted if it falls within a
programmable tolerance window. These windows can be tuned to achieve
the desired working point, balancing the muon identification
efficiency against the accepted background. To enable triggering on
cosmic muon candidates, the windows can be as large as a full DT
detector in order to also accept muons that are not pointing to the
interaction point. All possible station pairs are linked this way and
a track candidate is built. Then, the difference in azimuthal
positions of the two inner segments is translated into a transverse
momentum value, again using LUTs. Also the azimuthal and longitudinal
coordinates of the candidate are computed, while a quality code based
on the number and positions of the stations participating in the track
is generated. The hardware modules are VME 9U boards hosted in 6
crates with custom backplanes and VME access; there are 72 such track
finding boards, called sector processors (SP).  Each SP finds up to
two tracks from one DT sector. Two separate SPs analyze DTLTs from the
sectors of the central wheel, to follow tracks at positive or negative
pseudorapidity. Each SP receives also a subset of the DTLT information
from their neighboring SPs, through parallel electrical connections, in
order to perform track finding for tracks crossing detectors in
different sectors. SP from external wheels also receive track segments
from the CSC trigger.

The last stage of the DTTF system consists of the muon sorter
(MS)~\cite{DToffdetector}. First, a module called the wedge sorter (WS)
collects up to 12 track candidates from the 6 SPs of one ``wedge" (5 DT
sectors at the same azimuthal position) through parallel backplane
connections, and selects two based on the matched magnitude of the transverse
momentum and on their reconstruction quality. The resulting 24 muon
candidates from 12 wedge sorters are collected via parallel LVDS cables into the final
sorting module, called the barrel sorter (BS), which selects the final
four muon candidates to be delivered to the GMT. Both
the WS and BS perform ghost cancellation algorithms before the track
sorting, in order to remove duplicate tracks, \eg, multiple
track candidates originating from the same muon crossing from
neighboring SPs. Two WS modules are installed in each DTTF crate,
while the BS is located in a separate crate called central crate. Also
readout information (DTLT track segments and DTTF track candidates in
a $\pm1$ BX window) is provided by each DTTF module and concentrated in
a readout module (provided with serial link output and TTS inputs)
called a data concentrator card (DCC) and located in the central crate.
\subsubsection{Cathode strip chambers track finder}

The CSCTF  logic consists of pairwise comparisons of
track segments in different detector stations that test for the
compatibility in $\phi$ and $\eta$ of a muon emanating from the
collision vertex within certain tolerance windows. These comparisons
are then analyzed and built into tracks consisting of two or more
stations.
The track finding logic has the ability to accept segments in
different assigned bunch crossings by analyzing across a sliding time
window of programmable length (nominally 2 BX) every bunch
crossing. Duplicate tracks found on consecutive crossings are
canceled. The reported bunch crossing of a track is given by the
second arriving track segment. The reported \pt of a candidate muon is
calculated with
large static random-access memory (SRAM) LUTs that take
information such as the track type, track $\eta$, the segment $\phi$
differences between up to 3 stations, and the segment bend angle in
the first measurement station for two-station tracks.

In addition to identifying muons from proton collisions, the CSCTF
processors simultaneously identify and trigger on beam halo muons for
monitoring and veto purposes by looking for trajectories approximately
parallel to the beam line. A beam halo muon is created when a proton
interacts with either a gas particle in the pipe or accelerator
material upstream or downstream the CMS interaction point, and the
produced  hadrons decay.  The collection of halo muons is an
interesting initial data set; the muons' trajectory is highly parallel
to the beam pipe and hence also to parallel to the solenoidal
magnetic field; therefore, they are minimally deflected and their
unbent paths are a good tool for aligning different slices of the
detector disks.  Additionally, these muons are a background whose rate
need to be known as they have the potential to interact with multiple
detector subsystems. The halo muon trigger also allows monitoring of
the stability of the proton beam.

The CSCTF system is partitioned into sectors that correspond to a
$60^\circ$ azimuthal region of an endcap. Therefore 12 ``sector
processors'' are required for the entire system, where each sector
processor is a 9U VME card that is housed in a single crate. Three 1.6
Gbps optical links from each of five MPCs are received by each sector
processor, giving a total of 180 optical links for the entire
system. There is no sharing of signals across neighbor boundaries,
leading to slight inefficiencies. There are several FPGAs on each
processor, but the main FPGA for the track-finding algorithms is from
the Xilinx Virtex-5 family. The conversion of strip and wire positions
of each track segment to $\eta$, $\phi$ coordinates is accomplished
via a set of cascaded SRAM LUTs (each 512k$\times$16 bits). The final
calculation of the muon candidate \pt is also accomplished by SRAM
LUTs (each 2M$\times$16 bits). In the same VME crate there is also one
sorter card that receives over a custom backplane up to 3 muons from
each sector processor every beam crossing and then sorts this down to
the best four muons for transmission to the GMT. The crate also
contains a clock and control signal distribution card, a DAQ card with
a serial link interface, and a PCI-VME bridge~\cite{CSCAcosta,Trig-TDR}.

\subsubsection{Resistive plate chambers trigger system}
\label{sec:rpc}

The RPCs provide a complementary, dedicated triggering detector system
with excellent time resolution ($\mathcal{O}(1\mathrm{ns})$), to
reinforce the measurement of the correct beam-crossing time, even at
the highest LHC luminosities. The RPCs are located in both the barrel
and endcap regions and can provide an independent trigger over a large
portion of the pseudorapidity range ($\abs{\eta} < 1.6$). The RPCs are
double-gap chambers, operated in avalanche mode to ensure reliable
operation at high rates. They are arranged in six layers in the barrel
and three layers in the endcaps.  Details of the RPC chamber design,
geometry, gas mixtures used and operating conditions can be found in
Refs.~\cite{Chatrchyan:2008zzk,Chatrchyan:2012xi}.  The RPC trigger is based on
the spatial and temporal coincidence of hits in different layers.  It
is segmented into 25 towers in $\eta$ which are each subdivided into
144 segments in $\phi$. The pattern comparator trigger
(PACT)~\cite{pact} logic compares signals from all RPC chamber layers
to predefined hit patterns in order to find muon candidates. The RPCs
also assign the muon \pt, charge, $\eta$, and $\phi$ to the matched
pattern.

Unlike the CSCs and DTs, the RPC system does not form trigger
primitives, but the detector hits are used directly for muon trigger
candidate recognition. Analog signals from the chambers are
discriminated and digitized by front end boards (FEB), then assigned
to the proper bunch crossing, zero-suppressed, and multiplexed by a
system of link boards located in the vicinity of the
detector. They are then sent via optical links to 84
trigger boards in 12 trigger crates located in the underground
counting room. Trigger boards contain the complex PAC logic, which fits
into a large FPGA. The strip pattern templates to be compared with
the particle track are arranged in segments of approximately $0.1$ in
$\abs{\eta}$ and 2.5$^\circ$ (44\unit{mrad}) in $\phi$, called logical cones. Each
segment can produce only one muon candidate. The trigger algorithm
imposes minimum requirements on the number and pattern of hit planes,
which varies with the position of the muon. As the baseline, in the
barrel region ($\abs{\eta} \le$ 1.04), a muon candidate is created by at
least a 4-hit pattern, matching a valid template. To improve
efficiency, this condition is relaxed and a 3-hit pattern with at
least one hit found in the third or fourth station may also create a
muon candidate. In addition, low-\pt muons often do not penetrate
all stations.  Muon candidates can also arise when three
hits are found in four layers of the first and second station. In this case,
only low-\pt candidates will be reconstructed. In the endcap region
($\abs{\eta} > 1.04$) there are only 3 measurement layers available, thus
any 3-hit pattern may generate a muon candidate. A muon quality value
is assigned, encoded in two bits, that reflects the number of hit
layers (0 to 3, corresponding to 3 to 6 planes with hits).

Hits produced by a single muon may be visible in several logical cones
which overlap in space. Thus the same muon may be reconstructed,
typically with different momentum and quality, in a few segments. In
order to remove the duplicated candidates a special logic, called the
RPC ghost buster (GB), is applied in various steps during the
reconstruction of candidates. The algorithm assumes that among the
muon candidates reconstructed by the PACT there is the best one,
associated to the segment penetrated by a genuine muon. Since the
misreconstructed muons appear as a result of hit sharing between
logical cones, these muons should appear in adjacent segments. The
best muon candidate should be characterized by the highest number of
hits contributing to a pattern, hence highest quality. Among
candidates with the same quality, the one with highest \pt is
selected.  The muon candidates from all the PACTs on a trigger board
are collected in a GB chip. The algorithm searches for groups of
adjacent candidates from the same tower. The one with the best rank,
defined by quality and \pt, is selected and other candidates in the
cluster are abandoned. In the second step the selected candidate is
compared with candidates from the three contiguous segments in each of the
neighboring towers. In the last step, the candidates are sorted based
on quality criteria, and
the best ranked four are forwarded to the trigger crate sorter. After
further ghost rejection and sorting, the four best muons are sent to
system-wide sorters, implemented in two half-sorter boards and a
final-sorter board. The resulting four best muon candidates from the
barrel and 4 best muon candidates from the endcap region are sent to
GMT for subtrigger merging.

The RPC data record is generated on the data concentrator card that
receives data from individual trigger boards.

\subsubsection{Global muon trigger system}

The GMT fulfills the following functions: it synchronizes incoming
regional muon candidates from DTTF, CSCTF, and RPC trigger systems,
merges or cancels duplicate candidates, performs $\pt$ assignment
optimization for merged candidates, sorts muon candidates according to
a programmable rank, assigns quality to outgoing candidates and stores
the information about the incoming and outgoing candidates in the
event data. The GMT is implemented as a single 9U VME module with a
front panel spanning four VME slots to accommodate connectors for 16
input cables from regional muon trigger systems. Most of the GMT logic
is implemented in a form of LUTs enabling a high level
of flexibility and functional adaptability without changing the FPGA
firmware, \eg,  to adjust selection requirements, such as
transverse momentum, pseudorapidity, and quality, of the regional muon
candidates~\cite{Sakulin:687846}.

The input synchronization
occurs at two levels. The phase of each input with respect to the on-board clock can be adjusted
in four steps
corresponding to a quarter of the 25\unit{ns} clock cycle to latch correctly the incoming data.
Each input can be then delayed by up to 17 full
clock cycles to compensate for latency differences in regional systems such
that the internal GMT logic receives in a given clock cycle
regional muon candidates from the same bunch crossing.

The muon candidates from different regional triggers are then matched
geometrically, according to their pseudorapidity and azimuthal angle
with programmable tolerances, to account for differences in
resolutions.  In addition, the input $\eta$ and $\pt$ values are
converted to a common scale and a sort rank is assigned to each
regional muon candidate. The assignment of the sort rank is
programmable and in the actual implementation it was based on a
combination of input quality and estimated transverse momentum.

The matching candidates from the DT and barrel RPC and similarly from
the CSC and endcap RPC triggers are then merged. Each measured
parameter ($\eta$, $\phi$, $\pt$, charge, sort rank) is merged
independently according to a programmable algorithm. The $\eta$, charge, and rank were taken from
the either the DT or CSC. For $\pt$ merging, the
initial setting to take the lowest $\pt$ measurement was optimized
during the data taking to become input quality dependent in certain
pseudorapidity regions. In case of a match between DT and CSC,
possible in the overlap region ($0.9<\abs{\eta}<1.2$), one of the candidates
is canceled according to a programmable logic, dependent, for example,
on an additional match with RPC.

Each of the output candidates is assigned a three-bit quality value
which is maximal for a merged candidate. If the candidate is not
merged, its quality depends on the input quality provided by the
regional trigger system and on the pseudorapidity. The quality
assignment is programmable and allows for flexibility in defining
looser or tighter selection of muon candidates in GT
algorithms. Typically, muon candidates in double-muon triggers were
allowed to have lower quality.

The final step in the GMT logic is the sorting according to the sort
rank. Sorting is first done independently in the barrel and in the
endcap regions and four candidates in each region with the highest
rank are passed to the final sort step. Four candidates with the
highest rank are then sent to the GT.

Since the GMT module and the GT system are located in the same VME
crate, the two systems share a common readout. The data recorded from
GMT contains a complete record of the input regional muon candidates,
the four selected muon candidates from the intermediate barrel and
endcap sorting steps, as well as the complete information about the
four output candidates. This information is stored in five blocks
corresponding to five bunch crossings centered around the trigger
clock cycle.

\subsection{The L1 global trigger system}
\label{sec:global_trigger_desc}

The GT is the final step of the L1 Trigger system. It consists of several VME boards mounted in a VME 9U crate together with the GMT and the central trigger control system (TCS)~\cite{GTref1,GTref2}.

For every LHC bunch crossing, the GT decides to reject or accept a
physics event for subsequent evaluation by the HLT. This decision is
based on trigger objects from the L1 muon and calorimeter systems,
which contain information about transverse energy \ET or transverse
momentum \PT, location (pseudorapidity and azimuthal angle), and quality. Similarly, special trigger signals delivered by various subsystems are also used to either trigger or veto the trigger decision in a standalone way (``technical triggers'') or to be combined with other trigger signals into logical expressions (``external conditions''). These technical triggers (up to 64) are also used for monitoring and calibration purposes of the various CMS sub-detectors including L1 trigger system itself.

The trigger objects received from the GCT and GMT, and the input data
from the other subsystems are first synchronized to each other and to
the LHC orbit clock and then sent via the crate backplane to the global
trigger logic (GTL) module, where the trigger algorithm calculations
are performed. For the various trigger object inputs of each type
(four muons, four non-isolated and four isolated $\Pe/\Pgg$
objects, four central and four forward jets, four tau jets) certain
conditions are applied such as \ET or \pt being above a certain
threshold, pseudorapidity and/or azimuthal angle being within a
selected window, or requiring the difference in pseudorapidity and/or
azimuthal angle  between two particles to be within a certain range. In addition, ``correlation conditions'' can be calculated,
\ie, the difference in pseudorapidity and azimuthal angle between two objects of different kinds. Conditions can also be applied to the trigger objects formed
using energy sums such as  \ETm and \HT.

Several conditions are then combined by simple combinatorial logic (AND-OR-NOT) to form up to 128 algorithms. Any condition bit can be used either as a trigger or as a veto condition. The algorithm bits for each bunch crossing are combined into a ``final-OR'' signal by the final decision logic (FDL) module, where each algorithm can also be prescaled or blocked. An arbitrary number of
sets of prescales can be defined for the algorithms in a given
logic firmware version. A set of 128 concrete algorithms
form an L1 menu which together with the set of prescales
completely specifies the L1 trigger selection. The algorithms and the
thresholds of the utilized input objects (such as transverse momentum
or spatial constraints) are defined and hard-coded in firmware and are only changed by loading another firmware version.
Different prescale settings allow  adjustment of the trigger rate
during a run by modifying the prescale values for identical
copies of algorithms differing only in input thresholds.
In case of a positive ``final-OR'' decision and if triggers are not blocked by trigger rules or detector deadtime, the TCS sends out an L1A signal to
trigger the readout of the whole CMS detector and forward all data to the HLT for further scrutiny.

Trigger rules are adjustable settings to suppress trigger requests coming too soon after one or several triggers, as in this case subsystems may not be ready to accept additional triggers~\cite{Varela:687458}. Sources of deadtime can be subsystems asserting ``not ready'' via the trigger throttling system~\cite{DAQ-TDR}, the suppression of physics triggers for calibration cycles, or the trigger rules described above.

The GT system logs all trigger rates and deadtimes in a database to allow for the correct extraction of absolute trigger cross sections from data. The trigger cross section is defined as $\sigma = R/\mathcal{L}$, where $R$ is the trigger rate and $\mathcal{L}$ is the instantaneous luminosity.

Over the years of CMS running, the GT system has proved to be a highly flexible tool: the trigger logic implemented in the firmware of two ALTERA FPGAs
(the L1 menu) was frequently updated to adapt to changing beam conditions, increasing data rates, and modified physics requirements
(details in Section~\ref{sec:triggermenus}). Additional subsystems (\eg, the TOTEM detector~\cite{TOTEMJINST}) have also been configured as a part of the L1 trigger system.

\subsection{Beam position timing trigger system}
\label{sec:bptx}

The two LHC beam position monitors closest to the interaction point
for each LHC experiment are reserved for timing measurements and are
called the Beam Pick-up Timing eXperiment (BPTX) detectors. For CMS,
they are located at a distance of approximately 175~m on either side
of the interaction point (BPTX+ and BPTX-).

The trigger selects valid bunch crossings using the digitized BPTX
signal by requiring a coincidence of the signals from the detectors on
either side (``BPTX\_AND", logical AND of BPTX+ and BPTX-).

To suppress noise in triggers with high background, a coincidence with
BPTX\_AND is required. Another important application has been the
suppression of pre-firing from the forward hadron calorimeter caused
by particles interacting in the photomultiplier anodes, rather than
the detector itself. As the LHC was mostly running with a bunch
spacing of 50~ns and thus there was at least one 25~ns gap without
proton collisions between two occupied bunch crossings, the trigger
discarded pre-firing events by vetoing the trigger for the ``empty
bunch crossing" before a valid bunch crossing. This is achieved by
advancing the BPTX\_AND signal by one bunch crossing (25~ns time unit)
and using this signal to veto the L1 trigger signal (dubbed ``pre-BPTX
veto"). This solution also improved the physics capabilities of the L1
trigger by enabling a search for heavy stable charged particles
(Sec.~\ref{sec:HSCP} for details).

\subsection{High-level trigger system }
\label{sec:HLTDAQ}
The event selection at the HLT is performed in a similar way to that used in the offline
processing. For each event, objects such as
electrons, muons, and jets are reconstructed and identification
criteria are applied in order to select only those events which are of
possible interest for data analysis.

The HLT hardware consists of a single processor farm composed of
commodity computers, the event filter farm (EVF), which runs Scientific Linux.
The event filter farm consists of filter-builder units. In the builder
units, individual event fragments from the detector are assembled to
form complete events. Upon request from a filter unit, the builder
unit ships an assembled event to the filter unit. The filter unit in
turn unpacks the raw data into detector-specific data structures and
performs the event reconstruction and trigger filtering. Associated
builder and filter units are located in a single multi-core machine
and communicate via shared memory. In total, the EVF executed on
approximately 13,000 CPU cores at the end of 2012. More information
about the hardware can be found elsewhere~\cite{daqhlt2012}.

The filtering process uses the full precision of the data from the
detector, and the selection is based on  offline-quality
reconstruction algorithms.
With the $2011$ configuration of the EVF, the CPU power available
allowed L1 input rates of $100$\unit{kHz} to be sustained for an average HLT
processing time of up to about 90 ms per event. With increased CPU
power available in 2012, we were able to accommodate a per-event time
budget of 175~ms per event.
Before data-taking started, the HLT was commissioned extensively using
cosmic ray data~\cite{cmsCollaboration:2010gj}. The HLT design
specification is described in detail in~\cite{HLT-2006}.

The data processing of the HLT is structured around the concept of a
\emph{HLT path}, which is a set of algorithmic processing steps run in
a predefined order that both reconstructs physics objects and makes
selections on these objects. Each HLT path is implemented as a
sequence of steps of increasing complexity, reconstruction refinement,
and physics sophistication. Selections relying on information from the
calorimeters and the muon detectors reduce the rate before the
CPU-expensive tracking reconstruction is performed. The
reconstruction modules and selection filters of the HLT use the
software framework that is also used for offline reconstruction and
analyses.

Upon completion, accepted events are sent to another software process,
called the storage manager, for archival storage. The event data are
stored locally on disk and eventually transferred to the CMS Tier-0
computing center for offline processing and permanent storage. Events
are grouped into a set of non-exclusive streams according to the HLT
decisions. Most data are processed as soon as possible; however, a
special ``parked'' data stream collected during 2012 consisted of
lower-priority data that was collected and not analyzed until after
the run was over~\cite{parked}. This effectively increased the amount
of data CMS could store on tape, albeit with a longer latency than
regular, higher-priority streams. Example physics analyses enabled by
the parked data stream include generic
final states created via vector boson fusion, triggered by four
low-momentum jets ($\ET> 75,55,38,20\GeV$, for the four jets) and
parton distribution function studies via Drell--Yan events at low
dimuon mass, triggered by two low-\pt muons ($\pt>17,8\GeV$, for the
two muons.)

\begin{figure}[tbp]
  \centering
  \includegraphics[width=\textwidth]{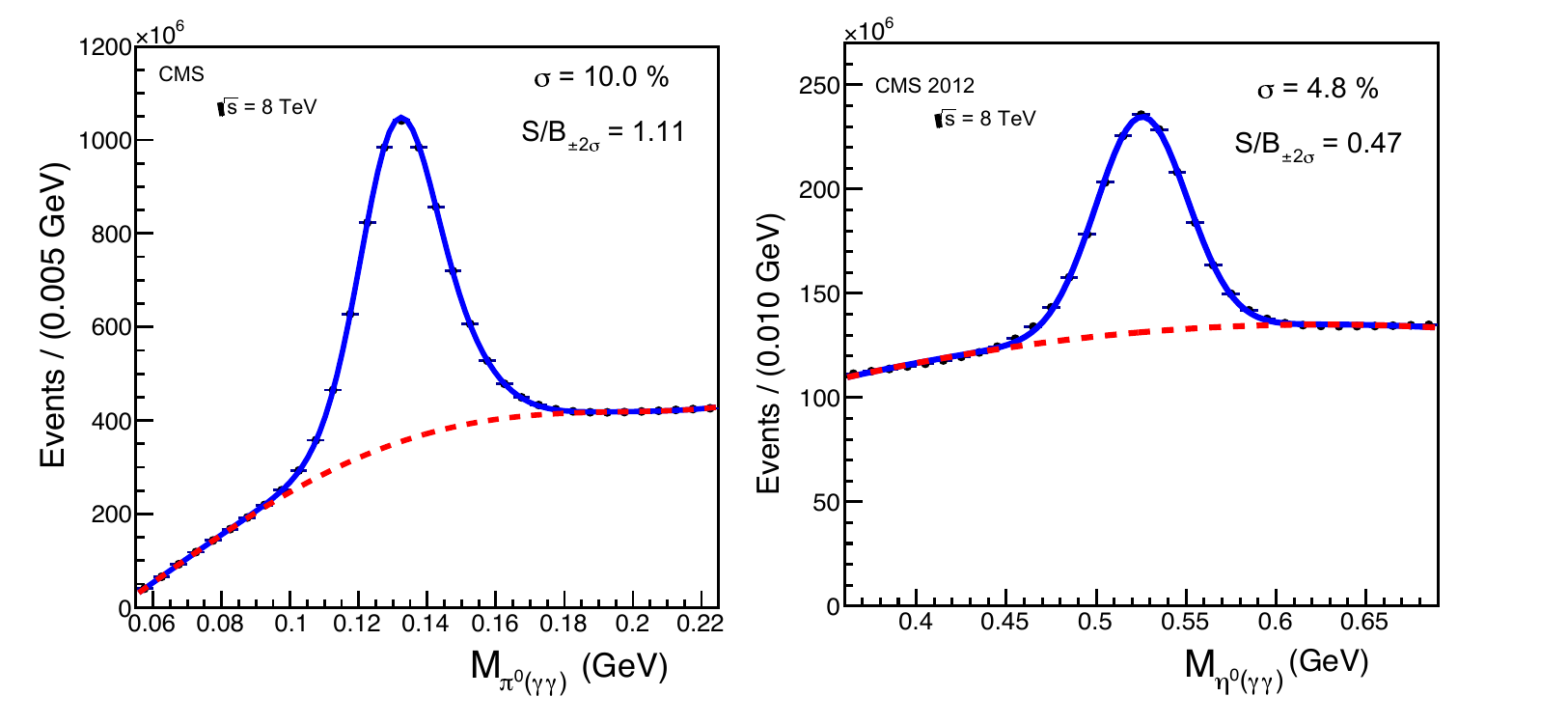}
    \caption{Neutral pion (left) and $\eta$ (right) invariant mass
      peaks reconstructed in the barrel with 2012 data. The spectra
      are fitted with a combination of a double (single) Gaussian for
      the signal and a 4th (2nd) order polynomial for the
      background. The entire 2012 data set is used, using special
      online $\pi^0/\eta$ calibration streams. The sample size is
      determined by the rate of this calibration stream. Signal over
      background (S/B) and the fitted resolution are indicated on the
      plots. The fitted peak positions are not exactly at the nominal
      $\pi^0/\eta$ mass values  mainly due to the effects of
      selective readout and leakage outside the $3{\times}3$ clusters
      used in the mass reconstruction; however, the absolute mass
      values are not used in the inter-calibration.}
  \label{fig:calib_etapi}
\end{figure}

Globally, the output rate of the HLT is limited by the size of the
events and the ability of the downstream systems (CMS Tier-0) to
process the events. In addition to the primary physics stream,
monitoring and calibration streams are also written. Usually these
streams comprise triggers that record events with reduced
content, or with large prescales in order to avoid saturating the
data taking bandwidth. One example is the stream set up for
calibration purposes. These streams require very large data samples
but typically need information only from a small portion of the
detector, such that their typical event size is around 1.5\unit{kB}, while the
full event size is around 0.5~MB. Among the triggers that
define the calibration stream, two select events that are used for the
calibration of the ECAL. The first one collects minimum bias events
and only the ECAL energy deposits are recorded. By exploiting the
$\phi$ invariance of the energy deposition in physics events, this
sample allows inter-calibration of the electromagnetic calorimeter
within a $\phi$ ring. The second ECAL calibration trigger reconstructs
$\Pgpz$ and $\Pgh$ meson candidates decaying into two photons. Only
the ECAL energy deposits associated with these photons are kept. Due
to the small event size, CMS was able to record up to 14\unit{kHz} of
$\Pgpz/\Pgh$ candidates in this fashion~\cite{Chatrchyan:2013dga}. Figure~\ref{fig:calib_etapi}
shows the reconstructed masses for $\Pgpz$ and $\Pgh$ candidates
obtained from these calibration triggers during the 2012 run.

\label{sec:hltintro}
\section{Object identification}
\label{sec:objid}
In this section, the L1 and HLT selection of each object is discussed
as well as the related main single- and double-object triggers using
those objects. The event selection at the HLT is performed in a
similar manner to that used in the offline event processing. For each event,
objects such as electrons, muons, or jets are reconstructed and
identification criteria are applied in order to select those
events which are of possible interest for data analysis.

The object reconstruction is as similar as possible to the offline
one, but has more rigorous timing constraints imposed by the
limited number of CPUs. Section~\ref{sec:hpa} describes how these objects
are used in a
representative set of physics triggers.

We emphasize the track reconstruction in particular as it is used in
most of the trigger paths, either for lepton isolation or for particle-flow
(PF) techniques~\cite{CMS-pf1,CMS-pf2}.
\subsection{Tracking and vertex finding}
\label{sec:HLTTrack}
Tracking and vertex finding
is very important for reconstruction at the HLT.
A robust
and efficient tracking algorithm can help the reconstruction of
particles in many ways, such as improving the momentum resolution of
muons, tracking-based isolation, and $\rm b$-jet tagging. Since
track reconstruction is a CPU-intensive task, many strategies have
been developed to balance the need for tracks with the increase in CPU
time. In this section we describe the algorithm for reconstructing the
primary vertex of the collision in an efficient and fast manner using
only the information from the pixel detector, as well as the algorithm
for reconstructing HLT tracks.
More details about the tracking
algorithm used in CMS, both online and offline, can be found
elsewhere~\cite{Chatrchyan:2014fea}.

It is worth emphasizing that since the tracking detector data in not
included in the L1 trigger, the HLT is the first place that charged
particle trajectories can be included in the trigger.

\subsubsection{Primary vertex reconstruction}
\label{sec:primvtx}

In many triggers, knowledge of the position of the primary vertex is
required. To reconstruct the primary vertex without having to run the full (and
slow) tracking algorithm, we employ a special track reconstruction
pass requiring only the data from the pixel detector.
With these tracks, a simple gap-clustering algorithm is used for
vertex reconstruction~\cite{Chatrchyan:2014fea}. All tracks are ordered by the $z$ coordinate of
their point of closest approach to the $\Pp\Pp$ interaction point.  Wherever two
neighboring elements in this ordered set of $z$ coordinates has a gap
exceeding a distance requirement $z_\text{sep}$, tracks on either side
are split into separate vertices. In such an algorithm, interaction
vertices separated by a distance less than $z_\text{sep}$ are merged.
\begin{figure}[tbph]
  \centering
  \includegraphics[width=0.6\textwidth]{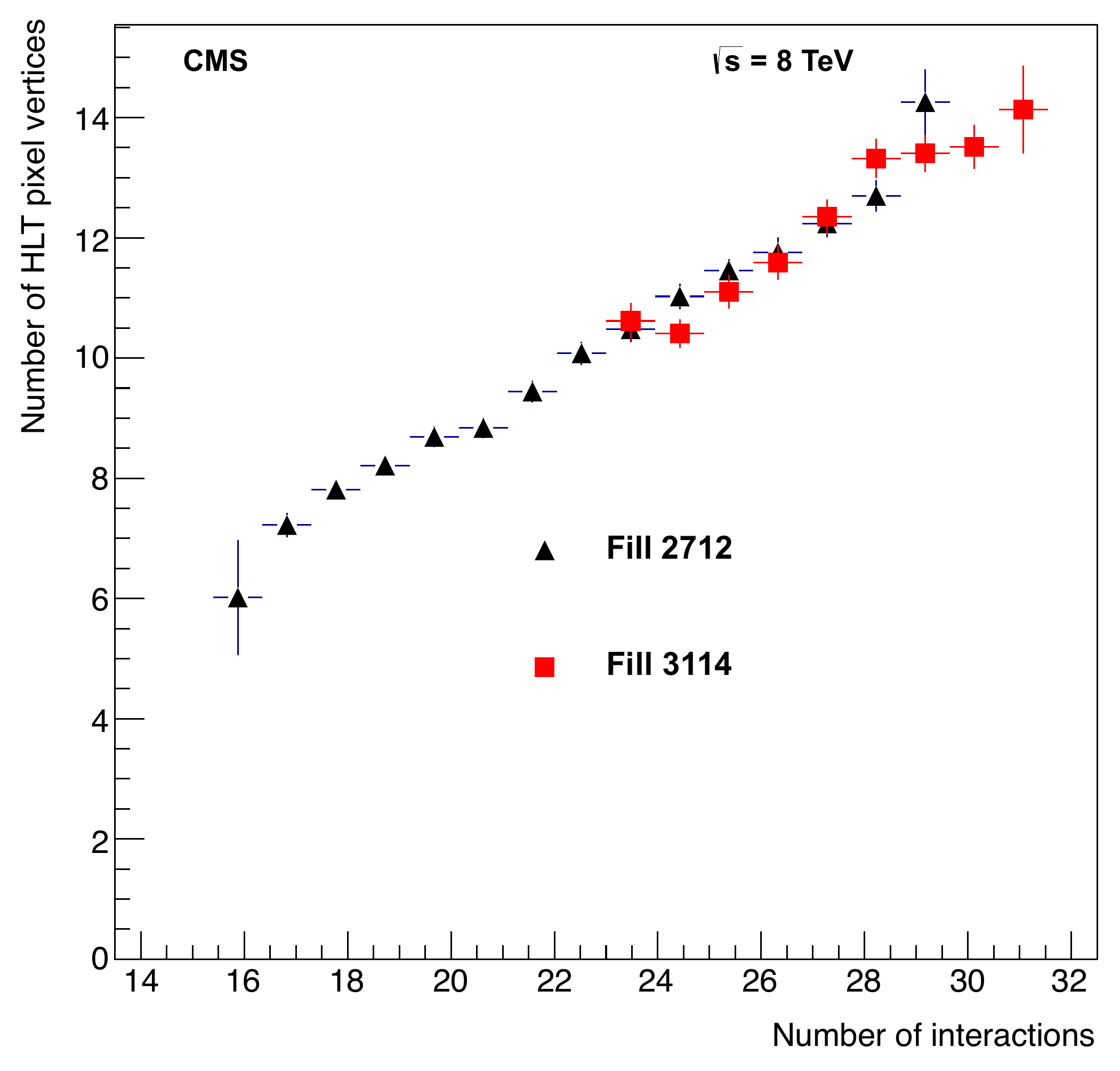}
  \caption{Number of vertices as a function of the number of
    $\Pp\Pp$ interactions as measured by the forward calorimeter, for fills
    taken in two different periods of the 2012 $\Pp\Pp$ run. A linear relation
    can be seen between the two quantities, demonstrating good
    performance of the HLT pixel vertex algorithm.}
\label{Fig:vtx_pu}
\end{figure}
Figure~\ref{Fig:vtx_pu} represents the estimated number of interactions
versus the number of reconstructed pixel vertices for two periods,
with different pileup conditions. The number of interactions is
measured using the information from the HF,
which covers the pseudorapidity range $3 < \abs{\eta} < 5$. The method
used is the so-called ``zero counting'', which relies on the fact that
the mean number of interactions per bunch crossing ($\mu$)
has a probability density described by the Poisson
distribution. The average fraction of empty HF
towers is measured and then $\mu$ is calculated by inverting the
Poisson zero probability. Figure~\ref{Fig:vtx_pu} shows that in the 2012 data,
where the number of interactions per bunch crossing reached 30, the
number of reconstructed vertices depends linearly on the
number of pileup events for a wide range of values,
demonstrating no degradation of performance due to pileup.

With increasing number of pileup collisions, we observed that the
CPU time to reconstruct pixel tracks and pixel vertices increased
nonlinearly. For a few HLT paths, the CPU time usage is largely
dominated by the pixel track and vertex reconstruction time and
it is prohibitive to use the primary-vertex finding algorithm
described above.

A second method, called fast primary vertex finding, was implemented
to reduce the CPU time usage. This method initially finds a coarse
primary vertex and reconstructs only pixel tracks in jets associated
to this vertex. The pixel tracks are then used to find the online
primary vertex using the standard method described above. The coarse
vertex is found as follows: initially,
jets with $\pt > 40\GeV$ are considered. Pixel clusters in the $\phi$
wedges corresponding to the jets are selected and projected to the
beam axis using the jet pseudorapidity. The projections are then
clustered along the $z$ axis. If a vertex exists, the clusters will
group around the $z$ position of the vertex.

Roughly 5\% of the time, the coarse vertex is not found. In these
cases, the standard vertex reconstruction is run. The coarse vertex
has a resolution of 0.4 cm. By using the fast primary vertex finding,
the overall CPU time needed to reconstruct the vertex is reduced by a
factor 4 to 6, depending on the HLT path. The reduced CPU time requirement
allowed some additional paths to use b-tagging techniques than would not
have been possible with the standard algorithm.
The two methods have similar performance in reconstructing the online primary
vertex. The efficiency of the reconstruction relative
to offline is about 92\% within the vertex resolution.
The pixel tracks are also used in other reconstruction steps as
described in the following subsections.

\subsubsection{HLT tracking}

Given the variety of the reconstructed objects and the fast changes in
the machine conditions, it has been impossible to adopt a unique full
silicon track reconstruction for all the paths. Different objects
ended up using slightly different tracking configurations, which had
different timing, efficiencies, and misreconstruction rates. All configurations use
a combinatorial track finder (CTF) algorithm,  which consists of four
steps:
\begin{enumerate}
\item The seed generation provides initial track candidates using a
  few (two or three) hits and the constraint of the  $\Pp\Pp$ interaction point
  position. A seed defines the initial estimate of the trajectory,
  including its parameters and their uncertainties.
\item The next step is based on a global Kalman
  filter~\cite{Fruhwirth:1987fm}. It extrapolates the seed
  trajectories along the expected flight path of a charged particle,
  searching for additional hits that can be assigned to the track
  candidate.
\item The track fitting stage uses another Kalman filter and smoother
  to provide the best possible estimate of the parameters of each
  trajectory.
\item Finally, the track selection step sets quality flags and
  discards tracks that fail minimum quality requirements.
\end{enumerate}

Each of these steps is configurable to reduce the time at the cost of
slightly degraded performance. As an example, when building track
candidates from a given seed, the offline track reconstruction retains
at most the five partially reconstructed candidates for extrapolation
to the next layer, while at HLT only one is kept. This ensures little
time increase in the presence of large occupancy events and high
pileup conditions. As another example, the algorithm stops once a
specified number of hits have been assigned to a track (typically
eight). As a consequence, the hits in the outermost layers of the
tracker tend not to be used. The different tracking configurations
can be divided into four categories:
\begin{itemize}
\item Pixel-only tracks, \ie, tracks consisting of only three
  pixel hits. As stated above, the pixel-based tracking is
  considerably faster than the full tracking, but pixel tracks have
  much worse resolution and are mostly used to build the primary
  vertex and are also used in parts of the b- and
  $\tau$-identification stages. These tracks are also used to build
  the seeds for the first iteration of the iterative tracking.
\item Iterative tracking, \ie, a configuration which is as similar as
  possible to that used offline. This is used as input to the
  PF reconstruction.
\item Lepton isolation, \ie, a regional one-step-tracking used in
  paths with isolated electrons and muons. On average, higher-\pt tracks are
  reconstructed in comparison to the iterative tracking method and as a result
  this variant is somewhat more time consuming than the iterative tracking.
\item b tagging, \ie, a regional one-step-tracking similar to the one
  used for lepton isolation.
\end{itemize}
The iterative tracking approach is designed to reconstruct tracks in decreasing order of complexity. In the early iterations, easy-to-find tracks, which have high \pt and small impact parameters, are reconstructed. After each iteration hits associated with found tracks are removed, and this reduces combinatorial complexity and allows for more effective searching for lower-\pt or highly displaced tracks. For data
collected in 2012, the tracking consisted of five iterations, similar
(but not identical) to those run offline. The main difference between
each iteration lies in the configuration of the seed generation and
final track selection steps.

The first iteration is seeded with three pixel hits. Each pixel track
becomes a seed. The seeds in this iteration are not required to be
consistent with the primary vertex position. For the other iterations,
only seeds compatible with the primary vertex $z$ position are used.
In the first iteration, we attempt to reconstruct tracks across the
entire detector. For speed reasons, later iterations are seeded
regionally, \ie, only seeds in a given $\eta$-$\phi$ region
of interest are considered. These regions are defined using the
$\eta$-$\phi$ direction of jets from tracks reconstructed in the
previous iterations. Unfortunately, due to hit inefficiency in the
pixel detector and the requirement of hits in each of the three pixel
layers in this step, 10--15\% of isolated tracks may be lost.
This leads to an efficiency loss for one-prong $\tau$ lepton decays, which
is recovered by adding extra regions based on the $\eta$-$\phi$
direction of isolated calorimeter jets. Finally, after the five
iterations, all tracks are grouped together (adding the separately
reconstructed muon tracks), filtered according to quality criteria and
passed to the PF reconstruction.

\begin{figure}[tbph]
  \centering
  \includegraphics[width=0.6\textwidth]{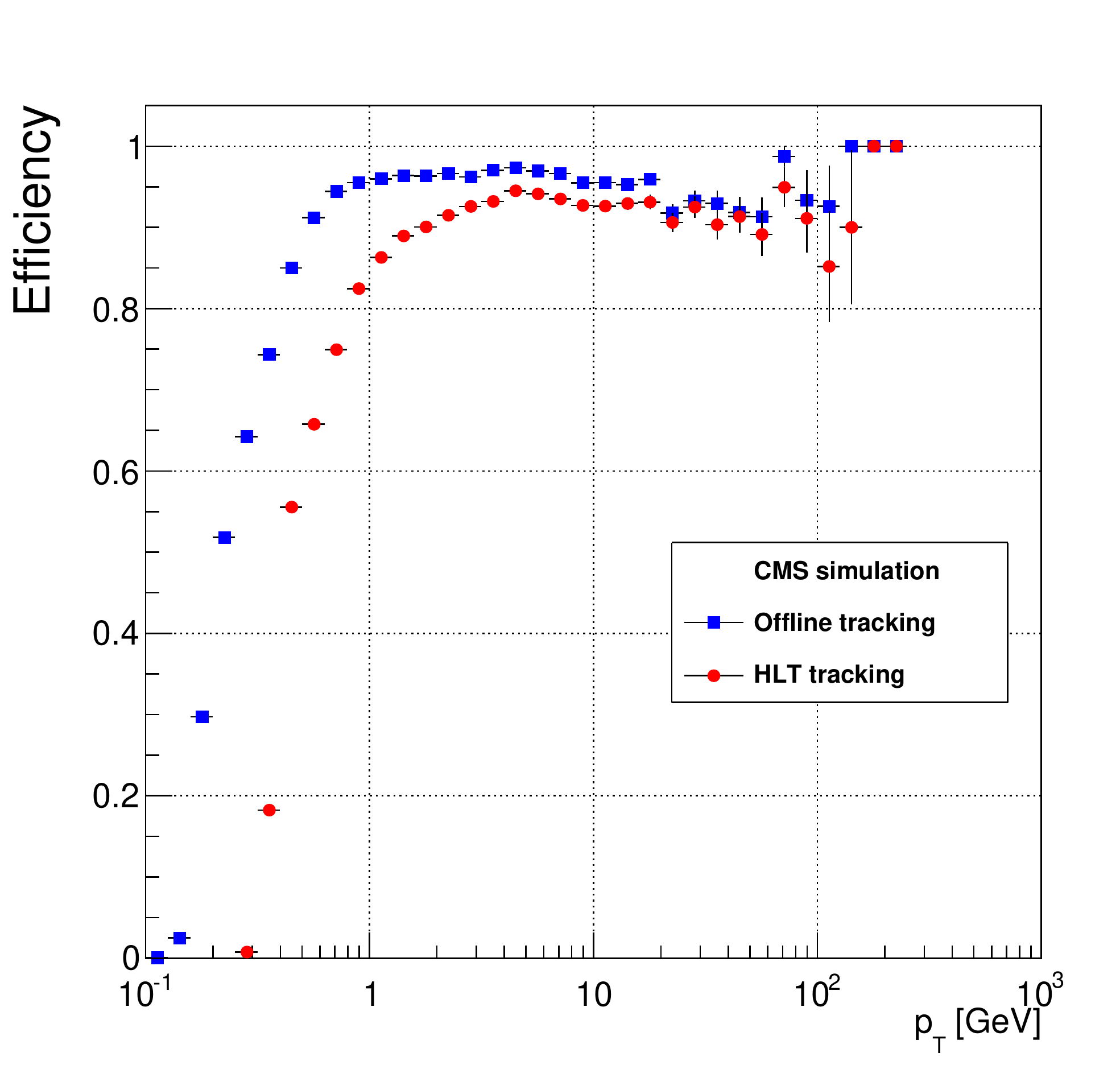}
  \caption{ Tracking efficiency as a function of the momentum of the
    reconstructed particle, for the HLT and offline tracking, as
    determined from simulated \ttbar events. Above 0.9\GeV,
    the online efficiency is above 80\% and plateaus at around 90\%.}
\label{fig:trackingEfficiency}
\end{figure}
Figure~\ref{fig:trackingEfficiency} shows the offline and online track
reconstruction efficiency on simulated top-antitop ($\PQt\PAQt$) events. Online
efficiencies are above 80\% for track \pt above 0.9\GeV.

\begin{figure}[tbph]
  \centering
  \includegraphics[width=0.6\textwidth]{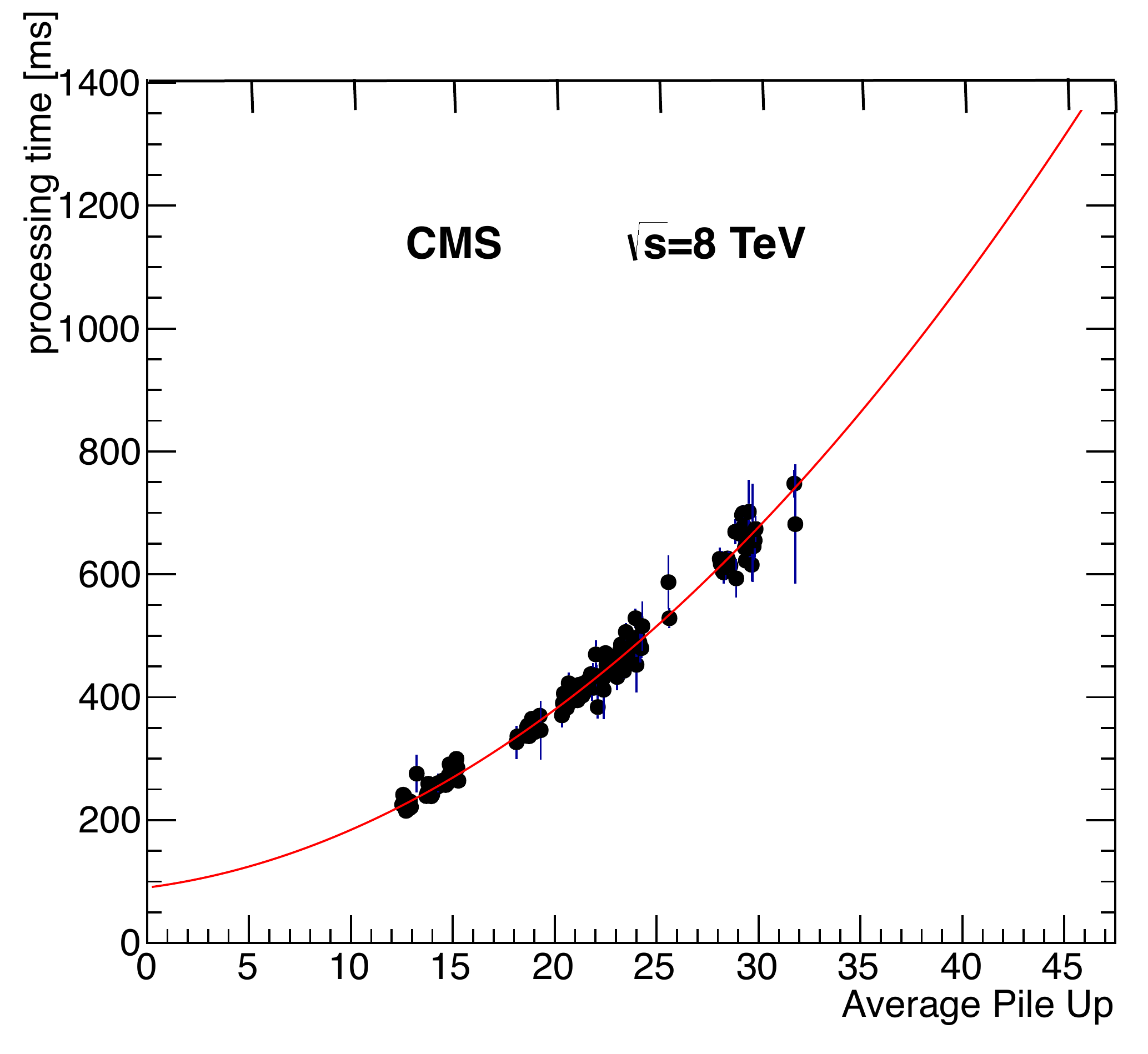}
  \caption{The CPU time spent in the tracking reconstruction as a function
    of the average pileup, as measured in pp data taken during the
    2012 run. The red line shows a fit to data with a second-order polynomial. On average, about 30\% of the total CPU time of the HLT
    was devoted to tracking during this run. }
\label{iterativetrackingtiming}
\end{figure}
Figure~\ref{iterativetrackingtiming} shows the time taken by the
iterative track reconstruction as a function of the average pileup.
As already discussed, the time spent in tracking is too high to allow the
use of the tracking on each L1-accepted event.
To limit the computing time, HLT tracking was only run on a subset of events that pass a set of filters,
reducing it to about 30\% of the
total HLT CPU time.

\subsection{Electron and photon triggers}
\label{sec:egamma_overview}
The presence of high-\pt leptons and photons is a strong
indicator for interesting high-$Q^2$ collisions and consequently much
attention has been devoted to an efficient set of triggers for these processes.
Electrons and photons (EG or ``electromagnetic objects'') are reconstructed
primarily using the lead-tungstate electromagnetic calorimeter.
Each electromagnetic object deposits its energy primarily in this detector, with little energy deposited in the hadron calorimeter. The transverse shower size is of
the order of one crystal. Electrons and photons are distinguished from
one another by the presence of tracks pointing to electrons
and lack thereof for photons. At L1,
only information from the calorimeter is available and no distinction
can be made between $\Pe$ and $\Pgg$. At the HLT level, tracks are
used to resolve this ambiguity.
\subsubsection{L1 electron / photon identification}

\paragraph{L1 electron / photon trigger performance}

\subparagraph{The L1 electron trigger resolution}

\begin{figure}
\centering
\includegraphics[width=0.467\textwidth]{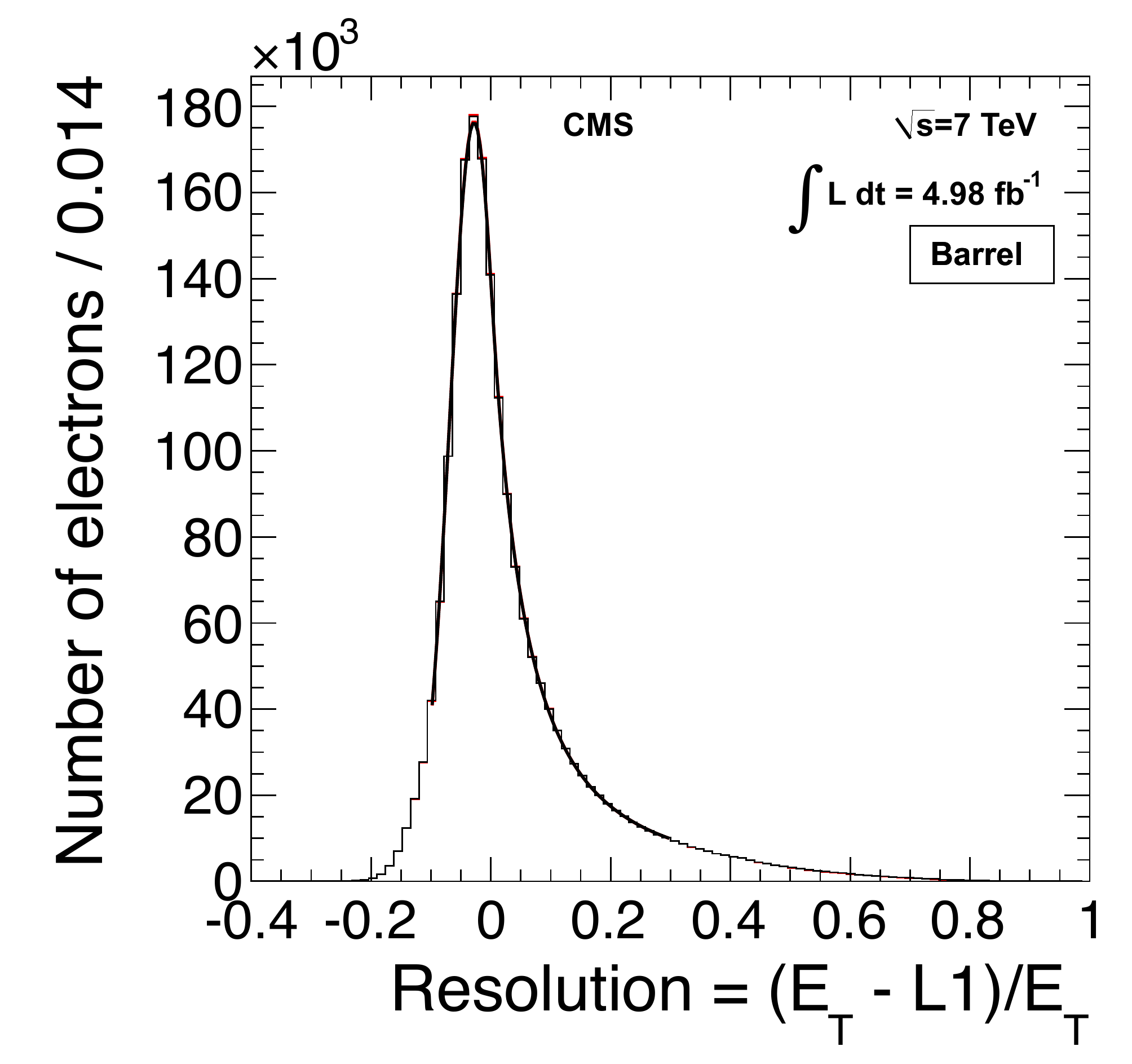}
\includegraphics[width=0.48\textwidth]{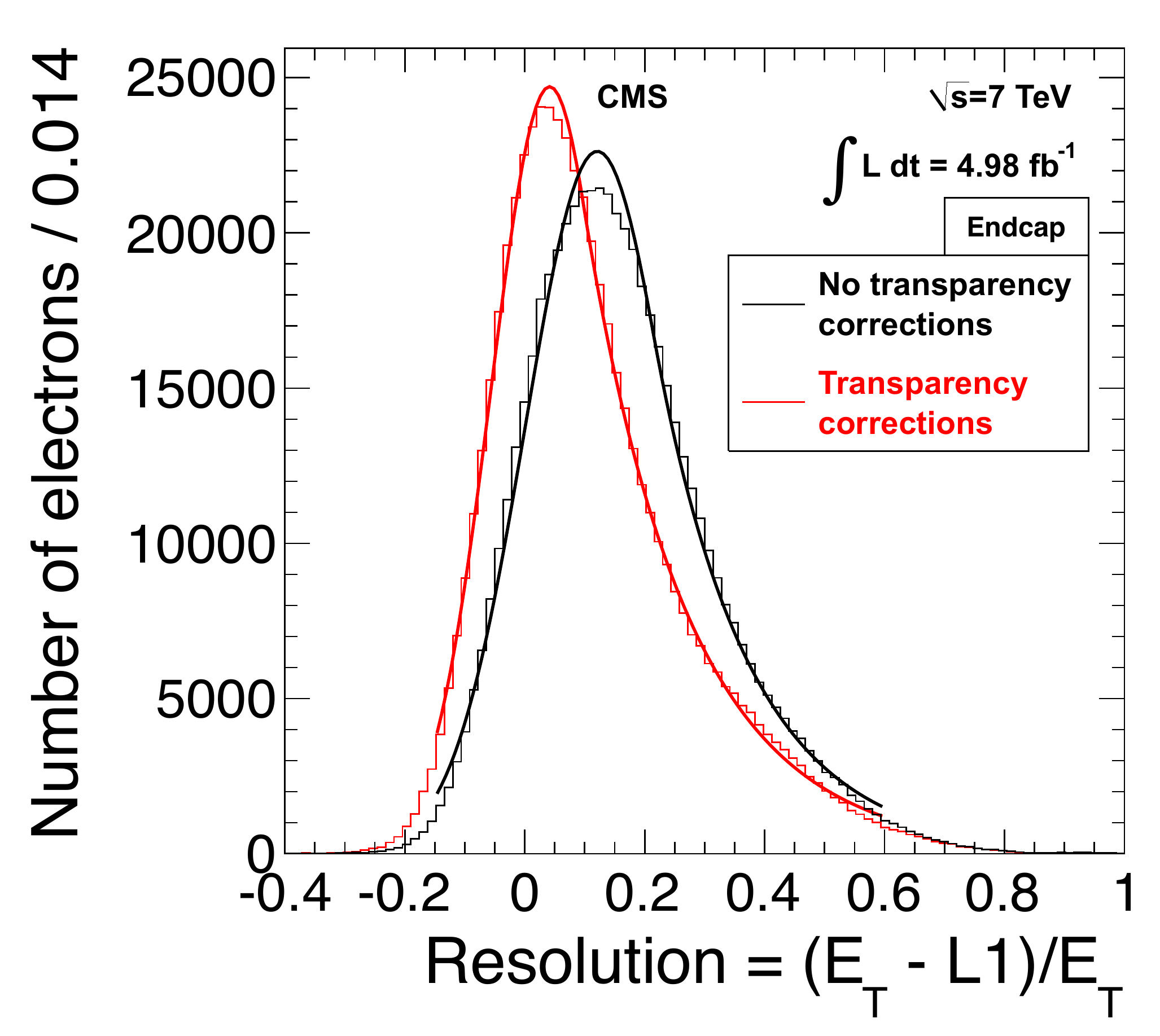}
\caption{The L1 EG resolution, reconstructed offline \ET minus L1 \ET divided by reconstructed offline \ET,  in the
barrel (left) and endcap (right) regions. For both distributions, a
fit to a Crystal Ball function is performed. On the right curve, the red solid
line shows the result after applying the transparency corrections (as
discussed in Sec.~\ref{sec:trans}) For EB, the resolution after transparency correction is unchanged.}
\label{fig:l1reso}
\end{figure}

\begin{figure}
\centering
\includegraphics[width=0.48\textwidth]{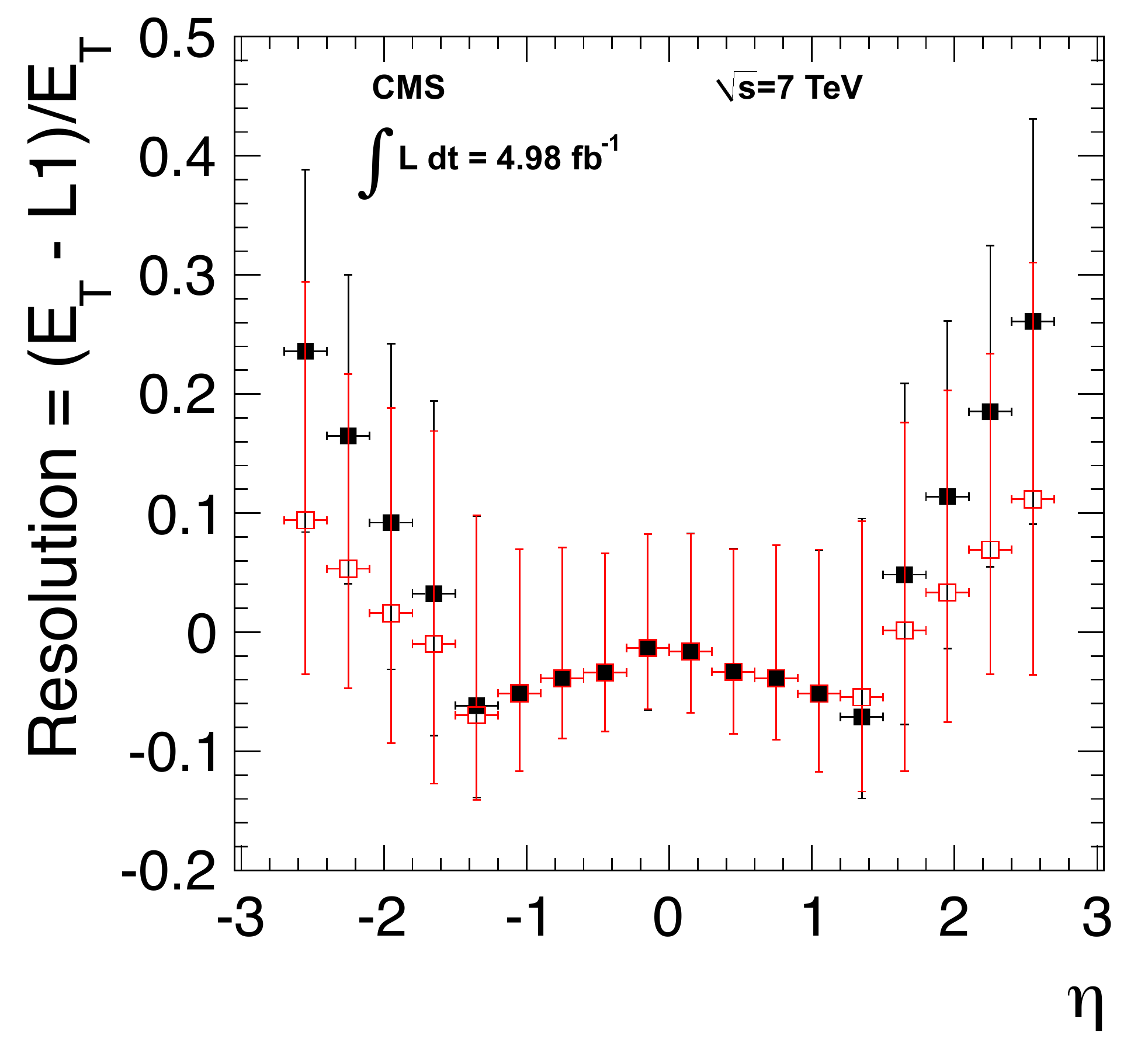}
\caption{
  The L1 EG resolution  for all
  electron \pt as a function of pseudorapidity $\eta$. For each $\eta$
  bin, a fit to a Crystal Ball function was used to model the data distribution.
  The vertical bars on each point represent
  the sigma of each fitted function which is defined as the width of
  the 68\% area.
  The red points show the improved resolution after
  applying transparency corrections (as discussed in
  Sec.~\ref{sec:trans}).
}
\label{fig:l1resoETeta}
\end{figure}

Offline reconstructed electrons are matched to L1 EG candidates by
looking for the RCT region which contains the highest energy trigger tower (TT)
within the electron supercluster (SC)~\cite{cms-e7,cms-e8}. In order to extract the resolution, the supercluster transverse energy reconstructed offline is compared to the
corresponding L1 candidate \ET. Figure~\ref{fig:l1reso} shows the distribution of the L1 EG
trigger resolution, offline reconstructed \ET minus L1 \ET divided by offline reconstructed \ET,  in the barrel and endcap regions. The
same observable is displayed as a function of the electron offline
supercluster \ET and $\eta$ in Fig.~\ref{fig:l1resoETeta}. Above 60\GeV, the resolution starts to degrade as the L1
saturation is reached.\footnote{The ECAL trigger primitives saturate at 127.5\GeV
and RCT EG candidates at 63.5\GeV.}

The resolution of L1 EG candidates (Fig.~\ref{fig:l1reso})
is reasonably well described by a fit to a Crystal Ball function~\cite{Oreglia:1980cs}. An electron supercluster can spread
its energy over a large region of the calorimeter due to the emission
of photons from bremsstrahlung. The L1 EG algorithm only aggregates
energy in 2 trigger towers (Section~\ref{sec:ecaltpg}). For this reason, the probability to
trigger is  reduced for electrons propagating across
a significant amount of material. This effect increases with the
pseudorapidity and peaks in the transition region between the EB and
the EE. Figure~\ref{fig:l1resoETeta} illustrates this effect by showing the L1 EG resolution as function of $\eta$. Further
effects such as the transparency change of ECAL crystals with time
certainly degrades the resolution further (see
Sec.~\ref{sec:trans}). The resolutions shown in
Figs.~\ref{fig:l1reso} and \ref{fig:l1resoETeta} were obtained after
correcting for this effect.

\subparagraph{L1 electron trigger efficiency}

The electron trigger efficiency was measured with electrons from
$\Z\to\Pe\Pe$ events, using a tag-and-probe method~\cite{cms-wz}.
The data collected in 2011 and 2012 were used. Both the tag and
the probe are required to pass tight identification requirements in order to
reduce significantly the background contamination. The tag electron
must also trigger the event at L1, while the probe electron is used for
the efficiency studies. The invariant mass of the tag-and-probe system
should be consistent with the \Z boson mass ($60 < M_{\Pe\Pe} <
120\GeV$), resulting in a pure unbiased electron
data sample. The trigger efficiency is given by the fraction of probes
above a given EG threshold, as a function of the probe \ET. In order
to trigger, the location of the highest energy TT within the electron
supercluster must match a corresponding region of an L1 candidate in
the RCT.

\begin{figure}[h]
\centering
\begin{minipage}{0.47\textwidth}
\includegraphics[width=\textwidth]{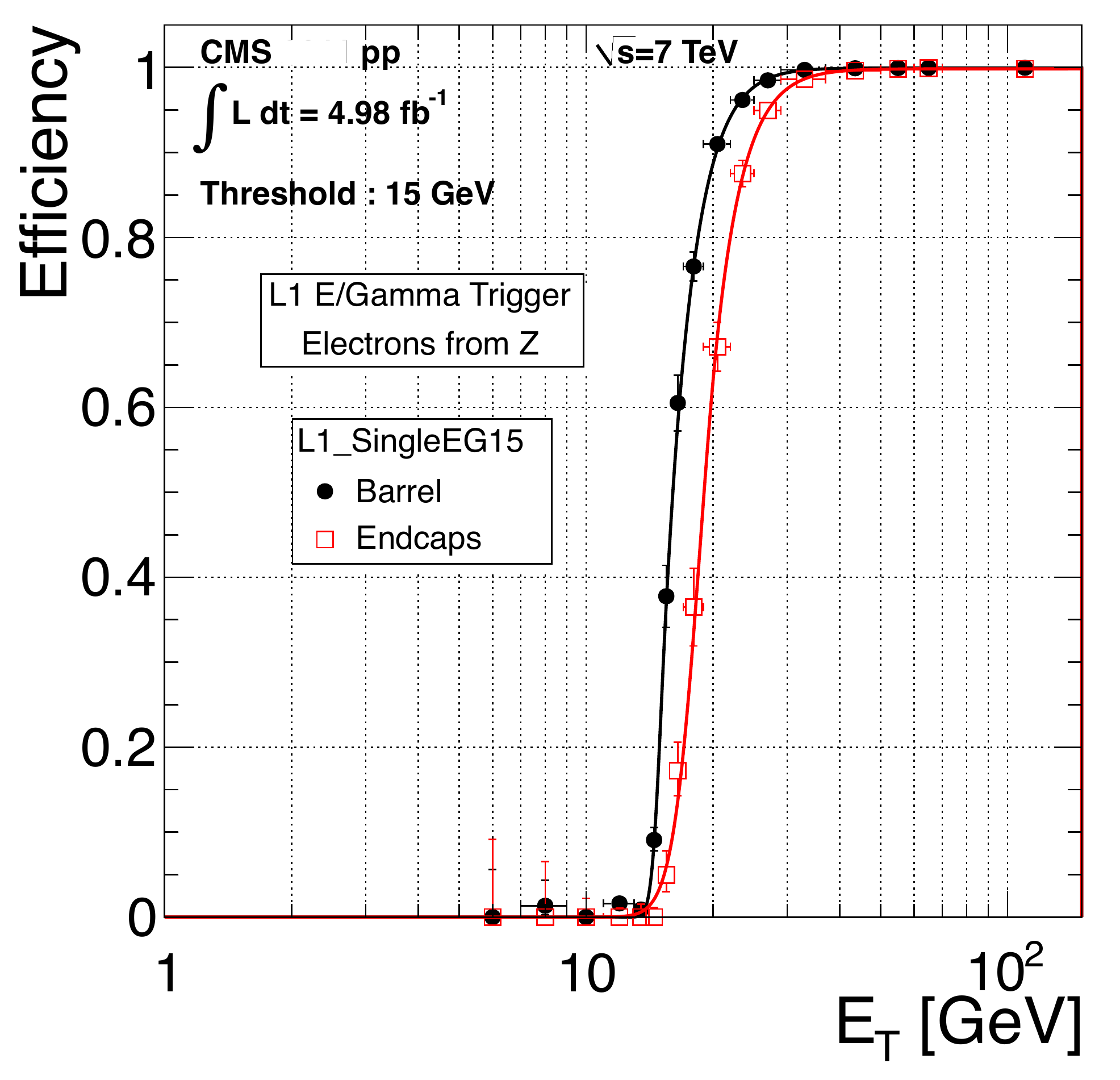}
\caption{\label{fig:l1eg15perf} The electron trigger efficiency at L1 as a
  function of offline reconstructed \ET for electrons in the EB (black dots) and EE (red
  dots), with an EG threshold: $\ET=15\GeV$. The curves show unbinned
  likelihood fits.}
\end{minipage}\hfill
\begin{minipage}{0.47\textwidth}
\includegraphics[width=\textwidth]{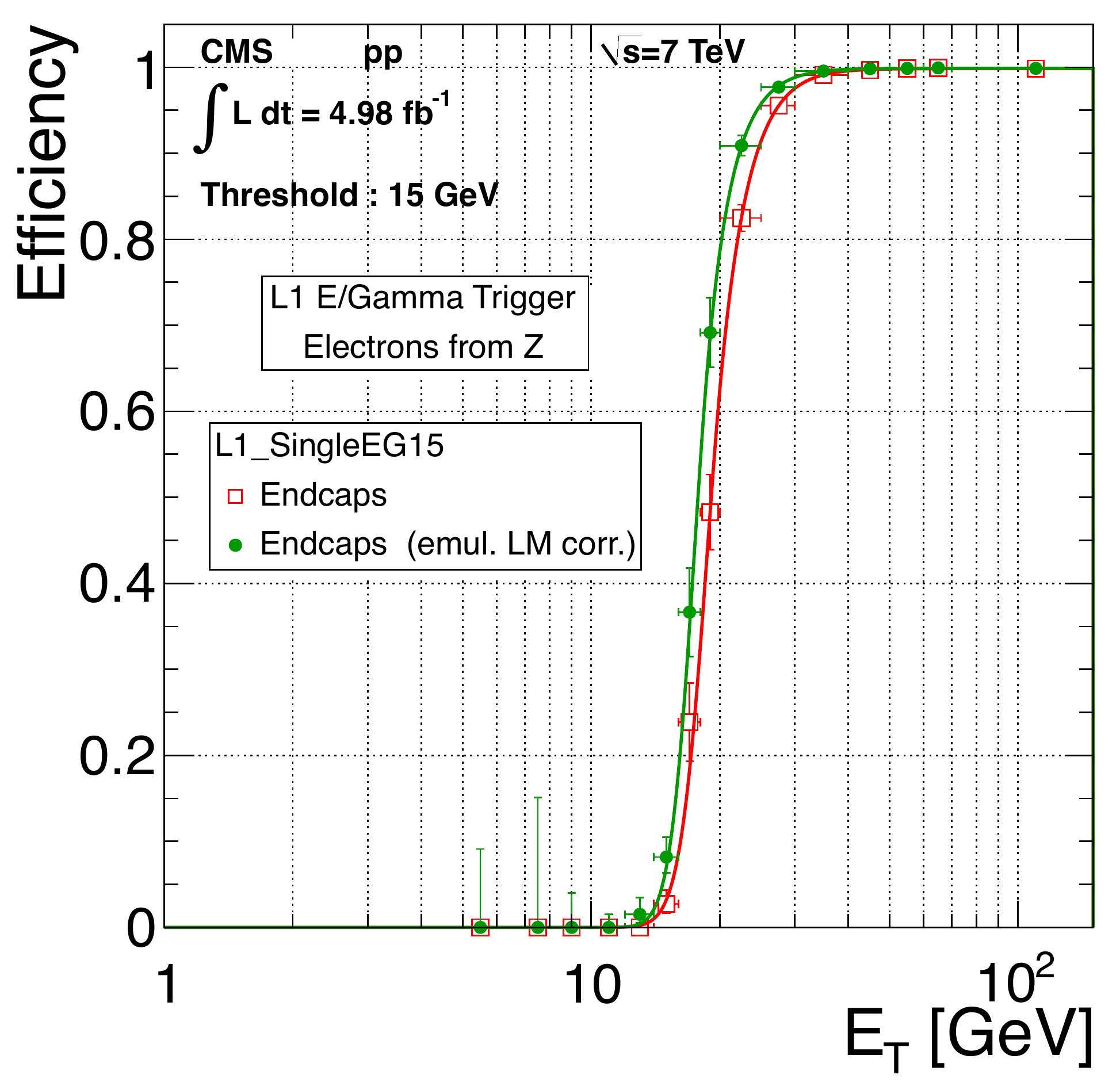}
\caption{\label{fig:transl1}
     The EE L1 electron trigger efficiency
     as a function of offline reconstructed \ET
    before (red) and after (green)  transparency corrections are
    applied at the ECAL TP level. The curves show unbinned likelihood fits.}
\end{minipage}
\end{figure}

The trigger efficiency curves are shown in Fig.~\ref{fig:l1eg15perf}
for an EG threshold of 15\GeV. The \ET on the $x$ axis is obtained
from the fully reconstructed offline energy. In the EE this includes
the pre-shower energy that is not available at L1. As a consequence, the
trigger efficiency turn-on point for the EE is shifted to the right
with respect to the EB. For both EB and EE, corrections for crystal
transparency changes were not included at L1 in 2011, which further
affects the turn-on curve (Sec.~\ref{sec:trans}). The width of the
turn-on curves is partly determined by the coarse trigger granularity,
since only pairs of TTs are available for the formation of L1
candidates, which leads to lower energy resolution at L1. An unbinned
likelihood fit was used to derive the efficiency
curves. Parameters of the turn-on curves are given in
Table~\ref{tab:fitbox}. Table~\ref{tab:fitboxtrans} summarizes the parameters of the turn-on
curves and compares them with the actual EE turn-on curve in 2011
(Fig.~\ref{fig:l1eg15perf}).

\begin{table}[h!]
\centering
\begin{minipage}[t]{0.47\textwidth}
\topcaption{\label{tab:fitbox} The L1 electron trigger turn-on curve
 parameters. This table gives the electron \ET thresholds for which
 an efficiency of 50\%, 95\% and 99\% are reached for EB and EE
 separately. The last entry corresponds to the efficiency obtained at
 the plateau of each curve shown in Figure~\ref{fig:l1eg15perf}.}
 \vfill
{\renewcommand{\arraystretch}{1.2}
\resizebox{\textwidth}{!}{
\begin{tabular}{|lll|}
\hline
  \multicolumn{1}{c}{EG15} & \multicolumn{1}{c}{EB} & \multicolumn{1}{c}{EE} \\
\hline
  50\% & $16.06^{+0.01}_{-0.01}$\GeV & $19.11^{+0.03}_{-0.06}$\GeV  \\
  95\% & $22.46^{+0.04}_{-0.05}$\GeV & $27.05^{+0.01}_{-0.01}$\GeV \\
  99\% & $28.04^{+0.07}_{-0.10}$\GeV & $34.36^{+0.01}_{-0.01}$\GeV \\
\hline
  100\GeV & $99.95^{+0.01}_{-0.88}$ \% & $99.84^{+0.06}_{-0.60}$ \% \\
\hline
\end{tabular}
}}
\end{minipage}\hfill
\begin{minipage}[t]{0.47\textwidth}
\topcaption{\label{tab:fitboxtrans} The EE L1 electron trigger turn-on curve
 parameters. This table gives the electron \ET thresholds for which
 an efficiency of 50\%, 95\% and 99\% are reached before and after
transparency  corrections are applied. The last entry corresponds to the
 efficiency obtained at the plateau of each curve shown in
 Figure~\ref{fig:transl1}.}
{\renewcommand{\arraystretch}{1.2}
\resizebox{\textwidth}{!}{
\begin{tabular}{|lll|}
\hline
  \multicolumn{1}{c}{EG15} & \multicolumn{1}{c}{EE} & \multicolumn{1}{c}{EE (corr)}\\
\hline
  50\% &  $19.11^{+0.03}_{-0.06}$\GeV &  $17.79^{+0.03}_{-0.06}$\GeV \\
  95\% &  $27.05^{+0.01}_{-0.01}$\GeV &  $24.46^{+0.10}_{-0.23}$\GeV \\
  99\% &  $34.36^{+0.01}_{-0.01}$\GeV &  $30.78^{+0.21}_{-0.48}$\GeV \\\hline
  100\GeV &  $99.84^{+0.06}_{-0.60}$ \% & $99.89^{+0.01}_{-0.67}$ \% \\
\hline
\end{tabular}
}}
\end{minipage}
\end{table}

\begin{figure}
\centering
\resizebox{0.48\textwidth}{!}{\includegraphics{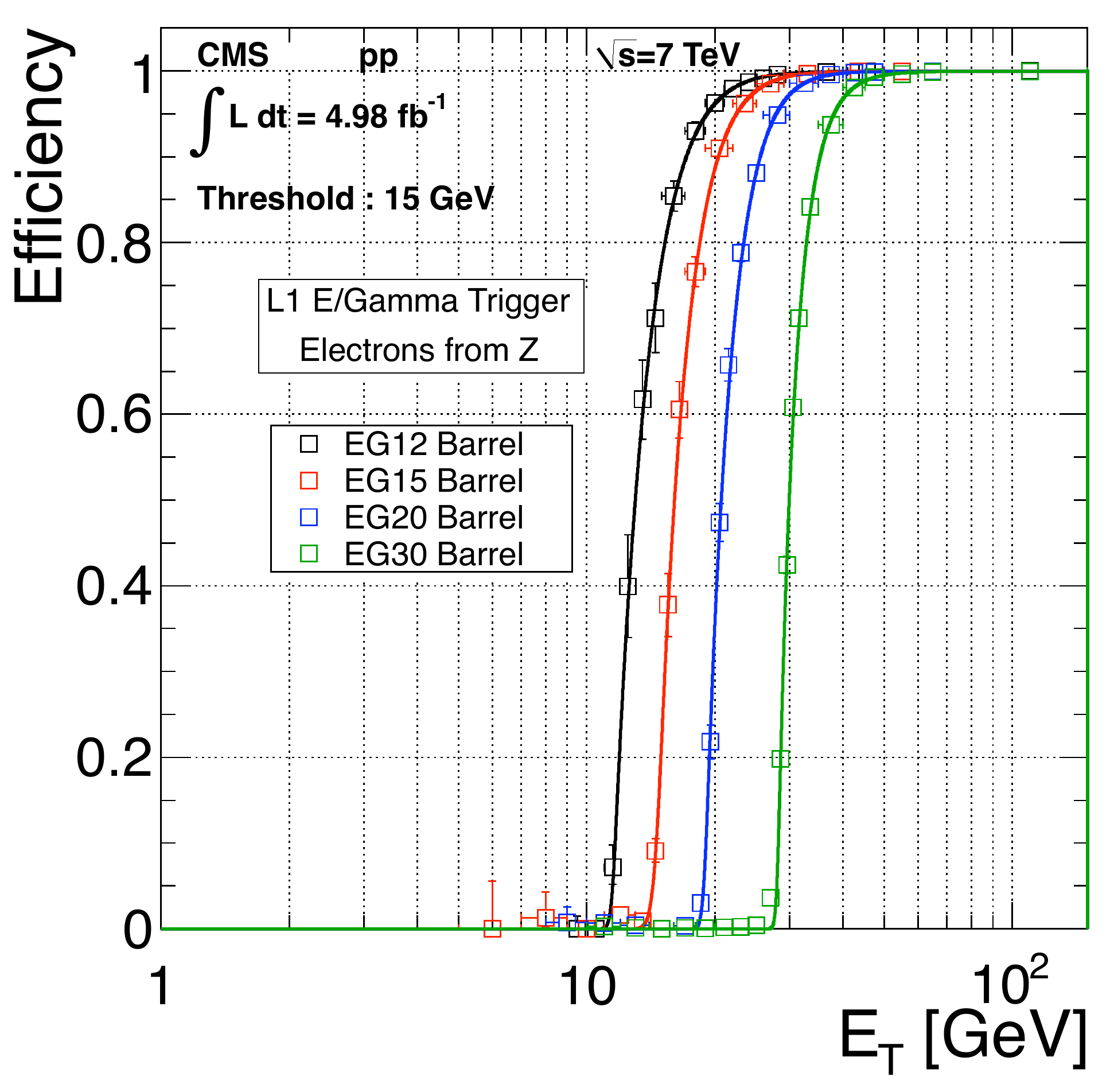}}
\resizebox{0.48\textwidth}{!}{\includegraphics{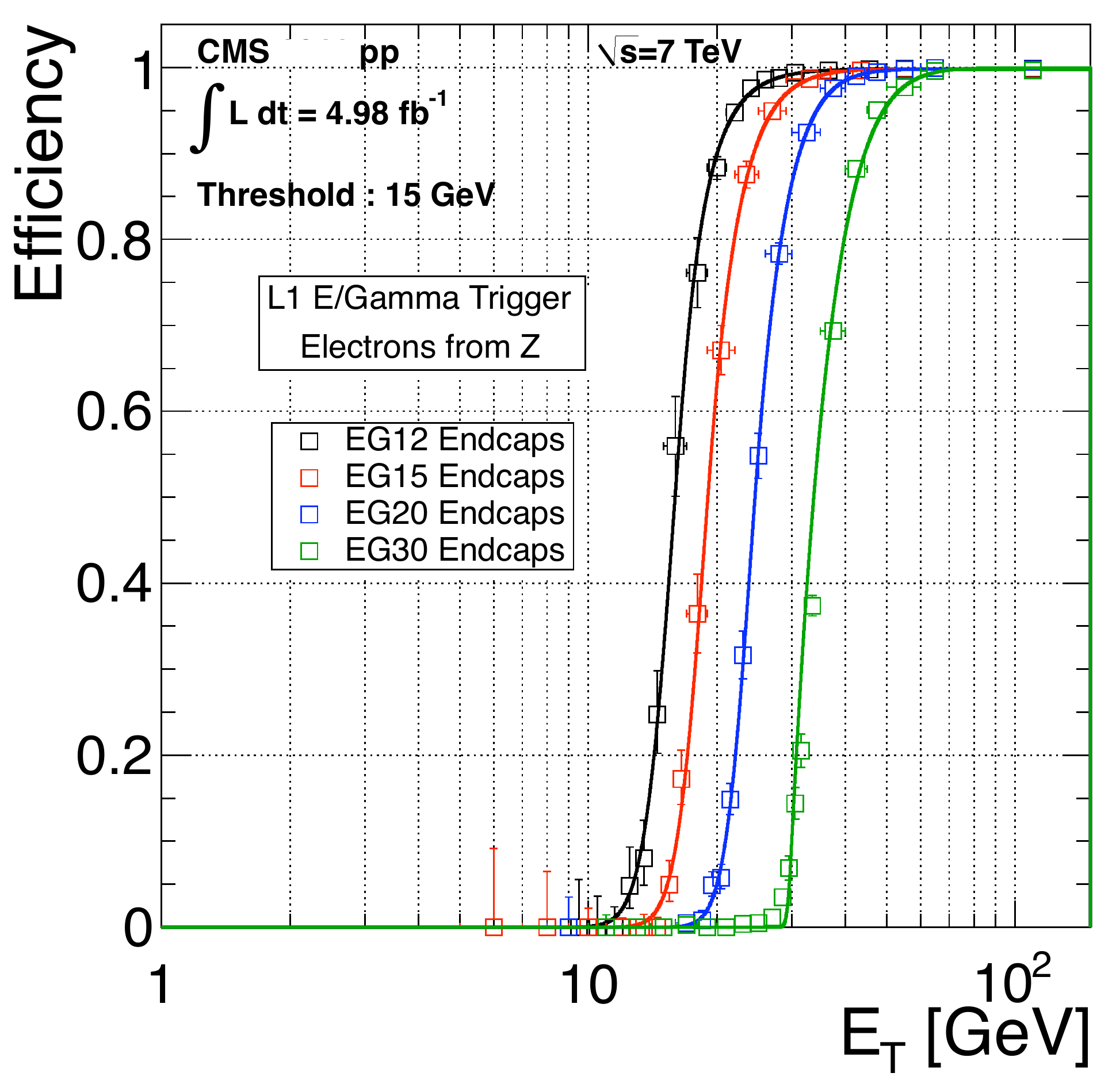}}
\caption{The L1 electron triggering efficiency as a function of the
  reconstructed offline electron \ET for barrel (left) and endcap
  (right).  The efficiency is shown for the EG12, EG15, EG20 and EG30
  L1 trigger algorithms. The curves show unbinned likelihood fits.}
\label{fig:turnoncurves}
\end{figure}

In the EE, the material in front of the detector causes more
bremsstrahlung, which together with the more complex TT geometry,
causes the turn-on curve to be wider than that for the EB. Some masked
or faulty regions (0.2\% in EB and 1.3\% in EE) result in the plateaus
being slightly lower than 100\% (99.95\% in EB and 99.84\% in EE) as
shown in Table~\ref{tab:fitbox}. The effect on efficiency of the L1
spike removal~\cite{cms-spike}, described in Sec.~\ref{sec:spikes},
is negligible, but will require further optimization as the number of
collisions per bunch crossing increases in the future.  Turn-on curves
for various EG thresholds are shown in Fig.~\ref{fig:turnoncurves}, and
Table~\ref{tab:turnoncurves} gives their turn-on points, \ie, the \ET value where the curve attains 50\% efficiency.

\begin{table}
\centering
  \topcaption{Turn-on points for the EG12, EG15, EG20, and EG30 L1
    trigger algorithms shown in Fig.~\ref{fig:turnoncurves}.}

\begin{tabular}{|lcccc|}
  \hline
  EG Threshold (\GeVns{}) & 12 & 15 & 20 & 30 \\
  \hline
  EB turn-on \ET (\GeVns{}) & 12 & 16.1 & 20.7 & 29.9 \\
  EE turn-on \ET (\GeVns{}) & 13 & 19.1 & 24.6 & 33.7 \\
  \hline
\end{tabular}

\label{tab:turnoncurves}
\end{table}

\begin{figure}[h]
\begin{minipage}[t]{0.49\textwidth}
\includegraphics[width=\textwidth]{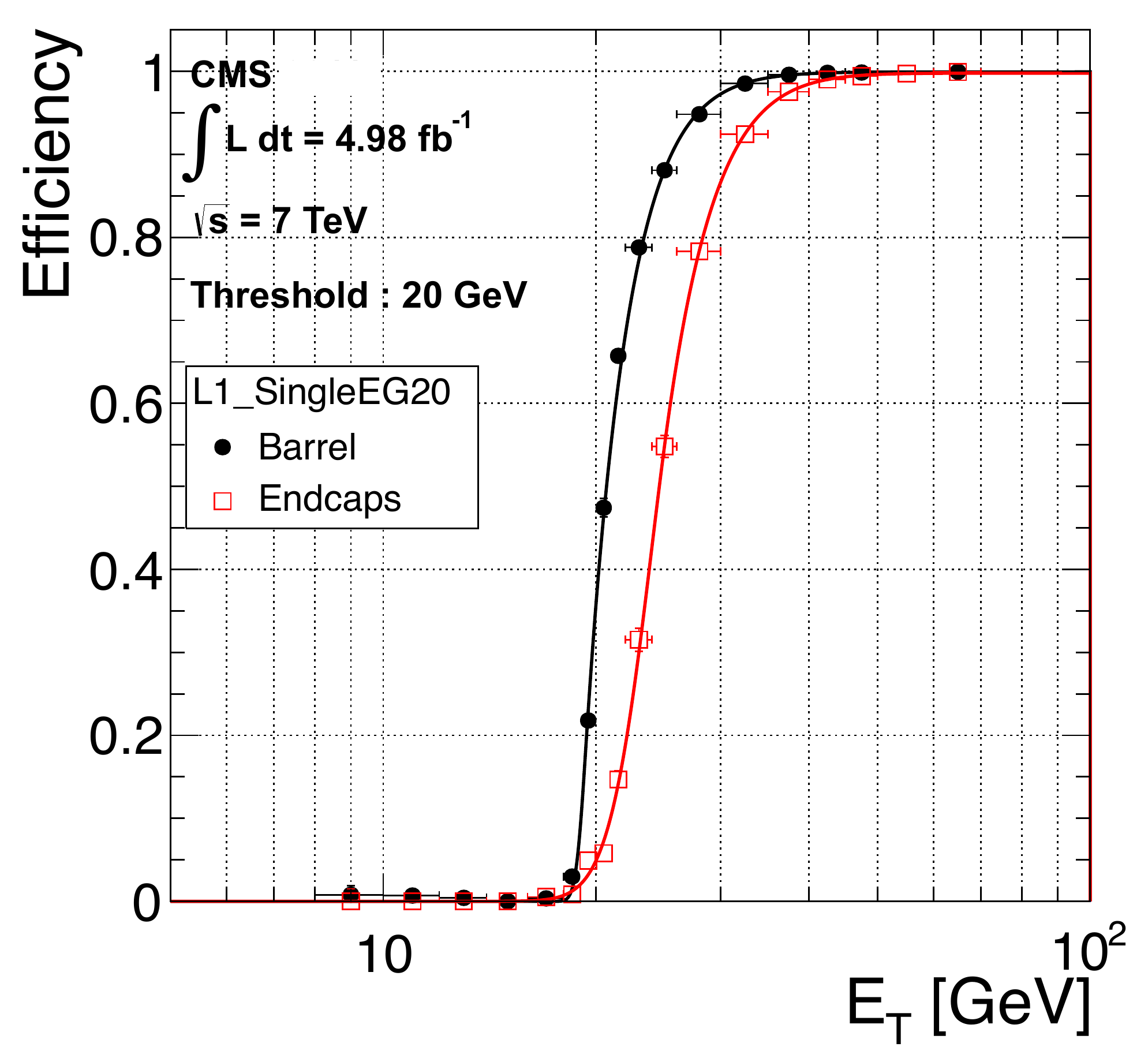}
\caption{\label{fig:finalEGeff2011} Electron trigger efficiency at L1, as a function of offline reconstructed \ET for
electrons in the EB (black dots) and EE (red squares) using the 2011
data set (EG threshold: $\ET=20\GeV$). The curves show unbinned
likelihood fits.}
\end{minipage}\hfill
\begin{minipage}[t]{0.47\textwidth}
\includegraphics[width=.995\textwidth]{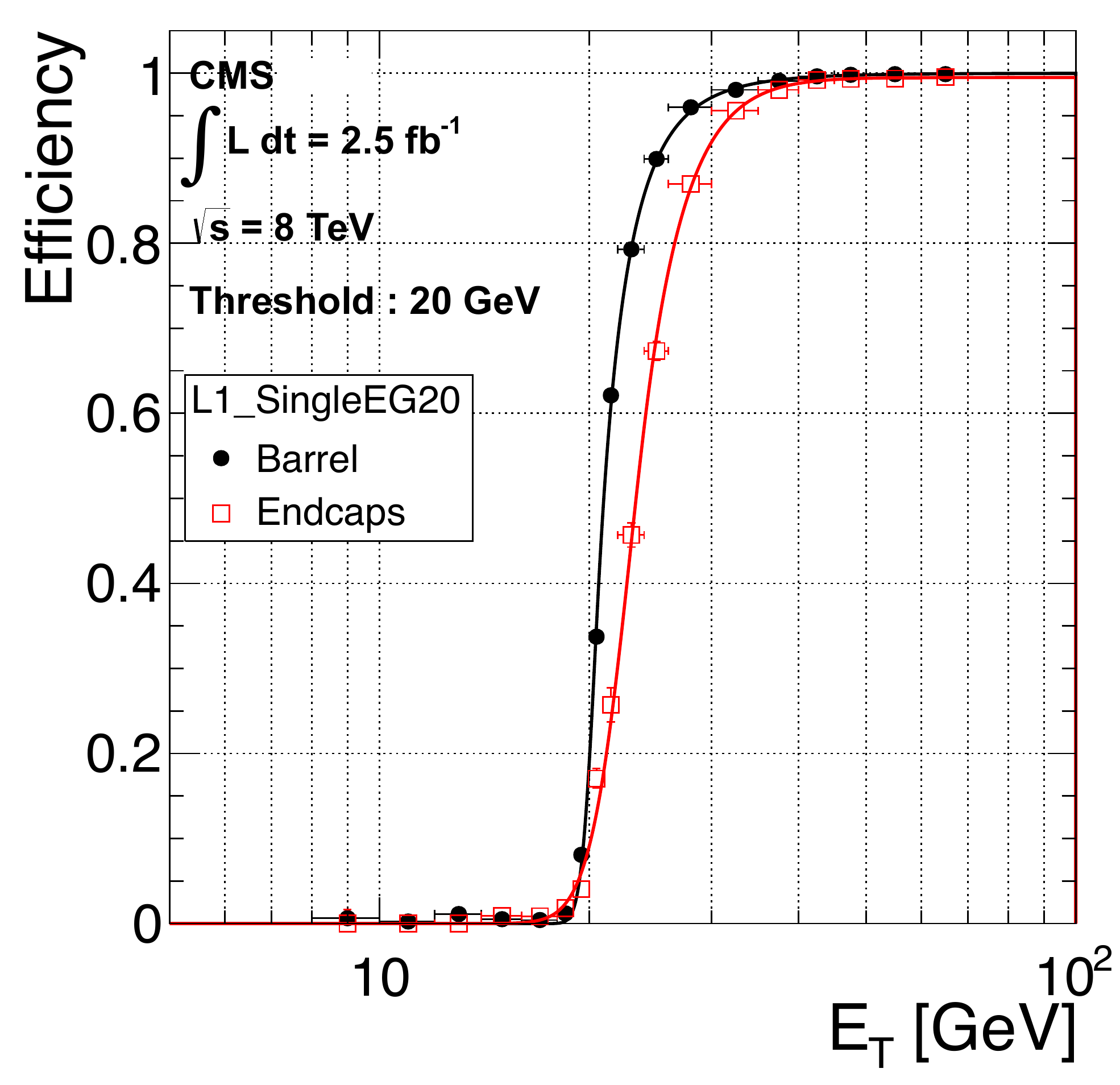}
\caption{\label{fig:finalEGeff2012} Electron trigger efficiency at L1
 as a function of offline reconstructed \ET for
electrons in the EB (black dots) and EE (red squares) using the 2012
data set (EG threshold:  $\ET=20\GeV$). The curves show unbinned
likelihood fits.}
\end{minipage}
\end{figure}

Figures~\ref{fig:finalEGeff2011} and \ref{fig:finalEGeff2012} show the
comparison of the EG20 algorithm performance obtained in 2011 and
2012. In the latter, the turn-on curve in EE is closer to that in EB. The
optimizations of the ECAL trigger primitive generation (spike killing
procedure and ECAL crystal transparency corrections) and RCT
calibration allowed the retention of the lowest possible unprescaled trigger to be used during physics runs.

\paragraph{L1 EG trigger rates}
The EG trigger rates were obtained from the analysis of a dedicated data stream,
containing only L1 trigger information, that was collected at high rate  on the
basis of L1 decision only.

For the study, events were selected using BPTX\_AND trigger
coincidences.  This selection provides unbiased information about the
L1 EG trigger response.  In this fashion, it was possible to apply
requirements related to the presence of L1 EG candidates with a given
\ET threshold and pseudorapidity acceptance region within the
analysis.

\begin{figure}[tbph]
\centering
  \includegraphics[width=0.6\textwidth]{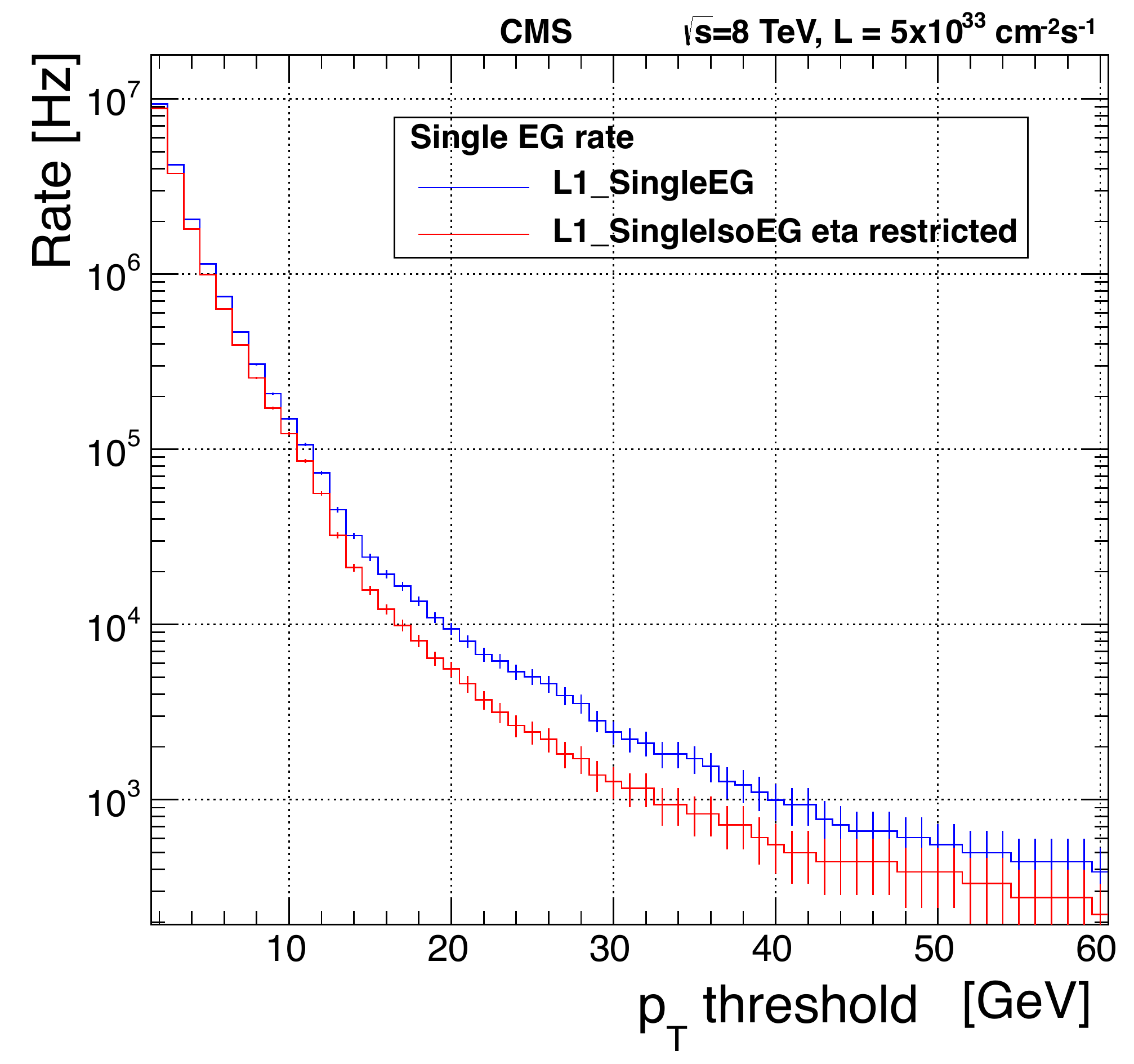}
  \caption{Rates of the isolated and nonisolated versions of the
    single-EG trigger versus the transverse energy threshold rescaled
    to an instantaneous luminosity of $5 \times 10^{33}\percms$.
    Isolated EG rates are computed within a pseudorapidity range of
    $\abs{\eta}<2.172$ to reflect the configuration of the L1 isolated EG
    algorithms used in 2012. }
  \label{figure:L1SingleEGRates}
\end{figure}

Rates of isolated and nonisolated single-EG
triggers are presented in Fig.~\ref{figure:L1SingleEGRates}. During
the 2012 run, isolated EG trigger algorithms were
restricted to $\abs{\eta}<2.712$ at the GT level.
Rates were calculated using
data collected with luminosities between $4.5$ and $5.5 \times
10^{33}\percms$ (for an average luminosity of $4.94 \times
10^{33}\percms$), and rescaled to a target instantaneous luminosity
of $5 \times 10^{33}\percms$. Uncertainties stemming from this
small approximation
are well within the fluctuations caused by data acquisition deadtime
variations.

\subsection{Online anomalous signals and their suppression}
\label{sec:spikes}
Anomalous signals were observed in the EB shortly after collisions
began in the LHC: these were identified as being due to direct
ionization within the APDs, thus producing spurious isolated signals
with high apparent energy. These {\it spikes} can induce large trigger
rates at both L1 and HLT if not removed from the trigger decision. On
average, one spike with $\ET >$ 3\GeV is observed per 370 minimum bias
triggers in CMS at $\sqrt{s}$ = 7\TeV. If untreated as many 60\% of
trigger objects containing only ECAL energy, above a threshold of
12\GeV, would be caused by spikes. At high luminosity these would be
the dominant component of the 100\unit{kHz} CMS L1 trigger rate
bandwidth~\cite{spike-petyt}. Spike identification and removal
strategies were developed, based on specific features of these
anomalous signals. In the ECAL the energy of an electromagnetic (EM)
shower is distributed over several crystals, with up to 80\% of the
energy in a central crystal (where the electron/photon is incident)
and most of the remaining energy in the four adjacent crystals. This
lateral distribution can be used to discriminate spikes from EM
signals. A topological variable $s=1-E_4/E_1$ ($E_1$: \ET of the
central crystal; $E_4$: summed \ET of the four adjacent crystals)
named ``Swiss-cross'' was implemented offline to serve this purpose. A
similar topological variable was also developed for the on-detector
electronics, a strip fine grain veto bit (sFGVB). Every
TP has an associated sFGVB that is set to 1 (signifying a true EM
energy deposit) if any of its 5 constituent strips has at least two
crystals with \ET above a programmable trigger  sFGVB threshold,
of the order of a few hundred \MeVns{}. If the sFGVB is set to zero, and
the trigger tower \ET is greater than a trigger  killing
  threshold, the energy deposition is considered spike-like. The
trigger tower energy is set to zero and the tower will not contribute
to the triggering of CMS for the corresponding event.

As the sFGVB threshold is a single value, the electron or photon
efficiency depends upon the particle energy: the higher the threshold,
the more low-energy genuine EM deposits would be flagged as
spikes. However, these spurious spikes may not pass the killing
threshold so they would still be accepted. With a very low sFGVB
threshold, spikes may not be rejected due to neighboring crystals
having noise. A detailed emulation of the full L1 chain was developed
in order to optimize the two thresholds to remove as large a fraction
of the anomalous signals as possible whilst maintaining excellent
efficiency for real electron/photon signals. In order to determine the
removal efficiency, data were taken in 2010 without the killing
thresholds active. Using the Swiss-cross method, spike signals were
identified offline. Those signals were then matched to L1
candidates in the corresponding RCT region and the emulator used to
evaluate the fraction of L1 candidates that would have been
eliminated. In a similar fashion the efficiency for triggering on
genuine electrons or photons could be estimated.

\begin{figure}[h]
\centering
\includegraphics[width=0.48\textwidth]{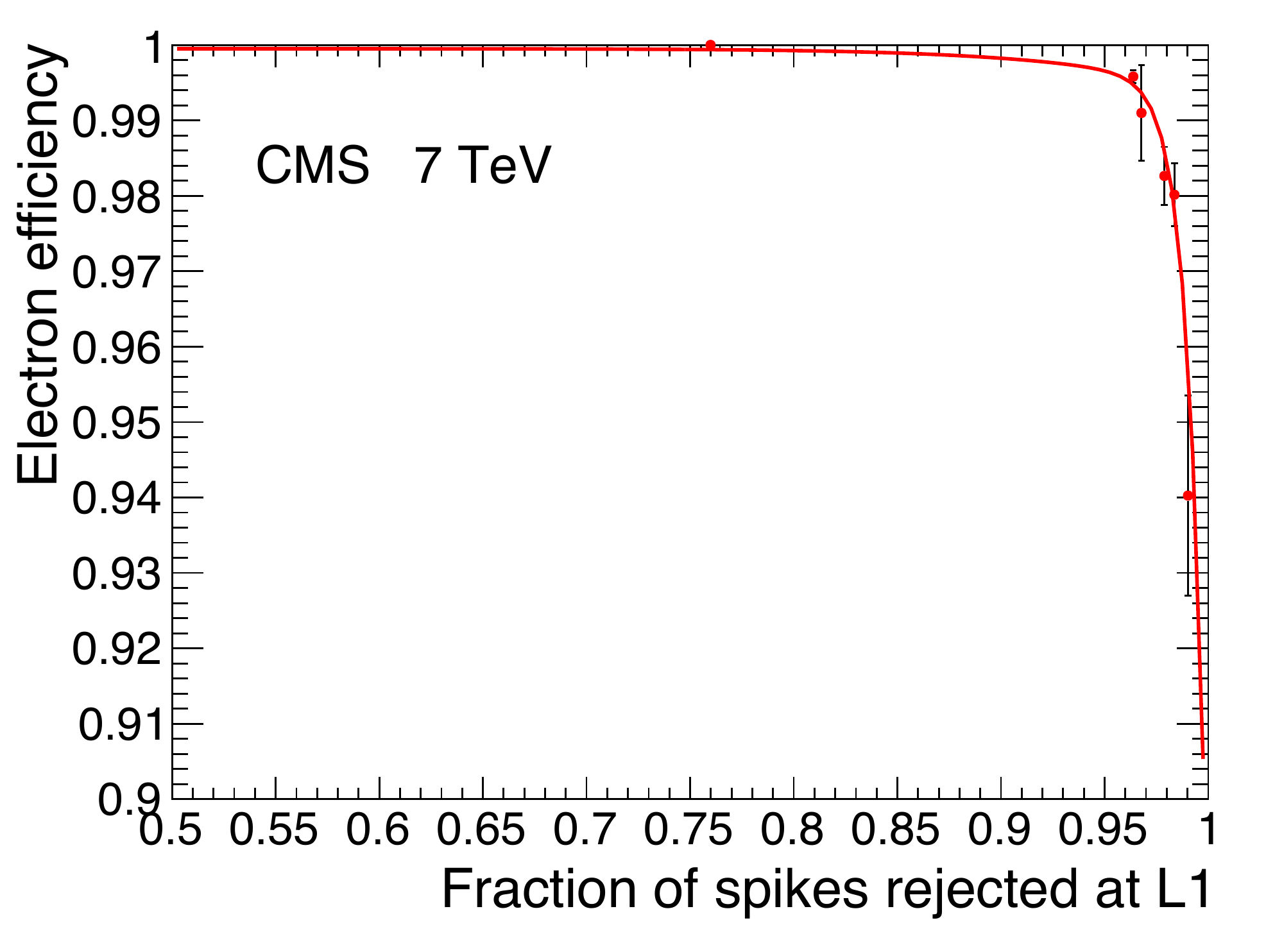}
\caption{\label{fig:spikeperf}Electron trigger efficiency as a
function of the spike rejection at L1. Each point corresponds to a
different spike removal trigger  sFGVB threshold. The trigger  killing
threshold is set to 8\GeV. The data were taken in 2010.}
\end{figure}

Three killing thresholds were emulated ($\ET = 8$, 12, and 18\GeV), combined with six sFGVB thresholds (152, 258, 289, 350, 456, 608\MeV). Figure~\ref{fig:spikeperf} shows the electron efficiency
(fraction of electrons triggered after spike removal) versus the L1
spike rejection fraction, for all sFGVB thresholds mentioned above
(one point for each threshold value) and a killing threshold of 8\GeV. The optimum configuration was chosen to be an sFGVB threshold of
258\MeV and a killing threshold of 8\GeV. This corresponds to a
rejection of 96\% of the spikes, whilst maintaining a trigger
efficiency for electrons above 98\%. With these thresholds the
efficiency for higher energy electrons is even larger: 99.6\% for
electrons with $\ET >20$\GeV.

\begin{figure}[h]
\centering
\includegraphics[width=0.48\textwidth]{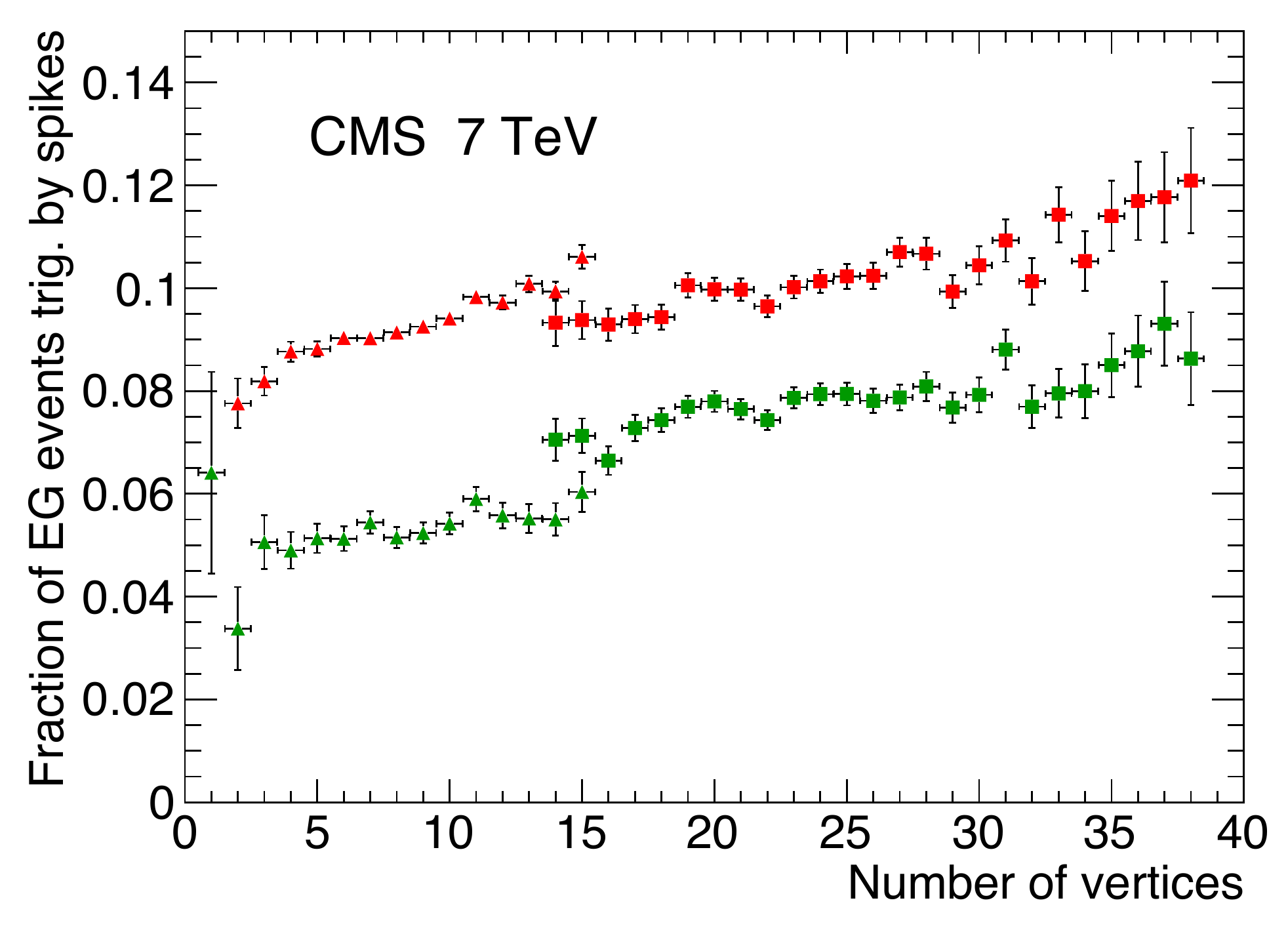}
\caption{\label{fig:spikepileup}Fraction of spike-induced EG triggers
    as a function of the number of reconstructed vertices.
The red points represent the spike removal working point used in 2011, and
    the green points the optimized working point for 2012.
The squares (triangles) correspond to higher (lower) pileup data.}
\end{figure}

Table~\ref{tab:spikerate} summarizes the rate reduction factors
obtained for L1 EG algorithms considering the working point discussed
above. This optimized configuration was tested online at the beginning
of 2011. It gave a rate reduction factor of about 3 (for an EG
threshold of 12\GeV), and up to a factor of 10 for \ET sum triggers
(which calculate the total EM energy in the whole calorimeter system).

\begin{table}[h]
  \caption{\label{tab:spikerate}Rate reduction factors obtained for
    L1 EG algorithms (considering a 258\MeV sFGVB threshold
    and an 8\GeV killing threshold on the ECAL Trigger Primitives) for various EG thresholds.}
\centering
\begin{tabular}{|lllll|}
\hline
EG Threshold (\GeVns{}) & 12 & 15 & 20 & 30\\
\hline
Rate reduction factors & 3.4 & 4.3 & 6.0 & 9.6 \\
\hline
\end{tabular}
\end{table}

At the end of 2011 the average pileup had peaked at 16.15, and in 2012
the highest average pileup was 34.55. Efficient identification of EM
showers at trigger level became more and more challenging. As pileup
events act as noise in the calorimeter, they degraded trigger object
resolution and reduced the probability of observing isolated
spikes. The fraction of spike-induced EG triggers was measured as a
function of the number of vertices (roughly equivalent to the number
of pileup events) in Fig.~\ref{fig:spikepileup}. The fraction of
spike-induced EG triggers reaches 10\% for collisions including more
than 20 pileup events (red points). Using the L1 trigger emulator, a
more efficient working point (sFGVB threshold~=~350\MeV, killing
threshold = 12\GeV) for the spike removal algorithm reduces this
fraction to 6\% (green points), but still preserves the same high
trigger efficiency for genuine electrons and photons.
\subsubsection{HLT electron  and  photon identification}
\label{sec:egammaHLT}

The HLT electron and photon identifications begin with a regional
reconstruction
of the energy deposited in the ECAL crystals around the L1 EM
candidates. This is followed by the building of the supercluster using
offline
reconstruction algorithms~\cite{cms-e8}.

Electron and photon candidates are initially selected based on the
\ET of the supercluster and on criteria based on properties of the
energy deposits in the ECAL and HCAL subdetectors. Selection
requirements include a cluster shape variable
$\sigma_{mathrm{i}\eta\mathrm{i}\eta}$ (the root-mean-square of the
width in $\eta$  of the shower)~\cite{cms-e8} and an isolation requirement
that limits the additional energy deposits in the ECAL in a cone
around the EM candidate with outer cone size of
$\DR\equiv\sqrt{\smash[b]{{\Delta\phi}^2 + {\Delta\eta}^2}}=0.3$, and inner
cone radius corresponding to the size of three ECAL crystals
($\DR=0.05$ in the barrel region.)  The energy deposits in channels that
are found in a strip along $\phi$ centered at the ECAL position of the
EM candidate with an $\eta$-width of 3 crystals are also not
considered.  Candidates are then required to satisfy selection
criteria based on the ratio of the HCAL energy in a cone of size
$\DR=0.3$ centered on the SC, to the SC energy.

These requirements typically reduce the trigger rate by a factor of 3--4, reaching
10 for the tightest selection used in 2012. The thresholds are
such that, after this set of calorimetric criteria, the rates of
electron candidates are about 1\unit{kHz}.
The previously described steps are common to electron and photon selection.
In addition, photon candidate selection imposes an additional
isolation requirement based on tracks reconstructed in a cone around
the photon candidate. In some trigger paths extra requirements are
needed to keep the rate at an acceptable level. The
$\RNINE \equiv E_{3{\times}3}/E_{SC}$ variable, where $E_{3{\times}3}$
denotes the energy deposited in a small window of $3{\times}3$ crystals
around the most energetic crystal in the SC, is very effective in
selecting good unconverted photons
even in the presence of large pileup. Finally,  to
distinguish electrons from photons, a nearby track is required, as
described later in this section.

An improvement deployed in the $\Pe/\Pgg$ triggers in 2012 was the use
of corrections for radiation-induced changes in the transparency of
the crystals in the endcap ECAL~\cite{Chatrchyan:2013dga}.
A new set of
corrections was deployed weekly.
\begin{figure}[tbp]
   \centering
   \includegraphics[width=0.5\textwidth]{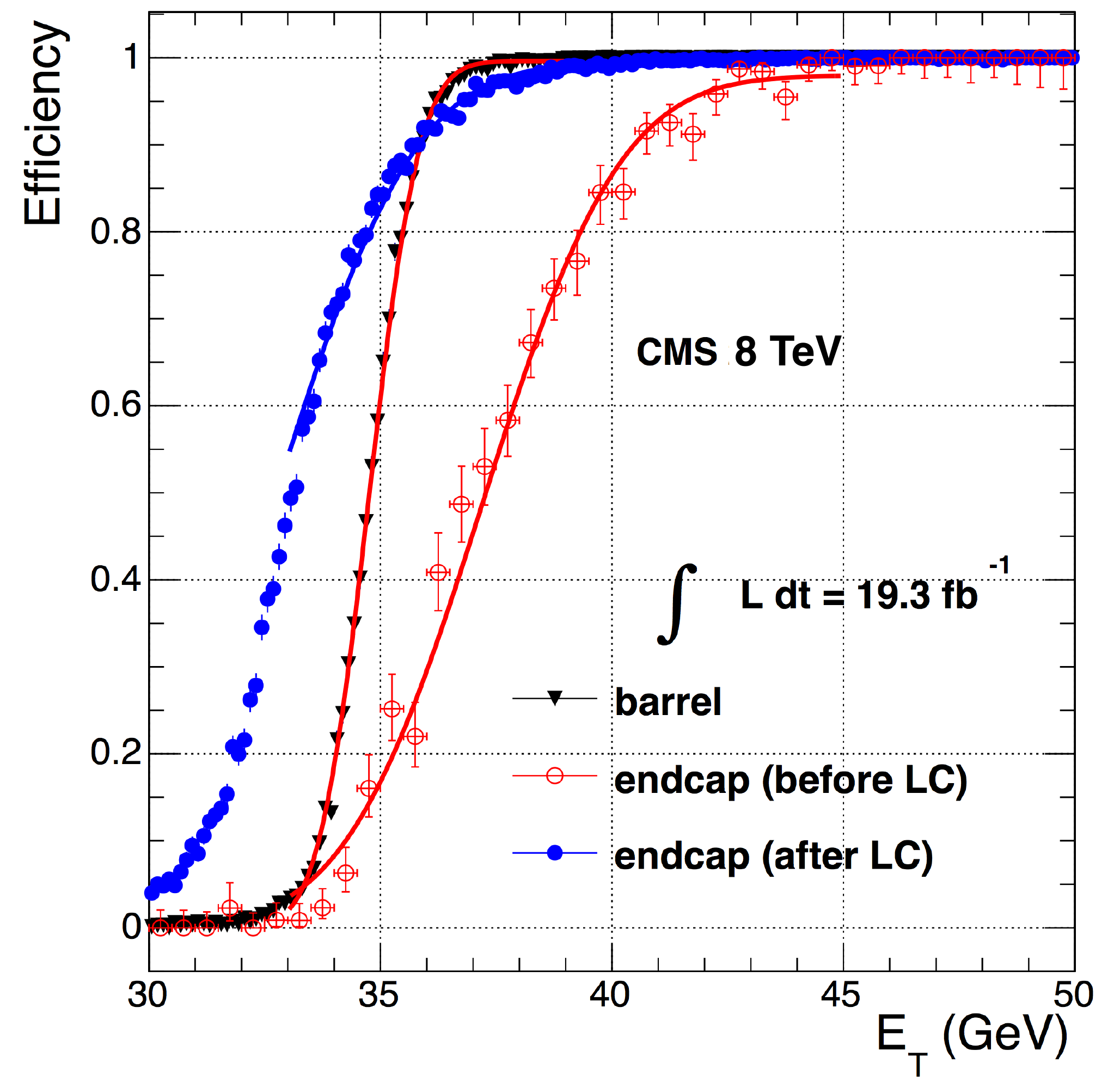}
   \caption{Efficiency of the online \ET selection as a function of the
     offline electron \et, in barrel and endcap regions, before
     and after the deployment of online transparency corrections. The data
     depicts the results of a double-electron trigger requiring
     $\pt>33\GeV$ for  both legs, and shows that applying the corrections
     causes a significant improvement of the online turn-on curve.}
   \label{fig:laser_corrections}
\end{figure}
Figure~\ref{fig:laser_corrections} shows that the introduction of
these corrections in the trigger significantly improved the
performance of the electron trigger in the endcap. The turn-on curve
refers to a double-electron trigger requiring  a 33\GeV threshold
for both legs.

\paragraph{Double-photon trigger efficiency.}
The tag-and-probe method with $\Z  \to\Pe\Pe$ events is used to
measure trigger efficiencies from the data. For photon triggers, the
probe electron is treated as a photon and the electron SC is required
to pass photon selection requirements. Events are selected from the
double-electron data set with the loosest prescaled tag-and-probe
trigger path. Since this path requires only one electron passing the
tight HLT selection for the leading leg of the trigger, the other electron,
which is only required to pass a very loose filter on its SC
transverse energy, is sufficiently unbiased such that it is suitable
for our measurement. We then require at least one offline electron to
match the HLT electron leg, and at least two offline photons to match
the HLT electron and the HLT SC leg, respectively. The two offline
photons are required to have an invariant mass compatible with the \Z
boson (between 70\GeV and 110\GeV), and to pass offline \pt threshold of
30\GeV and 22.5\GeV, respectively. Finally the event is required to
pass offline photon and event selections, \eg, for the \HGG\xspace
measurement.

The photon matched to the HLT electron leg is also required to match
to an L1 $\Pe/\Pgg$ isolated object with $\ET >22\GeV$. This photon is
considered to be the tag, while the other one is the probe. Each trigger step
is measured separately and, to account for the fact that electrons and
photons have different \RNINE distributions, each electron pair used
for the trigger efficiency measurement is weighted so that the \RNINE
distribution of the associated SCs matches the one of a simulated
photon. The net effect is an increase of the measured efficiency due
to the migration of the events towards higher \RNINE values.
\begin{figure}[tbh]
   \centering
   \includegraphics[height=150pt,trim={0 .2cm 0 0},clip]{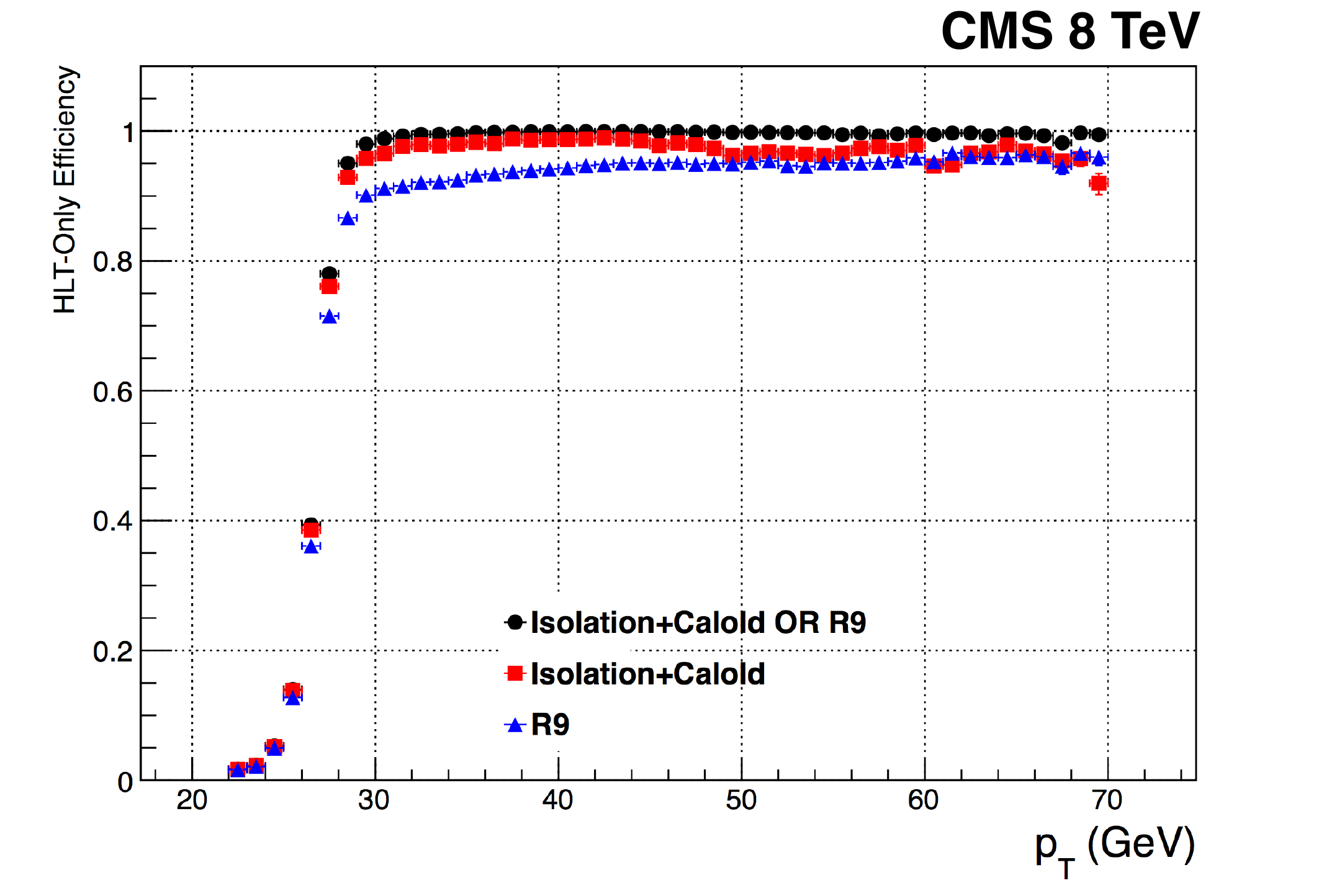}
   \includegraphics[height=151pt]{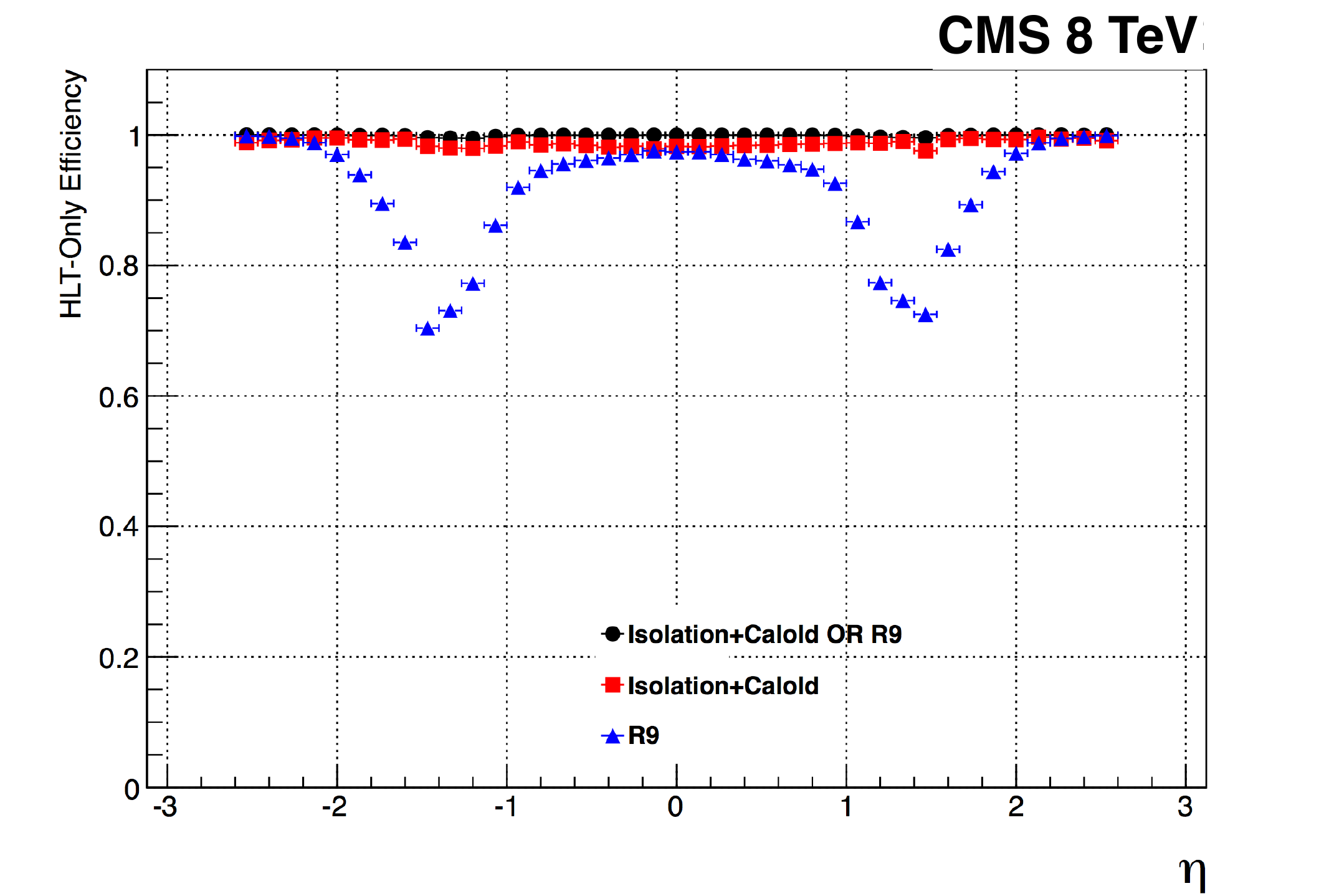}
   \caption{Efficiencies of the leading leg for the double-photon
     trigger as a function of the photon transverse energy (left) and
     pseudorapidity (right), as described in the text. The red symbols
     show the efficiency of the isolation plus calorimeter
     identification requirement, and the blue symbols show the
     efficiency of the \RNINE selection criteria. The black symbols
     show the combined efficiency.}
   \label{fig:hlt_26_pt_eta}
\end{figure}

\begin{figure}[tbh]
   \centering
   \includegraphics[width=0.5\textwidth]{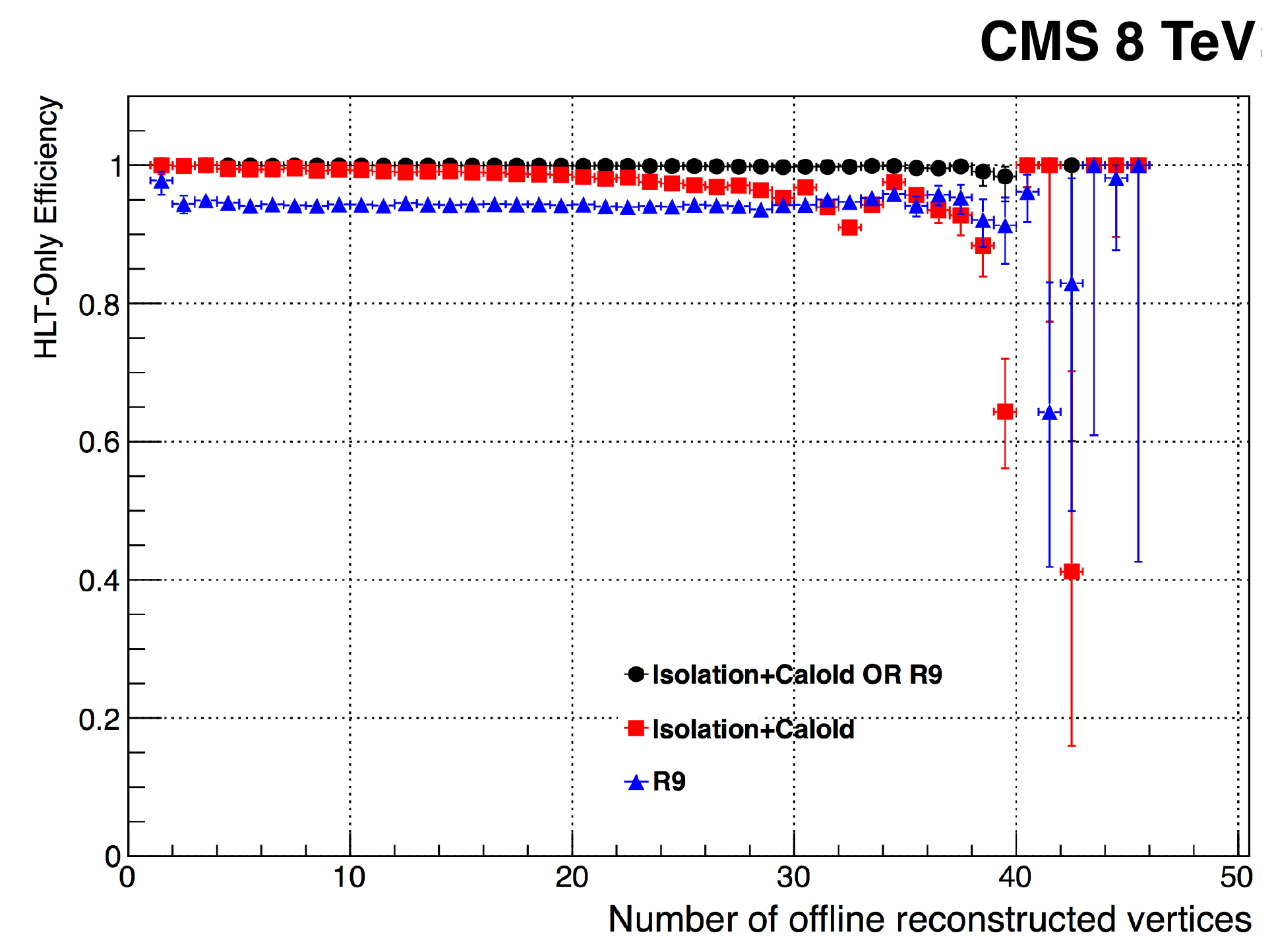}
   \caption{Efficiencies of the leading leg of the double-photon
     trigger described in the text as a function of the number of
     offline reconstructed vertices. The red symbols show the
     efficiency of the isolation plus calorimeter identification
     requirement, and the blue symbols show the efficiency of the
     \RNINE selection. The black symbols show the combined
     efficiency.}
   \label{fig:hlt_26_nvtx}
\end{figure}
Figures~\ref{fig:hlt_26_pt_eta} to~\ref{fig:hlt_26_nvtx} show the
efficiency of the leading leg selection as a function of the photon
transverse energy, pseudorapidity, and number of offline reconstructed
vertices (N$_\mathrm{vtx}$).

The double-photon trigger is characterized by a steep turn-on
curve. The loss of efficiency shown in Fig.~\ref{fig:hlt_26_pt_eta}
(right) for the \RNINE selection follows the increase of the tracker
material in the region around $\abs{\eta}{\approx}1.2$, where is more likely
to find converted photons with a smaller \RNINE value. The flat
efficiency versus N$_\text{vtx}$ curve demonstrates that the path is quite
insensitive to the amount of pileup events, although some small
dependence is noticeable for N$_\text{vtx} > 30$.

\paragraph{Electron selection.}
In order to distinguish between electron and photon candidates, the
presence of a reconstructed track compatible with the SC is
required. Hence, after the common selection described above, the
selection of online electron candidates follows with selections involving
the tracker. The first step is the so called ``pixel-matching'', which
uses the energy and position of the SC to propagate a hypothetical trajectories through the magnetic field under each charge hypothesis to search for compatible hits in the pixel detector. Full silicon
tracks are then reconstructed from the resulting pixel seeds. Timing
constraints prohibit the usage of the offline tracking algorithms and
a simple Kalman filter technique is used. Nevertheless, since 2012, it
is complemented by the Gaussian-Sum Filtering (GSF) algorithm, which better
parametrizes the highly non-Gaussian electron energy
loss. Due to the large CPU time requirements of the
algorithm, it was used only in paths where it is possible to achieve a
large reduction of the rate before the electron tracking (\eg,
in the path selecting two high-\ET electrons, where the transverse
energy requirement is of 33\GeV on each electron). The electron tracks
are required to have a measured momentum compatible with the SC
energy. Their direction at the last tracker layer should match the SC
position in $\eta$ and $\phi$. These selection criteria reduce the
rate of misidentified electrons by a factor of 10. Finally, isolation
requirements with respect to the tracks reconstructed around the
electron candidate are applied, if required for rate reasons. The
lowest-threshold inclusive single isolated electron path at the end of
the 2012 running (corresponding to instantaneous luminosities of
$7\times10^{33}\percms$) had a threshold of $\ET>27$\GeV, with a rate
of less than 50\unit{Hz}.
\begin{figure}[tbph]
   \centering
   \includegraphics[width=0.7\textwidth]{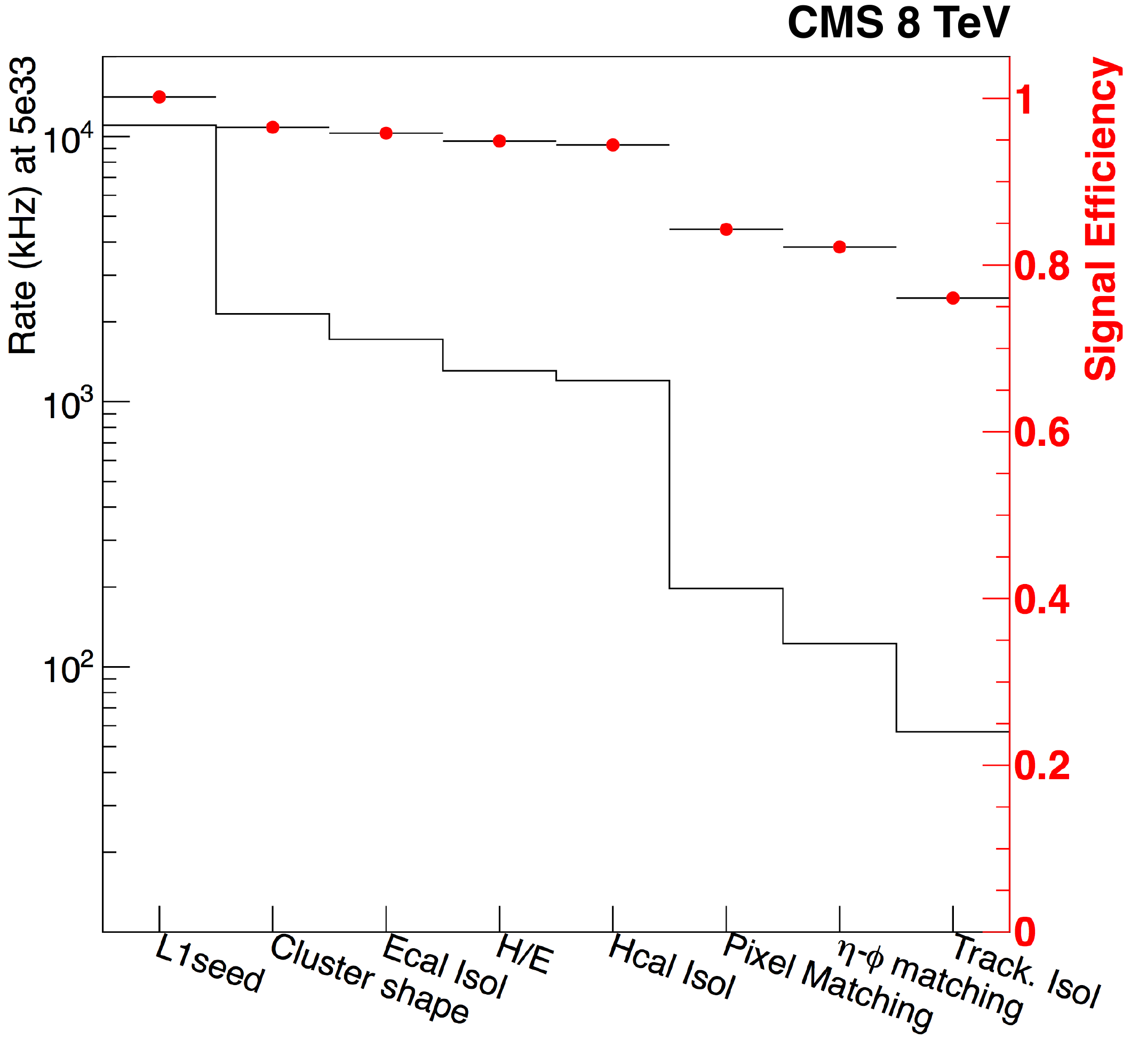}
   \caption{Performance of the internal stages of the lowest-\ET      unprescaled single-electron trigger. The rate is shown as the
     black histogram (left scale); the red symbols show the efficiency
     for electron selection (right scale).}
   \label{fig:single_ele}
\end{figure}
Figure~\ref{fig:single_ele} shows how the rate is gradually reduced by
the filtering steps of this trigger (black histogram), along with the
efficiency of electrons (red points).

\begin{figure}[tbph]
   \centering
   \includegraphics[width=0.49\textwidth]{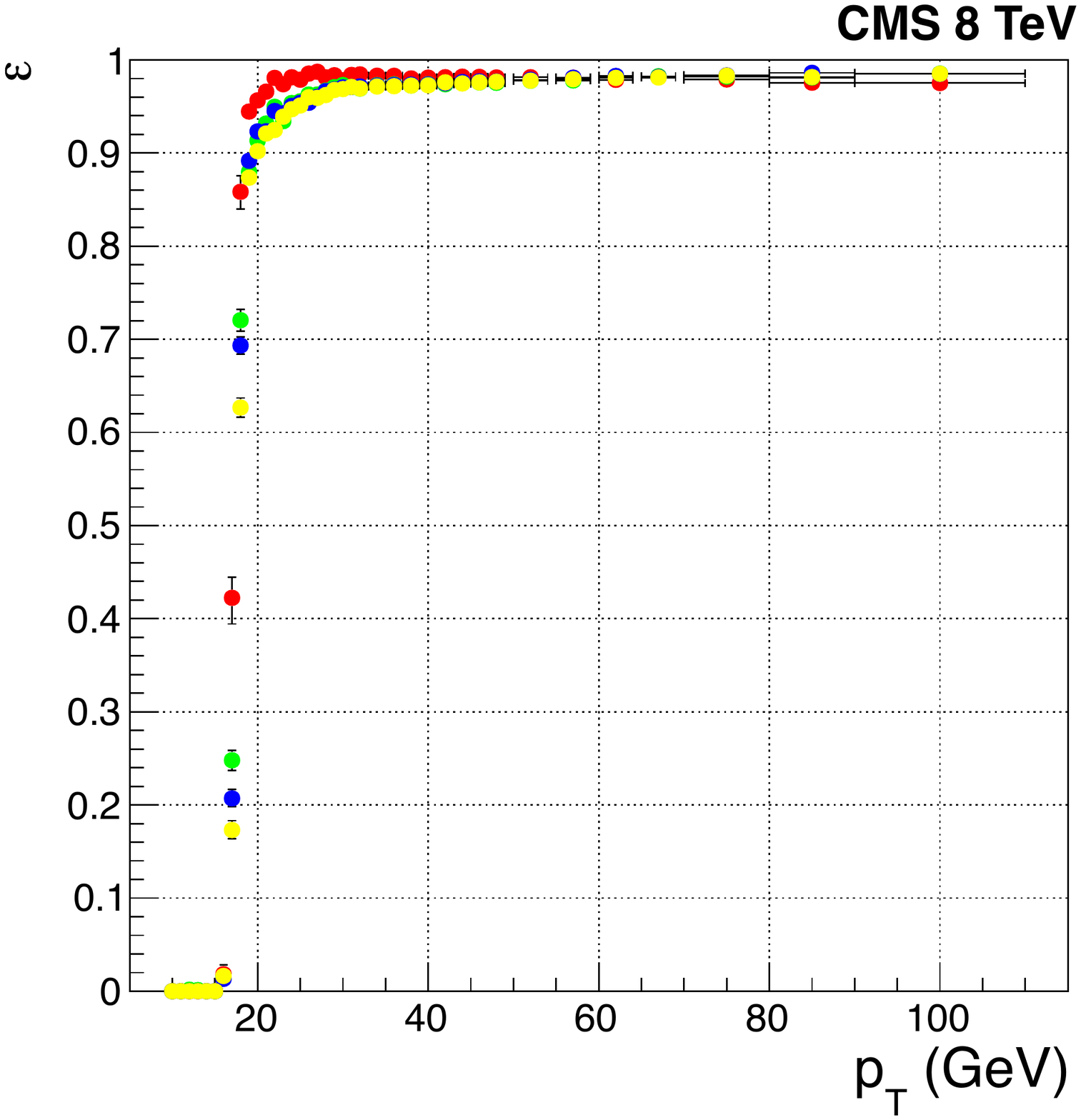}
   \includegraphics[width=0.49\textwidth]{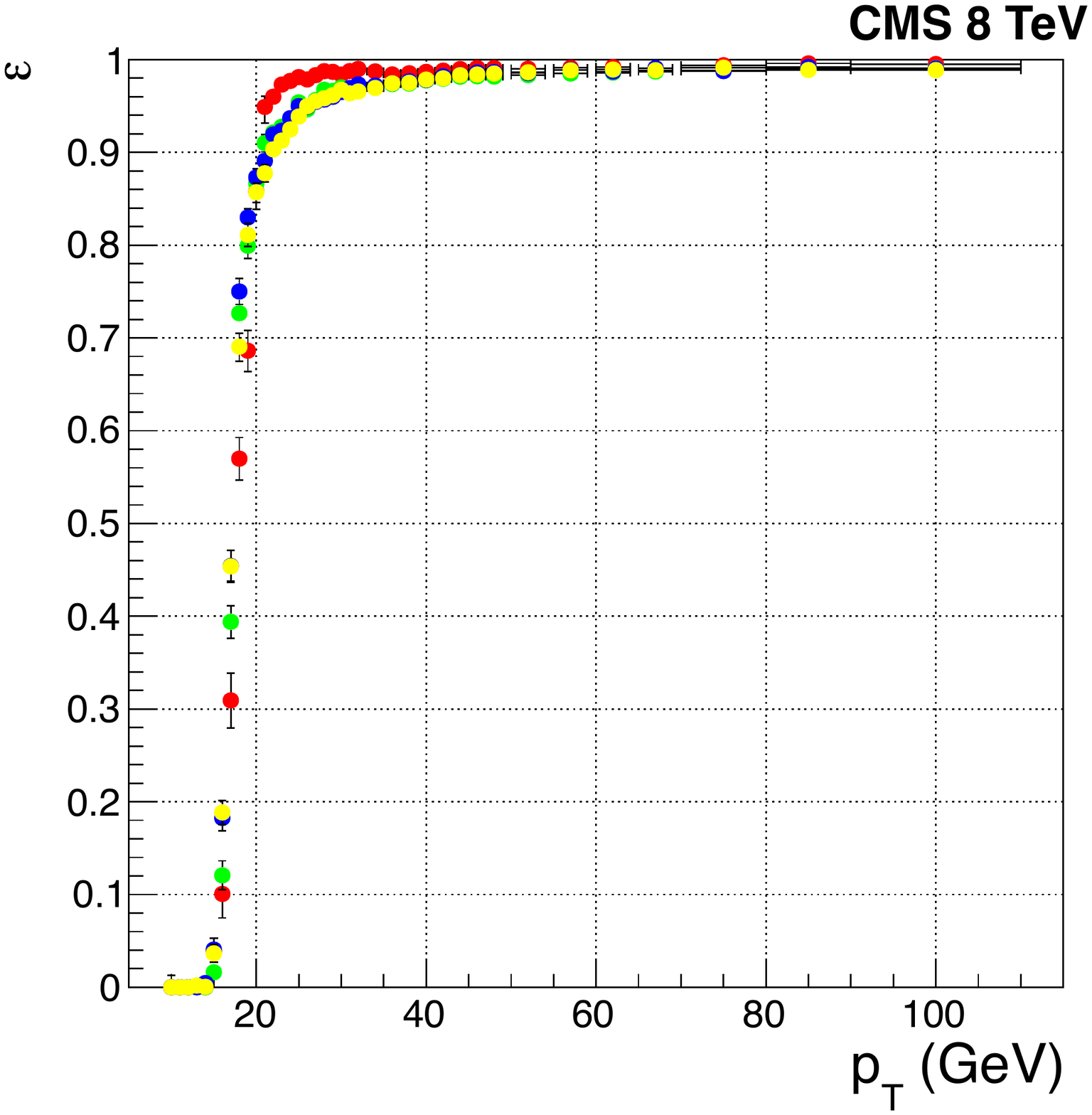}

   \caption{Efficiencies of the leading leg for the double-electron
     trigger described in the text as a function of the offline
     electron momentum. The trigger uses identical selection for both
     legs, so the other leg just has a different
     threshold. Efficiencies are shown for different running periods
     (red May, green June, blue August, and yellow November of 2012)
     and separately for electron reconstructed in barrel (left) and
     endcap (right).}  \label{fig:double_ele_pt}
\end{figure}
\begin{figure}[tbph]
   \centering
\includegraphics[width=0.49\textwidth]{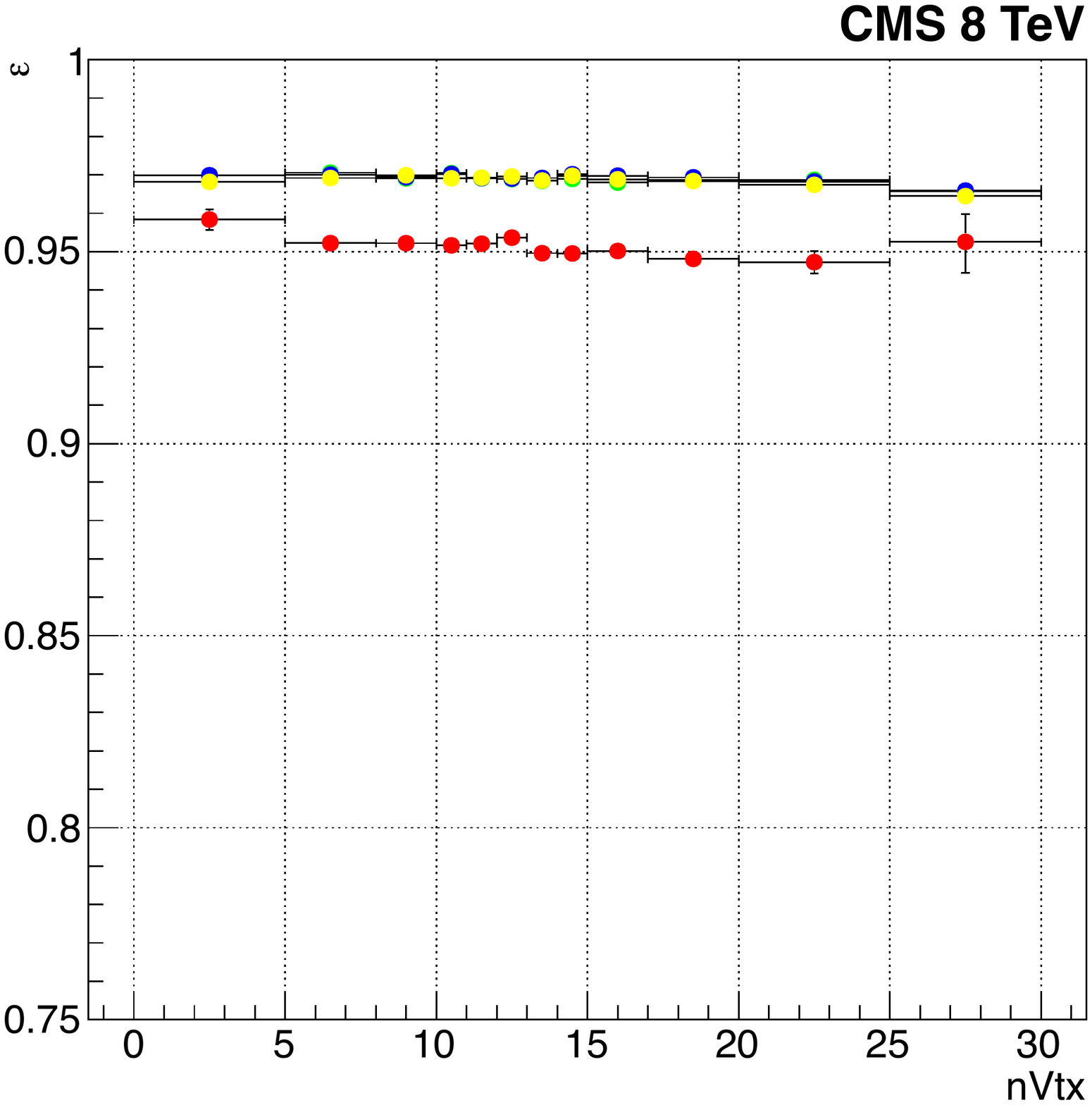}
\includegraphics[width=0.49\textwidth]{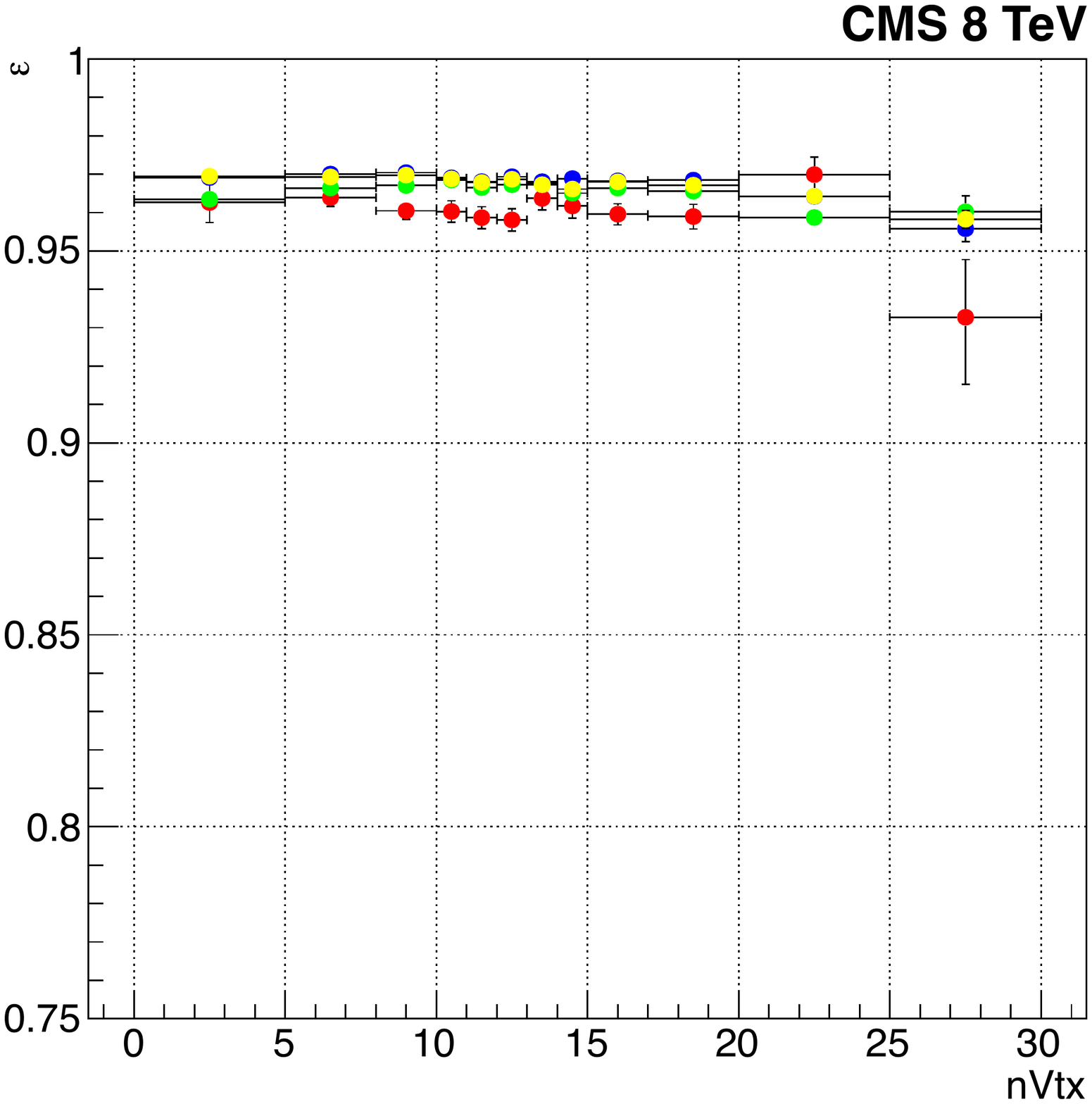}

   \caption{Efficiencies of the leading leg for the double-electron
     trigger described in the text as a function of the number of
     reconstructed vertices. The trigger uses identical selection for
     both legs, so the other leg just has a different
     threshold. Efficiencies are shown for different running periods
     (red May, green June, blue August, and yellow November of 2012)
     and separately for electron reconstructed in barrel (left) and
     endcap (right).}
     \label{fig:double_ele_nvtx}
\end{figure}
\paragraph{Double-electron trigger efficiency.}
\label{sec:hlt_egamma_doublee_eff}
Figures~\ref{fig:double_ele_pt} and~\ref{fig:double_ele_nvtx} show the
performance of the double-electron trigger. Efficiencies were measured
using a tag-and-probe technique similar to that described for the
photon path measurements and are computed with respect to a standard
offline selection.
The results are reported for various running periods; the different
results reflect the different pileup conditions.
Figure~\ref{fig:double_ele_nvtx} shows that the efficiency is only
loosely dependent on the pileup conditions.

\subsection{Muon triggers}
\label{sec:objid_muons}

\subsubsection{The L1 muon trigger performance}

The following sections report the performance of the L1 muon trigger
system described in Sec.~\ref{sec:l1muon}. Results concerning
efficiency, $\pt$ assignment resolution, rates, and timing are
presented.  At GT level, different GMT quality requirements are
required for single- and multi-muon algorithms. Therefore,
the performance for both the single- and multi-muon objects is
documented.

For most of the studies offline reconstructed muons are used as a reference to measure
the response of the L1 trigger. Muon identification requirements similar
to the ones used by CMS offline analysis are required. These are documented
in Ref.~\cite{Chatrchyan:2012xi}.

\paragraph{The L1 muon trigger efficiency}
The efficiency of the muon trigger was calculated by use of the tag-and-probe
method described in~\cite{Chatrchyan:2012xi}.
Events with two reconstructed muons having an invariant mass compatible with the one
of the \Z boson or of the \JPsi resonance were selected out of a sample of events
collected on the basis of single muon triggers.

\begin{figure}[tbph]
\centering
  \includegraphics[width=0.45\linewidth]{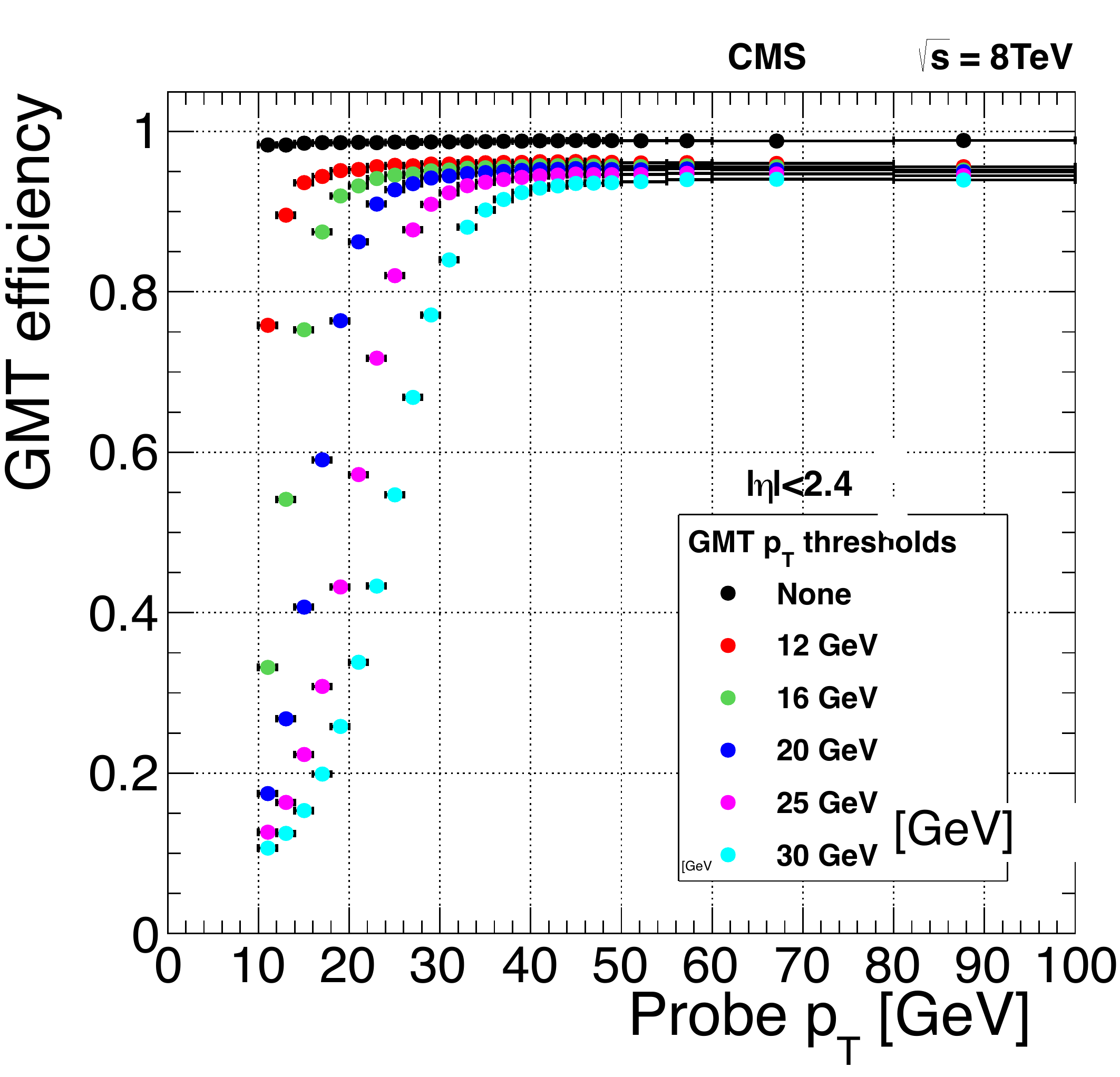}
  \includegraphics[width=0.45\linewidth]{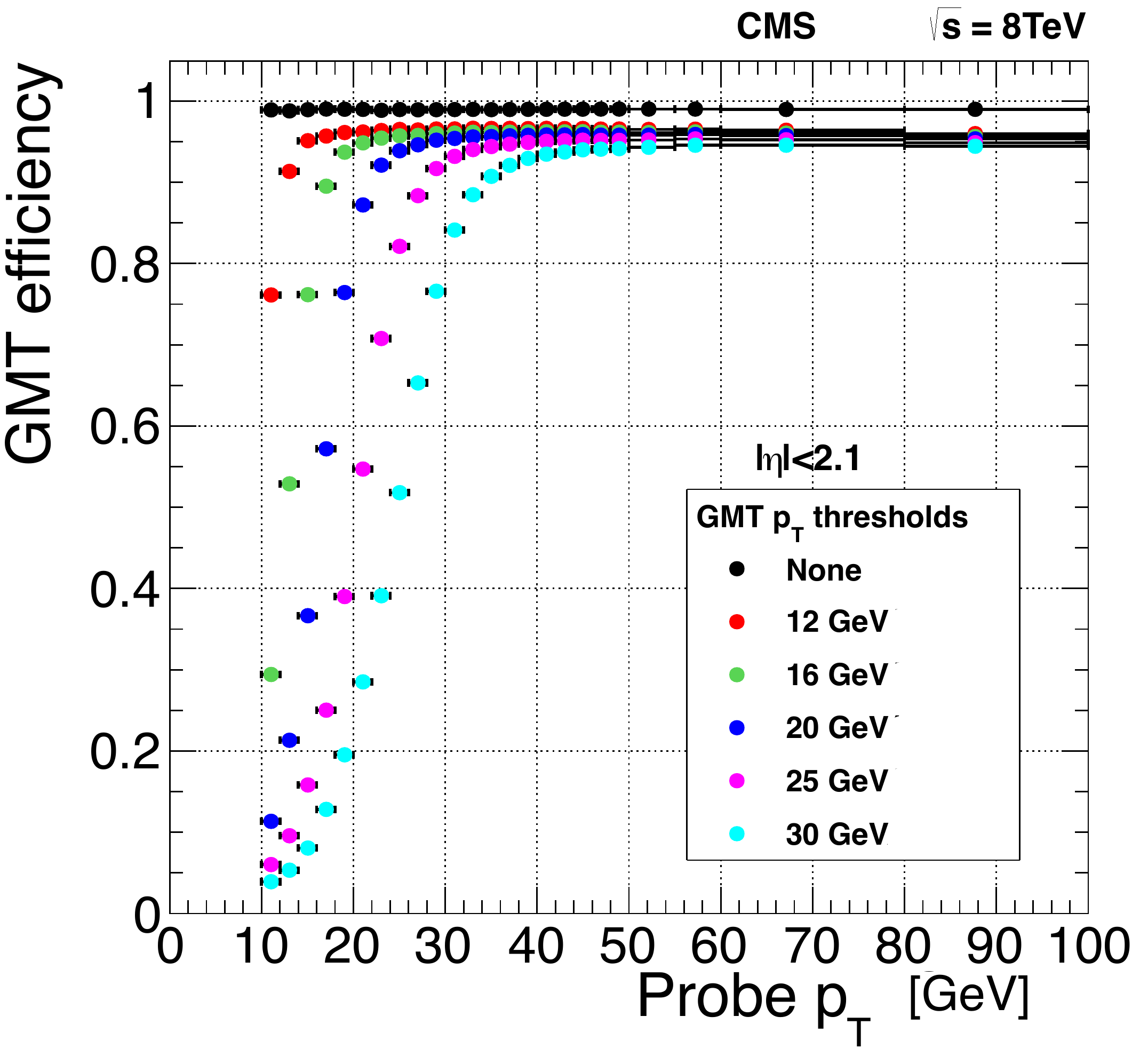}
  \caption{The efficiency of the single-muon trigger versus the
    reconstructed transverse momentum of the muon for different
    thresholds applied on the trigger candidate $\pt$ for the full
    pseudorapidity range $\abs{\eta} < 2.4$ (left), and limited to the
    range $\abs{\eta}< 2.1$ (right). The quality requirement used in
    the single-muon trigger algorithms (see text) was applied.
    Results are computed using the tag-and-probe method applied on a
    \Z boson enriched sample.}
  \label{figure:L1MuonSingleMuEfficiencyFromZ}
\end{figure}

\begin{figure}[tbp]
\centering
  \includegraphics[width=0.45\linewidth]{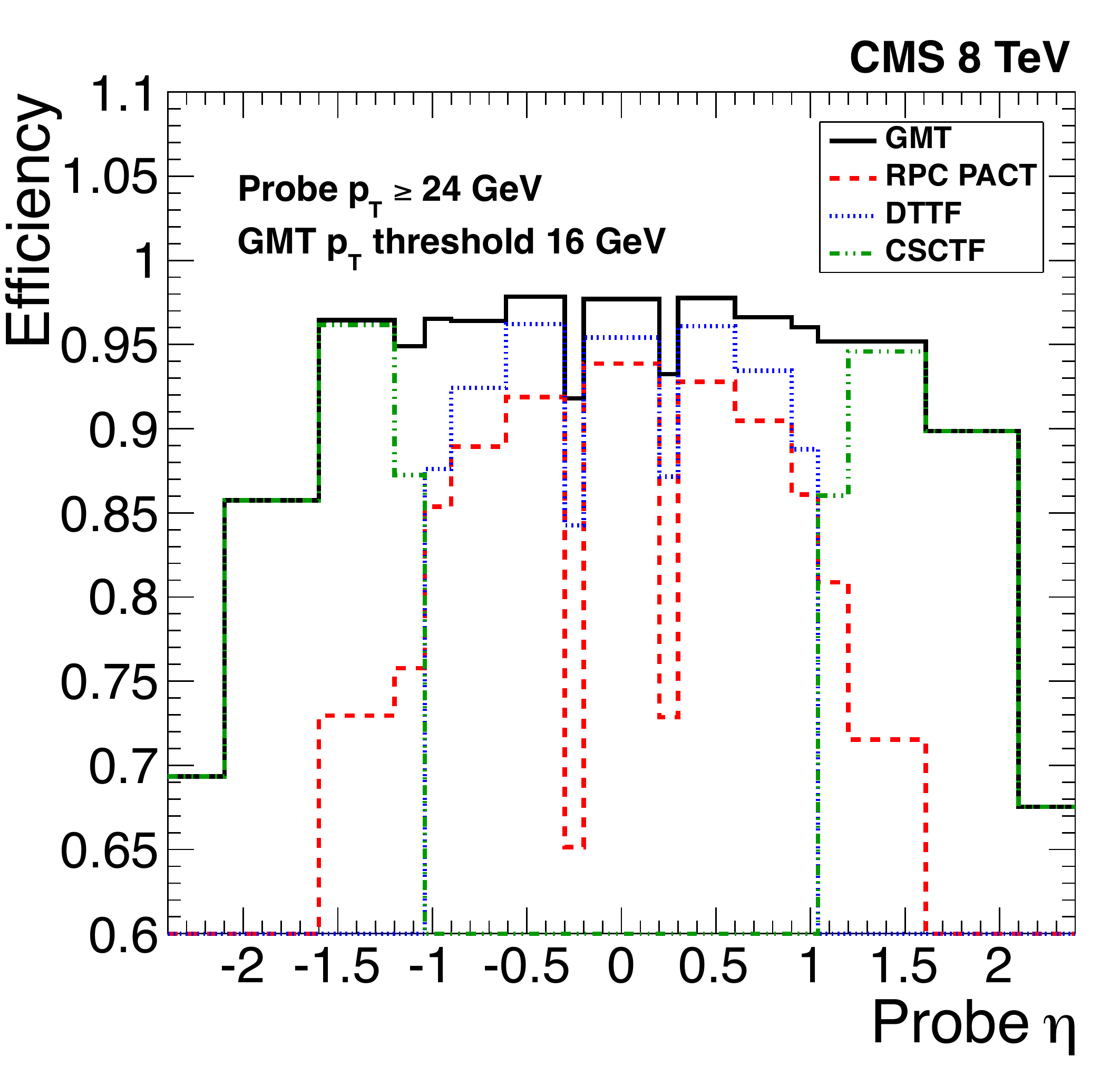}
  \caption{The efficiency of the single-muon trigger as a function of
  $\eta$ for the  threshold of 16\GeV (black) for muons with
  reconstructed $\pt > 24\GeV$.
  The contribution of the muon trigger subsystems to this efficiency is also
  presented: the red/green/blue points show the fraction of the GMT events based on
  the RPC/DTTF/CSCTF candidates, respectively. Results are computed using the
  tag-and-probe method applied to a \Z boson enriched sample.}
  \label{figure:L1MuonsEtaSubdetectors}
\end{figure}

\begin{figure}[tbp]
\centering
  \includegraphics[width=0.45\linewidth]{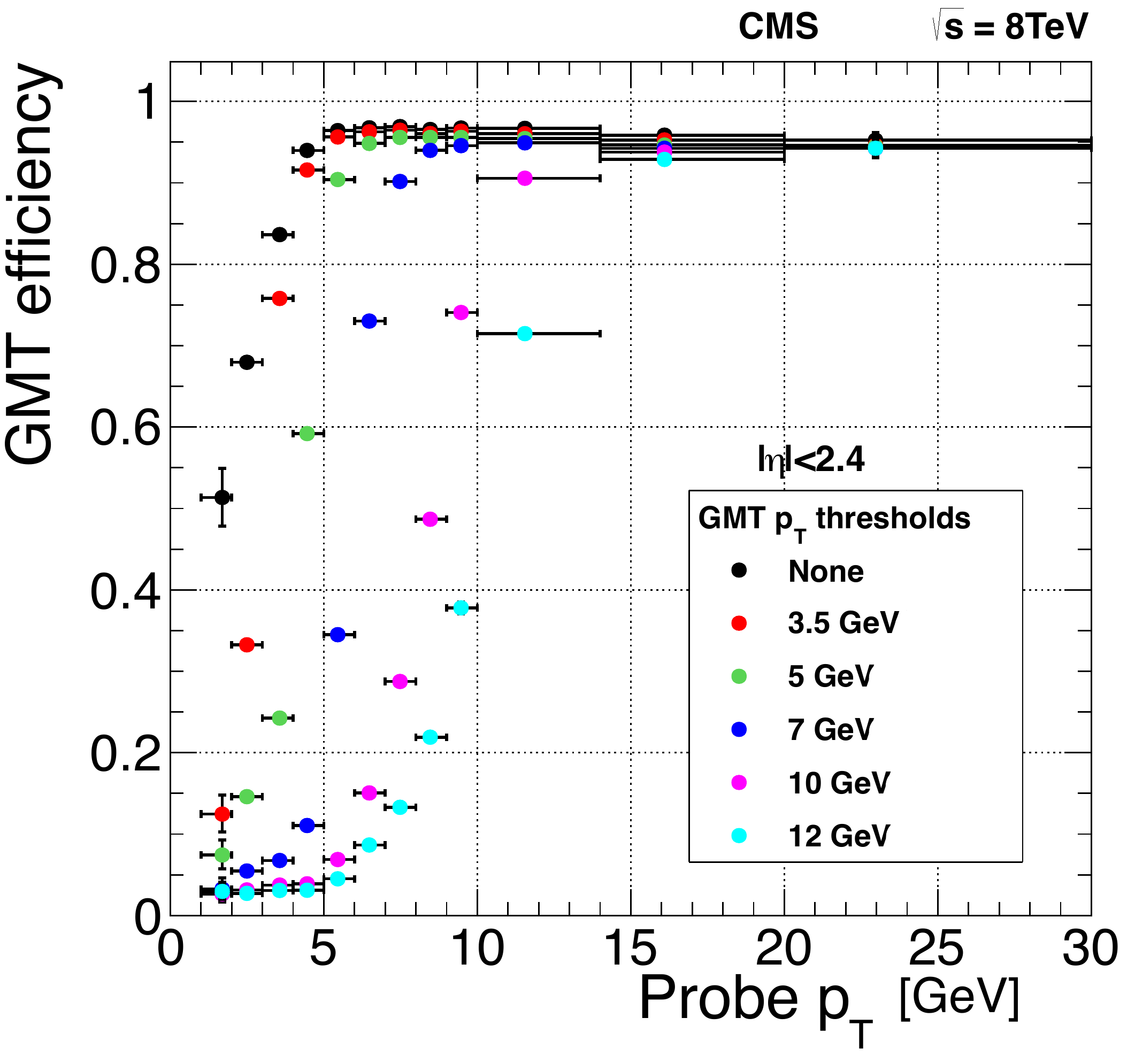}
  \caption{The efficiency of the double-muon trigger versus the reconstructed
   transverse momentum of the muon for different thresholds applied on the
   trigger candidate $\pt$. Results are computed using the tag-and-probe
   method applied to a \JPsi enriched sample.}
  \label{figure:L1MuonsJPsi0-2p4}
\end{figure}

Reconstructed tag muons were required to meet ``tight'' identification requirements
and to be matched to SingleMu HLT objects. This allowed the removal of trigger selection
biases.
Reconstructed probe muons had to be identified by either the ``tight'' or``loose''
identification criteria. The former selection matches the one used in most of the physics
analyses with single muons and was used to compute the efficiency for single L1 muon triggers
(Figs.~\ref{figure:L1MuonSingleMuEfficiencyFromZ} and
\ref{figure:L1MuonsEtaSubdetectors}), whereas the second is the muon identification
baseline for many analyses with multiple muons and it was used to compute efficiencies
for L1 double-muon triggers (Fig.~\ref{figure:L1MuonsJPsi0-2p4}).
The L1 muon trigger efficiency was calculated on the basis of probe muons geometrically
matched with L1 muon trigger candidates.

The L1 trigger candidates were matched to probes if the distance between the two was
found to be smaller than $\Delta\phi=0.15$ and $\Delta\eta=0.2$. If two L1 trigger
candidates were matched to a single probe the closest in $\phi$ was chosen.
Tag-and-probe muons were also required to be separated by
$\DR>0.5$
to exclude interference of the
two in the muon chambers.

The performance for different L1 $\pt$ requirements using a sample of dimuons satisfying a mass requirement around the \Z boson mass value is presented. Figure~\ref{figure:L1MuonSingleMuEfficiencyFromZ} shows the efficiency for
single L1 muon trigger GMT quality selections as a function of the reconstructed muon
$\pt$ for $\abs{\eta}<2.4$ and $\abs{\eta}<2.1$ acceptance regions, respectively.
Figure~\ref{figure:L1MuonsEtaSubdetectors} shows trigger efficiency
as a function of the reconstructed muon $\eta$. In this case a L1 $\pt >16\GeV$ is applied and probe muons are required to have a reconstructed
$\pt$ larger than 24\GeV.

The number of unbiased events recorded by CMS is not sufficient for a direct
and precise estimation of the overall L1 double-muon trigger efficiency.
In this case efficiency is
obtained using the tag-and-probe method on the \JPsi resonance.
Results imposing muon quality cuts as well as L1 $\pt$ requirements
from double-muon algorithms
are shown in Fig.~\ref{figure:L1MuonsJPsi0-2p4}.

The ability of CMS to trigger efficiently on dimuons at low \pt
allowed the CMS experiment to observe the rare $B^0_s\to\mu^+\mu^-$
decay at $4.3\sigma$ significance~\cite{Chatrchyan:2013bka}, where a
dimuon trigger with a \pt threshold of 4\GeV on each muon was applied
at the HLT. The decay was established definitively at $6.2\sigma$
significance with the combination of data from both the CMS and LHCb
experiments~\cite{bsmumu}.

\paragraph{The L1 muon trigger rates}
Muon trigger rate plots were obtained from the analysis of a dedicated data
stream, containing L1 trigger information alone, that was collected at high
rate on the basis of L1 decision only.
This stream provides unbiased information about the L1 trigger response,
which is ideal for L1 trigger rate studies.

For this analysis, events were selected on the basis of the loosest
possible L1 muon trigger algorithm. The latter implies no quality or
$\pt$ requirements on the L1 muon GMT candidates, therefore any further
selection (\eg, the $\pt$ threshold or quality requirements
corresponding to single- or double-muon triggers) was applied offline.

\begin{figure}[tbph]
\centering
  \includegraphics[height=200pt,trim={0 .3cm 0 0},clip]{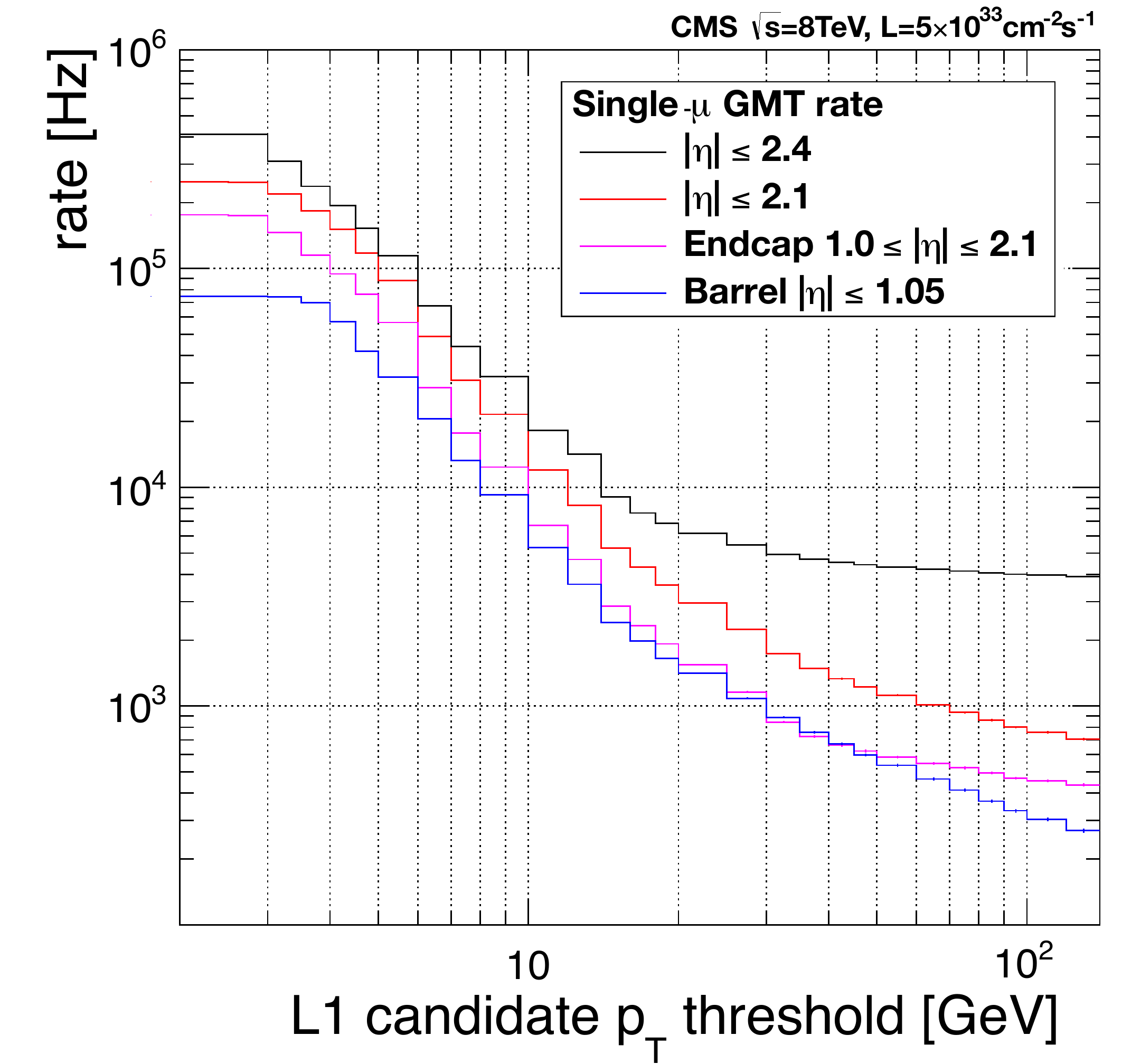}
  \includegraphics[height=201pt]{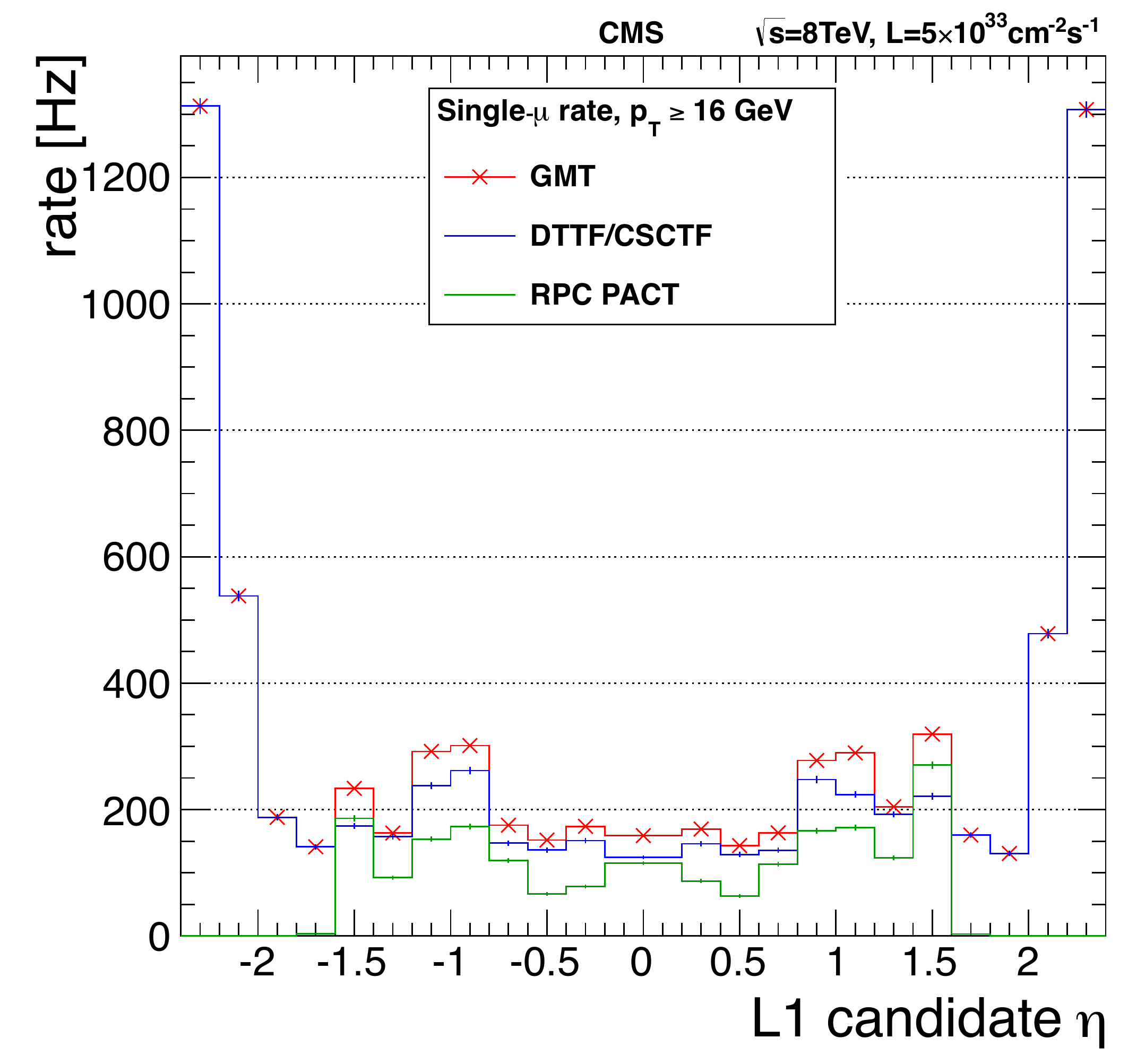}
  \caption{Left: rate of the single-muon trigger versus the transverse momentum
  threshold for the full pseudorapidity range $\abs{\eta}<2.4$ and for pseudorapidity
  limited to $\abs{\eta}<2.1$. Additionally the curves for pure endcap and barrel
  regions are presented.
  Right: the rate of the single-muon trigger GMT candidates as a function of
  $\eta$ for the $\pt$ threshold of 16\GeV (blue histogram). The contribution
  of the muon trigger subsystems to this rate is also presented: the green
  and blue histograms show how often the above GMT candidates built using
  RPC or DTTF/CSCTF candidates.
  On both plots the rates are rescaled to an instantaneous luminosity of
  $5 \times 10^{33}\percms$. The quality requirement used for single-muon trigger
  algorithms (see text) was applied.}
  \label{figure:L1MuonSingleMuRates}
\end{figure}

\begin{figure}[tbph]
\centering
  \includegraphics[scale=0.5]{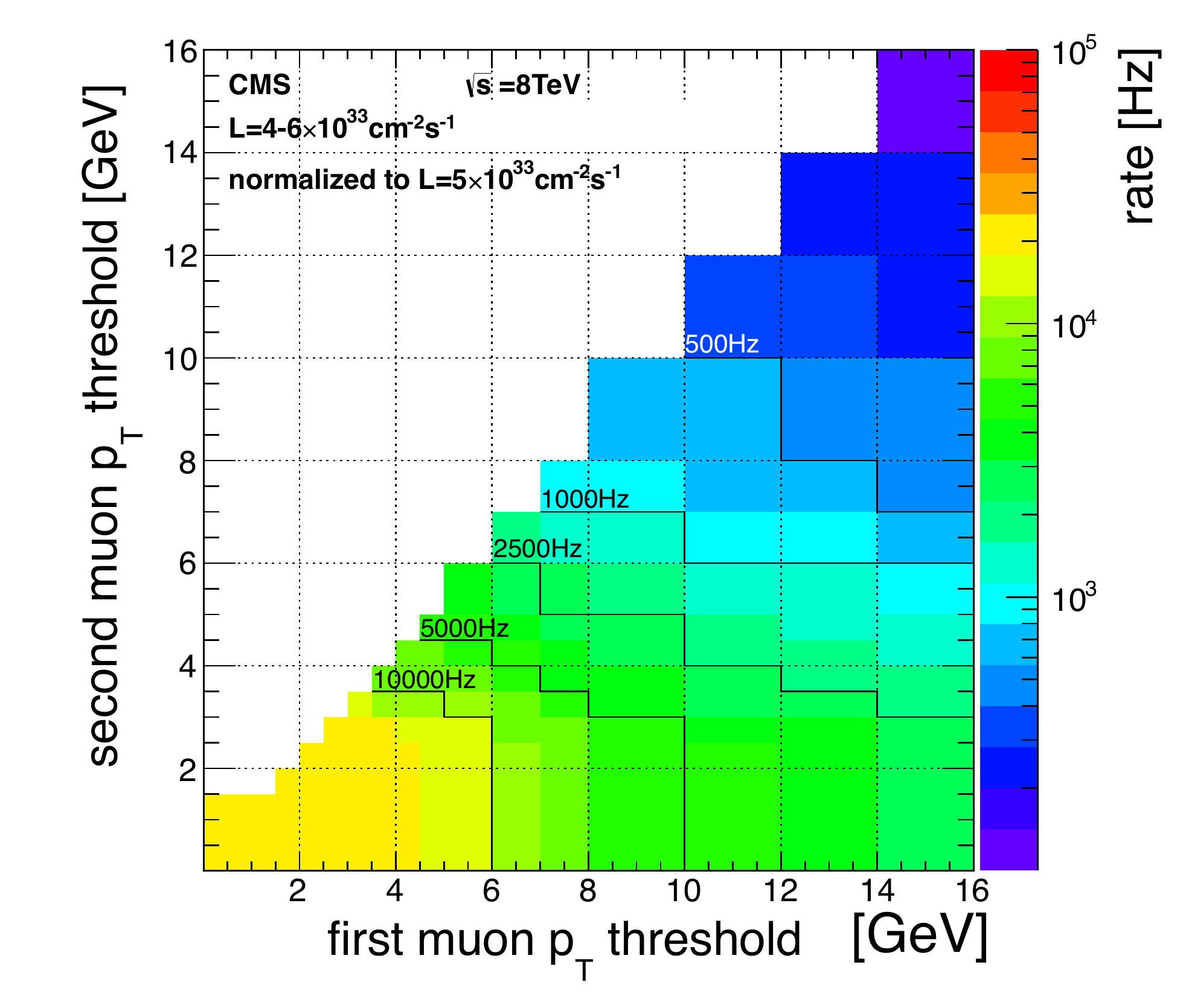}
  \caption{The rate of the double-muon trigger versus the threshold applied
  to the first and second muon. The rates are rescaled to the instantaneous
  luminosity $5 \times 10^{33}\percms$.}
  \label{figure:L1MuonDoubleMuRate}
\end{figure}

\begin{figure}[tbph]
\centering
  \includegraphics[width=0.45\linewidth]{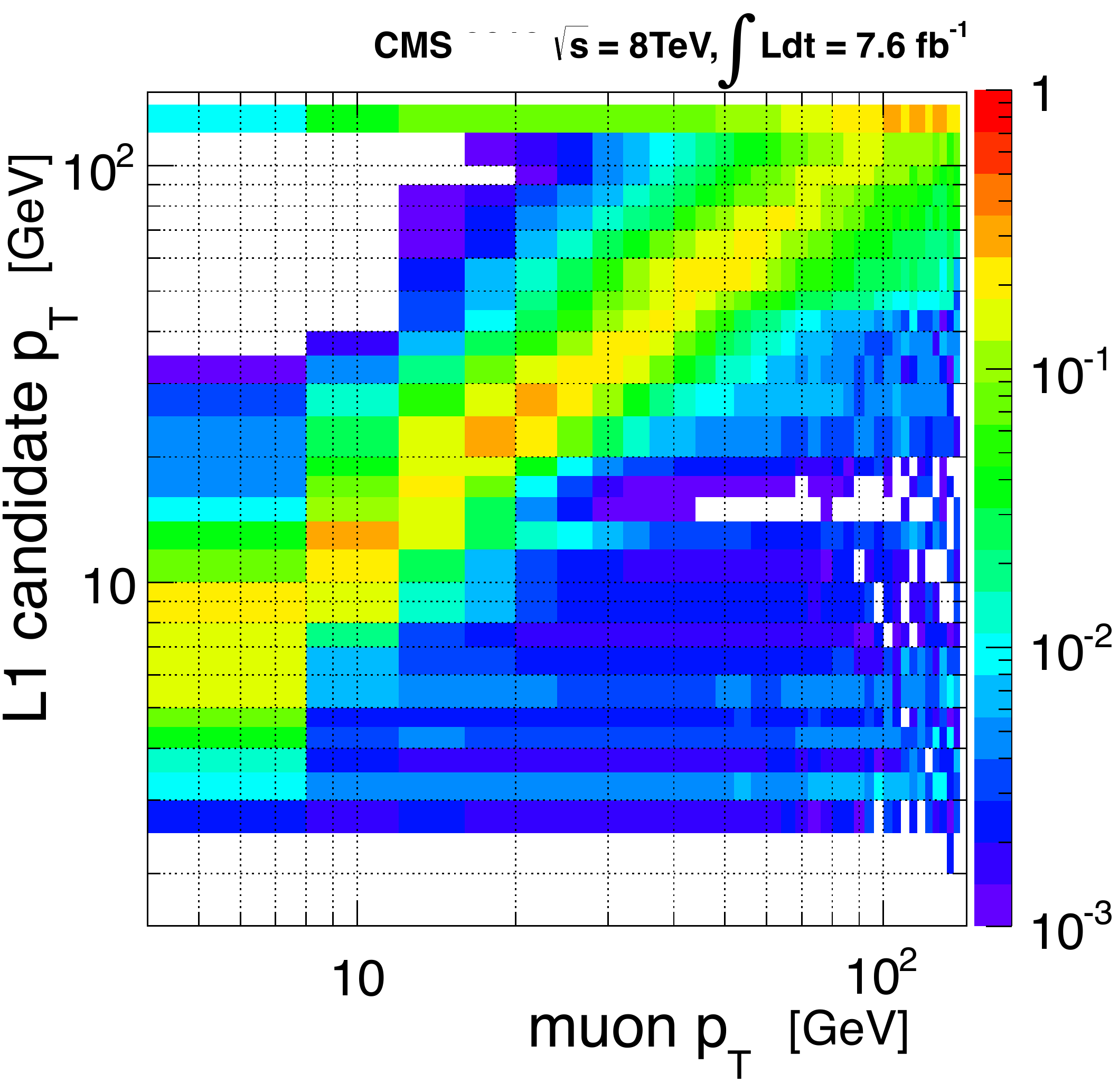}
  \caption{The distribution of the momentum of the L1 muon candidates versus the
   momentum of the corresponding reconstructed muon (``tight''identification
   criteria). Events with both \Z boson and \JPsi resonances contribute.
Offline
   muons in the full acceptance region ($\abs{\eta}<2.4$) are used.}
  \label{figure:L1MuonsPtL1vsPtRec}
\end{figure}

Results on the rates of single- and double-muon triggers are presented
in Figs.~\ref{figure:L1MuonSingleMuRates} and
\ref{figure:L1MuonDoubleMuRate}, respectively. The single-muon trigger
rate was calculated with a data recorded at instantaneous luminosities
up to $7.2 \times 10^{33}\percms$ and then rescaled to an
instantaneous luminosity of $5 \times 10^{33}\percms$. This
extrapolation was possible as the single-muon rate per instantaneous
luminosity (\ie, the trigger cross section)
is not a strong function of instantaneous luminosity.~\footnote{See
  Sec.~\ref{sec:muon_perf} and Fig.~\ref{fig:xsections}
  specifically. The variation is at the per-mille level.}
The left plot of
Fig.~\ref{figure:L1MuonSingleMuRates} shows a flattening of the slope
of the rate curve for single-muon triggers at high L1 $\pt$ threshold
values. The effect can be explained by studying the resolution of the
$\pt$ estimation of the L1 muon trigger computed with respect to
offline reconstructed ``tight muons''. The results of such a comparison
are presented in Fig.~\ref{figure:L1MuonsPtL1vsPtRec} and show that the muon trigger sometimes assigns very high \pt to muons with very low momentum. These candidates with overestimated transverse
momentum contribute significantly to the L1 muon trigger rate,
especially at high L1 $\pt$ thresholds.

In case of the double-muon triggers, the rate increases with
luminosity. The rates were calculated using data collected
with the luminosities in the range $4$--$6 \times 10^{33}\percms$ (for an
average luminosity of $4.9 \times 10^{33}\percms$), and rescaled to
a target instantaneous luminosity of $5 \times 10^{33}\percms$.
Errors from this small approximation are well within the fluctuations caused
by data acquisition deadtime variations ($\mathcal{O}(1\%)$).

\paragraph{The L1 muon trigger timing}

The muon trigger timing is a product of the timing performance of muon
trigger primitive generators and muon regional track-finders (DT, CSC,
RPC).  The GMT algorithm is executed independently for each BX. Thus
no further timing corrections on candidates generated by track finders
are performed at this stage.  Nevertheless, the GMT algorithm,
optimized for best momentum resolution and rejection of
misreconstructed double-muon candidates, can discard low quality
tracks, more prone to mistiming, affecting the overall L1 muon trigger
timing response as well.  This may result in the GMT accepting events
either in the earlier or later bunch crossing (pre- or post-firing).
Such errors do not currently cause incorrect L1 decisions since
triggers appearing in wrong LHC bunch crossings are suppressed at the
GT level by a BPTX veto.

\begin{figure}[tbph]
\centering
  \includegraphics[width=0.5\linewidth]{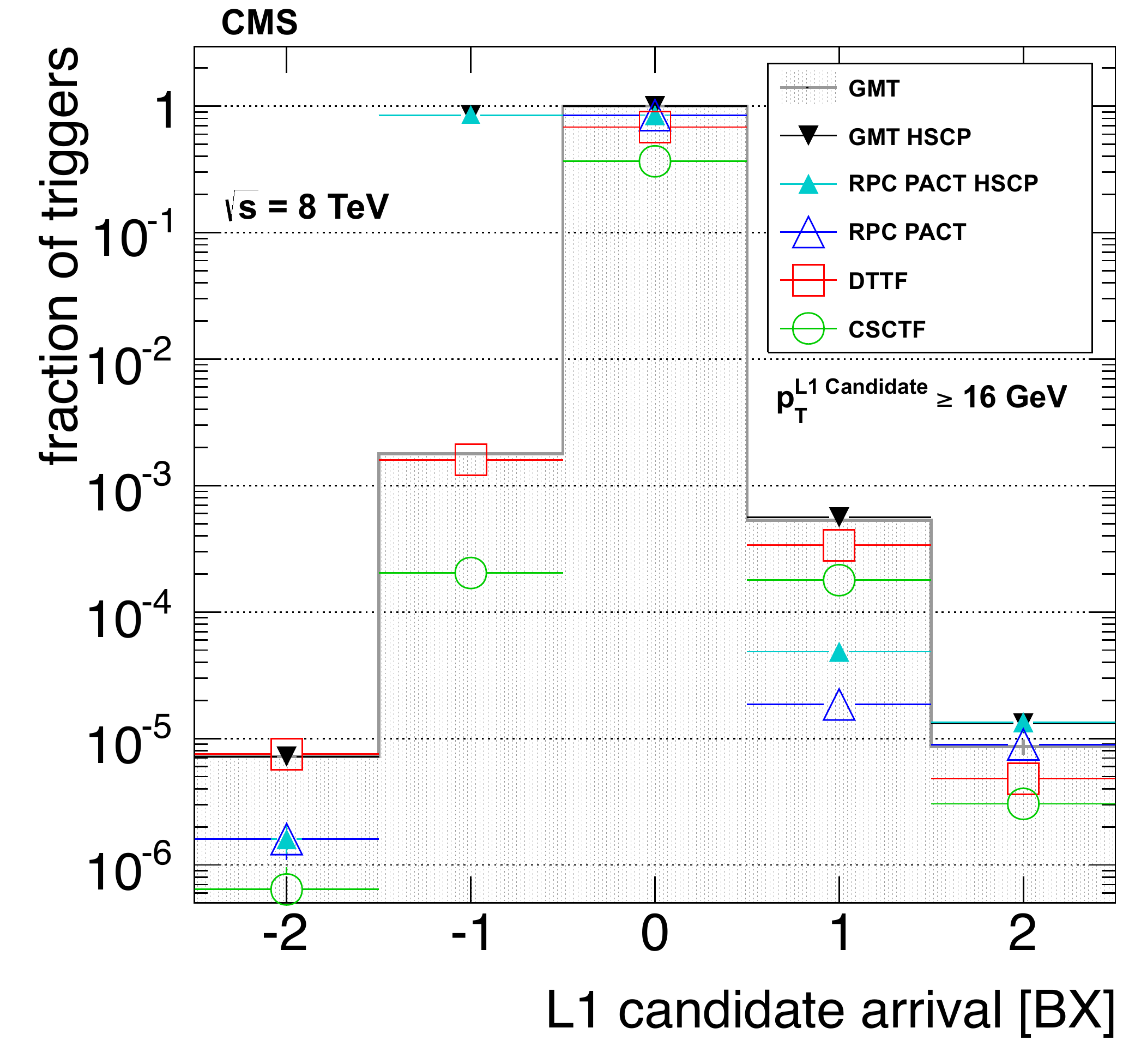}
  \caption{The overall timing distribution of L1 muon triggers. The
    distribution of GMT candidates is shown as a shaded histogram.
    The contributions from regional muon triggers (DT, CSC, RPC) are
    given. In addition, the GMT and RPC distributions for heavy stable
    charged particle trigger configurations are labeled separately.}
  \label{figure:L1MuonTimingL1BXDist}
\end{figure}

Ideally, the trigger timing logic assigns a muon trigger candidate to
the BX in which the actual muon was produced and reconstructed. In this
case the difference between trigger candidate LHC BX number and LHC BX
number of an event in which muon is reconstructed is 0, meaning that the
candidate arrives at (relative) $\mathrm{BX}=0$. To quantify the trigger timing
performance, the fraction of triggers appearing in a
given BX with respect to those with ideal timing is computed. This procedure
depends on an event selection used for muon reconstruction and the
underlying triggers. A typical distribution of L1 muon trigger
timing is shown in Fig.~\ref{figure:L1MuonTimingL1BXDist}.

The data of Fig.~\ref{figure:L1MuonTimingL1BXDist} come from a stream dedicated to the express monitoring of
muon reconstruction.  The event selection requires the
presence of a reconstructed muon with selections similar to the ones
used by the ``tight'' identification criteria.  To ensure a
correspondence between L1 muon trigger candidates and reconstructed
muons their position are requested to match within $\Delta R < 0.3$ of
each other.  No other reconstructed muons in the proximity of the one
matched with the trigger are allowed.  Since the most interesting
candidates are the ones that may affect the GT decision,
only events with $\pt$, $\abs{\eta}$, and quality requirements matching
the ones used for unprescaled L1 single-muon triggers in 2012 are
considered.

A L1 trigger is specifically implemented for heavy, stable
charged particles (HSCP) (Sec.~\ref{sec:HSCP}), which relies on
time extension of RPC signals in the RPC trigger logic. The typical
response to a prompt muon thus extends to two BXs.
It is therefore important that the presence of early or late signals
in the RPC and DT/CSC are not correlated.
Cases where both subsystem candidates respond in $\text{BX}=-1$
$(+1)$, therefore not providing a GMT candidate in $\text{BX}=0$, are
rare. It is therefore typical that GMT candidates to $\text{BX}=+1$
contribute as well to $\text{BX}=0$.

\begin{figure}[tbph]
\centering
  \includegraphics[width=0.5\linewidth]{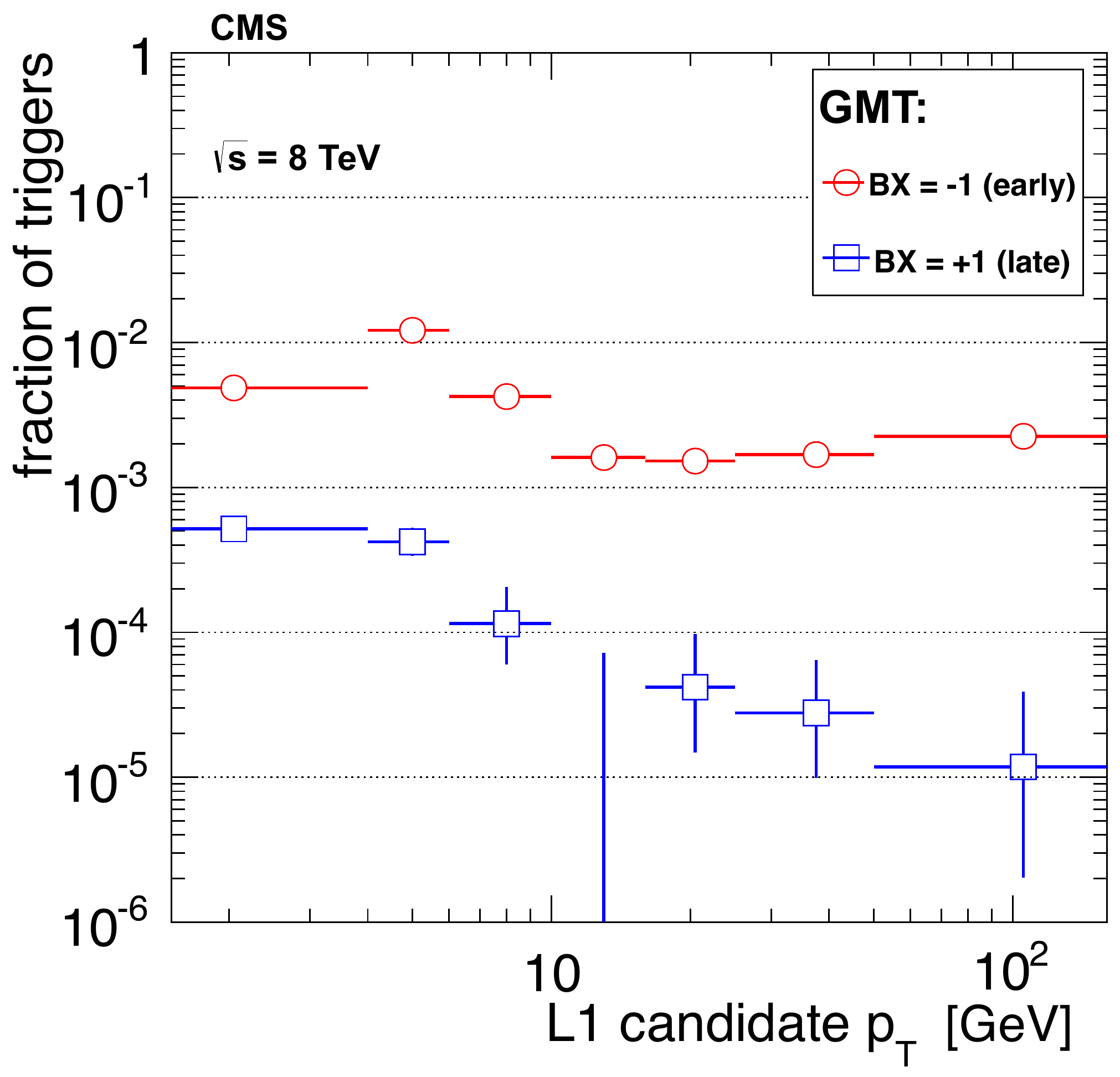}
  \caption{The fractions of GMT candidates in early and late bunch crossings as a function of L1 muon candidate transverse
           momentum.
           }
  \label{figure:L1MuonL1FirePrePost}
\end{figure}

A more detailed picture, derived from the same data set and illustrating
early and late GMT decisions, is given in
Fig.~\ref{figure:L1MuonL1FirePrePost}.
Here the fraction of events in $\text{BX}= +1$ and $-1$ is presented
as a function of GMT candidate transverse momentum. The low-$\pt$
behavior of the pre-firing curve follows the relative contribution of
DT and CSC candidates.  Event selection and trigger
rules affect trigger timing distributions.  In particular, a
trigger issued in an event suppresses possible triggers in the two
consecutive BXs. The above feature does not affect $\text{BX}=-1$
because triggers issued in non-colliding BXs are vetoed, but has an
impact on events triggered at $\text{BX}=+1$.  Therefore, in order to
extract the post-firing, only events with the first GMT candidate
appearing in $\text{BX}=+1$ are used.  To properly normalize
the plot, only events with a non-muonic additional trigger in the
event were selected.

\subsubsection{HLT muon identification}
\label{sec:muHLT}

The muon high-level triggers combine information from both the muon
and the tracker subdetectors to identify muon candidates and determine
their transverse momenta, $\pt$. The algorithm is composed
of two main steps: level-2 (L2), which
uses information from the muon system only, and level-3 (L3),
which combines measurements from both tracker and muon subdetectors.

\paragraph{Level-2.}
The reconstruction of a track in the muon spectrometer starts from an
initial state, called the \emph{seed}, built from patterns of DT and
CSC segments. The transverse momentum of the seed is parametrized as
$\pt = f(1/\Delta\phi)$, where $\Delta\phi$ is the azimuthal
angle between the two segments and $f$ is a first-order polynomial
function whose coefficients are determined using simulated CMS
data. Only seeds confirmed by the L1 decision are used.

Each seed is used to start the reconstruction of a track using
measurements (hits and segments) from all the muon detectors. Tracks
are built with the Kalman filter technique~\cite{Fruhwirth:1987fm}, a
recursive algorithm that performs pattern recognition and track
fitting. After all tracks were reconstructed, possible duplicates
of the same muon candidate are removed by checking that tracks do not share any
hits. The interaction point position is used to constrain the track parameters
to improve the transverse momentum resolution.

If one or more L2 muons are successfully reconstructed, their number
and parameters are used to filter the event. The main selection is
based on the L2 muon $\pt$. The number of muon chambers and
measurements used in the
track fit can also be used to suppress misreconstructed
muons.

\paragraph{Level-3.}

The L3 muon reconstruction exploits the excellent momentum and vertexing
resolution of the inner silicon tracker, and the larger lever arm of
the muon detector, to improve the momentum resolution at high
$\pt$ (greater than ${\approx}$200\GeV).
The L3 muon trigger algorithm consists of three main steps:
seeding of tracker reconstruction starting from L2 information,
track reconstruction in the tracker, and combined fit in the tracker
and muon systems.

Due to HLT timing and CPU constraints, the full tracker reconstruction
is not performed. Instead, tracks are seeded by L2 muon
candidates. Three different seeding algorithms are available:
\begin{enumerate}
\item
  the initial state (position, momentum) for track
  reconstruction is the L2-track state extrapolated to the outer
  surface of the tracker;
\item
  the initial state is the L2-track state extrapolated to
  the outer surface of the tracker, and updated with measurements
  found on the outermost layers of the silicon-strip detector; and
\item
  the initial state is defined by pairs of hits on adjacent
  layers of the silicon-pixel subdetector, in small rectangular
  $\eta$--$\phi$ regions around the L2 muon track.
\end{enumerate}
All these algorithms perform differently in different parts of
the detector. To optimize efficiency and timing, they are run in
reverse order of CPU time required: slower algorithms are only
called if the faster ones fail to reconstruct a L3 muon.
Starting from the initial seeds, tracks are reconstructed in the
silicon tracker using a Kalman filter. These tracks and the L2 muons
are propagated to a common surface (\eg, the innermost layer of
the muon system) and their compatibility is evaluated using several
criteria, such as their separation, directions, or relative
goodness-of-fit $\chi^2$. If a pair of compatible L2-tracker tracks is found, a final
refit of all the tracker and muon system measurements is performed.

If one or more L3 muons are successfully reconstructed, their number
and parameters are used to filter the event. The main selection is
based on the muon \pt. Other track parameters, such as $\chi^2$ and
impact parameter, can be used to suppress misreconstructed muons.

\paragraph{Isolation}

The isolation of L3 muons is evaluated combining information from the
silicon tracker, ECAL, and HCAL. Tracks are reconstructed in the
silicon tracker in a geometrical cone of size $\DR = 0.3$ around the
L3 muon. In the same cone, ECAL and HCAL deposits are summed. To
reduce the dependence of the isolation variable on the pileup of pp
collisions, the calorimeter deposits are corrected for the average
energy density in the event $\rho$~\cite{fastjetmanual}. A relative
isolation variable is defined as
\begin{linenomath*}
\begin{equation*}
I_\text{rel}  =  \frac{1}{\pt^{\mu}}
\biggl(\sum_i{p_{\mathrm{T,trk}}^{i}}  +
\max\Bigl[ 0, \textstyle{\sum_j}{E_{\mathrm{T,ECAL}}^{j}}  +
\textstyle{\sum_k}{E_{\mathrm{T,HCAL}}^{k}}  -
\pi (\DR)^2 \, \rho \Bigr] \biggr).
\end{equation*}
\end{linenomath*}
The standard selection is $I_\text{rel}<0.15$.

\paragraph{Double-muon triggers}

Double-muon triggers either require the presence of two L3 muons, as
described above, or one L3 muon and one
``tracker-muon''~\cite{Chatrchyan:2012xi}, \ie, a track in the silicon
tracker compatible with one or more segments in the muon detectors. The
latter class of triggers recovers possible inefficiencies of the L2
muon reconstruction (\eg, due to the muon detector
acceptance). Moreover, dropping the requirement of a fitted track in
the muon system allows reduction of the effective kinematic threshold,
making these triggers particularly suitable for quarkonia and B physics topologies.

The two legs of double-muon triggers are generally required to
originate from the same vertex to reduce the rate of misreconstructed dimuon
events. In specific quarkonia triggers, additional filtering is
applied to reduce the low-\pt background rate. This includes, for
example, mass requirements on the dimuon system and requirements on the
angle between the two muon candidates (Sec.~\ref{sec:bphbpag}.)

\paragraph{Performance of muon triggers.}
\label{sec:muon_perf}

This section describes the performance of the single- and double-
muon triggers during 2012 data taking at 8\TeV. The triggers are:
\begin{itemize}
\item a single-muon trigger seeded by a L1 requirement of $\pt>16$\GeV,
  and requiring a L2 track of $\pt > 16$\GeV and a L3 track of
  $\pt > 40$\GeV;
\item a single-muon trigger seeded by an L1 trigger of
  $\pt > 16$\GeV, and requiring a L2 track of $\pt > 16$\GeV and a
  L3 track of $\pt > 24$\GeV; the L3 track must also be isolated;
\item a double-muon trigger by a L1 trigger
  requiring two muon candidates of $\pt>10$ and 3.5\GeV,
  respectively; the L2 requirement is two tracks of $\pt > 10$ and
  3.5\GeV, and the L3 requirement is two tracks of $\pt > 17$ and
  8\GeV; the muons are required to originate from the same vertex;
  by imposing a maximum distance of 0.2~\cm between the points of
  closest approach of the two tracks to the beam line; and
\item a double-muon trigger seeded by a L1 trigger
  requiring two muon candidates of $\pt>10$ and 3.5\GeV,
  respectively; the L2 requires a track of $\pt > 10$\GeV, and the
  L3 a track of $\pt > 17$\GeV; in addition, a tracker muon of $\pt
  > 8$\GeV is required; the muons are required to come from the same
  vertex, by imposing a maximum distance of 0.2~\cm between the points
  of closest approach of the two tracks to the beam line.
\end{itemize}
Trigger efficiencies are measured with the tag-and-probe method,
using \Z bosons decaying to muon pairs. The tag must be identified as a
``tight muon''~\cite{Chatrchyan:2012xi} and triggered by the single-isolated-muon
path. The probe is selected either as a ``tight muon'' or a
``loose muon''~\cite{Chatrchyan:2012xi}, respectively,
for single- and double-muon efficiency studies. When measuring the efficiency of isolated
triggers, the probe is also required to be isolated.
The efficiency is obtained by fitting simultaneously the \Z resonance
mass for probes passing and failing the trigger in question.

\begin{figure}[tbh]
  \centering
    \includegraphics[width=0.477\textwidth]{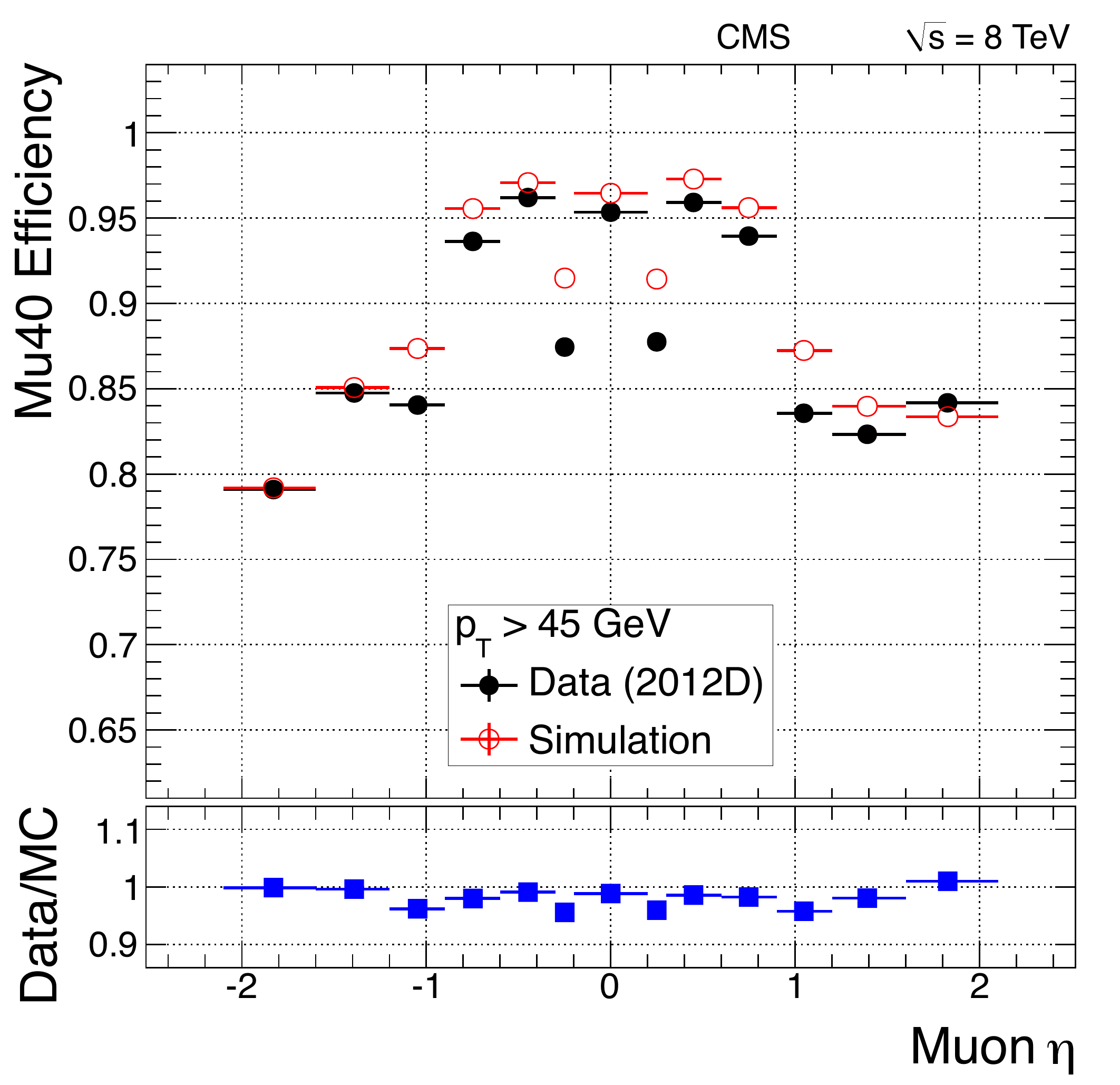}
    \includegraphics[width=0.477\textwidth]{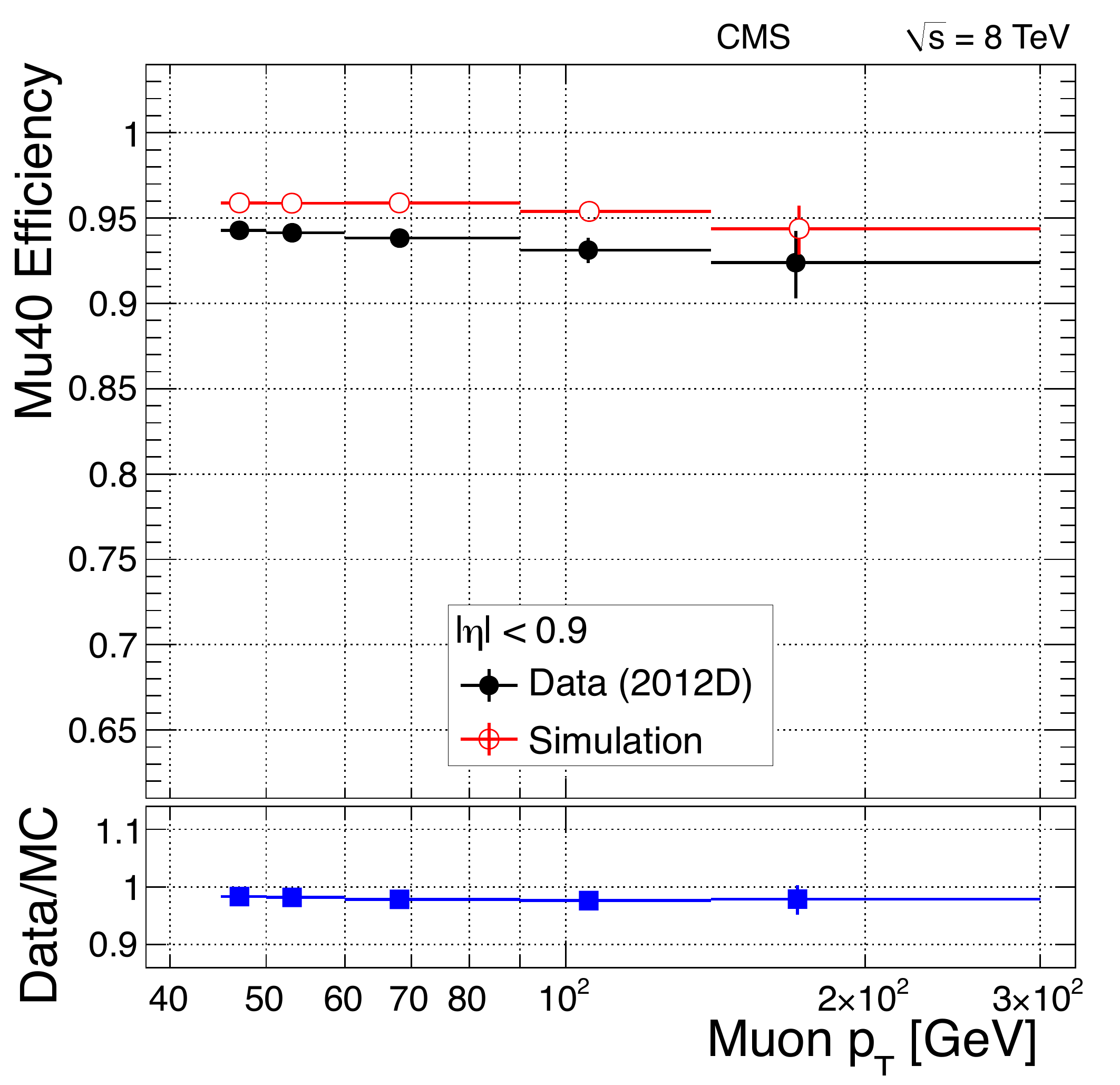}\\
    \includegraphics[width=0.477\textwidth]{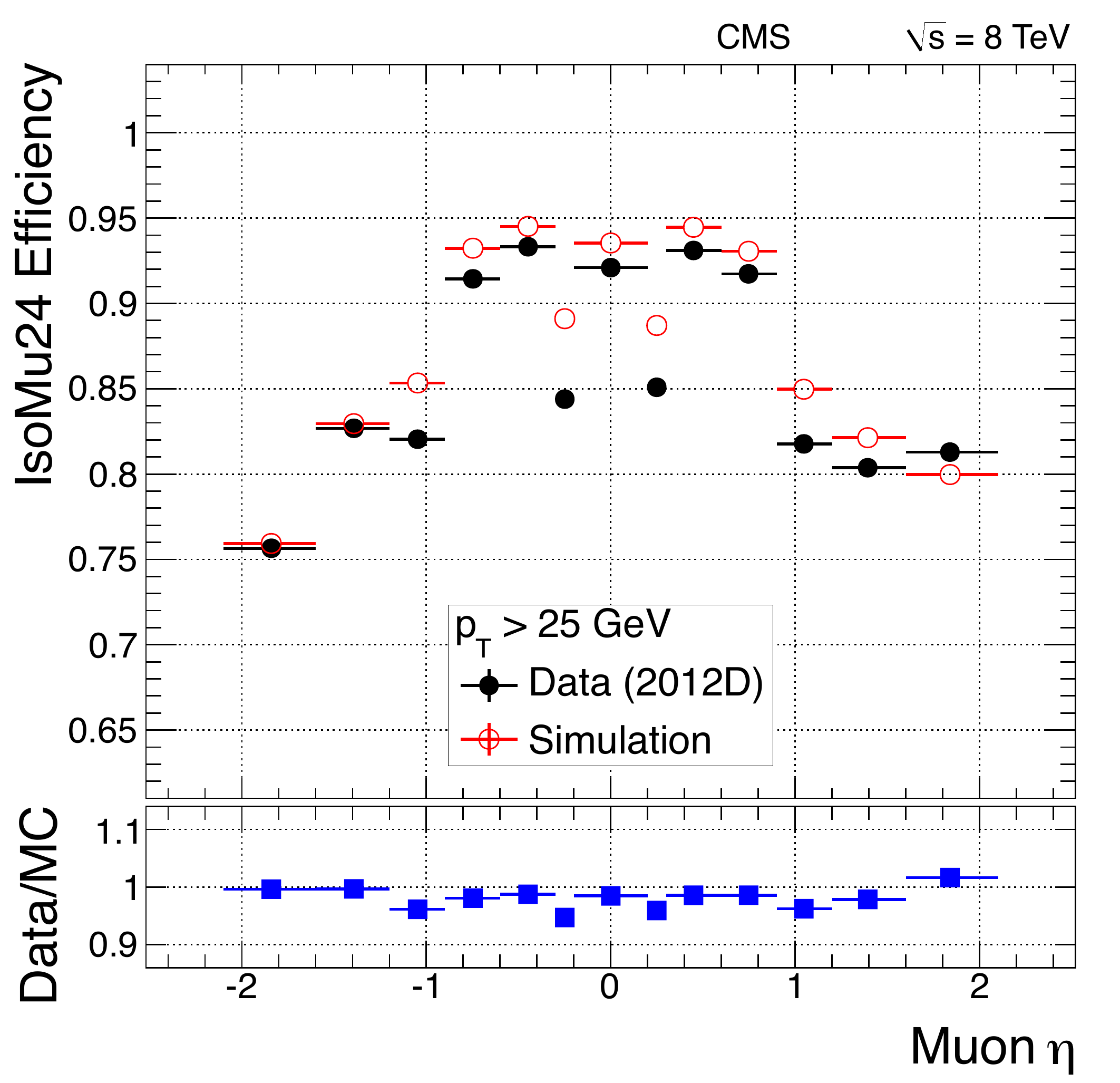}
    \includegraphics[width=0.477\textwidth]{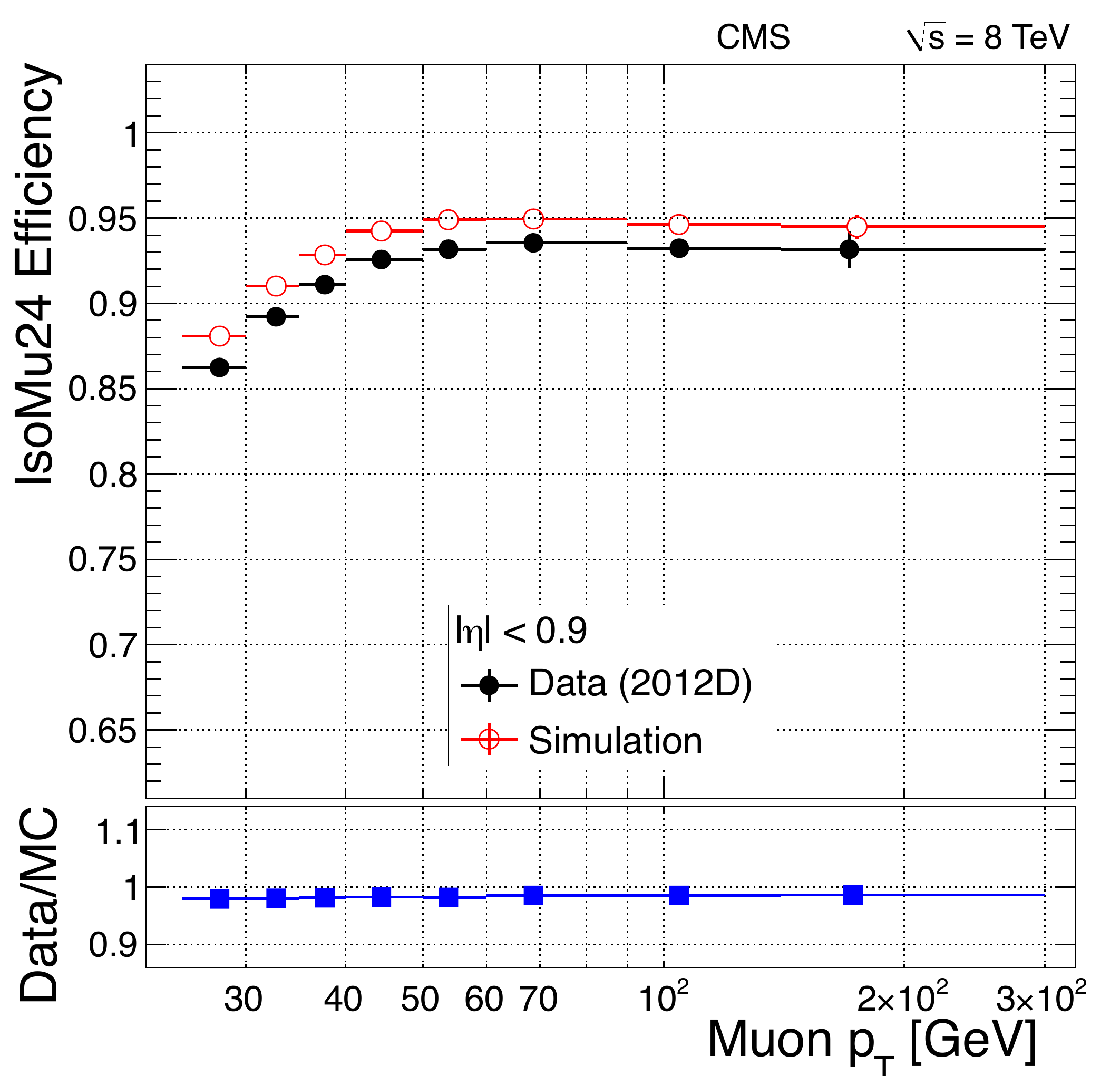}
  \caption{Efficiency of single-muon triggers without isolation
    (top) and with isolation (bottom) in 2012
    data collected at 8\TeV, as functions of $\eta$ (left) and
    $\pt$, for $\abs{\eta} < 0.9$ (right).}
  \label{fig:singlemuons}
\end{figure}

Figure~\ref{fig:singlemuons} shows the efficiencies of single-muon triggers with and without
isolation, as functions of $\eta$  and $\pt$ (for $\abs{\eta} < 0.9$), in 2012 data and in
simulation. The ratio between data and simulation is also shown. An
agreement of the level of 1--2\% is observed.

\begin{figure}[tbh]
  \centering
    \includegraphics[width=0.48\textwidth]{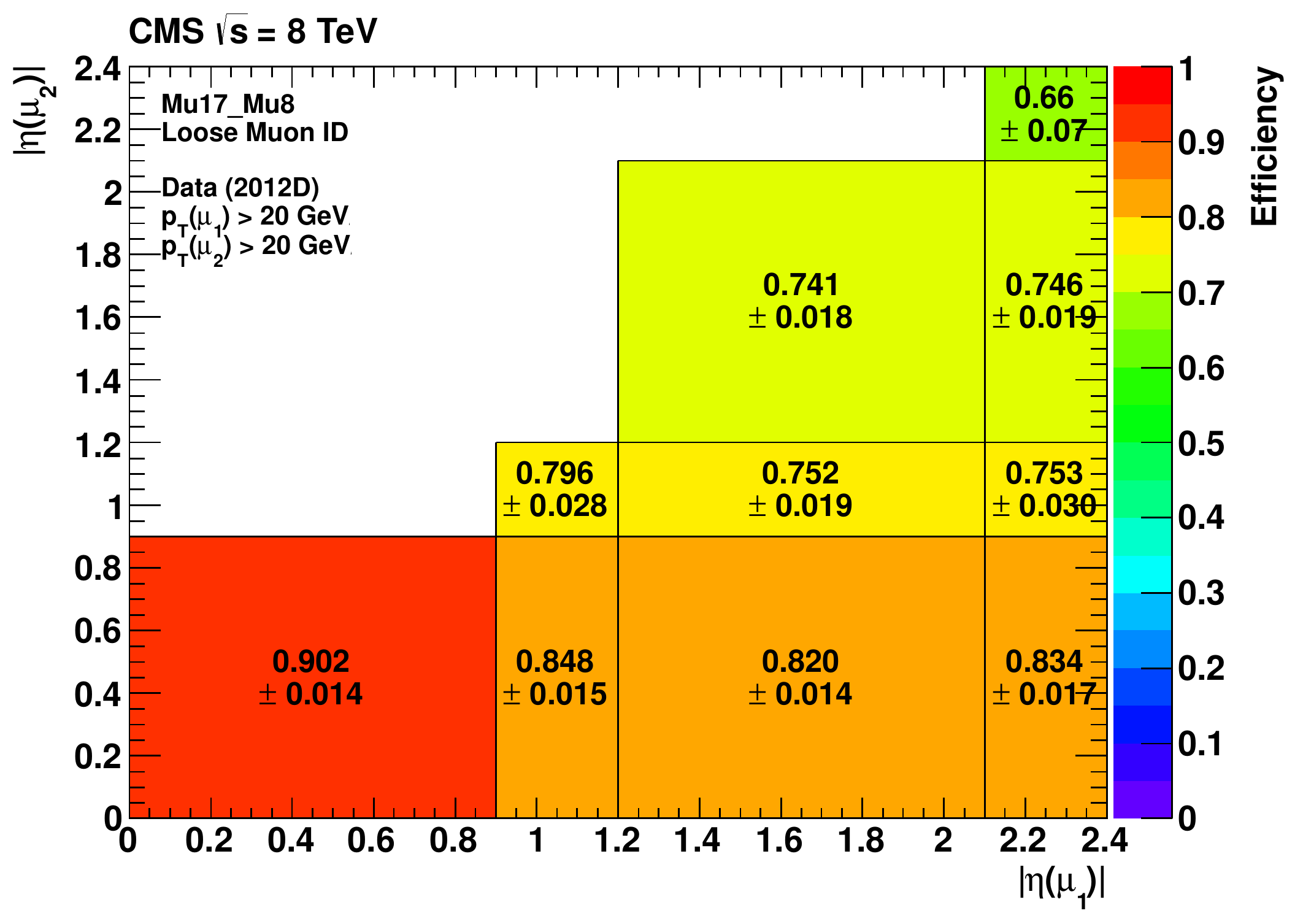}
    \includegraphics[width=0.48\textwidth]{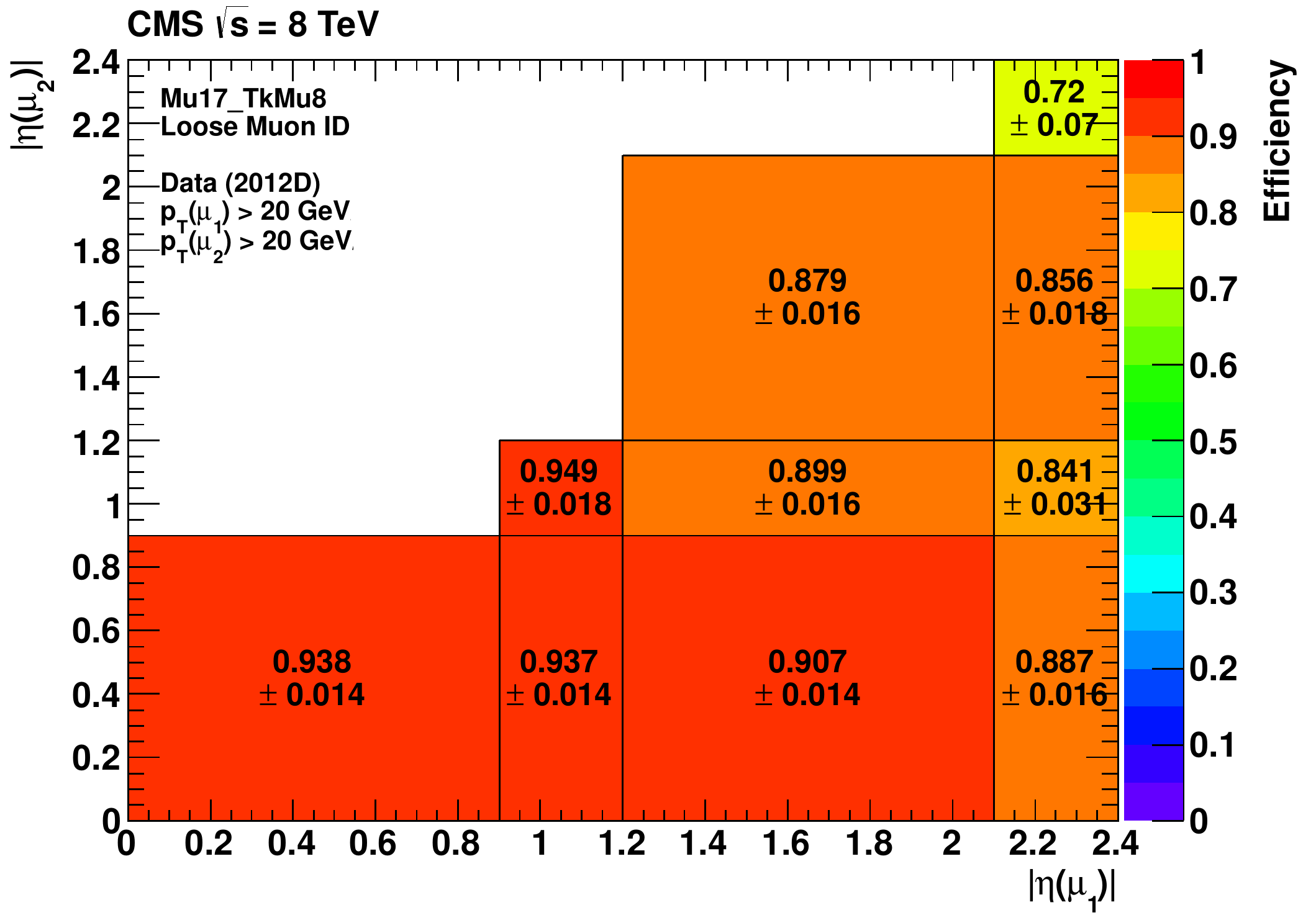}
  \caption{Efficiencies of double-muon triggers without
    (left) and with (right) the tracker muon requirement in 2012
    data collected
at 8\TeV as functions of the pseudorapidities $\abs{\eta}$ of the two muons,
for loose muons with $\pt > 20$\GeV.}
  \label{fig:doublemuons}
\end{figure}
Figure~\ref{fig:doublemuons} shows the efficiencies
for the double-muon triggers with and without the tracker muon requirement
 for tight muons of $\pt > 20$\GeV, as functions of $\eta$ of the two
muons. The total efficiency includes contributions from the efficiency
of each muon leg and from the dimuon vertex constraint.

\begin{figure}[tbh]
  \centering
    \includegraphics[width=0.6\textwidth]{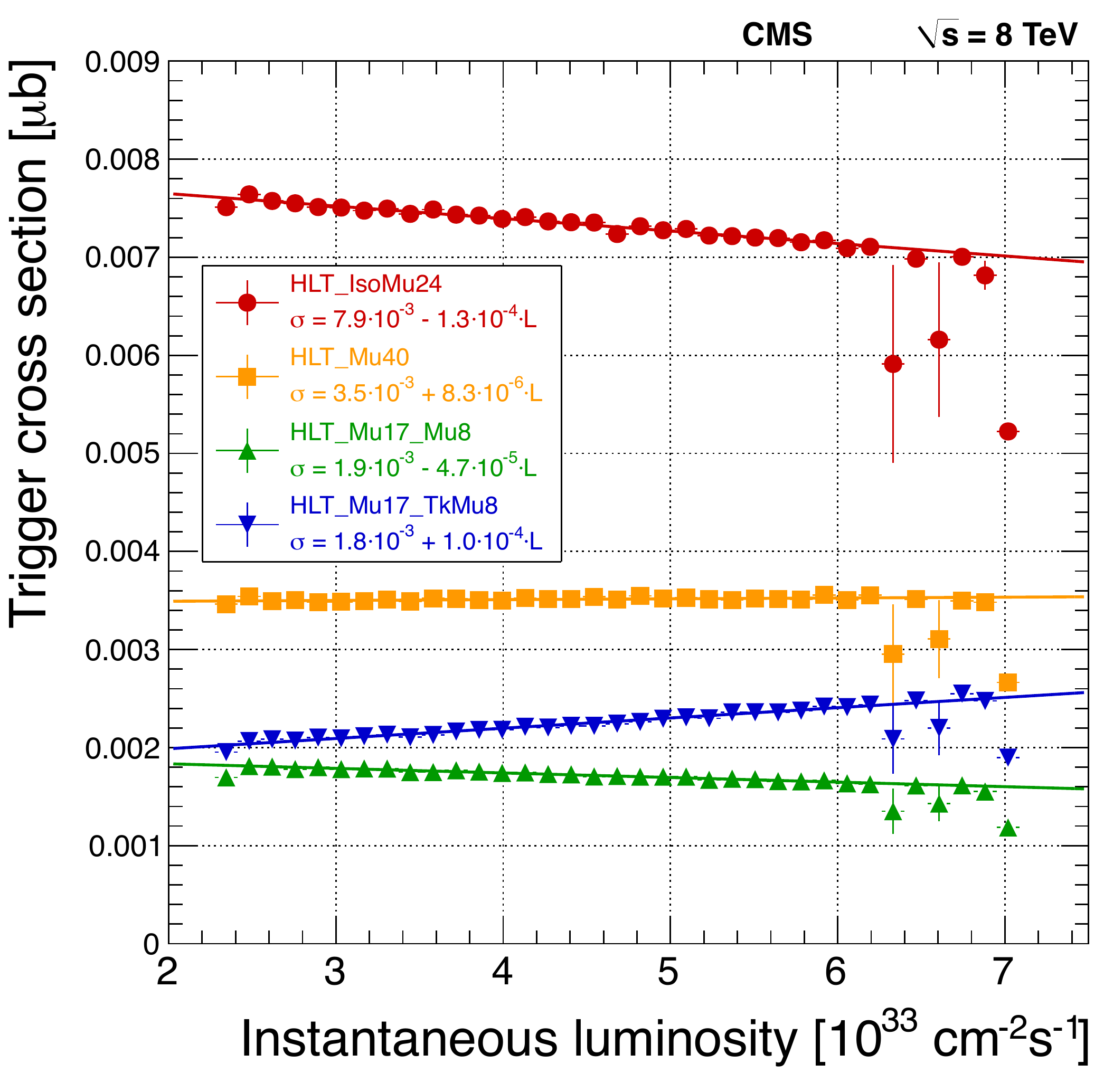}
  \caption{Cross sections of the  four main single- and double-muon
    triggers used in 2012 data taking, described in the text,
    as a function of the LHC instantaneous luminosity. Mild pileup
    dependencies are visible. }
  \label{fig:xsections}
\end{figure}
Figure~\ref{fig:xsections} shows the trigger cross sections of the four main
muon triggers in 2012 data taking, as functions of the LHC
instantaneous luminosity. As is shown in the Figure, during the 2012
run, a mild pileup-dependent inefficiency was observed for paths using
L3 reconstruction. This effect caused a drop in the cross section of
the isolated muon trigger at high
luminosity. Figure~\ref{fig:xsections} shows that this effect is not
visible in nonisolated triggers (such as the single-muon path with a
$\pt>40$\GeV requirement) as in those cases it is masked by a slight
luminosity-dependent cross section increase.

\subsection{Jets and global energy sums}
\label{sec:JetMET}

Triggers based on jet and missing transverse energy (\MET)
triggers play an important role for search for new physics.
Single-jet triggers are primarily
designed to study quantum chromodynamics  (QCD), but can also be used for many analyses, such
as searches for new physics using initial state radiation (ISR)
jets.
The dijet triggers are designed primarily for jet energy scale
studies. The \MET triggers are designed to search for new physics with
invisible particles, such as neutralinos  in supersymmetric models.

\subsubsection{The L1 jet trigger}
\label{sec:l1jet}

\begin{figure}[tbph]
  \centering
  \includegraphics{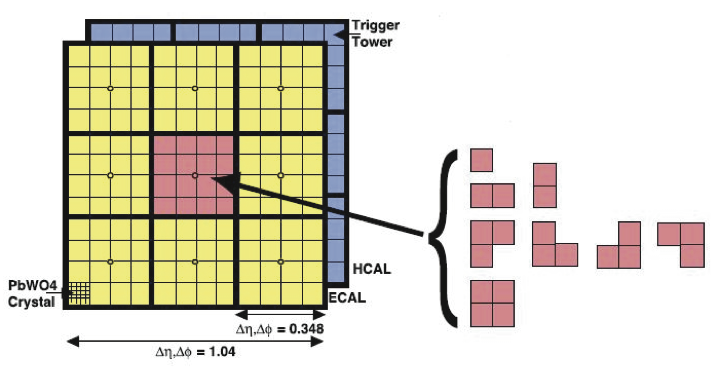}
  \caption{Illustration of the available tower granularity for the L1
    jet finding algorithm in the central region, $\abs{\eta} < 3$
    (left). The jet
    trigger uses a $3{\times}3$ calorimeter region sliding window
    technique which spans the full $(\eta, \phi)$ coverage of the
    calorimeter. The active tower patterns allowed for L1 $\tau$ jet candidates are shown on the right.}
  \label{fig:l1jetalgo}
\end{figure}

The L1 jet trigger uses transverse energy sums computed using both HCAL and ECAL in the central region ($\abs{\eta} < 3.0$) or HF in the forward region ($\abs{\eta} > 3.0$). Each central region is composed of a $4 \times 4$ matrix of trigger towers (Fig.~\ref{fig:l1jetalgo}), each spanning a region of $\Delta\eta \times \Delta\phi = 0.087 \times 0.087$ up to $\abs{\eta}{\approx}2.0$; for higher rapidities the $\Delta\phi$ granularity is preserved, while the  $\Delta\eta$ granularity becomes more coarse. In the forward region, each region consists of 4 or 6 HF trigger towers and has the same $\Delta\phi$ granularity of 0.384 as in the central region, with the $\Delta\eta$ granularity of 0.5. The jet trigger uses a ``sliding window" technique~\cite{Trig-TDR} based on a $3 \times 3$ regions (\ie, 144 trigger towers in the central region and up to 54 trigger towers in the forward region), spanning the full $(\eta,\phi)$ coverage of the CMS calorimeter. The L1 jet candidate is found if the energy deposits in the $3 \times 3$ window meet the following conditions: the central region of the $3 \times 3$ matrix must have the \ET higher than any of the eight neighbors, and this \ET must exceed a specific threshold (used to suppress the calorimeter noise). The L1 jets are characterized by the transverse energy \ET equal to the sum of transverse energies in the $3 \times 3$ regions of the sliding window centered on the jet. The L1 jet is labeled by the $(\eta,\phi)$ of its central region.

Jets with $\abs{\eta}>3.0$ are classified as forward jets, whereas those
with $\abs{\eta}<3.0$ are classified as central or $\tau$ jets, depending on the OR of the nine $\tau$ veto bits
associated with the 9 regions in the $3{\times}3$ window. To improve the detection efficiency for genuine L1 $\tau$ jets, a geometrical tower pattern is utilized for L1 $\tau$ jet candidates (Fig.~\ref{fig:l1jetalgo}).

The four highest energy central, forward, or central $\tau$ jets in the calorimeter are selected. After jets are found,
LUTs are used to apply a programmable $\eta$-dependent jet energy
scale correction.

The performance of the L1 jets is evaluated with respect to offline
jets, which are formed from the standard CaloJet reconstruction, as
well as PF jet reconstruction.
Jets are reconstructed using the anti-\kt algorithm and calibrated
for the nonlinearity of the calorimeter response and pileup effects
using a combination of studies based on simulation and collision data,
as detailed in Ref.~\cite{Chatrchyan:2012xx}.
A moderate level of noise rejection is applied to the offline jets by
selecting jets passing ``loose''~\cite{Chatrchyan:2012xx} identification criteria.

\paragraph{L1 jet trigger efficiency}

The L1 jet trigger efficiency was measured with a data sample from
the single-muon data set requiring an isolated muon with $\pt>24$\GeV
(HLT\_IsoMu24). Events from the muon paths are unbiased with respect
to the jet trigger paths.

The L1 jet efficiency is calculated relative to the offline
reconstructed jets. The efficiency is defined as the fraction of
leading offline jets that were matched to an L1 central,
forward, or central, $\tau$ jet above a certain trigger threshold, divided by the
number of offline (leading) jets that were matched to an L1
central, forward, or central $\tau$ jet above any threshold. This quantity is
then plotted as a function of the offline jet \pt, $\eta$, and
$\phi$. The efficiency is determined by matching the L1 and
reconstructed offline jets spatially in $\eta$-$\phi$ space. This is
done by calculating the minimum separation, \DR, between the
highest-\ET reconstructed jet (with \pt$>10$\GeV and $\abs{\eta}<3$)
and any L1 jet above a certain \et~threshold, and requiring it to
be less than 0.5. Should there be more than one jet satisfying this selection,
the one closest (in $\Delta R$) is taken as the matched jet.

\begin{figure}[tbp]
\centering
 \includegraphics[width=0.48\textwidth]{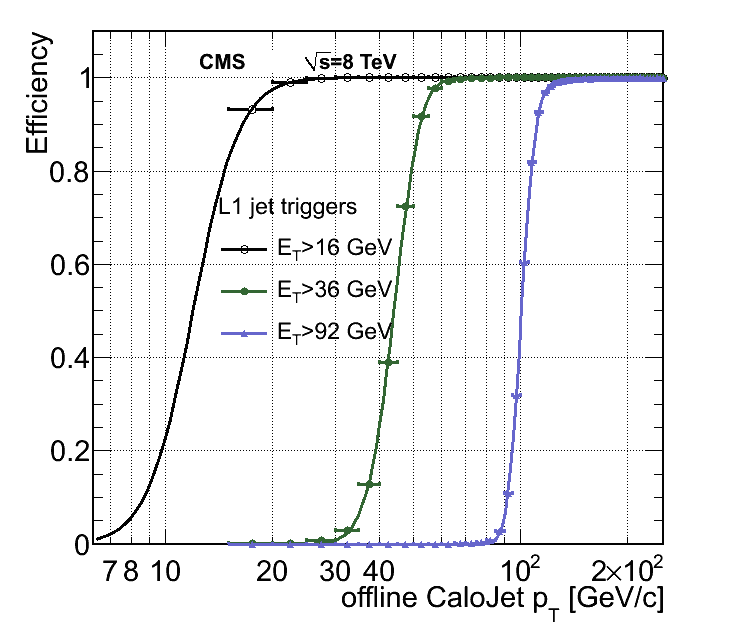}
 \includegraphics[width=0.48\textwidth]{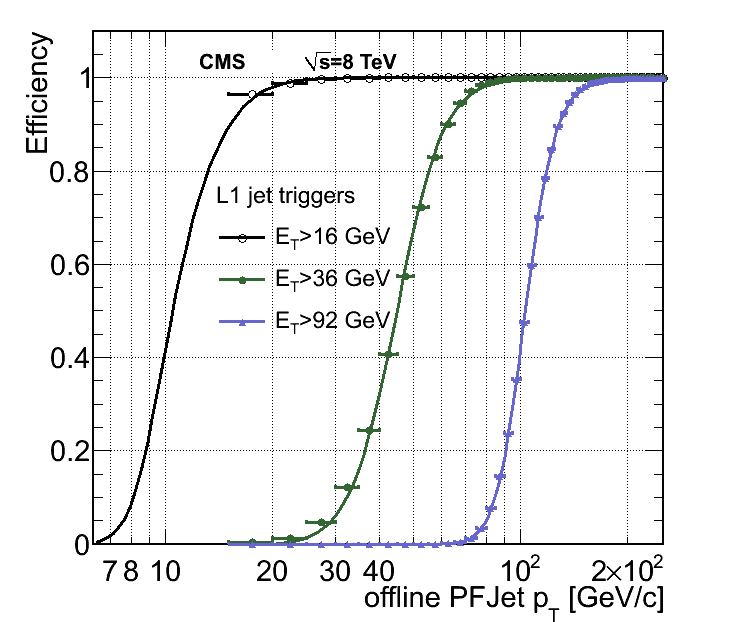}
 \caption[l1jets_eff]{Left: The L1 jet trigger efficiency as a
   function of the offline CaloJet transverse momentum. Right: The L1
   jet trigger efficiencies as a function of the PF jet transverse
   momentum. In both cases, three L1 thresholds ($\ET>16, 36, 92$\GeV)
   are shown.}
\label{fig:l1jets_eff}
\end{figure}

We evaluated the efficiency turn-on curves for various L1 jet
thresholds ($\ET>16$, $36$ and $92\GeV$) as a function of the
offline jet \pt. The efficiency is calculated with respect to offline
PF and CaloJet transverse energies (Fig.~\ref{fig:l1jets_eff}). Each
curve is fitted with a function that is the cumulative distribution
function of an exponentially modified Gaussian (EMG) distribution. In
this functional form, a parameter, $\mu$, determines the point of 50\%
efficiency and $\sigma$ represents the resolution.

\paragraph{Pileup dependence}
\label{pileup_dependence}

To evaluate the effect on the performance of the L1 triggers in
different pileup scenarios, the L1 jet efficiency is also benchmarked
as a function of pileup. The measure of the pileup per event is defined by the number
of `good' reconstructed primary vertices in the event, with each
vertex satisfying the following requirements

\begin{itemize}
	\item $N_\text{dof} > 4$;
	\item vertex position along the beam direction of $\vert z_\text{vtx} \vert < 24$\unit{cm};
	\item vertex position perpendicular to the beam of $\rho < 2$\unit{cm}.

\end{itemize}

\begin{figure}[tbp]
  \centering
  \includegraphics[scale=0.35]{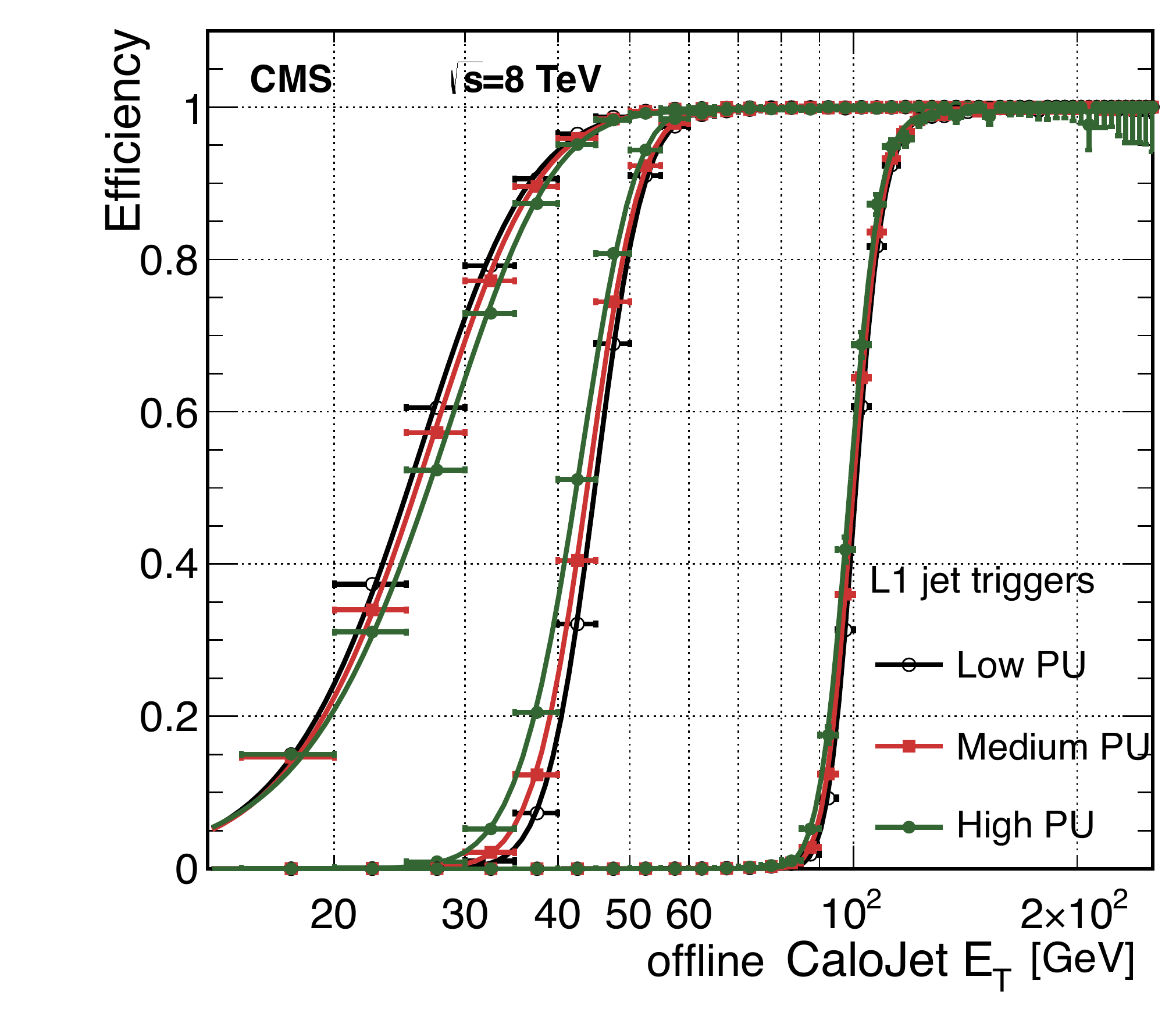}
  \includegraphics[scale=0.35]{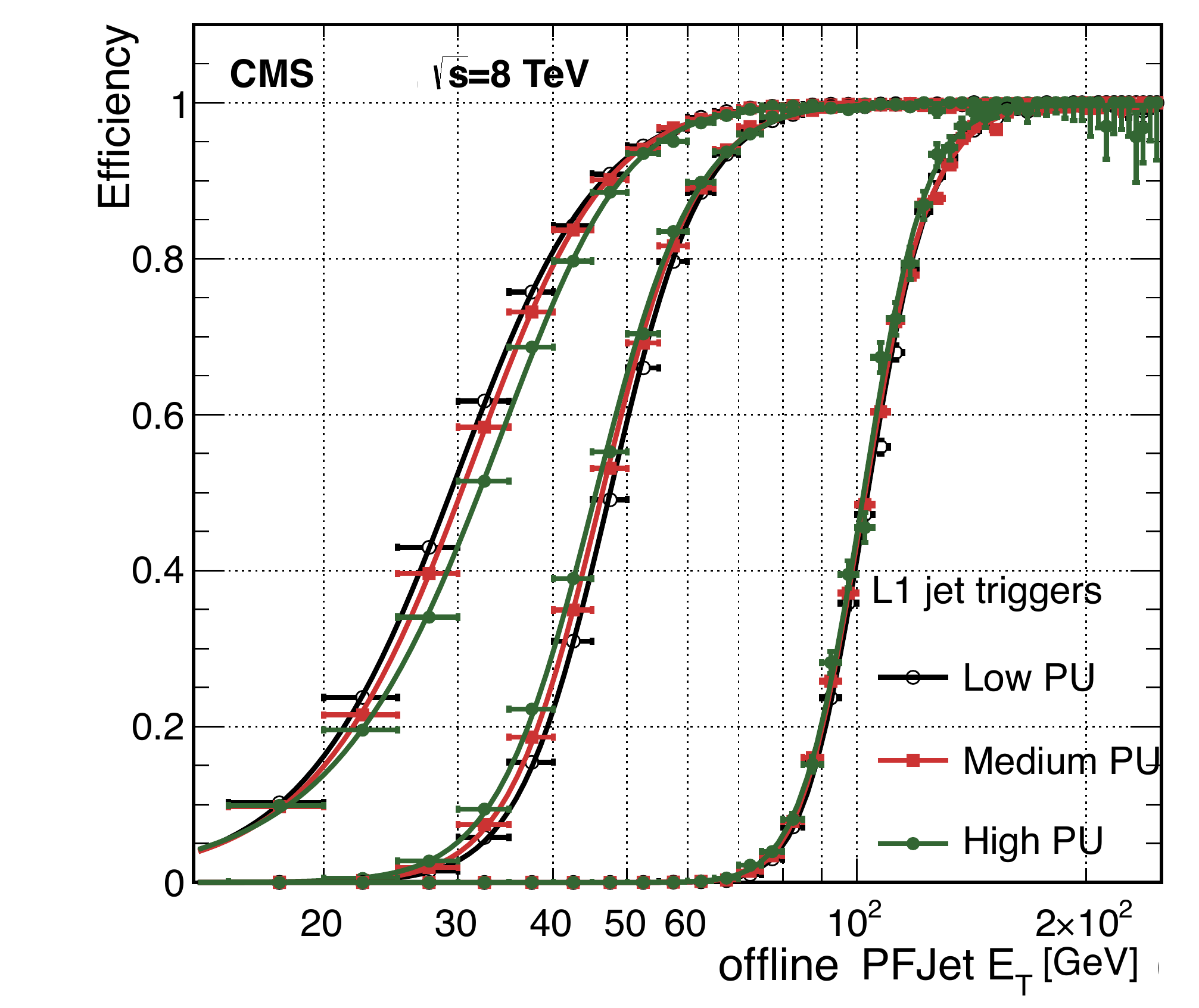}
  \caption{The L1 jet efficiency turn-on curves as a function of the leading
    offline CaloJet \ET (left) and as a function of the leading
    offline PF jet \ET (right), for low-, medium-, and high-pileup
    scenarios for three different thresholds: $\ET>16,36$, and $92\GeV$.}
  \label{fig:l1jet-pu}
\end{figure}

Three different pileup bins of 0--10, 10--20, and ${>}20$ vertices are
defined, reflecting the low-, medium-, and high-pileup running conditions
in 2012 for CaloJets and PF jets, respectively. The corresponding turn-on curves are shown in Fig.~\ref{fig:l1jet-pu}.

There is no significant change of the jet trigger efficiency observed in the
presence of a high number of primary vertices. The increase in
hadronic activity in high-pileup events, combined with the absence of
pileup subtraction within L1 jets, results in the expected observation
of a decrease in the $\mu$ value of the jet turn-on curves as a function of
pileup, while the widths ($\sigma$) of the turn-on curves are found to gradually
increase with increasing pileup.

\subsubsection{The L1 energy sums}
The GCT calculates the total scalar sum of \ET over the calorimeter
regions, as well as $\ETmiss$ based on individual regions.
In addition, it calculates the total scalar sum of L1 jet transverse
energies ($H_\mathrm{T}$) and the corresponding  missing transverse energy
$H_\mathrm{T}^\text{miss}$ based on L1 jet candidates.

\paragraph{Energy sum trigger efficiencies}
The performance of the various L1 energy sum trigger quantities is
evaluated by comparison with the corresponding offline quantities. The
latter are defined at the analysis level according to the most common
physics analysis usage. The following offline quantities are defined:
\begin{itemize}
\item Missing transverse energy, \MET , which is the standard
  (uncorrected) calorimeter-based \MET.
\item Total transverse jet energy, \HT (see Section 1).

\end{itemize}

\begin{figure}[tbp]
\centering
  \includegraphics[width=0.48\textwidth]{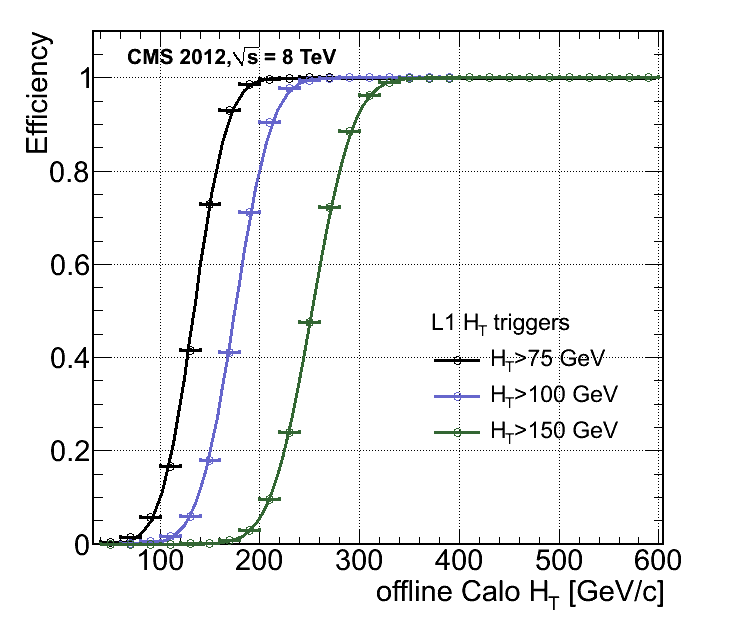}
  \centering
  \includegraphics[width=0.48\textwidth]{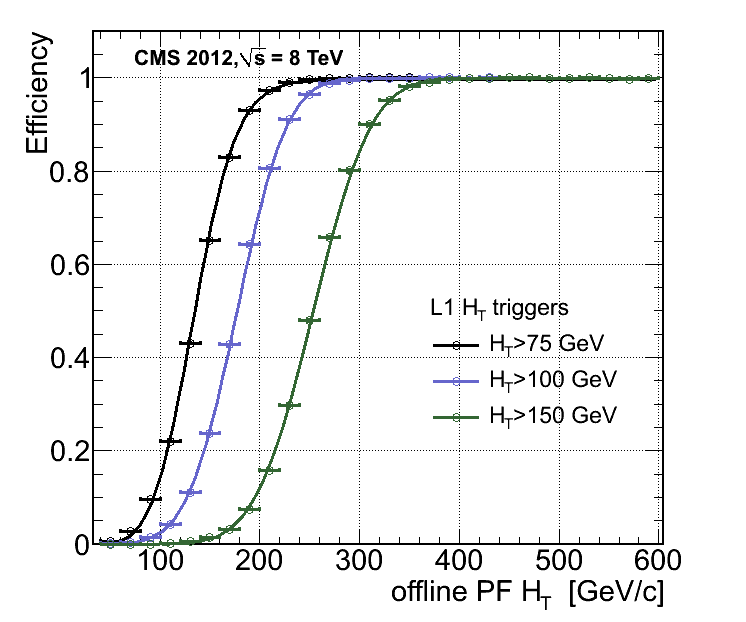}
\caption[esumcurves]{The L1 \HT ~efficiency turn-on curves as a function
  of the offline CaloJet (left) and PF (right) \HT, for three thresholds
  ($\HT>75,100,150$\GeV).}
\label{fig:esumcurves}
\end{figure}

\begin{figure}[tbp]
\centering
  \includegraphics[width=0.48\textwidth]{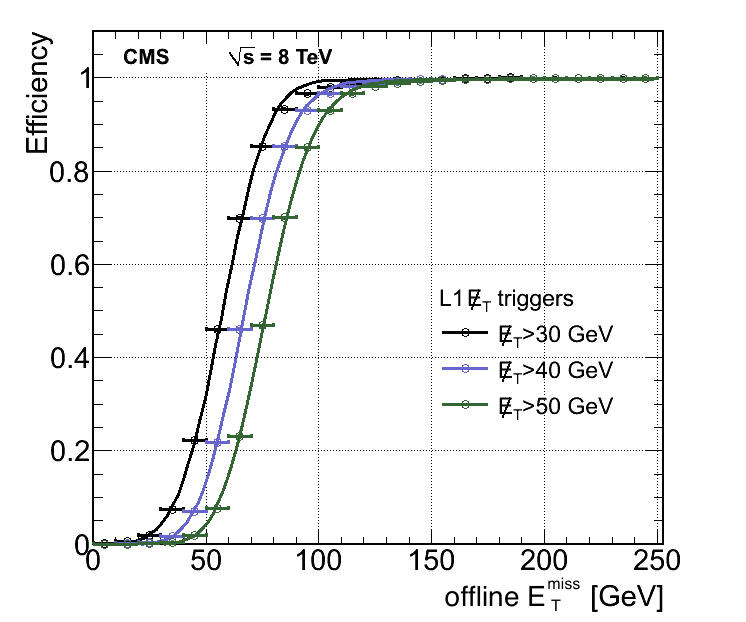}
\caption{The L1 \MET ~efficiency turn-on curve as a function of the
  offline calorimeter \MET, for three thresholds ($\MET>30,40,50$\GeV). }
\label{fig:metcurves}
\end{figure}

Figure~\ref{fig:esumcurves} show the L1 \HT efficiency turn-on curve
for three L1 \HT thresholds of 75, 100, and 150\GeV as a function
of offline CaloJet \HT (left), and PF \HT
(right). Figure~\ref{fig:metcurves} shows the
L1 \MET efficiency curve for three L1
\MET thresholds of 30, 40, and 50\GeV. The turn-on
points in all the efficiency curves are shown to be shifted towards larger
values than the corresponding L1 trigger thresholds, which is
explained by the fact that the quantities are defined in different way at
the trigger and offline levels; the trigger uses standard calorimeter reconstruction
based object definition, whereas offline uses the PF object
definition. The same  reasoning explains the slow turn-on curves observed in
the performance of the energy sum triggers versus the PF quantities,
with the resolution appearing to worsen when compared to the
performance obtained using the standard calorimeter reconstruction.
In both cases, the L1 \HT and L1 \MET efficiencies plateau at 100\%.

\subsubsection{L1 jet and energy sum rates}
\begin{figure}[tbp]
  \centering
  \includegraphics[width=0.48\textwidth]{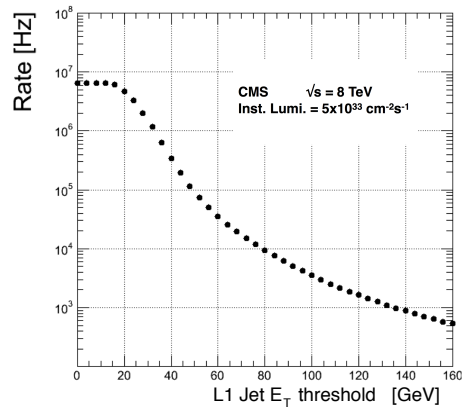}
  \caption{The rate of the L1 single-jet trigger as a function of the
    \ET threshold. The rates are rescaled to the
    instantaneous luminosity $5\times10^{33}\percms$. }
\label{fig:l1jetrate}
\end{figure}

\begin{figure}[tbp]
  \centering
  \includegraphics[width=0.45\linewidth]{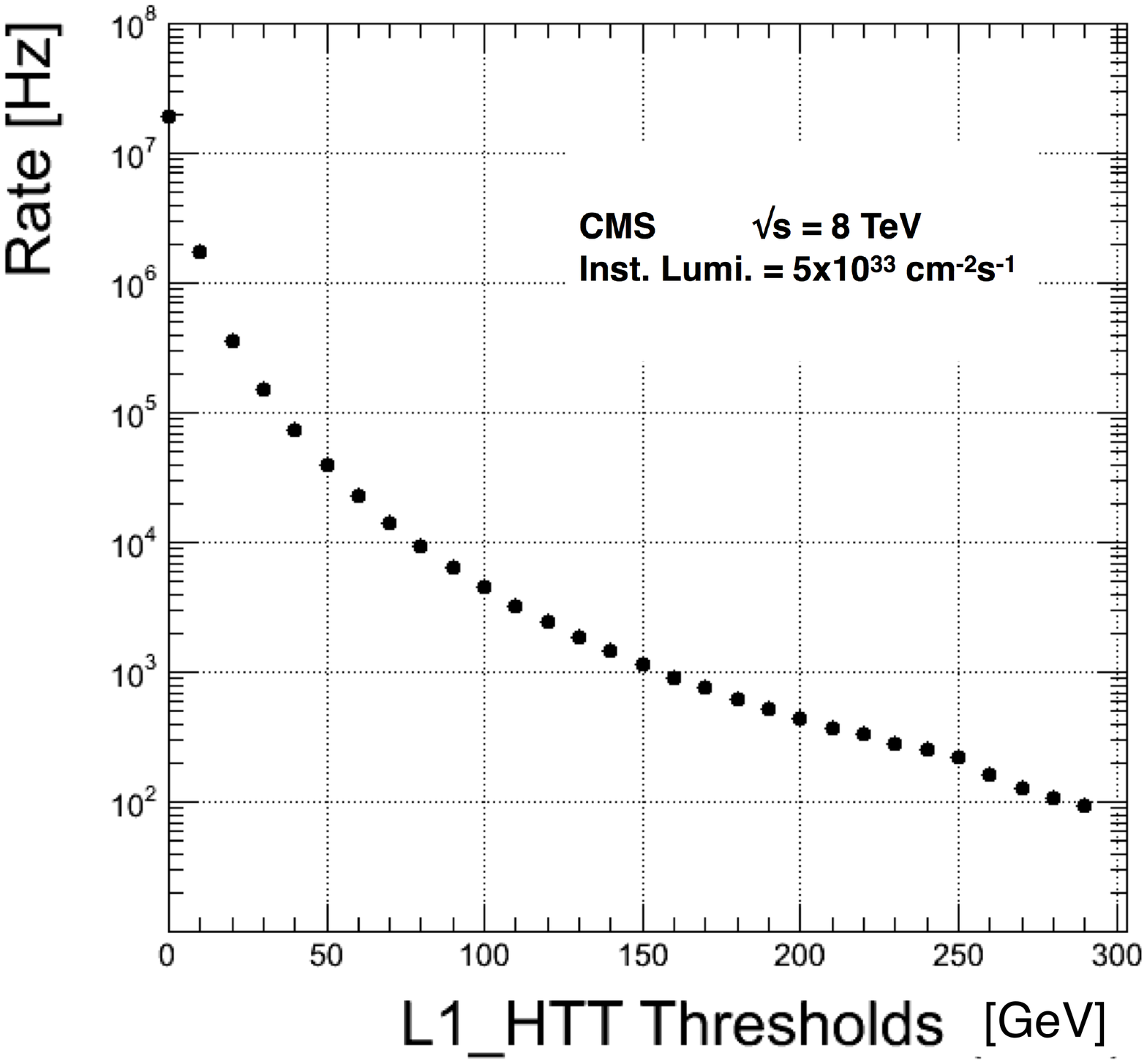}
  \includegraphics[width=0.45\linewidth]{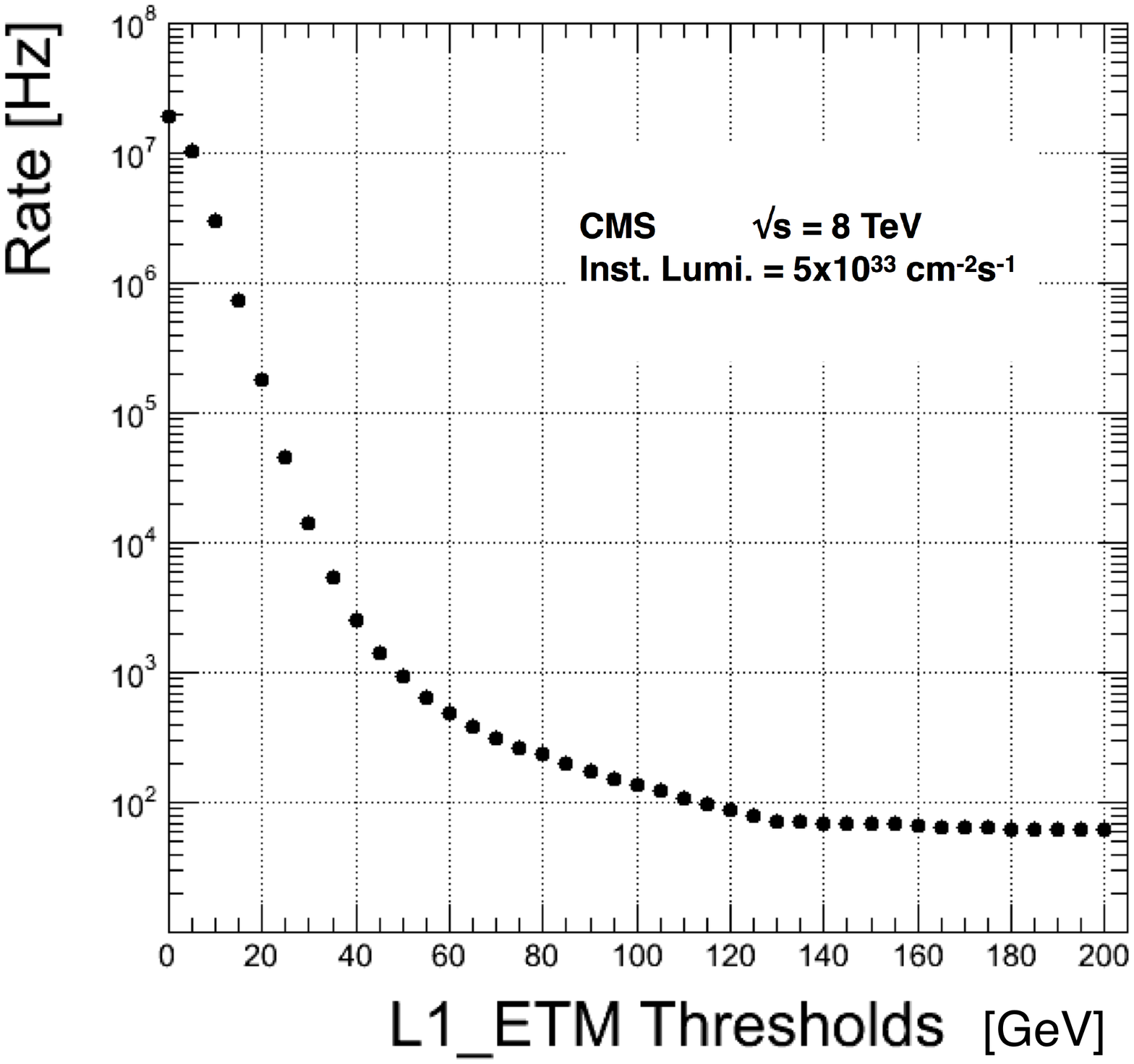}
  \caption{Left: Rate of the L1\_HTT trigger versus the
    L1\_HTT threshold. Right: Rate of the L1\_ETM missing
    transverse energy trigger  as a function of the
    L1\_ETM threshold. On both plots, the rates are rescaled to the
    instantaneous luminosity $5\times10^{33}\percms$. }
  \label{fig:esumrates}
\end{figure}

The L1 single jet trigger rates as a function of the L1 jet threshold were
also evaluated, using similar strategy to that described in the muon
identification section. We used data recorded in a special data set in
which only the essential needed information about the events was
stored, and further selected events without
any bias based on the trigger selection (\ie, zero bias triggered
events)
and correspond to an instantaneous luminosity of
$5 \times 10^{33}\unit{cm}^{-2}\unit{s}^{-1}$.
Figure~\ref{fig:l1jetrate} shows the L1
single-jet trigger rate as a function of the L1 jet threshold. Similarly, the
rates of the L1 energy sum triggers (L1\_HTT and L1\_ETM triggers
here) are shown in Fig.~\ref{fig:esumrates}.

\subsubsection{The HLT jet triggers }
\label{sec:JetHLT}
At the HLT, jets are reconstructed using the anti-\kt clustering
algorithm with cone size $R = 0.5$~\cite{fastjetmanual,Cacciari:2008gp}. The inputs
for the jet algorithm are either calorimeter towers (resulting in
so-called ``CaloJet'' objects), or the reconstructed particle flow
objects (resulting in ``PFJet'' objects). In 2012, most of the jet
trigger paths use PFJet as their inputs. As the PF
algorithm uses significant CPU resources, PFJet trigger paths have a
pre-selection based on CaloJets.   Matching between CaloJets and
PFJets is then required in single PFJet paths.

\begin{figure}[tbp]
\centering
  \includegraphics[width=0.48\textwidth]{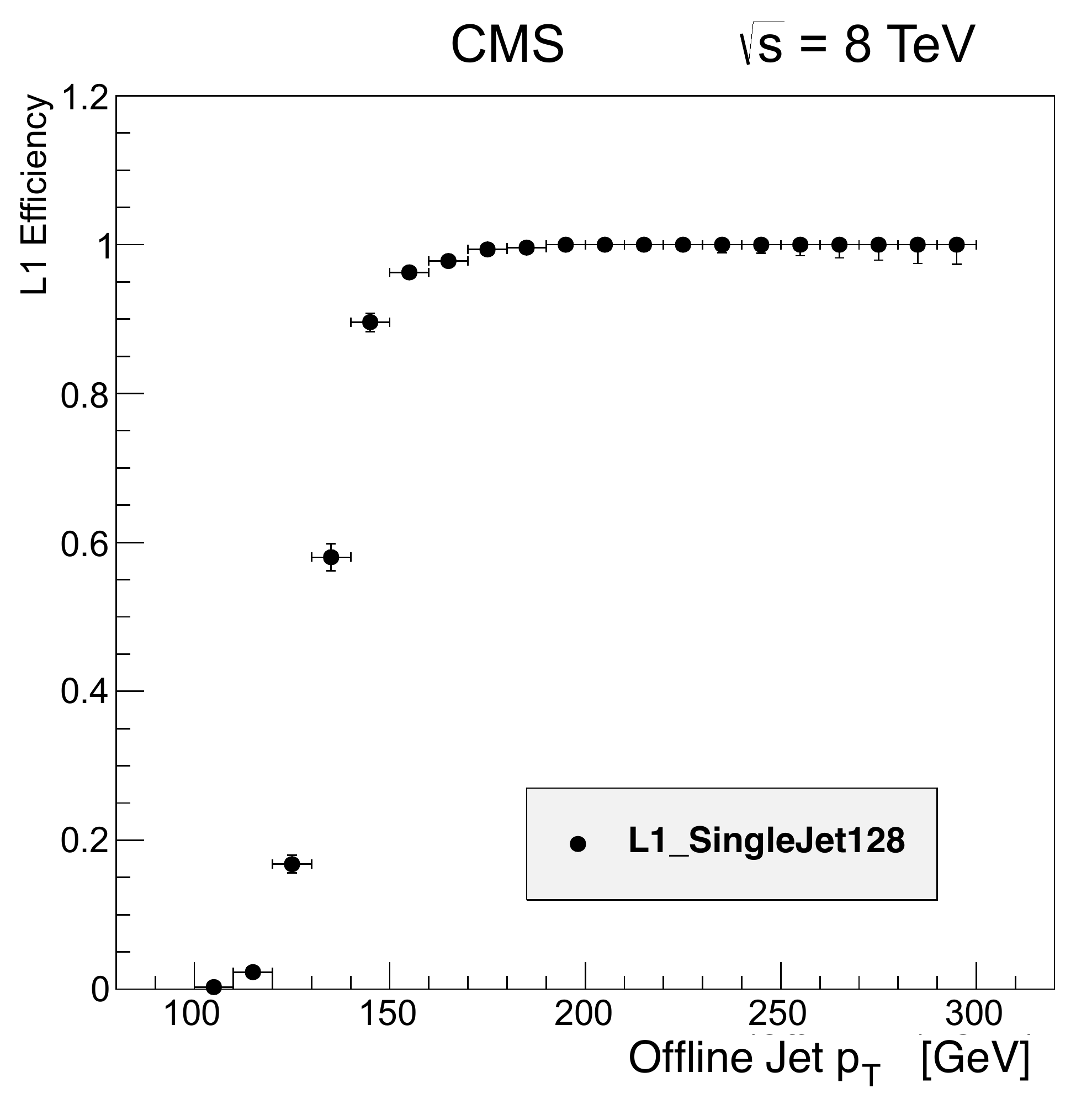}
  \includegraphics[width=0.48\textwidth]{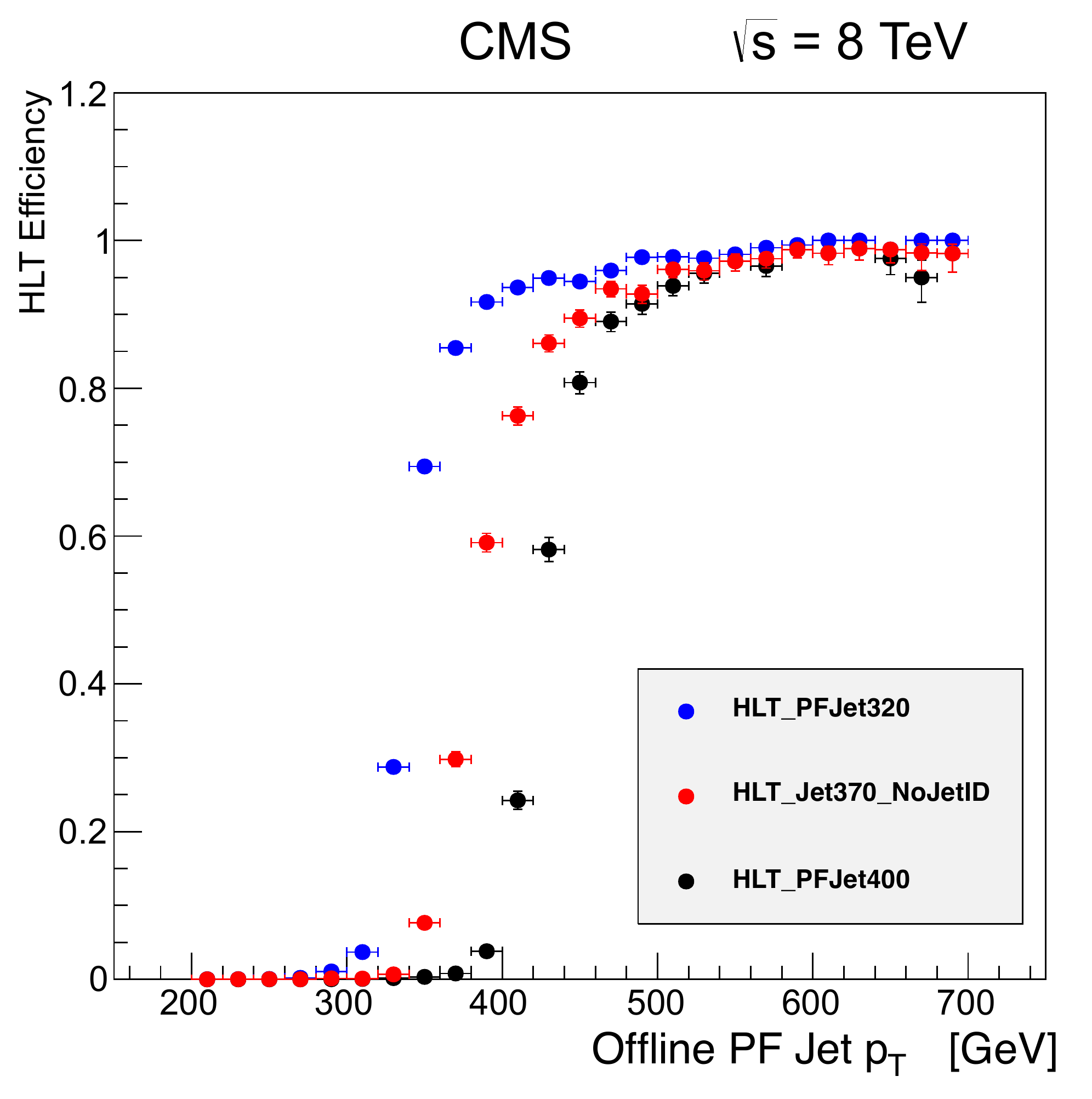}
  \caption[JetEff_RunComparison]{Left: Efficiency of the L1
single-jet trigger
with an \ET threshold of 128 \GeV as a function of
the offline jet transverse momentum.
Right:
    The HLT efficiencies as a function of transverse momentum
    for a calorimeter jet trigger with a 370\GeV threshold and no jet
identification requirements~\cite{CMS-PAS-JME-09-008}, and two PF jet triggers
with 320 and 400\GeV thresholds.}
\label{fig:JetEff_RunComparison}
\end{figure}

\paragraph{Single-jet paths}
The L1 thresholds for the single-jet paths were chosen such that the
L1 efficiency is at least 95\% at the corresponding HLT threshold. The
jet energy scale corrections (JEC) were applied to the single-jet
paths. The lowest threshold path was a L1 pass-through path that
simply requires a L1 jet in the event with $\pt > 16$\GeV. The single
PFJet trigger paths for $\lumi=7\times10^{33}\percms$ (pileup
${\approx}$32), along with the L1, prescales, and approximate rates are
listed in Table~\ref{tab:JetTrigger}. The trigger turn-on curves for
selected single PFJet paths as a function of transverse momentum
of the offline jet is
shown in Fig.~\ref{fig:JetEff_RunComparison}. The trigger efficiency
was calculated from an independent data sample collected using a
 single isolated muon trigger with a $\pt>24$\GeV
threshold. As in the L1 case (Section~\ref{sec:l1jet}), the
efficiency is evaluated in comparison to offline jets, in this case,
PF jets.

\begin{table}[tbp]
\centering
\topcaption{Single-jet triggers used for $\lumi=7\times10^{33}\percms$
(pileup ${\approx}$32),
  their prescales, and trigger rates at that instantaneous luminosity.}

\begin{tabular}{ | l  c  c  c  c | }
	\hline
	Path name & L1 seed & L1 prescale & HLT prescale & Approx. Rate (Hz) \\
	\hline \hline
	\texttt{HLT\_L1SingleJet16}   & \texttt{L1\_SingleJet16} & 200,000 & 55 & 0.9\\
	\texttt{HLT\_L1SingleJet36}   & \texttt{L1\_SingleJet36} & 6,000 & 200 & 1.8\\
	\hline
	\texttt{HLT\_PFJet40}   & \texttt{L1\_SingleJet16} & 200,000 & 5 & 0.2\\
	\texttt{HLT\_PFJet80}   & \texttt{L1\_SingleJet36} & 6,000 & 2 & 1.0\\
	\texttt{HLT\_PFJet140} & \texttt{L1\_SingleJet68} & 300 & 2 & 1.5 \\
	\texttt{HLT\_PFJet200} & \texttt{L1\_SingleJet92} & 60 & 2 & 1.2 \\
	\texttt{HLT\_PFJet260} & \texttt{L1\_SingleJet128} & 1 & 30 & 1.3\\
	\texttt{HLT\_PFJet320} & \texttt{L1\_SingleJet128} & 1 & 1 & 12.7\\
	\texttt{HLT\_PFJet400} & \texttt{L1\_SingleJet128}  & 1 & 1 & 3.7\\
	\hline
	\texttt{HLT\_Jet370\_NoJetID} & \texttt{L1\_SingleJet128}  & 1 & 1 & 6.7\\
	\hline
\end{tabular}

\label{tab:JetTrigger}
\end{table}

\paragraph{Dijet paths}
The dijet trigger is primarily used to collect data for
$\eta$-dependent energy corrections using a \pt-balance technique~\cite{Chatrchyan:2012xx}. This
correction removes any variation in the calorimeter response to a
fixed jet \pt as a function of jet $\eta$.

The dijet triggers require two HLT jets with an average transverse
energy greater than a given threshold. The lowest threshold path
requires two HLT jets with an average transverse energy greater than
40\GeV. The DiPFJet trigger paths for $\lumi=7\times10^{33}\percms$
(pileup ${\approx}$32), along with the L1 and HLT prescales and rates
are listed in Table~\ref{tab:DiJetTrigger}. The lowest transverse energy
unscaled path
has a threshold of 400\GeV.
\begin{table}[tbph]
\centering
\caption{Dijet-triggers used at $\lumi=7\times10^{33}\percms$
(pileup ${\approx}$ 32), their
  prescales, and trigger rates. The main purpose of these triggers is the
  $\eta$-dependent calibration of the calorimeter.}

\begin{tabular}{ | l  c  c  c  c | }
	\hline
	Path name & L1 seed & L1 prescale & HLT prescale & Rate (Hz) \\
	\hline
	\texttt{HLT\_DiPFJetAve40}   & \texttt{L1\_SingleJet16} & 200,000 & 1 & 0.51\\
	\texttt{HLT\_DiPFJetAve80}   & \texttt{L1\_SingleJet36} & 6,000 & 1 & 0.71\\
	\texttt{HLT\_DiPFJetAve140} & \texttt{L1\_SingleJet68} & 300 & 1 & 1.51 \\
	\texttt{HLT\_DiPFJetAve200} & \texttt{L1\_SingleJet92} & 60 & 1 & 1.36 \\
	\texttt{HLT\_DiPFJetAve260} & \texttt{L1\_SingleJet128} & 1 & 15 & 1.41\\
	\texttt{HLT\_DiPFJetAve320} & \texttt{L1\_SingleJet128} & 1 & 5 & 1.19\\
	\texttt{HLT\_DiPFJetAve400} & \texttt{L1\_SingleJet128}  & 1 & 1 & 1.44\\
	\hline
\end{tabular}
\label{tab:DiJetTrigger}
\end{table}

\subsubsection{The HLT \texorpdfstring{\MET}{Missing Transverse Energy} triggers}
\label{sec:METHLT}
In this section, triggers that exclusively place requirements on
missing transverse energy  are
described. Unscaled \MET triggers are of particular interest for
searches for new physics processes beyond the standard
model. Hypothetical  particles, such as the lightest
supersymmetric particle (LSP), graviton, or dark matter, will interact
weakly in the CMS detector before escaping. Their presence can be
inferred by a measured imbalance in the energy or momentum of the observed particles in the event.

\paragraph{The \texorpdfstring {\MET}{Missing Transverse Energy} algorithms}
The \MET at the HLT is calculated using the same algorithms as the offline
analysis. Two algorithms were used to reconstruct the \MET in the
HLT. The first algorithm, called CaloMET, calculated the \MET  by
summing all towers in the calorimeter,
\begin{equation}
\label{equ:CaloMET}
\MET = \sqrt{\Bigl( \sum_\text{towers} E_x\Bigr)^2 + \Bigl(\sum_\text{towers} E_y\Bigr)^2}.
\end{equation}
Another algorithm (PFMET) uses the negative of the vector sum
over transverse momenta of reconstructed anti-\kt PF
jets,
\begin{equation}
\label{equ:PFMET}
\mbox{PF }\MET = \sqrt{\Bigl( \sum_\text{PFJet} P_x\Bigr)^2 + \Bigl(\sum_\text{PFJet} P_y\Bigr)^2}.
\end{equation}
No minimum threshold requirement on jet $\pt$ was applied in this
algorithm at the HLT. As with the PFJet trigger paths, a pre-selection
based on the CaloMET is applied before the PFMET is calculated to
reduce the required CPU time of the PF
algorithm. Table~\ref{tab:METTrigger} shows the \MET triggers used for
$\lumi= 8\times10^{33}\percms$ in 2012, together with
prescale factors at L1 and HLT, and rate estimated using a 2012 dedicated data sample.

\begin{table}
\centering
\topcaption{The \MET triggers used for $\lumi=7\times10^{33}\percms$ (pileup
  ${\approx}$32), their prescales, and rates at that luminosity. Note
  that the L1 $\MET>36\GeV$ trigger (\texttt{L1\_ETM36}) was highly
  prescaled starting at this luminosity and hence the need to use an
  OR with the L1 $\MET>40\GeV$ trigger  (\texttt{L1\_ETM40}).  The
  parked HLT $\MET>80\GeV$ trigger (\texttt{HLT\_MET80\_Parked}) was
  also anticipated to be highly prescaled starting from
  $\lumi=8\times10^{33}\percms$. The \MET parking triggers were
  available at the end of 2012. ``Cleaned'' refers to application of
dedicated algorithms to remove noise events.
  }
\begin{tabular}{ | l  c  c  c | }
	\hline
Path name & L1 seed & HLT prescale & Rate (Hz) \\
	\hline \hline
	\multicolumn{4}{|c|}{Prompt triggers} \\ \hline
	\texttt{HLT\_MET80} & \texttt{L1\_ETM36 OR L1\_ETM40}  & 100 & 0.48\\
	\texttt{HLT\_MET120} & \texttt{L1\_ETM36 OR L1\_ETM40}  & 8 & 0.71\\
	\texttt{HLT\_MET120\_HBHENoiseCleaned} & \texttt{L1\_ETM36 OR L1\_ETM40} & 1 & 3.92 \\
	\texttt{HLT\_MET200} & \texttt{L1\_ETM70}  & 1 & 1.46 \\
	\texttt{HLT\_MET200\_HBHENoiseCleaned} & \texttt{L1\_ETM70} & 1 & 0.63\\
	\texttt{HLT\_MET300} & \texttt{L1\_ETM100}  & 1 & 0.47\\
	\texttt{HLT\_MET300\_HBHENoiseCleaned} & \texttt{L1\_ETM100}  & 1 & 0.15\\
	\texttt{HLT\_MET400} & \texttt{L1\_ETM100}   & 1 &0.19\\
	\texttt{HLT\_MET400\_HBHENoiseCleaned} & \texttt{L1\_ETM100}   & 1 & 0.05\\
	\texttt{HLT\_PFMET150} & \texttt{L1\_ETM36 OR L1\_ETM40}  & 1  & 3.05\\
	\texttt{HLT\_PFMET180} & \texttt{L1\_ETM36 OR L1\_ETM40}  & 1 & 1.92\\
	\hline
	\multicolumn{4}{|c|}{Parked triggers}  \\ \hline
	\texttt{HLT\_MET80\_Parked} & \texttt{L1\_ETM36 OR L1\_ETM40}  & 1 & 47.54 \\
	\texttt{HLT\_MET100\_HBHENoiseCleaned} & \texttt{L1\_ETM36 OR L1\_ETM40}  & 1 & 9.09\\	

	\hline
\end{tabular}
\label{tab:METTrigger}
\end{table}

\begin{figure}
\centering
  \includegraphics[width=0.6\textwidth]{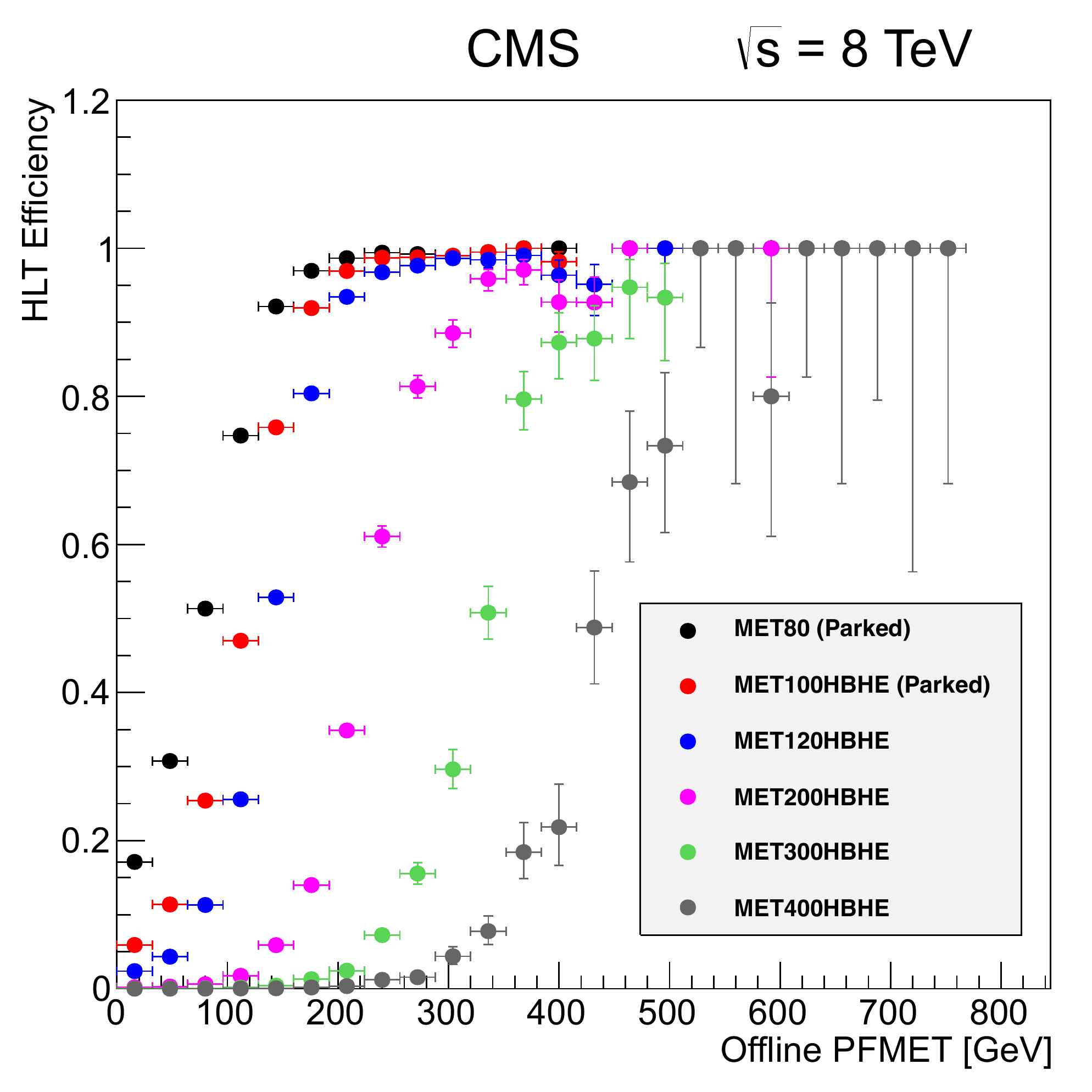}
  \caption[METEff_RunComparison]{
    The HLT efficiencies as a function of the offline PF\MET
    for different \MET thresholds ($\ETmiss = 80$--$400\GeV$). }
\label{fig:METEff_RunComparison}
\end{figure}

\paragraph{Efficiency of $\ETmiss$ triggers}
The trigger turn-on curves as a function of \MET are shown in
Figs.~\ref{fig:metcurves} and \ref{fig:METEff_RunComparison}. The
trigger efficiency is calculated from an independent data sample
collected using the lowest-\pt unscaled isolated single muon
trigger path, with $\pt>24$\GeV.

\subsection{\texorpdfstring{$\tau$}{Tau} lepton triggers}
\label{sec:tau}
The $\tau$-jet triggers are important for a wide variety of physics analyses
that use $\tau$ leptons decaying hadronically. In many models of new
physics, third-generation particles play a special role in
elucidating the mechanism for spontaneous symmetry breaking and
naturalness.
The $\tau$ leptons, as the charged leptons of the third generation,
constitute important signatures for $\mathrm{h} \to \tau\tau$
searches and certain new physics scenarios.
The tau triggers are designed to collect events with $\tau$ leptons
decaying hadronically. Hadronic decays make up more than 60\% of the tau
branching fractions, mostly via final states with either one or three
charged hadrons in a tightly collimated jet with little additional
activity around the central cone. Leptonic tau decays are
automatically collected by electron and muon triggers.
In what follows, we refer to
taus that decay hadronically as $\tau_\mathrm{h}$ and $\tau$ leptons that decay
to electrons (muons) as $\tau_{\Pe}$ ($\tau_{\mu}$).

\subsubsection[L1 tau identification]{The L1 $\tau$ lepton identification}
\label{sec:l1tau}

A common approach to separate $\tau$ leptons decaying to hadrons ($\tau_h$)
from quark and gluon jets is by using isolation criteria. This is a
challenging task to perform at the L1 trigger because of the given coarse
granularity of the L1 calorimeter readout (Fig.~\ref{fig:l1jetalgo}).
The L1 $\tau$ objects are mandatory, however, for analyses such as
$\mathrm{h} \to \tau\tau$, with both $\tau$
leptons decaying hadronically.

\begin{figure}[htbp]
\centering
  \includegraphics[width=0.75\linewidth]{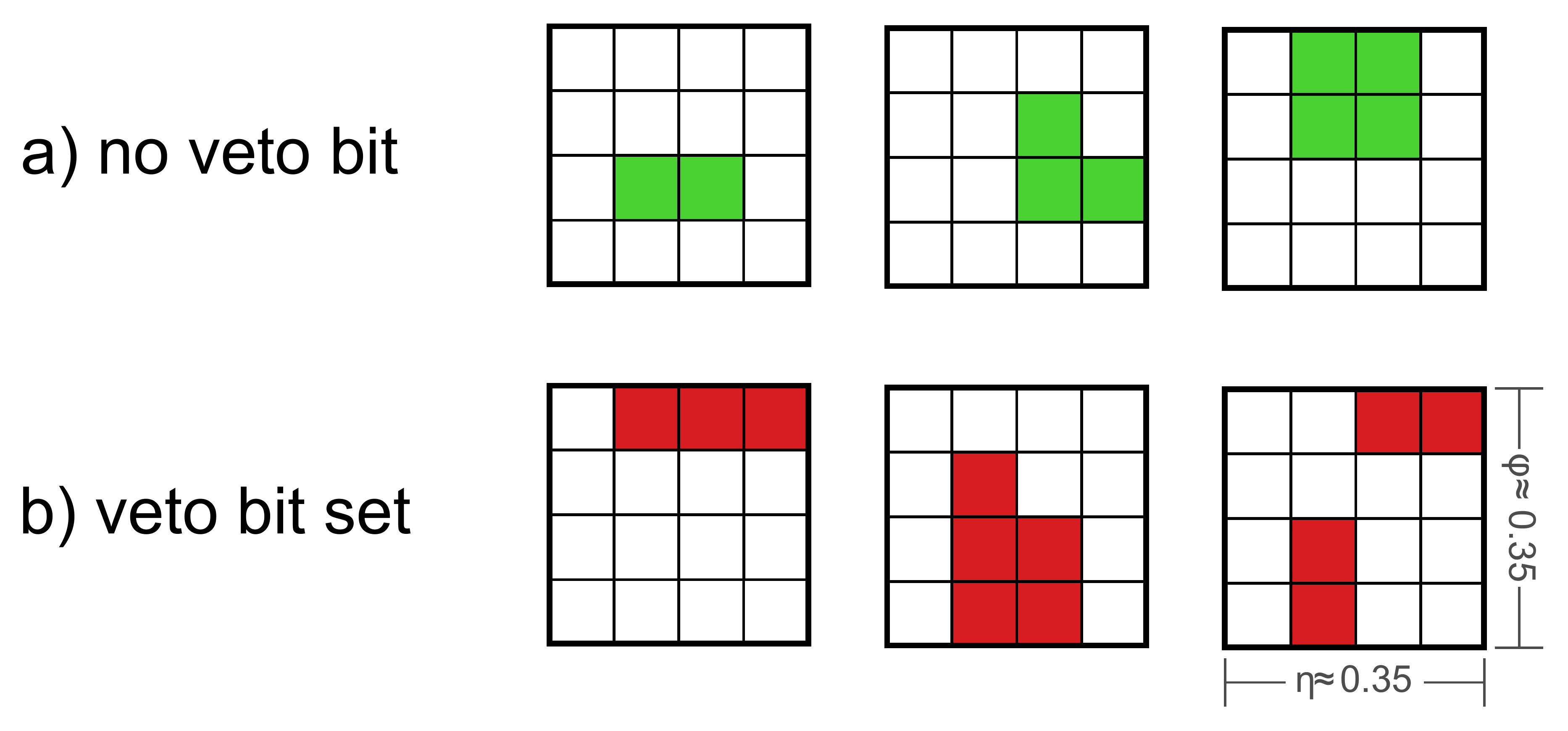}
  \caption{Examples of trigger regions, where trigger towers with
    energy deposits $\ET^\mathrm{ECAL}>4\GeV$ or $\ET^\mathrm{HCAL}>4\GeV$,
    are shown as shaded squares. The L1 $\tau$ veto bit is not set if
    the energy is contained in a square of $2{\times}2$ trigger towers
    (a). Otherwise, the $\tau$ veto bit is set (b). }
  \label{figure:L1TauVeto}
\end{figure}

The L1 $\tau_\mathrm{h}$ identification starts from previously identified L1 jet
objects (Section \ref{sec:l1jet}), which are further investigated
using an isolation variable and a $\tau$ veto bit. We require that seven
out of the eight noncentral trigger regions contain small
energy deposits ($\ET < 2\GeV$). This acts as an isolation
requirement. In addition, for each trigger region a $\tau$ veto bit is
set if the energy deposit is spread over more than $2\times2$ trigger
towers (Fig.~\ref{figure:L1TauVeto}). The L1 $\tau$ objects are
required to have no $\tau$ veto bit set in all nine trigger regions,
further constraining the energy spread within the two most energetic
 trigger regions.
If either the isolation or the $\tau$ veto bit requirement
fails, the object is regarded as an L1 central jet.

\begin{figure}[tbph]
\centering
  \includegraphics[width=0.6\linewidth]{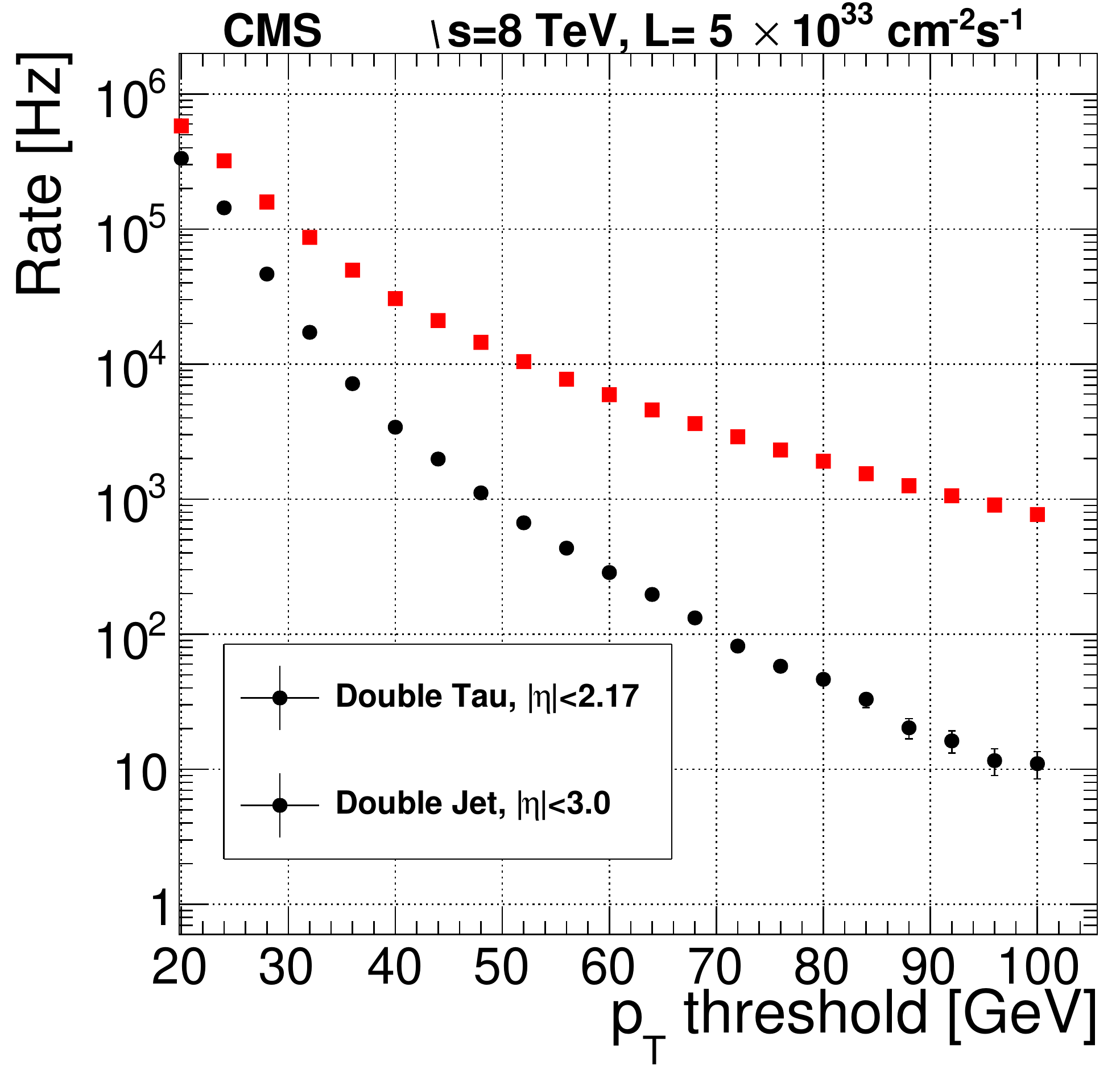}
  \caption{Rate of L1 double-$\tau$ and double-jet seeds as a function of the
  $\pt$ threshold on the two objects. The double-$\tau$ objects are restricted to
  $\abs{\eta}<2.17$, while the double-jet requires two seed objects (either $\tau$ or jet)
  within $\abs{\eta}<3.0$. The given rates are scaled to an instantaneous luminosity of
  $5 \times 10^{33}\percms$.}
  \label{figure:L1TauRate}
\end{figure}

\begin{figure}[tbph]
\centering
  \includegraphics[width=0.45\linewidth]{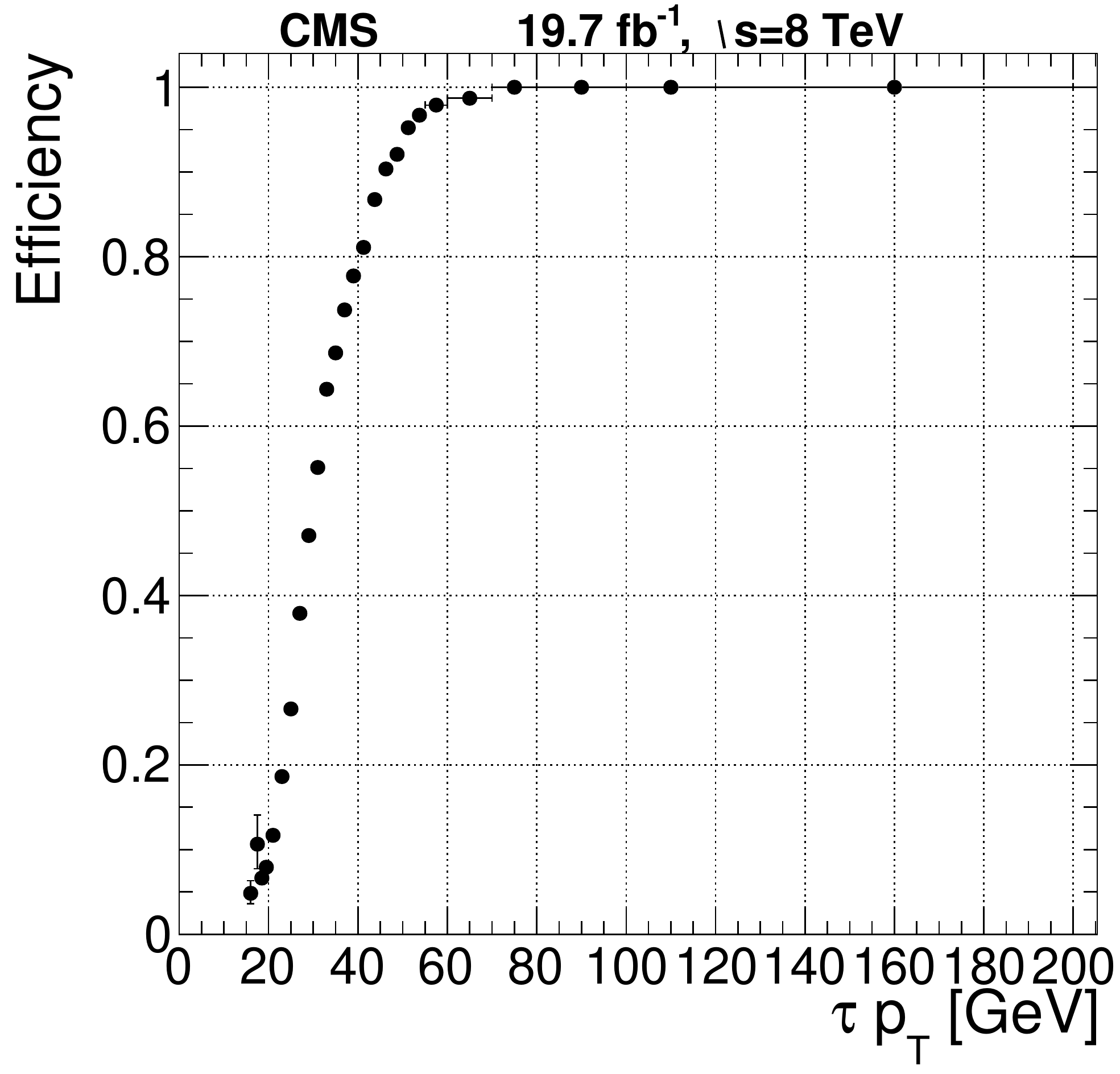}
  \includegraphics[width=0.452\linewidth]{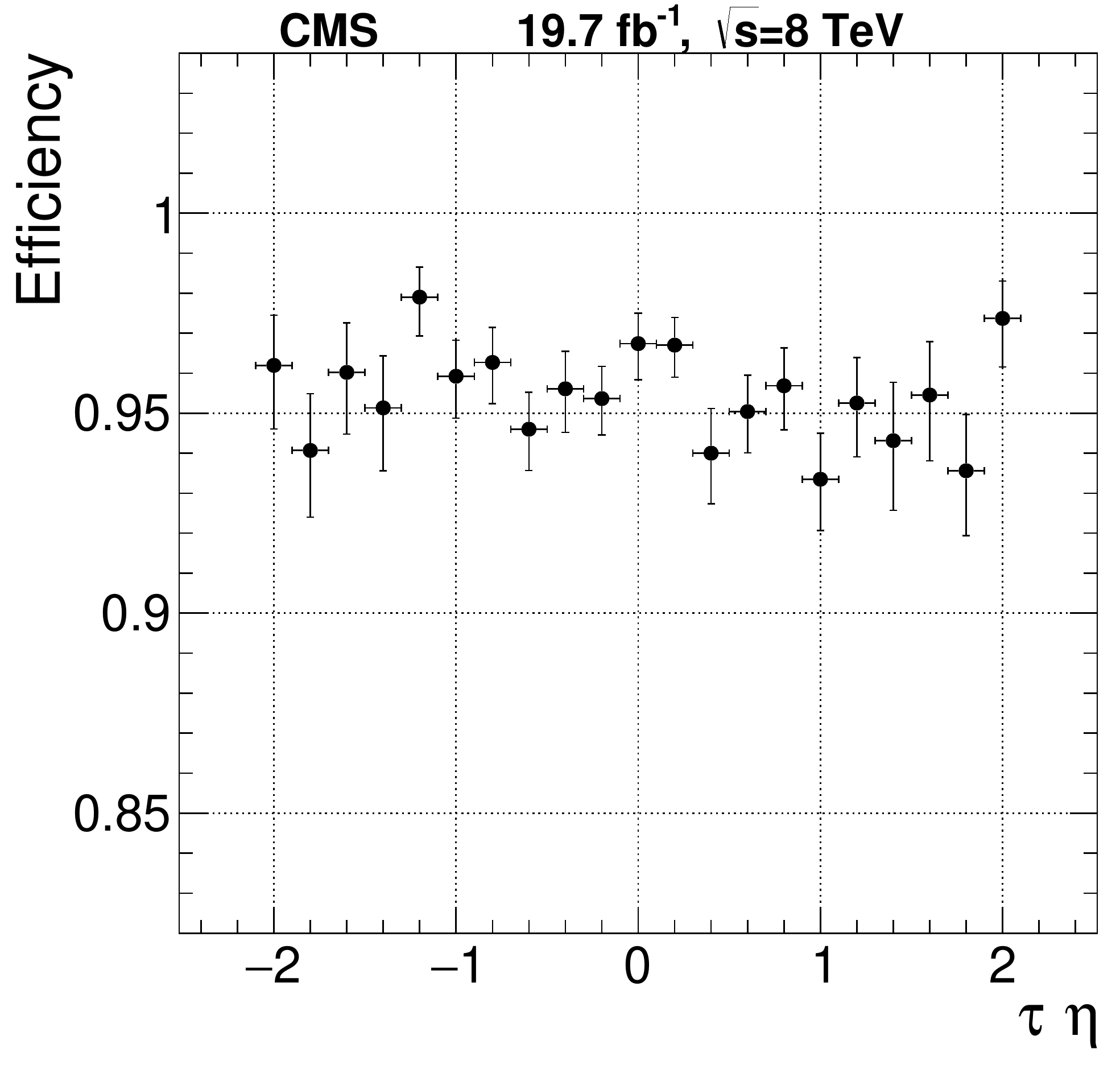}
  \caption{Efficiency of the double-$\tau_\mathrm{h}$ L1 trigger with a
    threshold of 44 and 64\GeV on the L1 $\tau$ and jet objects,
    respectively. Presented is the efficiency of one $\tau$ lepton candidate as
    a function of transverse momentum (left) and pseudorapidity
    (right).}
  \label{figure:L1TauEfficiency}
\end{figure}

The $\mathrm{h} \to \tau_h\tau_h$
search~\cite{Chatrchyan:2014nva} uses
an L1 seed requiring two L1 $\tau$ objects with $\pt>44\GeV$ and
$\abs{\eta} < 2.17$. For large $\tau$ energies, the isolation
criteria introduce an inefficiency for genuine $\tau$ leptons. This is
recovered by also allowing events with two L1 jets (central or
$\tau$) with $\pt>64\GeV$ and $\abs{\eta} < 3.0$ to be
selected. Figure~\ref{figure:L1TauRate} shows the rate of these
L1 seeds as a function of the applied $\pt$ threshold on the two
objects.
The measured efficiency of this L1 seed reaches a plateau of 100\% at
$\pt \approx70$\GeV, as shown in
Figure~\ref{figure:L1TauEfficiency}. The efficiency as function of the
pseudorapidity is obtained using $\tau$ leptons with $\pt
>45\GeV$.
This requirement emulates the \pt requirement used in the $\mathrm{h} \to \tau_\mathrm{h}\tau_\mathrm{h}$
search.

\subsubsection[HLT tau lepton identification]{The HLT $\tau$ lepton identification}
\label{sec:tauHLT}

The $\tau$-jet triggers identify and select events with hadronic
decays of the $\tau$ leptons; leptonic decays are
selected as prompt electrons and muons.
There are three levels of the $\tau$ HLT; each is designed to reduce
the rate before running the more complex subsequent step. The first
step we call the level-2 (L2) $\tau$ trigger; it is built with
CaloJets. The second step is referred to as level-2.5 (L2.5); this
step requires isolation for matching tracks reconstructed from the pixel
detector. The last step, called level-3 (L3), uses the PF
algorithm to build $\tau$ lepton candidates using information from all major
subdetectors. Offline $\tau$ reconstruction with
CMS is described in more detail elsewhere~\cite{Chatrchyan:2012zz}. The HLT $\tau$
paths come in two distinct varieties. The first is for $\tau_\mathrm{h}$
candidates triggered with the L1 trigger. These $\tau$ lepton triggers
have a L2 and L2.5 step to reduce the rate before running the more
advanced L3 $\tau$ reconstruction. The second type of $\tau$ trigger
path is triggered at L1 by a lepton or other event quantity
such as \MET. These triggers have HLT electron, muon or missing
energy selections to reduce the rate before running the L3 $\tau$
algorithm.

The L2 $\tau$-jet trigger reconstruction is entirely based on calorimeter
information. The CaloJets are built with a cone of radius equal to 0.2
seeded by L1 $\tau$ jets (Section~\ref{sec:l1tau}) or L1 central
jets. The only selection applied is a \pt threshold on the jet transverse
energy.

The L2.5 step consists of a track-based isolation applied on the L2
$\tau$ candidates that are above the \pt threshold. The isolation starts
by reconstructing the pixel tracks and selecting those coming from the
primary vertex and matched to  the L2 $\tau$ candidate. A L2 $\tau$ is
considered to be isolated if there is no pixel track from the same vertex
with transverse momentum greater than 1.2\GeV in an isolation annulus
between $0.2 <\DR< 0.4$ around the $\tau$ candidate.

Finally, the L3 $\tau$ reconstruction uses the PF
algorithm. The online reconstruction uses a so-called \emph{fixed cone} $\tau$
algorithm with a signal cone of $\DR = 0.18$, which contains the
$\tau$ decay products,  and an isolation annulus of
$0.18<\DR< 0.45$. The trigger uses tracker-only isolation built using
tracks from a vertex compatible with the primary vertex of the
$\tau$ to minimize pileup dependence. There are two isolation
working points: loose and tight. A loose $\tau$ is considered
isolated if no tracks with $\pt>1.5\GeV$
are found in the isolation annulus. A $\tau$ candidate is considered to be ``tight'' if it has no
tracks with $\pt>1.0\GeV$ with a signal/isolation cone boundary at
0.15.

Trigger efficiencies are measured individually in each step. For the
double-$\tau$ trigger a per-leg efficiency is measured. A sample of
$\Z\to\tau\tau$ events selected by a single-muon trigger is
used for the measurement, with one $\tau$ decaying hadronically and
the other to muon and neutrinos. The $\tau_\mathrm{h}\tau_\mu$ candidates are
selected and discriminated against multijet and W boson backgrounds
using muon isolation, charge requirements, and low transverse mass
$M_\mathrm{T}$ to
achieve a $\tau_\mathrm{h}$ purity of approximately 50\%.

\begin{figure}[tbph]
  \centering
    \includegraphics[width=0.45\textwidth]{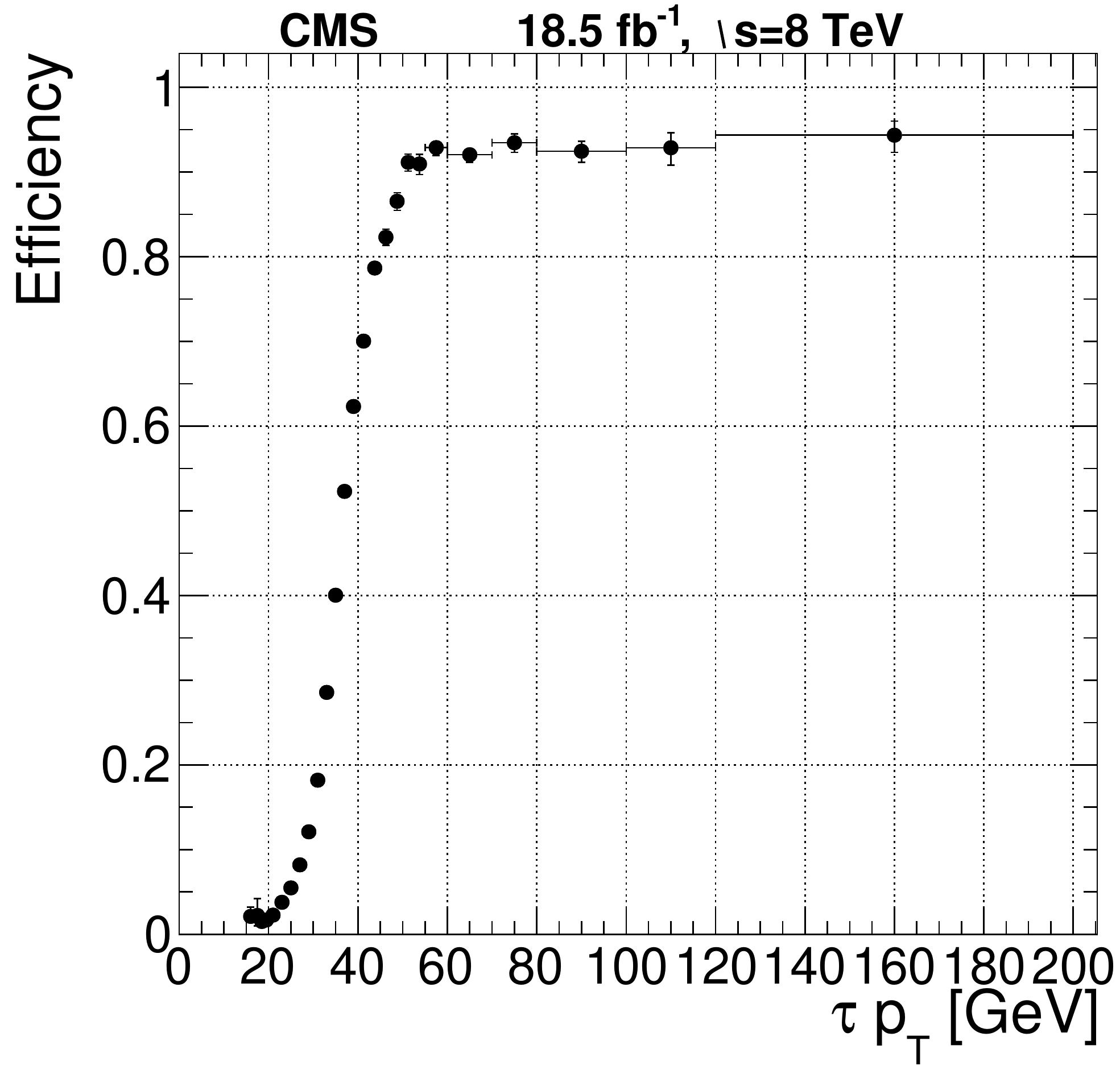}
    \includegraphics[width=0.45\textwidth]{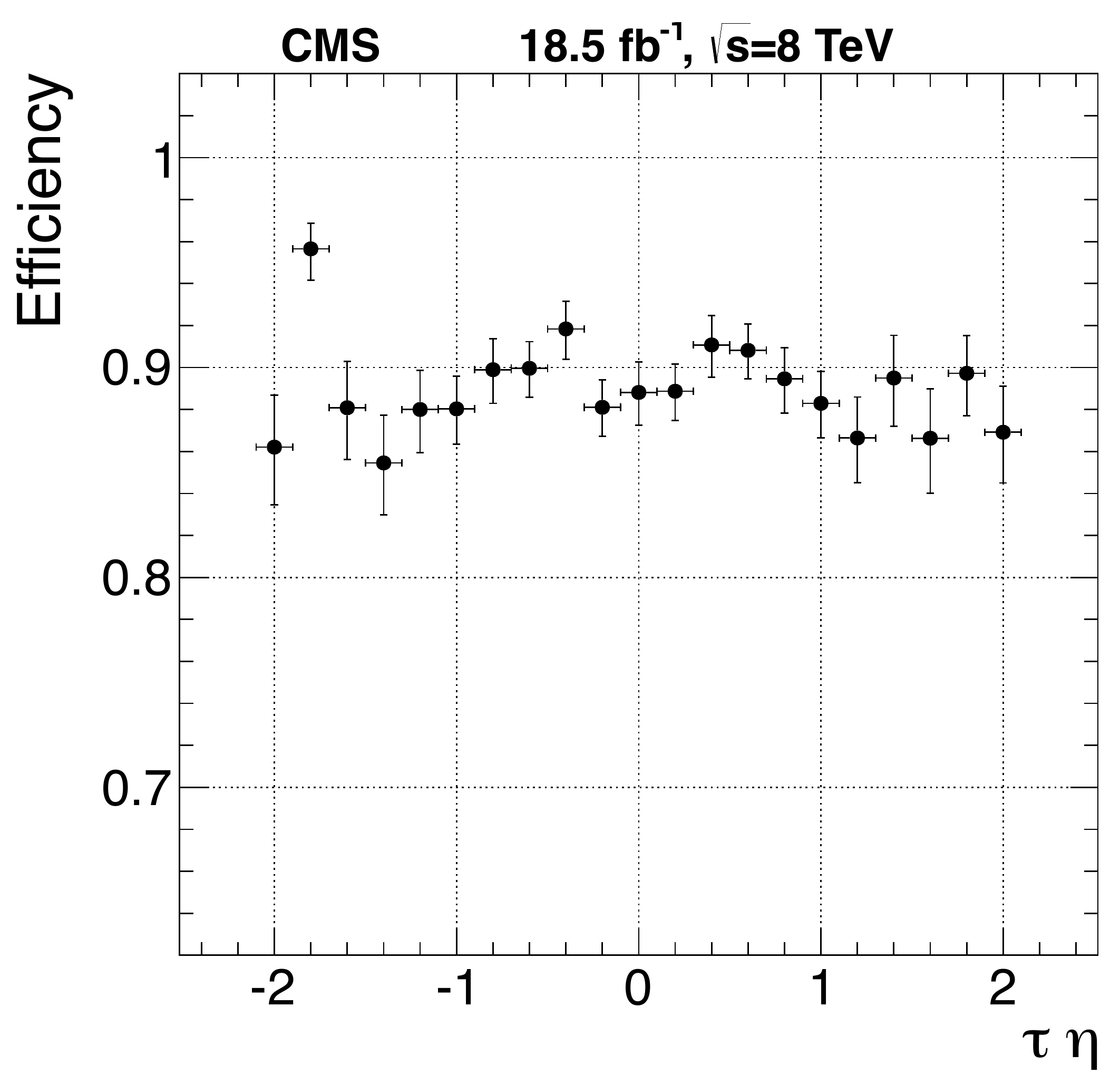}
    \caption{Efficiency of the L2 and L2.5 $\tau$ trigger with a
      35\GeV threshold as a function of the offline reconstructed
      $\tau$ transverse momentum (left) and pseudorapidity (right). }
    \label{figure:L2TauEff}
\end{figure}

The efficiency for the L2/L2.5 stages of the $\tau$ trigger with a
transverse momentum threshold of 35\GeV is shown in
Fig.~\ref{figure:L2TauEff}. The efficiency reaches a plateau of
$93.2$\% at 55\GeV.

\begin{figure}[tbph]
  \centering
    \includegraphics[width=0.32\textwidth]{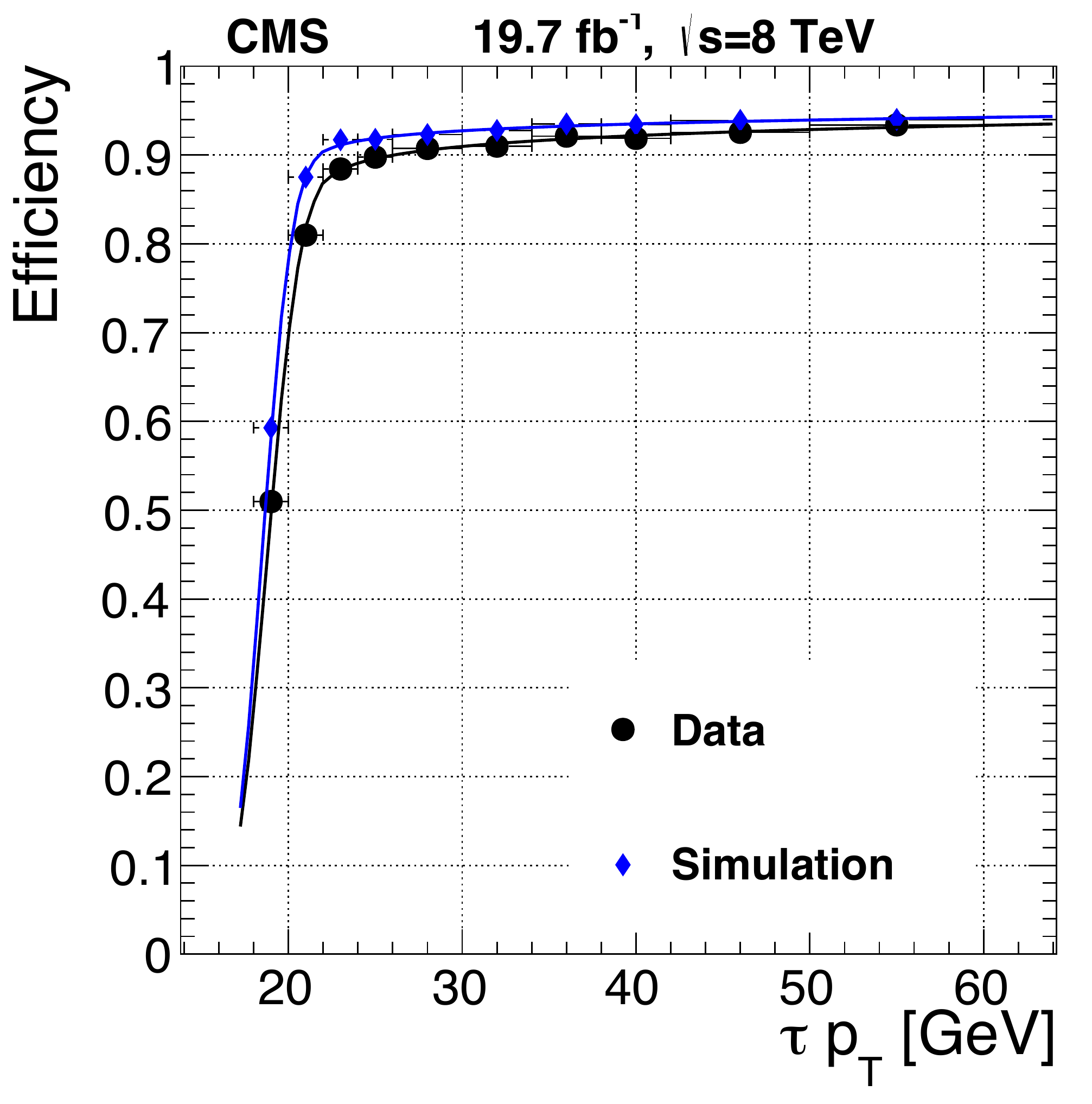}
    \includegraphics[width=0.32\textwidth]{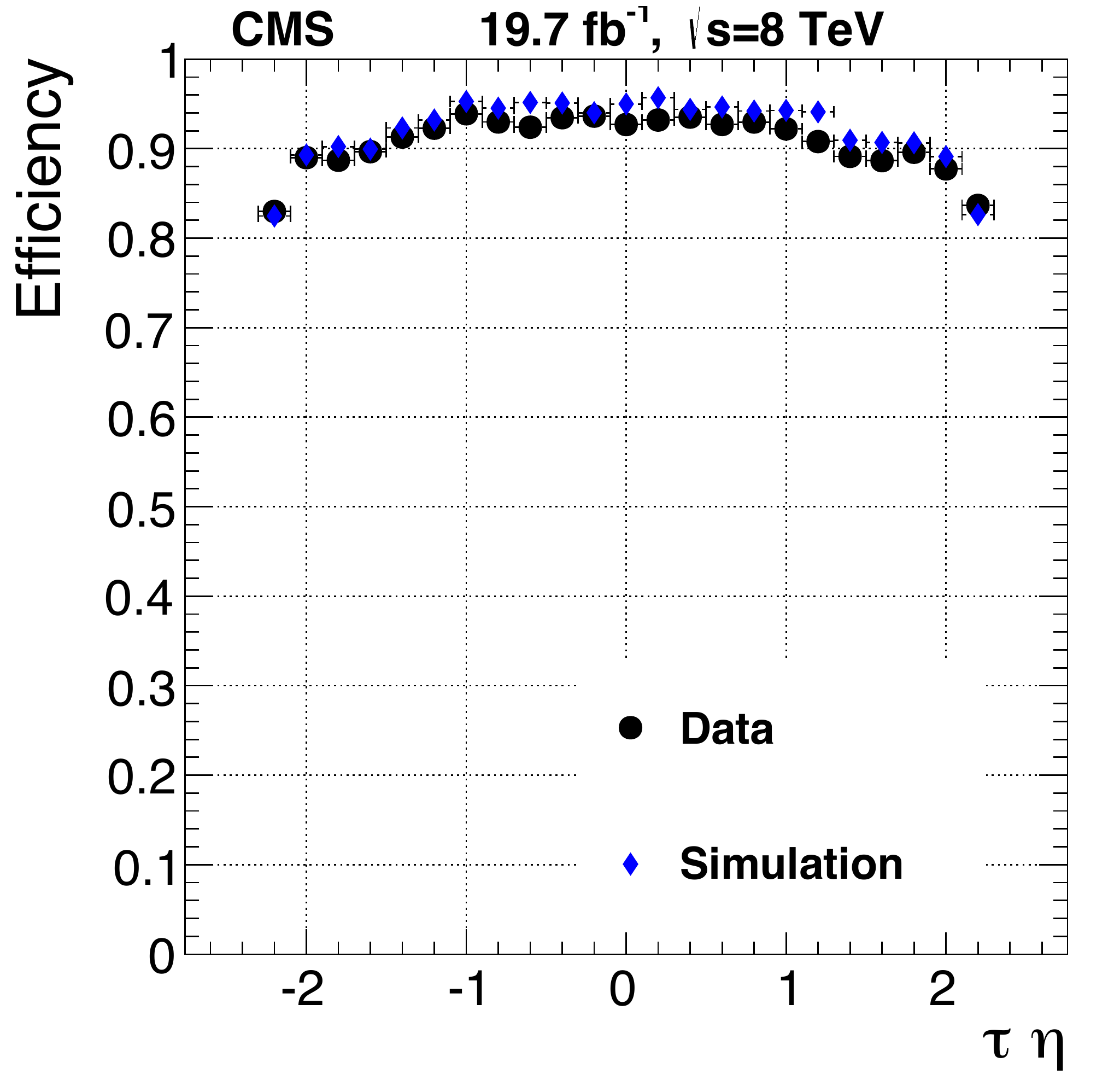}
    \includegraphics[width=0.32\textwidth]{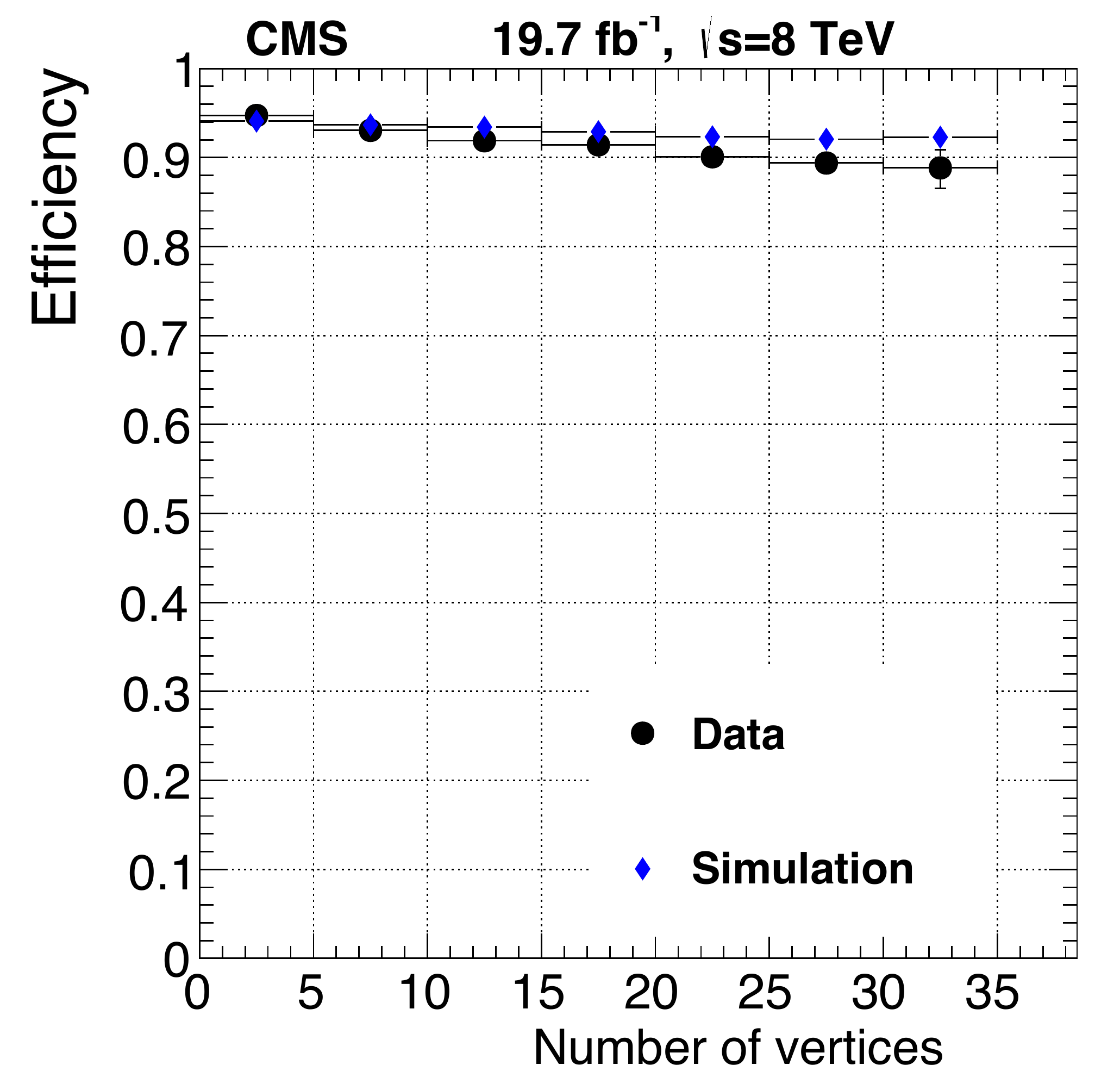}
    \caption{Efficiency of the loose L3 $\tau$ algorithm from the
      $\tau_h\tau_\mu$ events  plotted as a function of
      offline $\tau_h$ transverse momentum (left),
      pseudorapidity (center), and number of vertices (right). }
    \label{figure:L3TauEff}
\end{figure}

For the L3 efficiency measurement, a slightly different event
selection is applied: $\Z\to\tau\tau_\ell$ events
(with $\ell =  \Pe$ or $\mu$) are selected with a
muon-plus-$\ETmiss$ or a single-electron trigger.
Tight isolation on the
electron/muon and $M_\mathrm{T}<20\GeV$, measured between the
electron/muon and the missing energy, are also required. The purities
after this selection are 78\% and 65\% for
$\abs{\eta_{\tau_\mathrm{h}}}<1.5$ and $1.5<\abs{\eta_{\tau_\mathrm{h}}}<2.3$, respectively.
The event samples used to calculate the efficiencies in the simulation
are mixed with simulated \PW+jets events to produce a compatible purity.
The efficiency for the L3 $\tau$ trigger with a 20\GeV threshold is
shown in Fig.~\ref{figure:L3TauEff}. The efficiency reaches a plateau
of 90\% very quickly at about 22\GeV. The $\tau_\mathrm{h}\tau_\mathrm{h}$
triggers use the tight working point. This event topology is
dominated by multijet background. The tighter working point
substantially reduces the rate and provides an efficiency of 80\% on
the plateau. In offline analyses the efficiency of the simulation is
corrected as a function of the transverse momentum to match the
efficiency measured in data events.

In summary, the $\tau$ HLT is used in a variety of very important
physics analyses, including standard model Higgs boson searches. These
analyses combine the $\tau_h$ trigger algorithms described
above with other HLT objects, such as electrons, muons, and
missing transverse energy. These analysis have efficiencies as high
as 90\% while maintaining a manageable HLT rate.

\subsection{ b-quark jet tagging}
\label{sec:BTag}

Many important processes studied at the LHC contain jets
originating from b quarks. The precise identification of b jets
is crucial to reduce the large backgrounds. In CMS, this background
can  be suppressed at the HLT by using b tagging algorithms,
giving an acceptable trigger rate with large signal
efficiency.

The b tagging algorithms exploit the fact that B hadrons typically have
longer decay lifetimes than
 the hadrons made of light or charm quarks.
As a consequence, their decay product tracks and vertices
are significantly displaced from the
primary vertex. Similarly, B hadrons decay more frequently to final
states with leptons than their light-flavor counterparts.

The track counting (TC) and combined secondary vertex (CSV) algorithms
used for offline b tagging~\cite{CMS-PAS-BTV-12-001} are adapted to be
used at the HLT to trigger events containing jets originating from b
quarks. The TC algorithm uses the impact parameter
significance of the tracks in the jets to discriminate between jets
originating from b quark jets
from other flavors.
Combined information on impact parameter
significance of the tracks and properties of reconstructed secondary
vertices in the jets are combined in a multivariate discriminant in
the CSV algorithm.

The choice of which b tagging is used in a particular HLT path depends
on timing requirements. A compromise has to be found to keep the CPU
usage and trigger rates at low levels while keeping the trigger
efficiency as high as possible. Therefore, online b tagging techniques
were designed to be very flexible, allowing the use of not only
different algorithms, but also input objects, namely primary vertex and
tracks, reconstructed with different methods. The b tagging
algorithms depend on the primary vertices found via the fast primary
vertex algorithm described in Section~\ref{sec:primvtx}.

\subsubsection{Tracking for b tagging}
\label{subsec:BTagTrack}

Three tracking reconstruction methods are available at the HLT
(Section~\ref{sec:HLTTrack}) and are used for b tagging: pixel,
regional, and iterative tracking.

The reconstruction of pixel tracks is very fast; however, the
performance is limited. Thus, the pixel tracks are essentially only
used in online b tagging using TC algorithms with jets reconstructed
from energy deposits in the calorimeter at an intermediate step
(L2.5) of the trigger paths. At L2.5 the b tagging discriminant
thresholds are typically loose with the exclusive aim to reduce the
input rates to the slower, but better performing, reconstruction of
regional tracks. The regional tracks are used as input to b tagging
at a later step, called L3, of event triggering. Paths using online
PF jets have tracks reconstructed with the high-performance
iterative tracking, which can be used by both online algorithms.

\subsubsection{Performance of online b-tagging}
\label{subsec:BTagPerf}

\begin{figure}[tbhp]
  \centering
  \includegraphics[width=0.6\textwidth]{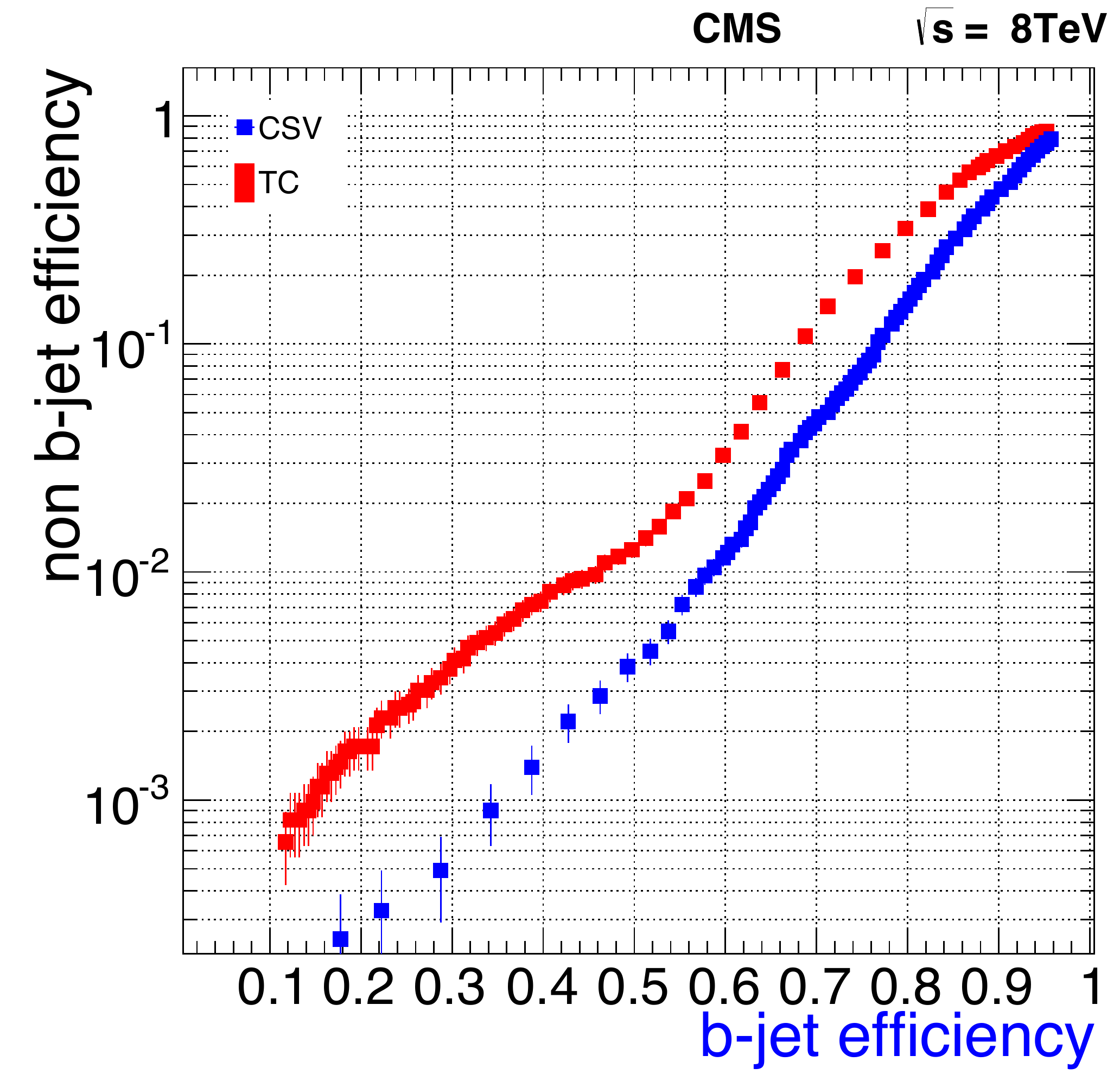}
    \caption{The efficiency to tag b quark jets versus the mistag rate,
      obtained from Monte Carlo simulations, for the track counting (TC)
      and  for the combined secondary vertex (CSV) algorithms. As expected from
      offline  studies, the CSV algorithm performs better than the
      TC algorithm.}
    \label{figure:btag_performance}
\end{figure}
\begin{figure}[tbp]
  \centering
    \includegraphics[width=0.6\textwidth]{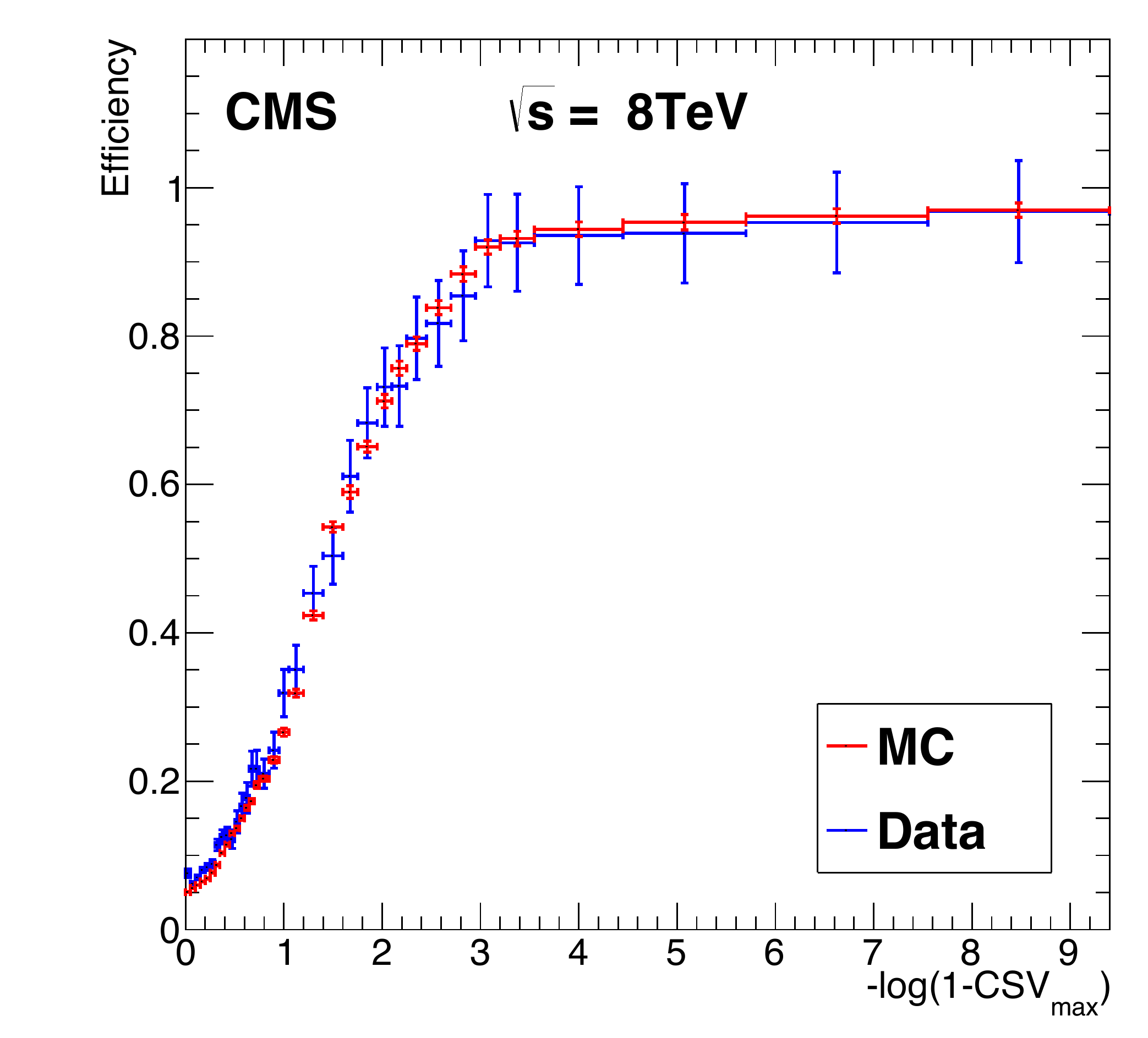}
    \caption{The efficiency of the online CSV trigger as a function of
      the offline CSV tagger discriminant, obtained from the
      data and from Monte Carlo simulations. Good agreement between
      the two is observed.}
    \label{figure:btag_efficiency}
\end{figure}
The performance of the online b tagging at the HLT is illustrated
in Figs.~\ref{figure:btag_performance} and
\ref{figure:btag_efficiency}. Figure~\ref{figure:btag_performance} shows
the efficiency to tag b quark jets versus the mistag rate, obtained from
Monte Carlo simulations, for both algorithms. As expected
from studies of the performance of the algorithms used offline, the
CSV algorithm performs better than the TC algorithm.

The efficiency of the online CSV trigger as a function of the offline
CSV tagger discriminant, obtained from the data, is shown in
Fig.~\ref{figure:btag_efficiency} for a trigger path with selections
on central PF jets with $\ET>30\GeV$ and $\MET>80\GeV$ relative to
an identical (prescaled) trigger path without the b tagging part.
The data are a \ttbar-enriched control region (requiring at least
three jets and at least one isolated lepton).
This defines the denominator
of the efficiency ratio.
The numerator additionally applies a requirement such that
$\epsilon_\text{CSV}>70\%$ for the b tagging discriminant. For the
simulation studies, a sample of \ttbar events is used with the same
selection. The choice to use $-\ln(1-\mathtt{CSV})$ in the $x$-axis
is because on the fact that the distribution of the CSV discriminant
is limited to the range between zero and unity, and  peaks at unity. This
choice makes it possible to visualize the turn-on behavior. A typical
requirement of $\text{CSV}>0.9$ corresponds to 2.3 on the $x$ axis.

\subsection{Heavy ion triggers}
\label{sec:hinobj}

The running conditions for PbPb collisions are significantly different
from the pp case. The instantaneous luminosity delivered by the LHC in
the 2010 (2011) PbPb running periods was $3\times 10^{25}$
($5\times 10^{26}$)\percms resulting in maximum interaction rates of
250\unit{Hz} (4\unit{kHz}), much lower than in pp running, with a negligible pileup
contribution and an inter-bunch spacing of 500~ns (200~ns). During the
pPb run in 2013 an instantaneous luminosity of
$10^{29}\percms$ was achieved, corresponding to an interaction
rate of 200\unit{kHz}, again with a very low pileup contribution.

In PbPb collisions, the number of produced particles depends strongly
on the geometrical overlap of the Pb ions at the time of the
collisions. The number of charged particles produced per unit of
pseudorapidity, $\rd{}N_\mathrm{ch}/\rd\eta$, varies from event to event from
${\approx}$10 for glancing collisions to ${\approx}1600$ for head-on
collisions. The large particle multiplicity of head-on collisions
leads to very high detector occupancies in the inner layers of the
silicon tracker. For such high detector occupancies the hardware-based
zero-suppression algorithm implemented in the front-end-drivers (FED)
of the tracker does not function reliably. As a consequence the
tracker had to be read out without hardware zero suppression and the
zero suppression was performed offline in 2010 and in the HLT in
2011. Table~\ref{tab:ionrunning} shows a summary of the conditions in
various heavy ion running periods.

A consequence of reading out the tracker without zero suppression is the
limited data throughput from the detector
due to the large event size. This limits the readout rate of the
detector to 3\unit{kHz} in PbPb collisions. The limit has to be taken into
account when setting up the
trigger menu for HI collisions.

\begin{table}[tbp]
\topcaption{Summary of the heavy ion running conditions in various
  data-taking periods.}
\label{tab:ionrunning}
\centering
\begin{tabular*}{0.75\textwidth}{@{\extracolsep{\fill}}|cccc|}
\cline{1-4}
Run period & Ion species ($\sqrt{s_{\rm NN}}$) & Max. collision rate & Zero suppression \\
\cline{1-4}
Winter 2010 & PbPb (2.76\TeV) & 200\unit{Hz} & Offline \\
Winter 2011 & PbPb (2.76\TeV) & 4500\unit{Hz} & HLT \\
Winter 2013 & pPb (5.02\TeV) & 200\unit{kHz} & FED \\
\cline{1-4}
\end{tabular*}
\end{table}

The HI object reconstruction is based on the pp HLT reconstruction
algorithms described in the previous sections. The physics objects or event
selection criteria used in the trigger menu are the following:
\begin{itemize}
\item Hadronic interactions (minimum bias);
\item Jets;
\item Photons;
\item Muons;
\item High-multiplicity events.
\end{itemize}
In the following we discuss the differences between the algorithms
used in pp running to those used offline, and the performance efficiencies of
these algorithms in the PbPb case.

\textbf{Hadronic interactions.} Since the interaction probability per
bunch crossing during HI data taking is only $\approx10^{-3}$,
it is necessary to deploy a dedicated trigger to select hadronic
interactions. This selection is based on coincidences between the
trigger signals from the $+z$ and $-z$ sides of either beam
scintillation counters (BSCs) or the
HF which cover a pseudorapidity range of $2.9<\abs{\eta}<5.2$. This
trigger has a selection efficiency of more than 97\% for hadronic
inelastic collisions and is thus also referred to as a ``minimum
bias'' trigger. The selection efficiency of this trigger was
determined using a MC simulation of HI events using the
{\textsc{hydjet}} event generator \cite{Lokhtin:2005px} and was
cross-checked with a control data sample
selected using the BPTX signal to identify crossing beam bunches. The
event sample selected this way is referred to as ``zero bias'' sample.
From the zero-bias sample, inelastic events can be selected by
requiring a charged-particle track consistent with originating from
the beam crossing region. The fraction of the zero bias sample selected using the minimum bias trigger is consistent with
the selection efficiency determined from simulated events.

\textbf{Jets.} The jet reconstruction algorithm used for HI data taking
closely follows the corresponding pp algorithm which reconstructs
calorimeter-based jets as described in Sections~\ref{sec:l1jet} and
\ref{sec:JetHLT}, with the addition of a step subtracting the
high-multiplicity underlying event using the iterative pileup
subtraction technique \cite{Kodolova:2007hd}.  During the 2010 and
2011 HI data-taking periods an iterative-cone type algorithm was used
for jet clustering.

\begin{figure}[tph]
  \centering
    \includegraphics[width=0.49\textwidth]{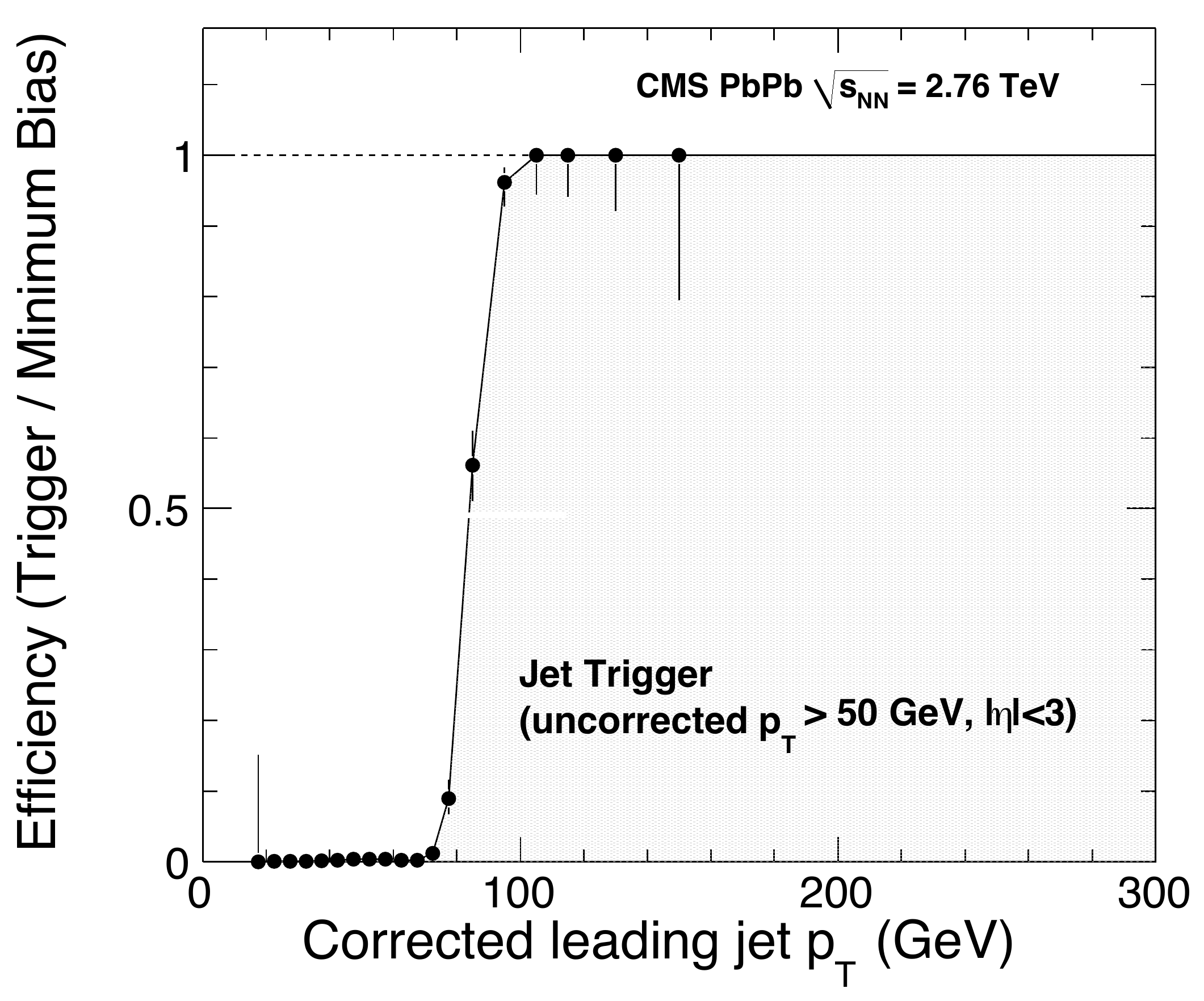}
    \includegraphics[width=0.49\textwidth]{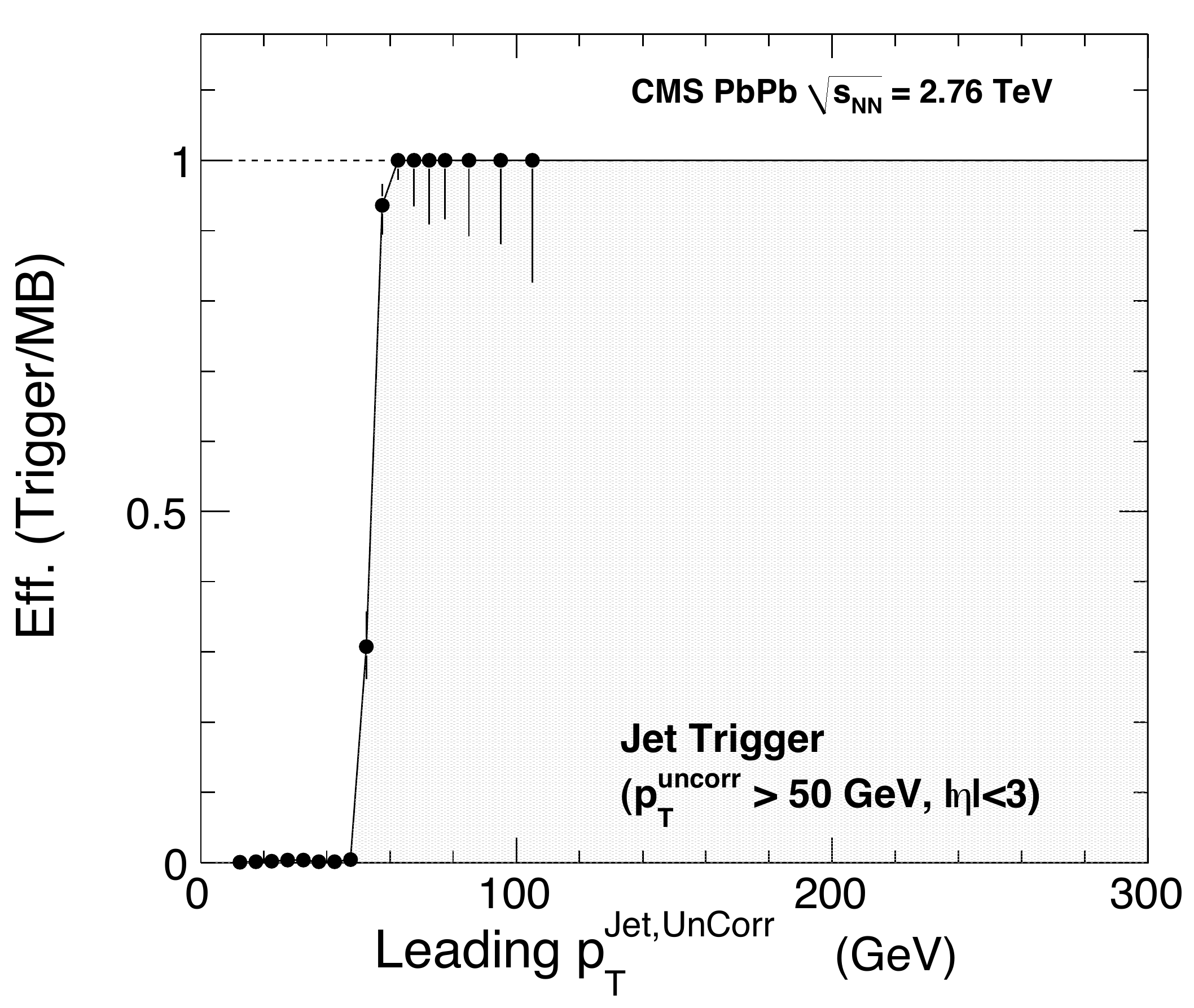}
    \caption{Efficiency curves for the Jet50U trigger in PbPb at $\sqrt{s_\mathrm{NN}}=2.76\TeV$, as a function of
the corrected (left) and uncorrected (right) leading jet transverse
momentum.}
    \label{fig:Jet50U}
\end{figure}

The efficiency of the jet triggers deployed for the 2010 PbPb
run is illustrated in Fig.~\ref{fig:Jet50U} by the efficiency turn-on
curve of the Jet50U trigger. This trigger was discriminating events
based on uncorrected jet energies. The efficiencies are given as a
function of leading-jet transverse momentum for offline-corrected (left)
and for uncorrected jets (right). The given efficiencies were determined
based on offline jets reconstructed using the iterative cone algorithm with pileup
subtraction in a data sample collected using a minimum bias trigger.
The efficiency is defined as the fraction of the minimum bias sample containing a leading jet of a given \pt that is selected by the jet trigger.

\begin{figure}[tph]
  \centering
    \includegraphics[width=0.49\textwidth]{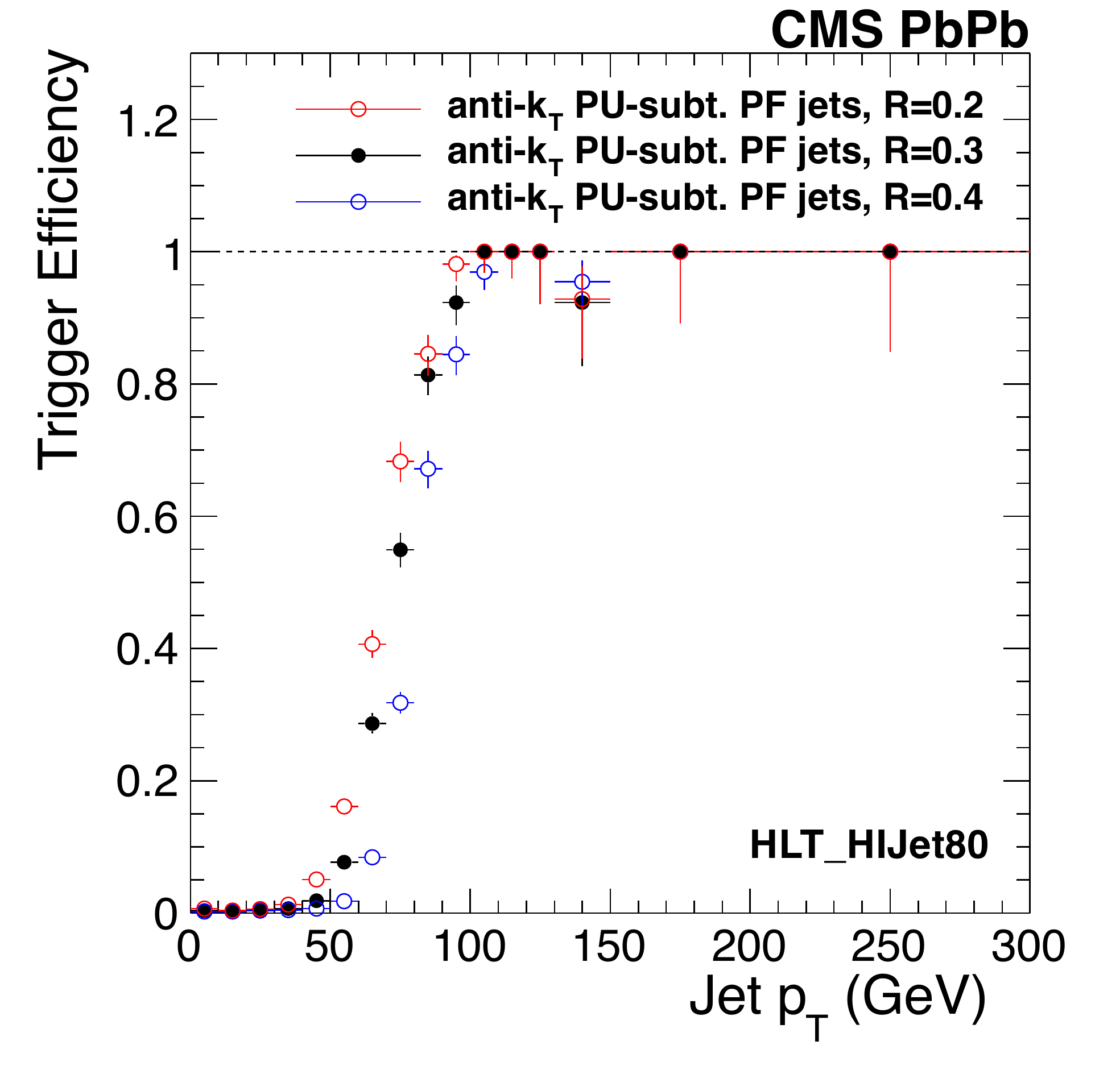}
    \caption{Efficiency curves for the Jet80 trigger in PbPb at
      $\sqrt{s_\mathrm{NN}}=2.76\TeV$, as a function of the leading jet
      transverse momentum in the $\abs{\eta}<2$ region evaluated from
      minimum bias sample. The red, black, and blue points correspond
      to anti-\kt jets with cone size of 0.2, 0.3, and 0.4,
      respectively. }
    \label{fig:Jet80}
\end{figure}

During the 2011 PbPb run, the jet triggers had energy corrections applied
at the HLT, leading to sharper turn-on curves, and thereby to more
efficient data taking. Figure~\ref{fig:Jet80} illustrates the
improvement by showing the efficiency of the Jet80 trigger as a function
of leading-jet transverse momentum in the $\abs{\eta}<2$ region.
The efficiencies are evaluated from a minimum bias sample, as in the 2010 case,
with the jet reconstructed using the anti-\kt algorithm based on PF objects and
also subtracting the underlying event using the iterative pileup subtraction
technique. The efficiencies are given for various cone radii.

\begin{figure}[tbph]
  \centering
    \includegraphics[width=0.407\textwidth]{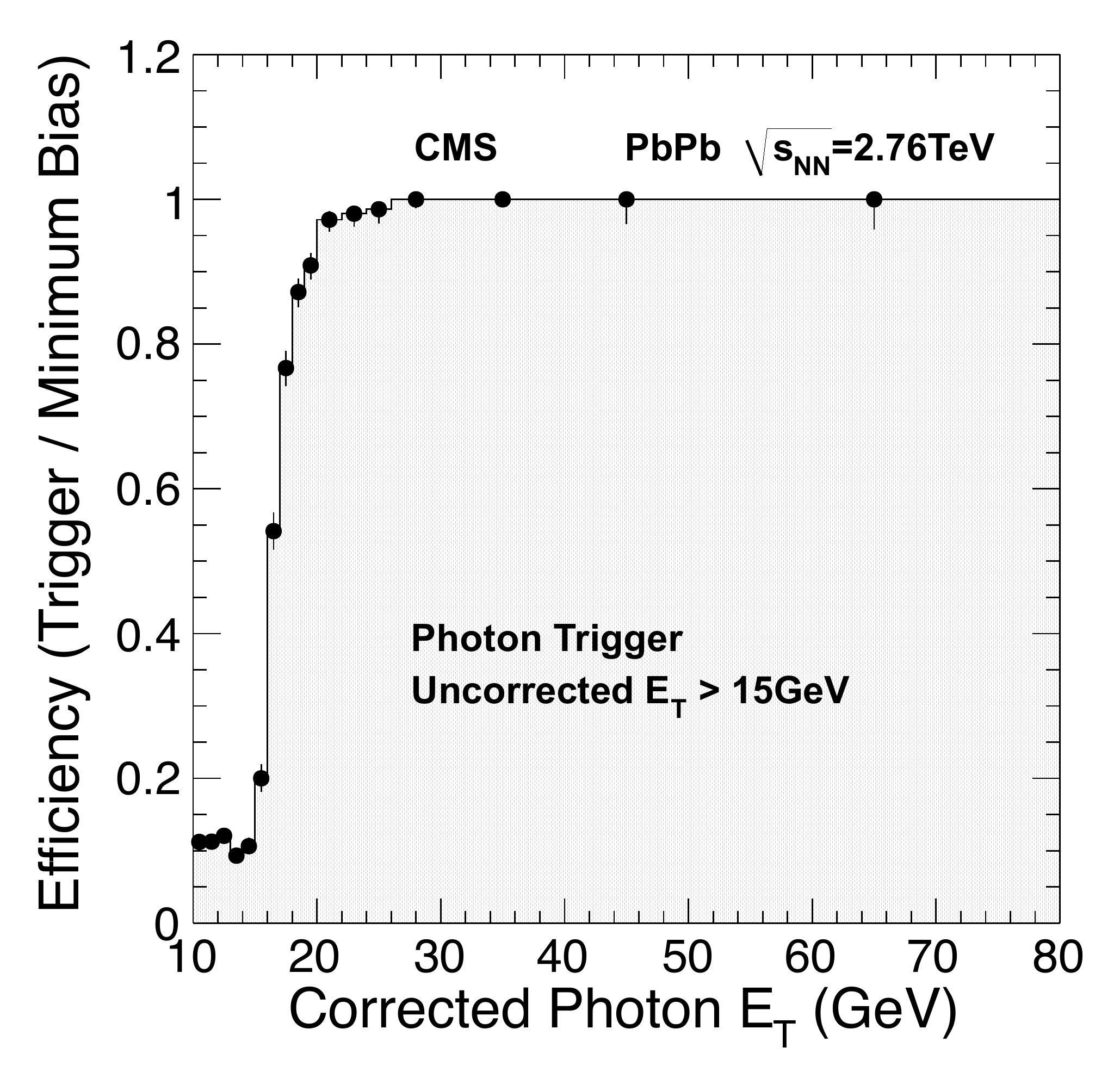}
    \includegraphics[width=0.584\textwidth,trim={0 .3cm 0 0},clip]{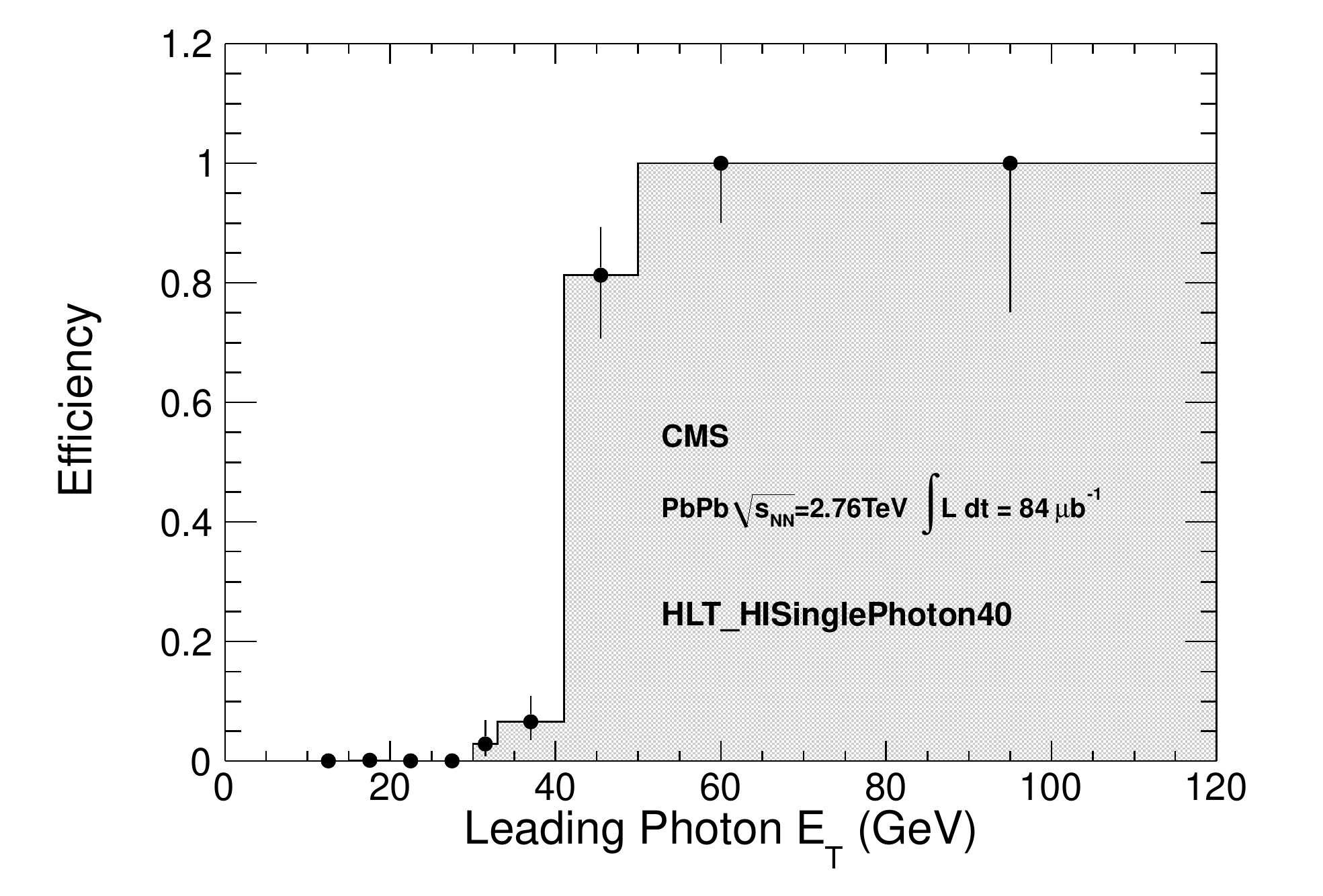}
    \caption{Trigger efficiency of the uncorrected Photon15 (left) and the corrected Photon40 (right)
triggers as a function of corrected offline photon transverse
momentum, in PbPb collisions at $\sqrt{s_{\rm NN}}= 2.76\TeV$.}
    \label{fig:PhotonTriggers}
\end{figure}

\textbf{Photons.} During the 2010 PbPb run the photon triggers employed at HLT were based on the energy
deposits in the ECAL reconstructed using the island clustering algorithm~\cite{cms-e7}. This is the same algorithm as
 used for offline analyses based on the 2010 data, but without energy correction already
applied at HLT. The trigger efficiency for the uncorrected Photon15 trigger for minimum bias events is
shown in the left  panel of Fig.~\ref{fig:PhotonTriggers}.

For the data taking of 2011, energy corrections were already applied in the HLT. The performance
of such corrected HLT photon paths is illustrated in the right panel of Fig.~\ref{fig:PhotonTriggers},
which shows the efficiency turn-on curve for the Photon40 trigger,
again determined with respect to minimum bias events.

\textbf{Muons.} Efficient triggering on high-\pt muons
is of primary importance for the HI physics program in CMS.
During data-taking both single- and double-muon triggers were
deployed to allow for maximal flexibility in event selection.

\begin{figure}[tbph]
  \centering
    \includegraphics[width=0.32\textwidth]{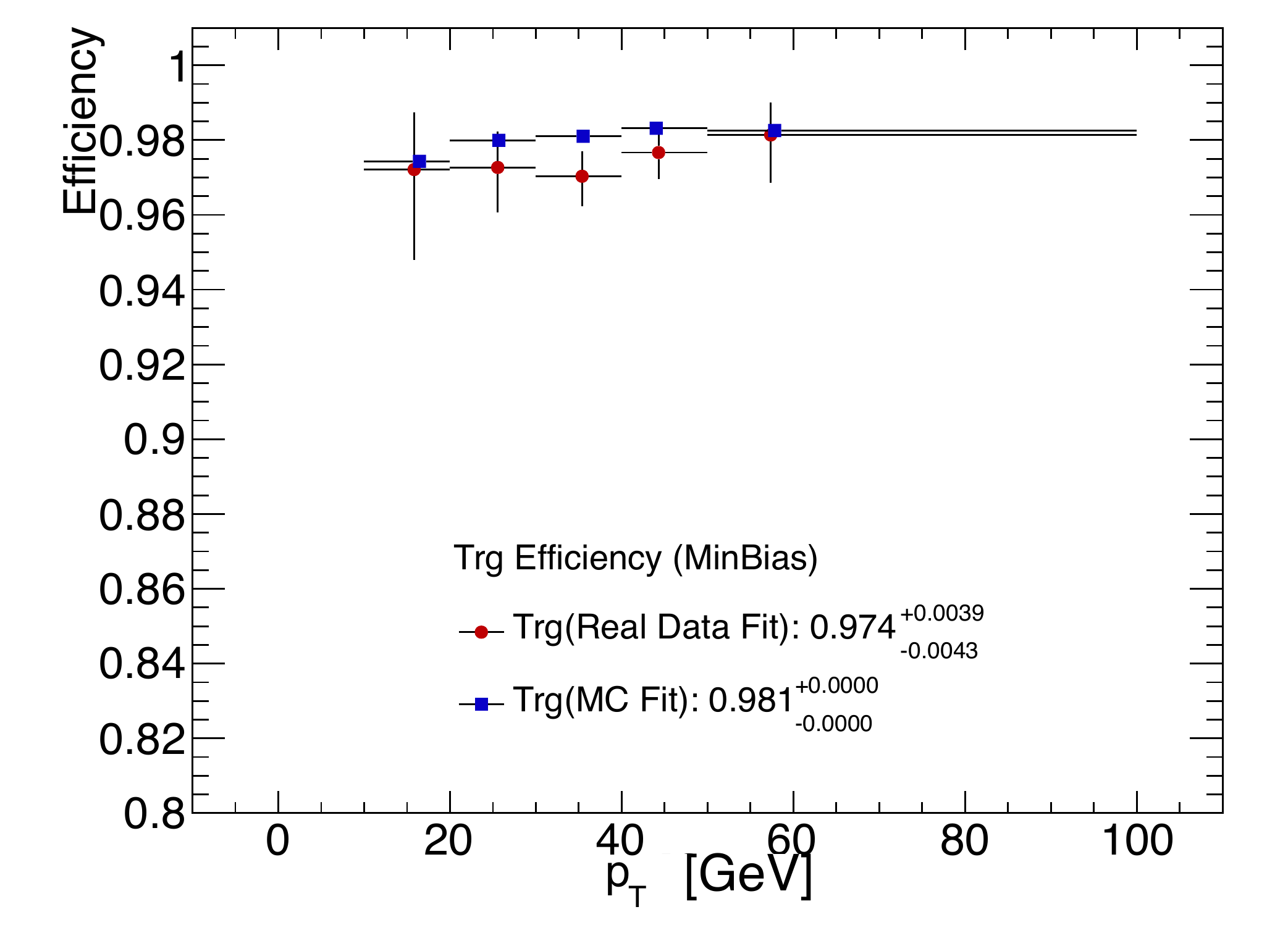}
    \includegraphics[width=0.32\textwidth]{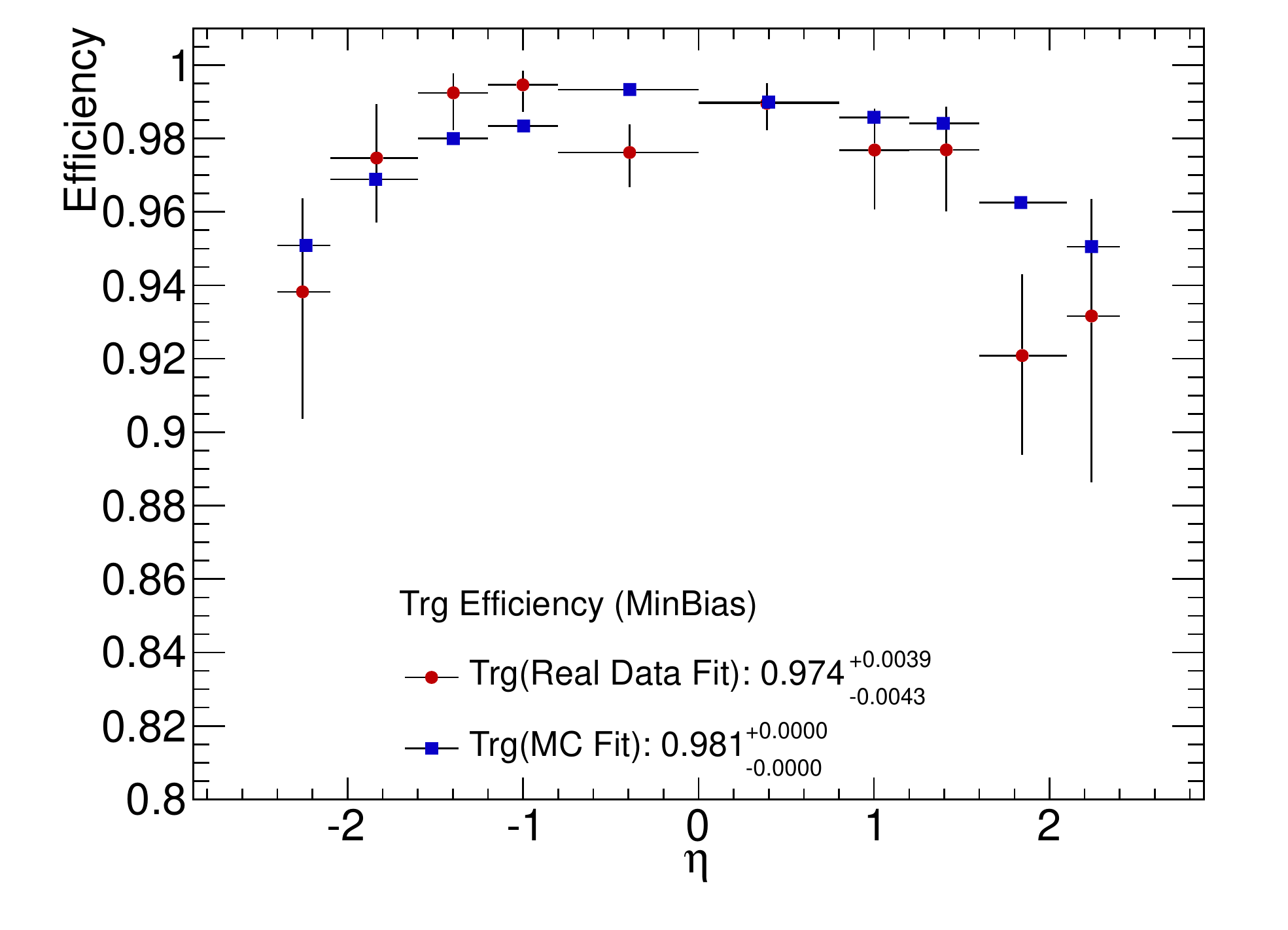}
    \includegraphics[width=0.32\textwidth]{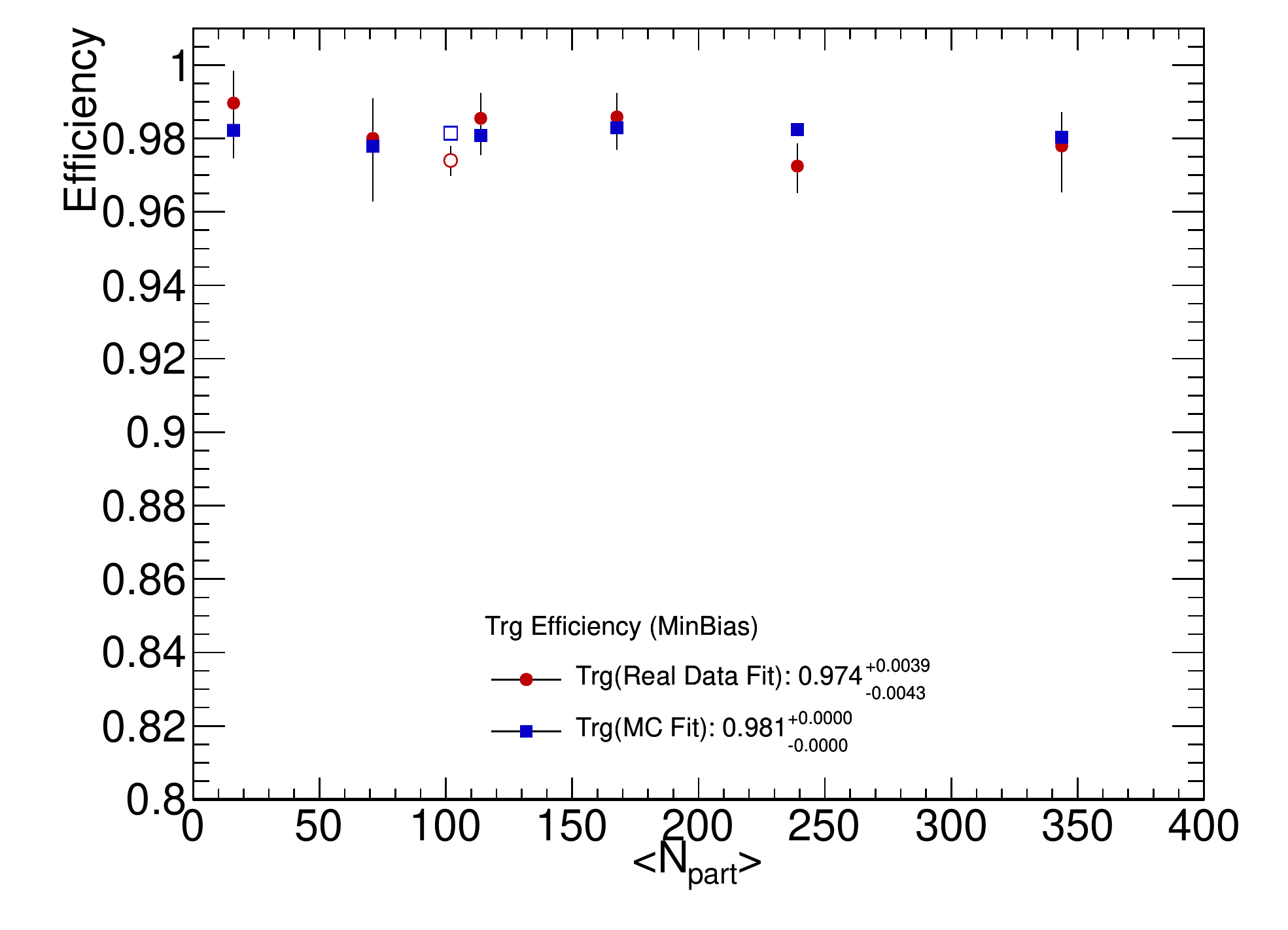}
    \caption{Per-muon triggering efficiency of the HLT HI double-muon
      trigger as a function of \pt (left), $\eta$ (center), and average
      number of participant nucleons (right).  \cPZ\xspace bosons in
      data (red) are compared to  simulated \cPZ\xspace bosons embedded in HI
      background simulated with \textsc{hydjet} (blue).}
    \label{fig:DoubleMu3}
\end{figure}

The per-muon trigger efficiency of the double-muon trigger (which
requires two muons with $\pt> 3\GeV$) in the 2011 PbPb data
determined by a tag-and-probe method is shown in
Fig.~\ref{fig:DoubleMu3}. The three panels show the efficiency as a
function of transverse momentum, pseudorapidity, and the overlap
between the two colliding nuclei, expressed by the ``number of
participants.'' Data are shown in red and simulated Z bosons
embedded in {\sc hydjet} background are shown in blue. On average, the
trigger efficiency is very good, reaching 98.2\% as obtained from
tag-and-probe with simulated data.

\begin{figure}[tbph]
  \centering
    \includegraphics[width=0.32\textwidth]{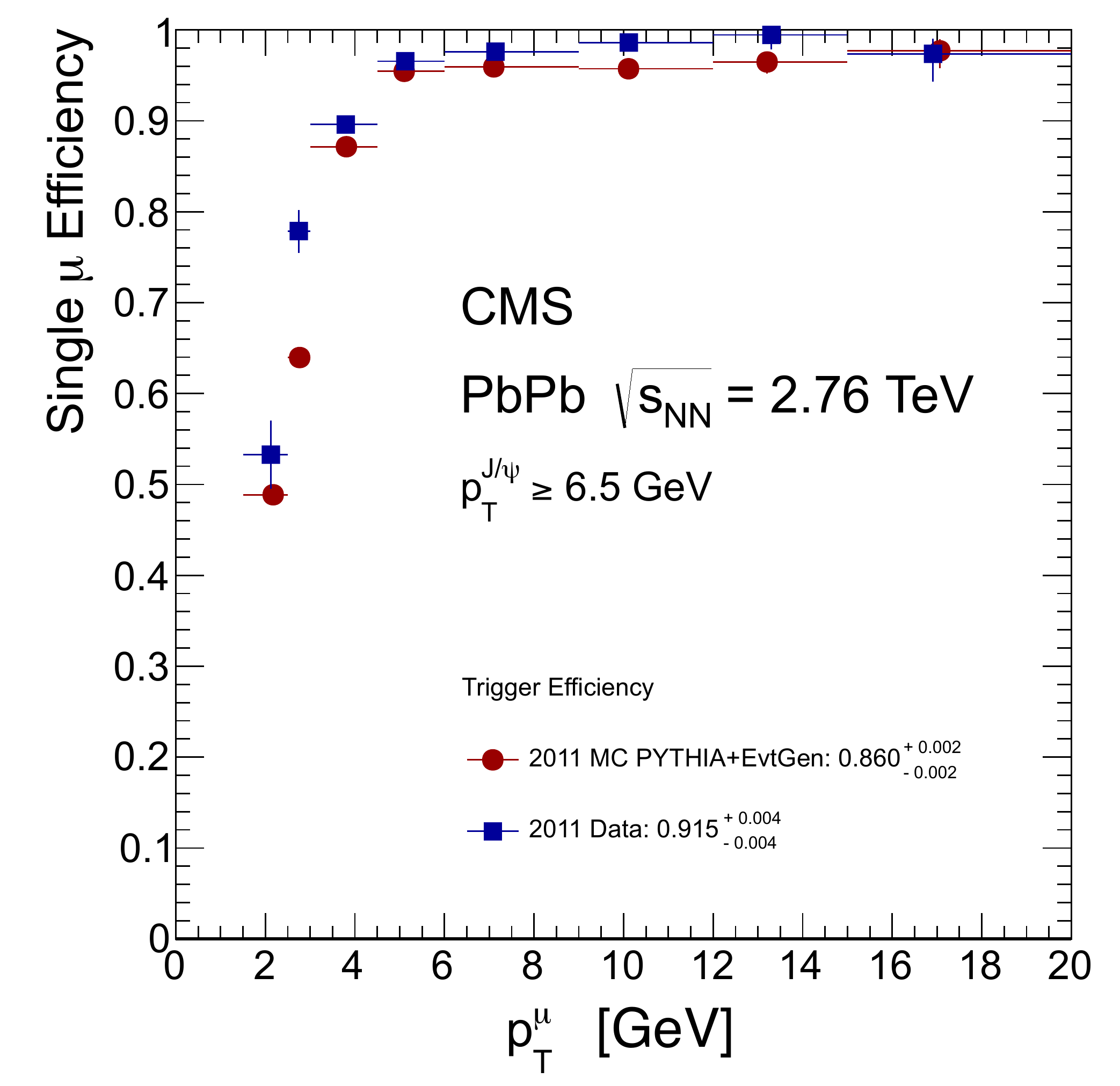}
    \includegraphics[width=0.32\textwidth]{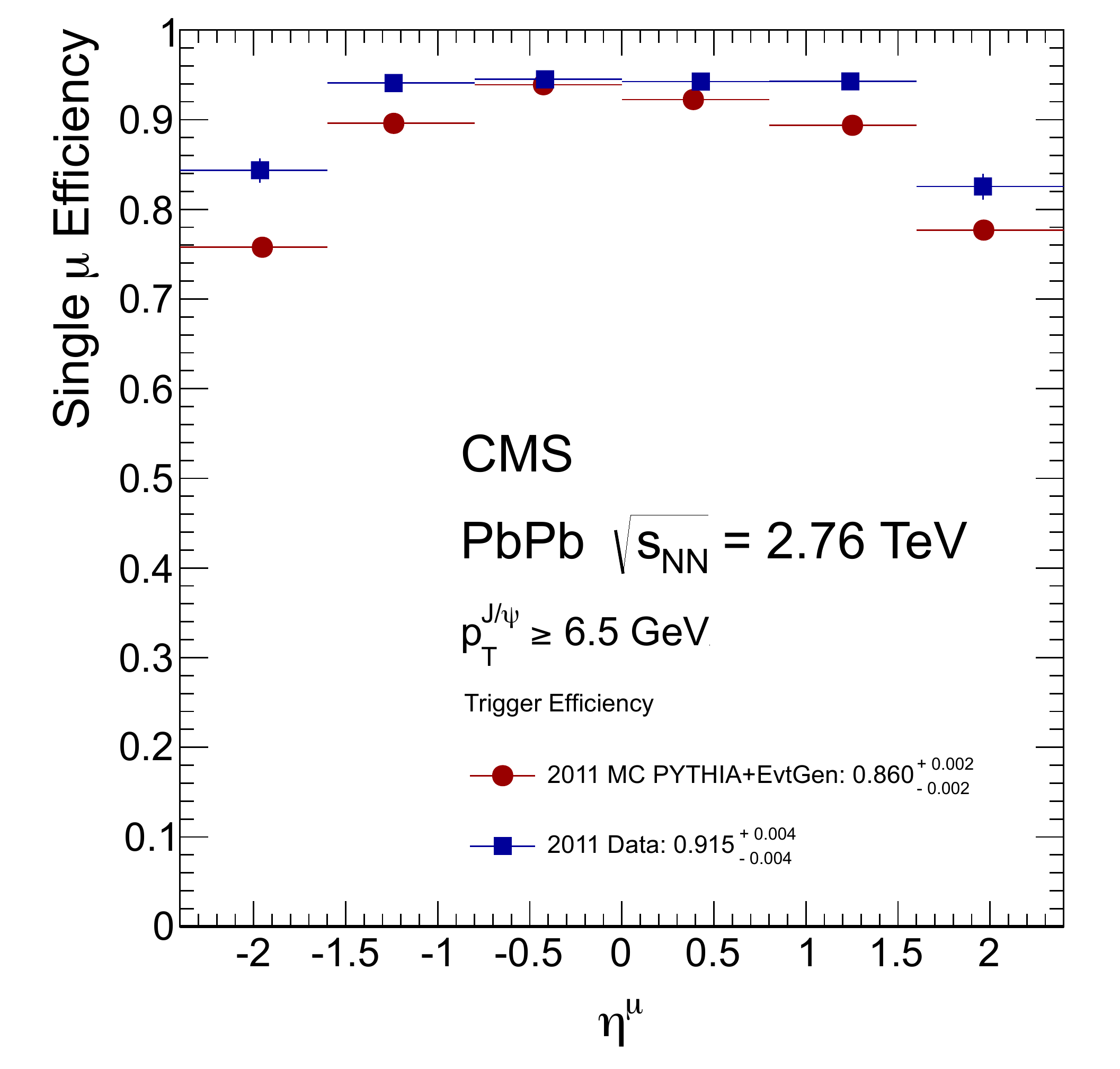}
    \includegraphics[width=0.32\textwidth]{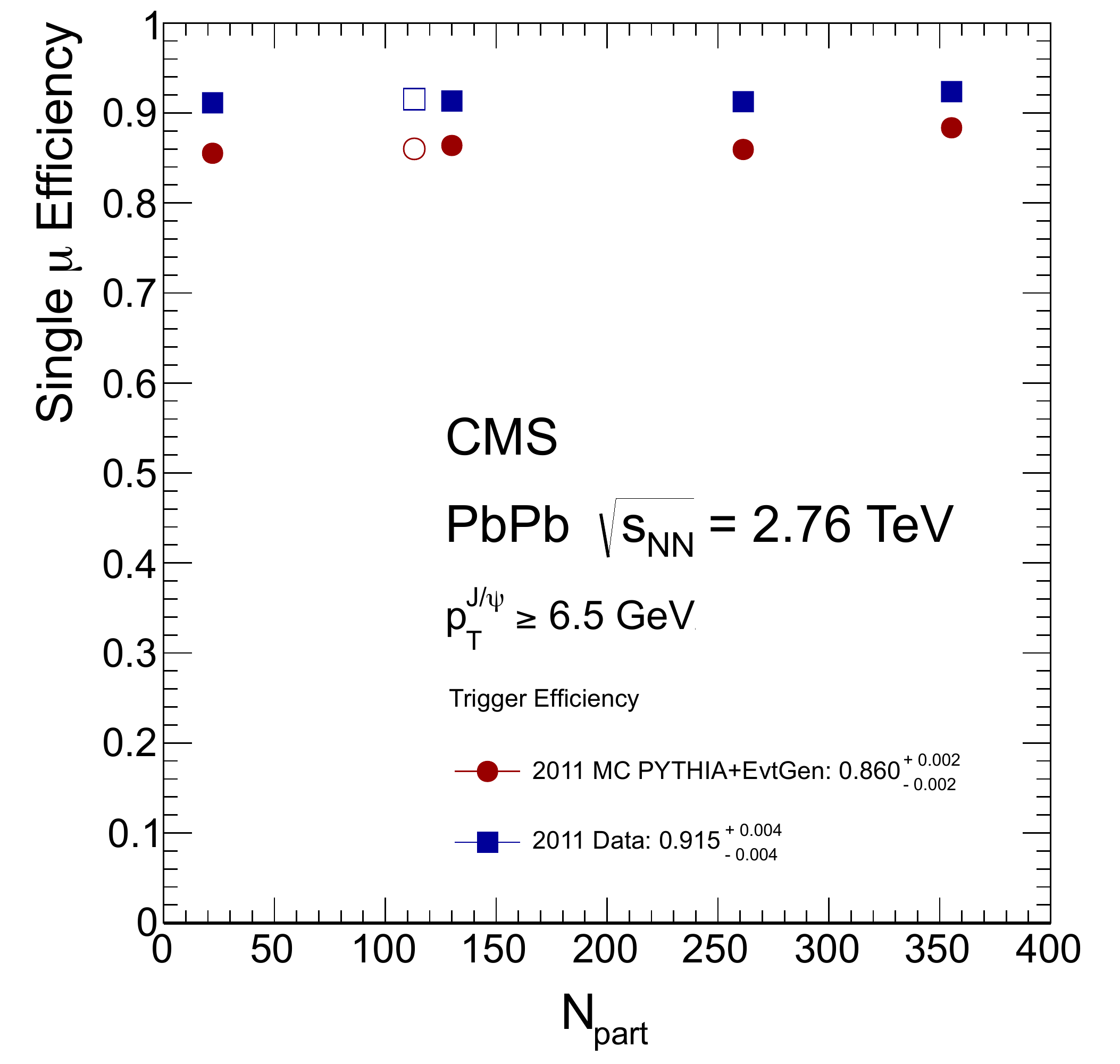}
    \caption{Single-muon trigger efficiencies as functions of probe
muon transverse momentum, pseudorapidity, and number of participants in the 2011 PbPb data.
Red full circles are simulation and blue full squares are data. The numbers quoted in the legends of the figures are the integrated efficiencies.}
    \label{fig:DoubleMuOpen}
\end{figure}

The single-muon trigger efficiencies for the daughters of $\JPsi$
mesons with $\pt >6.5$\GeV in the 2011 PbPb data as a function of
transverse momentum, pseudorapidity, and the number of participants
are shown in the various panels of Fig.~\ref{fig:DoubleMuOpen}.  The
\pt and $\eta$ integrated trigger efficiency is $86.0\pm0.2\%$ in MC,
and $91.5\pm0.4\%$ in data.  The trigger efficiency shows no
significant dependence on the number of participants, as expected, in
data or simulation (Fig.~\ref{fig:DoubleMuOpen}, right).

\begin{figure}[tbph]
  \centering
    \includegraphics[width=0.49\textwidth]{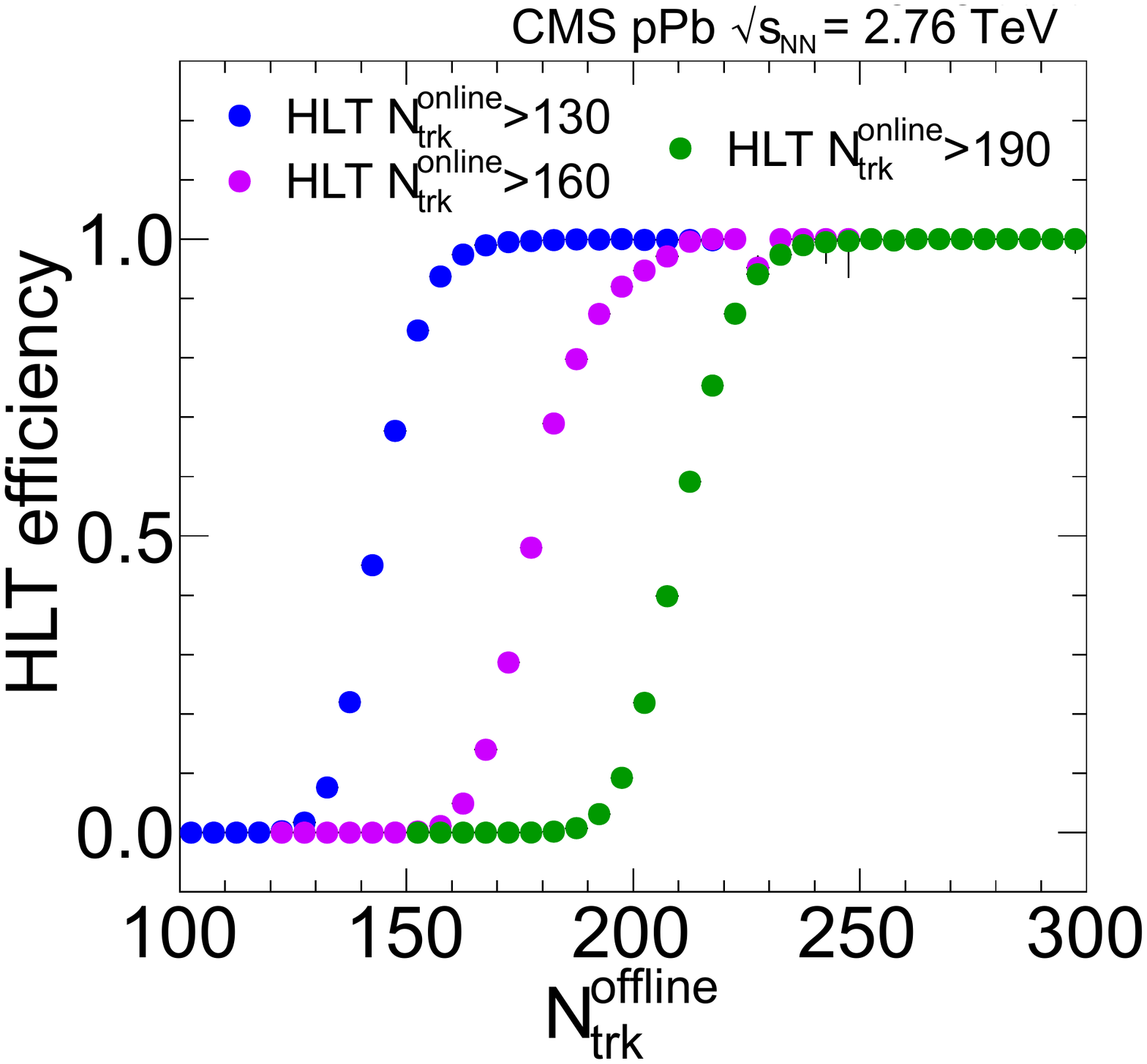}
    \includegraphics[width=0.49\textwidth]{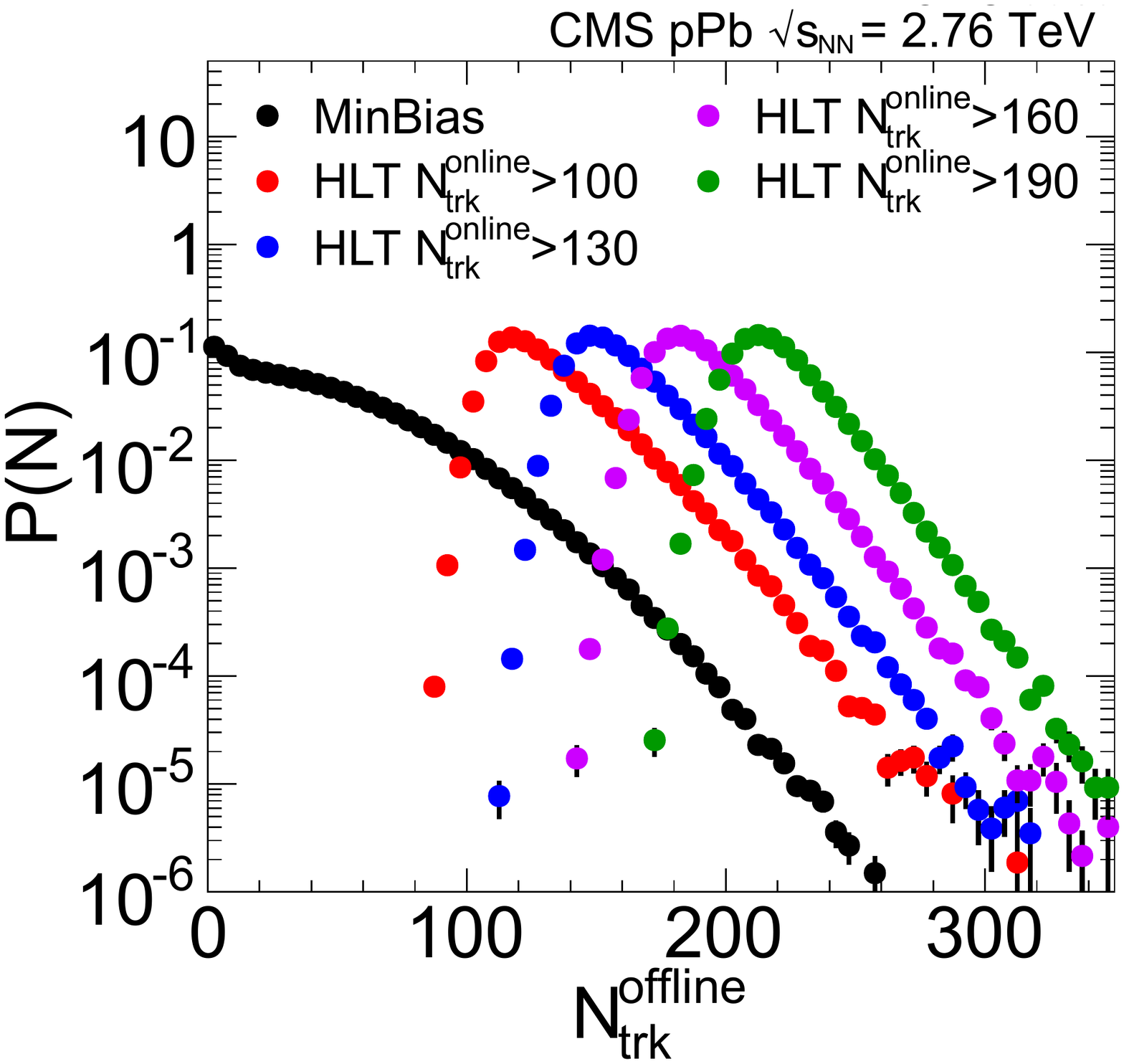}
    \caption{Left: Trigger efficiency as a function of the offline track multiplicity, for the three most
selective high-multiplicity triggers. Right: The spectrum of the offline tracks for minimum bias and
for all the different track-based high-multiplicity triggers in the 2013 pPb data.}
    \label{fig:highMult}
\end{figure}

\textbf{High-multiplicity events.} In order to trigger on
high-multiplicity events, several trigger paths were deployed during
the HI data-taking periods. Triggers based on energy deposits
in the calorimeter systems, signals in the BSC
detectors, as well as triggers based on track multiplicities were
employed and used in supplementary roles.  The efficiency of
high-multiplicity track triggers used during the 2013 pPb run is shown
in the left panel of Fig.~\ref{fig:highMult}. The histograms
correspond to different thresholds of the same kind as for track-based
triggers. The efficiencies are shown as a function of the offline track
multiplicity. The efficiencies are determined using either minimum bias
events or a lower threshold high multiplicity trigger as a
reference. The efficiency is defined as the fraction of events passing
a given trigger threshold in the reference sample and is shown as a
function of number of offline reconstructed tracks. The gain in the
number of high-multiplicity events is demonstrated in the right-hand
side panel of Fig.~\ref{fig:highMult}.

\section{Physics performance of the trigger}
\label{sec:hpa}

In the previous sections, we described the performance of the CMS
trigger system for single- and multi-object triggers. However, most
physics analyses published using the data taken in the first years of the LHC were performed using more complicated triggers. These triggers either take
advantage of different categories of objects, such as a mixture of
jets and leptons, or are topological triggers, which look at the event as
a whole and calculate quantities such as the scalar sum of jet transverse energy \HT in the event or the missing transverse energy. In this section, to illustrate the performance of the trigger system, we give specific examples of some high-priority analyses that CMS carried out based on data taken in 2012, at a center-of-mass energy $\sqrt{s}=8\TeV$.

\subsection{Higgs boson physics triggers}
\label{sec:higpag}

The observation of the Higgs boson~\cite{Chatrchyan:2012ufa,cms-higgs-long-paper} is the most important CMS
result in the first LHC run. In this section, we  discuss the
CMS trigger performance for Higgs boson physics. Single-object
triggers were discussed in Section~\ref{sec:objid}. In
this section, more complex triggers are described. The strategy of
combining different trigger paths to maximize the signal acceptance
for the Higgs boson measurements is also presented.

\subsubsection[Triggers for  Higgs boson diphoton analysis]{$\mathrm{h} \to\gamma\gamma$}

As already discussed in Section~\ref{sec:egammaHLT}, diphoton triggers
have been designed to efficiently collect \HGG\xspace events. To be
as inclusive as possible, any photon that passes the general
identification requirements described in Section~\ref{sec:egammaHLT}, and either
the isolation and calorimeter identification or the \RNINE requirement, is
accepted in the diphoton path. Asymmetric thresholds of 26\GeV on
the leading photon and 18\GeV on the subleading photon have been
applied together with a minimum invariant mass requirement on the diphoton system of
60\GeV. In the very late 2012 data-taking period, a similar path
with more asymmetric \ET requirements was added to the HLT menu to enhance the
discriminating power for the non-standard Higgs boson spin-0 and
spin-2 scenarios. The performance of the trigger was shown in
Figs.~\ref{fig:hlt_26_pt_eta} to~\ref{fig:hlt_26_nvtx}.

\subsubsection[Triggers for multi-lepton Higgs boson analyses]{$\rm H \to ZZ\to 4\ell$}
The four-lepton channel provides the cleanest experimental signature
for the Higgs boson search: four isolated leptons originating from a common
vertex. As the number of expected events is very low, it is necessary to preserve the highest possible signal efficiency. The
analysis performance therefore heavily relies on the lepton
reconstruction, identification efficiency, and, due to the low branching
fraction of the Higgs boson into $\Z\Z$, a robust trigger strategy to avoid any signal loss. The
events are selected requiring four leptons (electrons or muons)
satisfying identification, isolation, and impact parameter requirements
(Sections~\ref{sec:egammaHLT} and~\ref{sec:muHLT}).
The triggers described in this section were instrumental in the Higgs
boson discovery and in the studies of its
properties~\cite{Chatrchyan:2012ufa,Chatrchyan:2013mxa}.

In the following, we will describe the main triggers that are used to
collect  most of the data, as well as a set of utility
triggers used to measure the online selection efficiencies.
The main trigger selects $\PH\to \Z\Z\to 4\ell$
events with an efficiency larger than 95\% for
$m_\mathrm{h} = 125$\GeV, at a rate less than 10\unit{Hz} at an instantaneous luminosity of
$5 \times 10^{33}\percms$. This trigger has loose isolation and identification requirements applied, and these are
critical for proper background estimation. To improve the absolute
trigger efficiency, a combination of single-electron and dielectron
triggers was used. This combination achieved a 98\% overall trigger
efficiency.

For the $\PH\to \Z\Z\to 4\ell$ analysis, a basic set
of double-lepton triggers is complemented by the triple-electron paths
in the 4e channel, providing an efficiency gain of 3.3\% for signal
events with $m_{\PH} = 125$\GeV. The minimum momenta of the first and
second lepton are 17 and 8\GeV, respectively, for the double-lepton
triggers, while they are 15, 8 and 5\GeV for the triple-electron
trigger. The trigger paths used in 2012 are listed in
Table~\ref{tab:anatriggers}, where ``CaloTrk'' stands for calorimeter-
and tracker-based identification and isolation requirements applied
with very loose criteria, while the ``CaloTrkVT'' name denotes
triggers that make use of the same objects as discriminators, but with more stringent requirements placed on them.

\begin{figure}[tbph]
\centering
\includegraphics[width=0.45\linewidth]{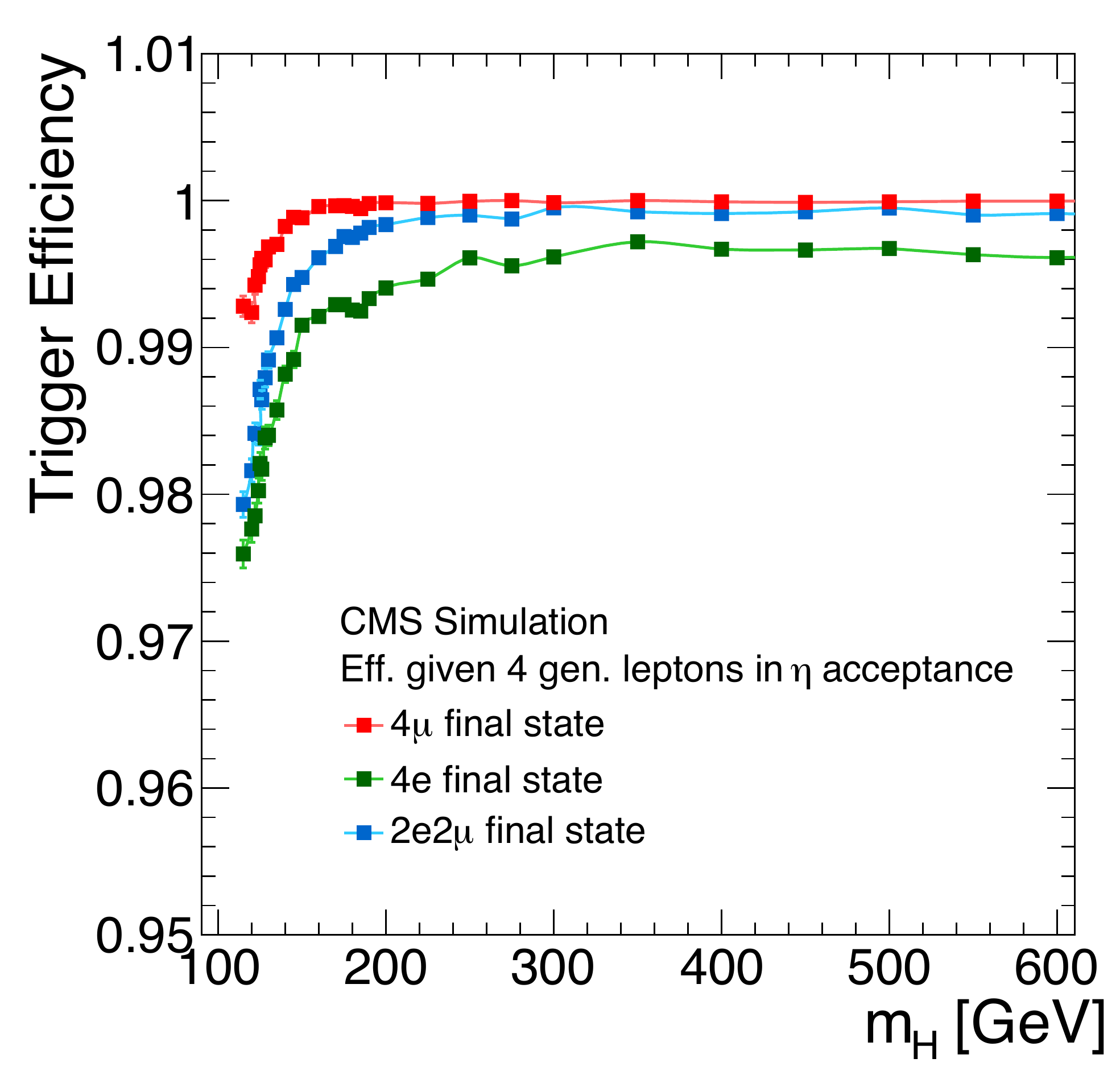}
\includegraphics[width=0.45\linewidth]{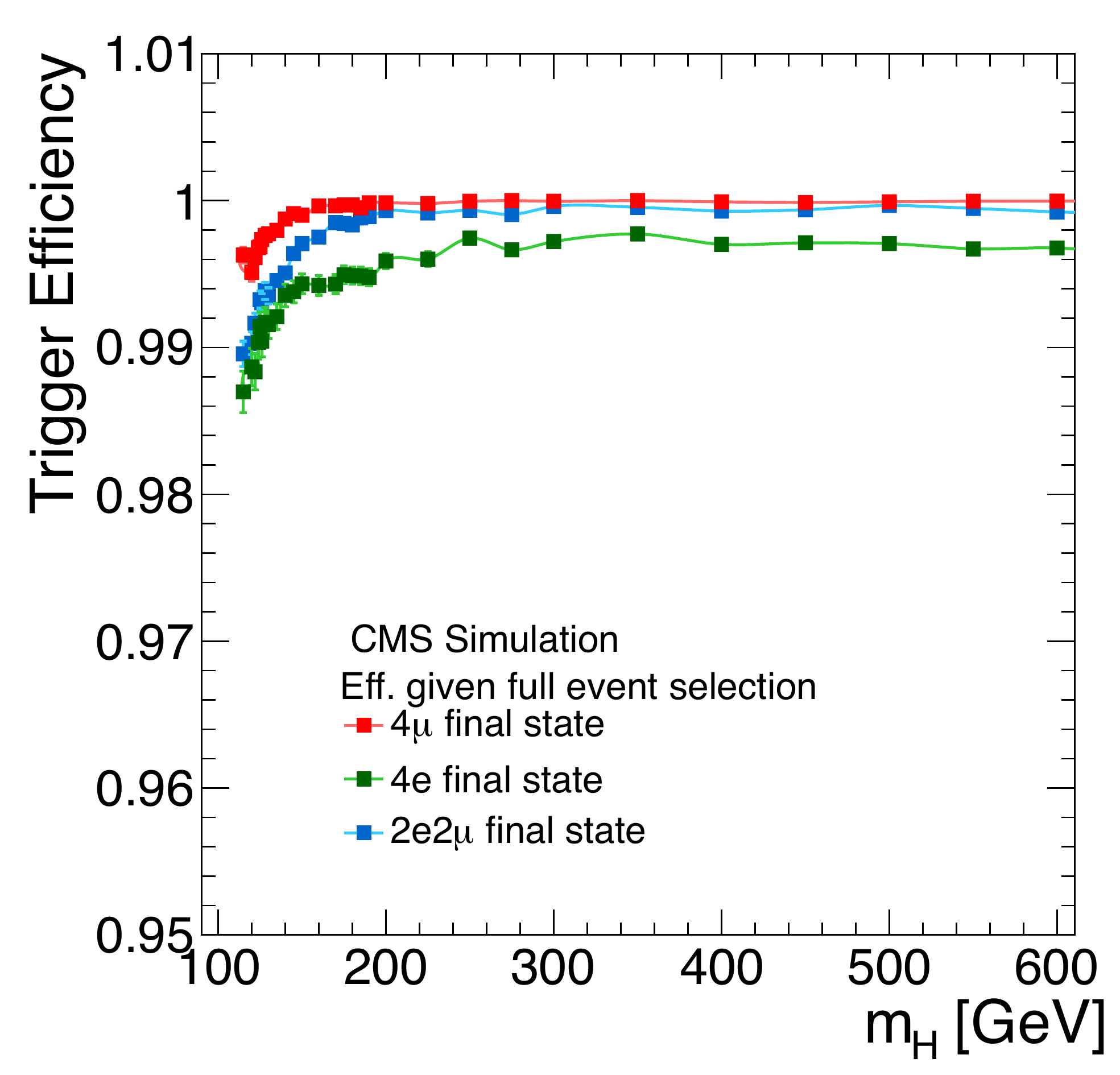}\\
\caption{Trigger efficiency for simulated signal events with four
  generated leptons in the pseudorapidity acceptance (left), and
  for simulated signal events that have passed the full
  $\PH\to4\ell$ analysis
  selection (right).
}
\label{fig:triggerplots4l}
\end{figure}

\begin{table}[tbh]
    \centering
    \topcaption{
    Triggers used in the $\PH\to4\ell$ event selection (2012 data and
    simulation). No prescaling is applied to these triggers.
    }
    \label{tab:anatriggers}
    \resizebox{\textwidth}{!}{
    \begin{tabular}{|lll|}
\hline
Channel & \multicolumn{1}{c}{HLT path}                      							   & \multicolumn{1}{c}{L1 seed}  \\ \hline
	4e        & \texttt{ HLT\_Ele17\_CaloTrk\_Ele8\_CaloTrk }   				   & \texttt{ L1\_DoubleEG\_13\_7 }    	  \\
                                 & \texttt{ OR HLT\_Ele15\_Ele8\_Ele5\_CaloIdL\_TrkIdVL  	}	               & \texttt{ L1\_TripleEG\_12\_7\_5 }  	  \\
	4$\mu$    & \texttt{ HLT\_Mu17\_Mu8  }   					           & \texttt{ L1\_Mu10\_MuOpen   }          \\
	             & \texttt{ OR HLT\_Mu17\_TkMu8   }  				           & \texttt{ L1\_Mu10\_MuOpen   }          \\
      2e2$\mu$    & \texttt{ HLT\_Ele17\_CaloTrk\_Ele8\_CaloTrk   } 				   & \texttt{ L1\_DoubleEG\_13\_7 }    	  \\
	               & \texttt{ OR HLT\_Mu17\_Mu8  	}					   & \texttt{ L1\_Mu10\_MuOpen   }    	  \\
	                & \texttt{ OR HLT\_Mu17\_TkMu8  }					           & \texttt{ L1\_Mu10\_MuOpen   }    	  \\
	              & \texttt{ OR HLT\_Mu8\_Ele17\_CaloTrk	}				   & \texttt{ L1\_MuOpen\_EG12   }    	  \\
               & \texttt{ OR HLT\_Mu17\_Ele8\_CaloTrk	}				   & \texttt{ L1\_Mu12\_EG6      }    	  \\
\hline
    \end{tabular}}
\end{table}
Figure~\ref{fig:triggerplots4l} shows the efficiency of the trigger
paths described above as a function of the Higgs boson mass, for signal
events with four generated leptons in the pseudorapidity acceptance and for those that have passed
the analysis selection, as determined from simulation.  With these trigger paths, the trigger
efficiency within the acceptance of this analysis is greater than 99\%
for a Higgs boson signal with $m_\mathrm{H} > 120\GeV$.

The tag-and-probe method is used to measure the per-lepton
efficiency for double-lepton triggers, as described in
Section~\ref{sec:egammaHLT} for electrons, and in Section~\ref{sec:muHLT} for muons.
The performance in data and simulation of the per-leg efficiencies of
the double-lepton triggers
are shown in those sections.
The position and the steepness of the turn-on curve of the trigger
efficiency as a function of the lepton \pt measured on data is in good
agreement with the expectations from simulation for all the triggers
considered. A measurement of the trigger efficiency on the plateau
reveals generally lower efficiency in data, compared to simulation,
by about 1--2\%. The effect of this inefficiency is mitigated,
however, by the fact that multiple leptons in the event can pass the trigger requirements, and so no correction factor is applied. A systematic
uncertainty of 1.5\% in the expected signal yields is included to
allow for this difference in trigger performance between data and
simulation.

In Table~\ref{tab:effitriggers}, the trigger paths used to select the
tag-and-probe pairs for the efficiency measurements are listed. In
case of muons, the prescaled double-muon triggers in the \JPsi mass
window are used to select a low-\pt muon probe to measure the
identification and isolation efficiency for muons with $\pt <
15\GeV$.

\begin{table}
    \centering
    \topcaption{
    Triggers used for tag-and-probe (T\&P) efficiency measurements of
    four-lepton events in 2012 data and
    simulation:  CaloTrk = CaloIdT\_CaloIsoVL\_TrkIdVL\_TrkIsoVL,
    CaloTrkVT = CaloIdVT\_CaloIsoVT\_TrkIdT\_TrkIsoVT
    }
    \resizebox{\textwidth}{!}{
    \begin{tabular}{|llllc|}
\hline
Channel     & Purpose & HLT path                      							   & L1 seed            & prescale  \\ \hline
4e and 2e2$\mu$ & \Z T\&P         & \texttt{ HLT\_Ele17\_CaloTrkVT\_Ele8\_Mass50 }			   & \texttt{ L1\_DoubleEG\_13\_7 }   & 5 	  \\
4e and 2e2$\mu$ & \Z T\&P low \pt  & \texttt{ HLT\_Ele20\_CaloTrkVT\_SC4\_Mass50\_v1  }			   & \texttt{ L1\_SingleIsoEG18er } & 10  	  \\
4$\mu$ and 2e2$\mu$ & \Z T\&P     & \texttt{ HLT\_IsoMu24\_eta2p1}				           & \texttt{ L1\_SingleMu16er }    &  	  \\
4$\mu$ and 2e2$\mu$ & \JPsi T\&P & \texttt{ HLT\_Mu7\_Track7\_Jpsi	}				   & 				 & 	  \\
		    &            & \texttt{ HLT\_Mu5\_Track3p5\_Jpsi  }				   &				 & 	  \\
		    &		 & \texttt{ HLT\_Mu5\_Track2\_Jpsi  	}				   &				 & 	  \\
\hline
    \end{tabular}}
    \label{tab:effitriggers}
\end{table}

\subsubsection[Triggers for the di-tau Higgs boson analysis]{$\PH\to\tau\tau$}
\label{sec:HTauTau}

The triggers used for the Higgs boson $\PH\to\tau\tau$ analysis in the
$\tau_{\mu}\tau_\mathrm{h}$ and $\tau_{\Pe}\tau_\mathrm{h}$ channels require both an
electron or muon and a hadronic tau object.
The electron or muon is required to be isolated; the energy in the isolation cone is corrected for the effects of the pileup~\cite{fastjetmanual}.
The tracks for the $\tau_\mathrm{h}$ candidate and the tracks used to compute the isolation were required to come from a
vertex compatible with the electron/muon origin. The efficiencies are
measured using \Z$\to\tau\tau$ events with a muon-plus-\MET or
a single-electron trigger. The events are selected by requiring the electron/muon to pass the tight isolation criteria, and also to have a transverse mass $M_\mathrm{T}<20\GeV$ measured
between the electron/muon and the missing transverse momentum vector. The purities after
this selection are 78\% and 65\% for
$\abs{\eta (\tau_\mathrm{h})} <  1.5$ and
$1.5 < \abs{\eta (\tau_\mathrm{h})} < 2.3$,
respectively. The event samples used to calculate the efficiencies
are mixed with \PW+jets simulated events to produce a
compatible purity. The $\tau$-leg trigger efficiencies are discussed
in detail in Section~\ref{sec:tauHLT}.

\subsubsection[Triggers for ZH to 2 neutrinos + b jets
analysis]{$\Z(\PGn\PAGn)\PH(\bbbar)$ }
The production of the Higgs boson in association with vector bosons is
the most effective way to observe the Higgs boson in the $\PH\to\bbbar$
decay mode~\cite{cms_zhbb}.
In this section, we report on the trigger performance for the 2012
data taking period.
\begin{table}[tbp]
  \topcaption{List of L1 and HLT used for 2012 data for
    the $\Z(\PGn\PAGn)\PH(\bbbar)$ channel. We use PF \MET. All triggers are combined
    to maximize acceptance. In all cases, an OR of the L1 $\MET>36\GeV$
    and $\MET>40\GeV$ are used as the L1 seed.}
\label{tab:trgs2012ZnnHbb}
\centering
\begin{tabular}{|lc|} \hline
HLT  & Run Period \\ \hline
 $\MET > 150\GeV$   & 2012  \\ 				
 $\MET > 80\GeV$ and 2 central jets with $\pt > 30\GeV$ & early 2012 \\
 $\MET > 100\GeV$ and 2 central jets and $\Delta \phi$ requirement   & late 2012\\
 $\MET > 100\GeV$ and 2 central jets with $\pt > 30\GeV$ and at least one b tag  &late 2012 \\
\hline
\end{tabular}
\end{table}
Table~\ref{tab:trgs2012ZnnHbb} summarizes these triggers. The main trigger requires $\MET>150\GeV$ and was active during
the entire year.
\begin{figure}[tbph]
  \centering
    \includegraphics[width=0.32\textwidth]{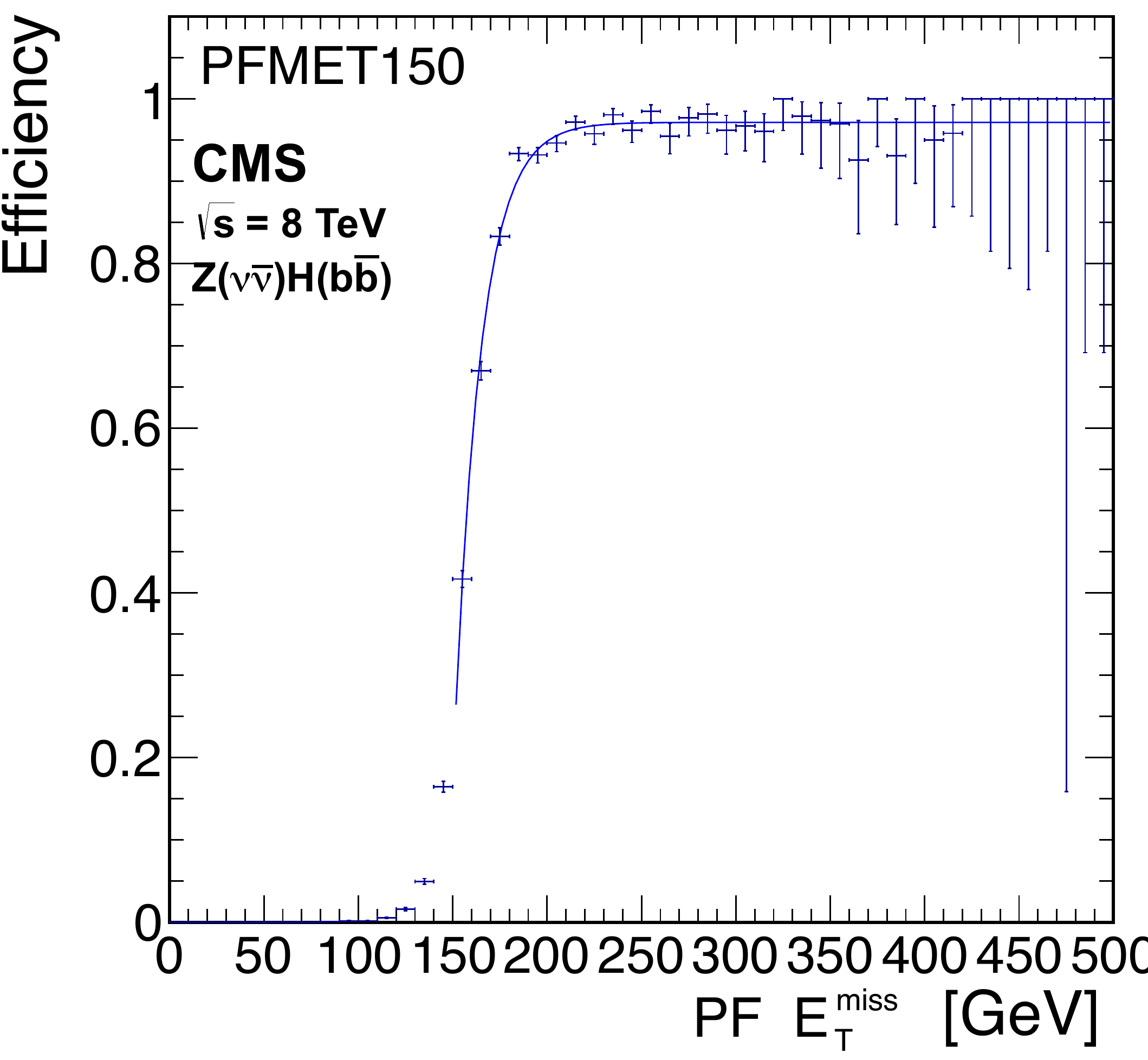}
    \includegraphics[width=0.32\textwidth]{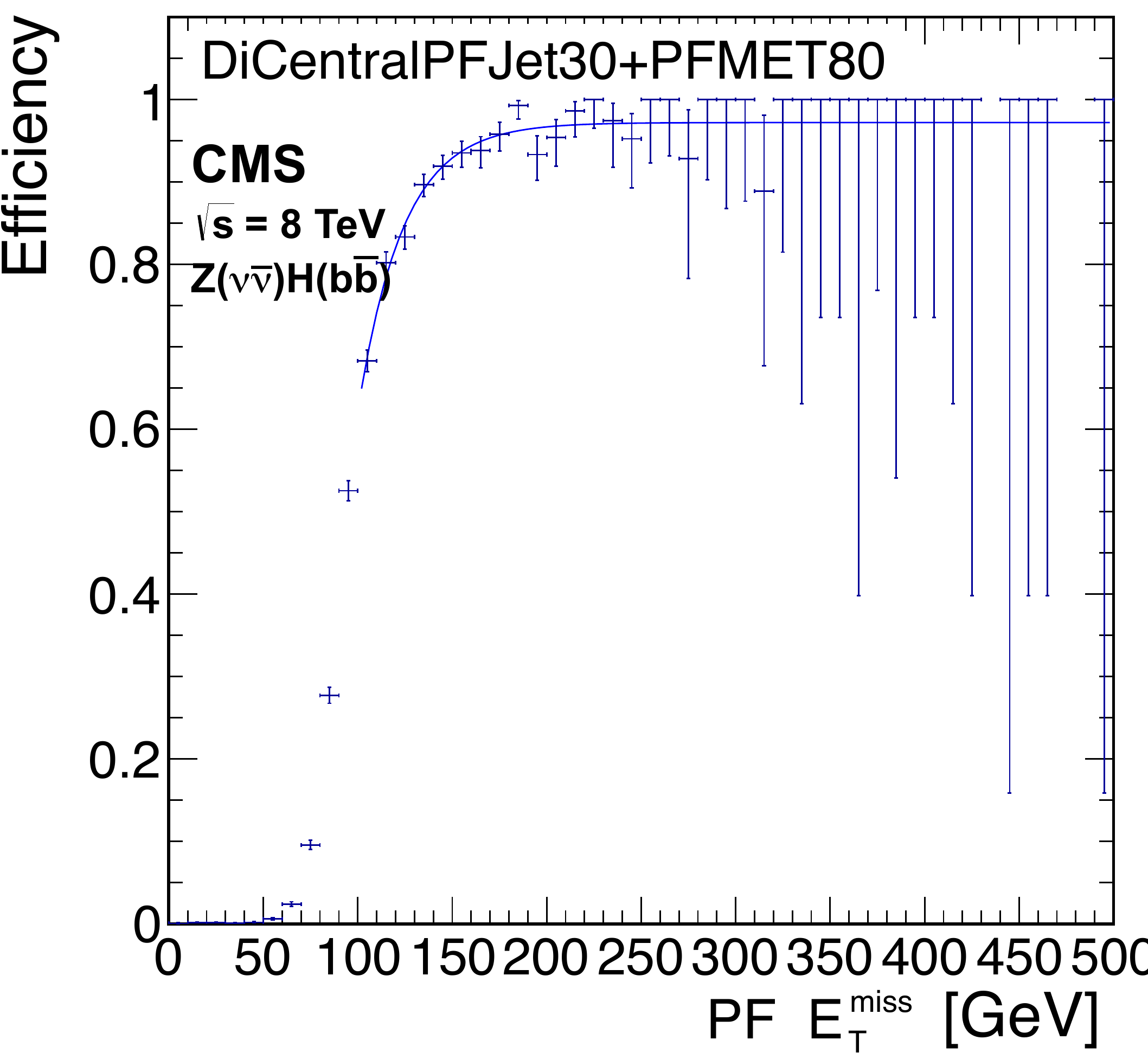}
    \includegraphics[width=0.32\textwidth]{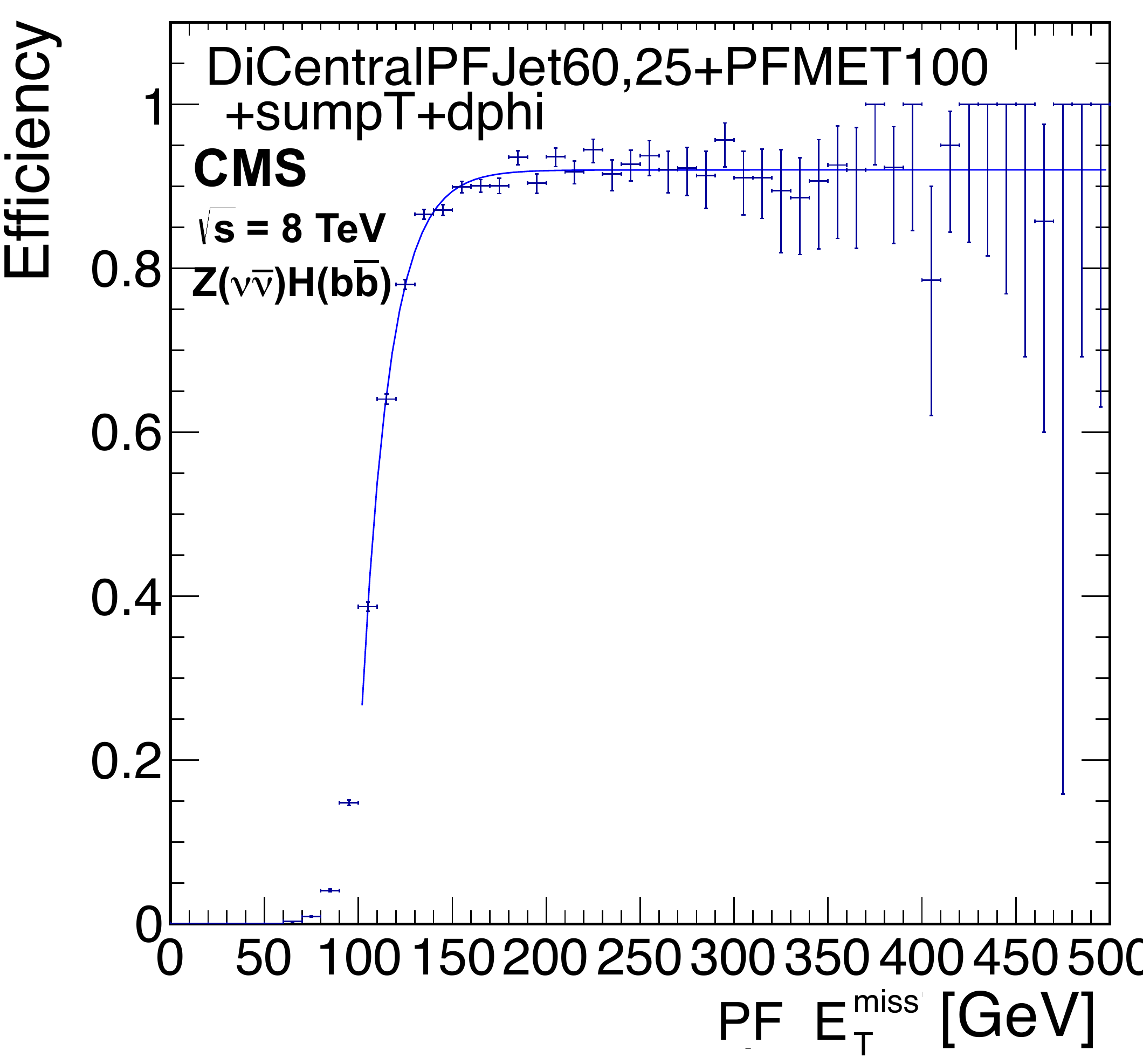}
    \caption{Trigger efficiencies for the $\Z(\PGn\PAGn)\PH(\bbbar)$ analysis,
      as a function of offline PF \MET for the pure $\MET>150\GeV$
      trigger (left) using late 2012 data, dijet and \MET trigger
      (center) using early 2012 data, and dijet, \MET and
      $\Delta \phi$ requirement trigger (right) using 2012 late data, as
      described in the text. }
    \label{fig:ZnunuHbbTrigEff2012}
\end{figure}
This trigger, however, attains an efficiency of 95\% at
$\MET{\approx}190\GeV$, as shown in Fig.~\ref{fig:ZnunuHbbTrigEff2012} (left).
To accept events with lower $\MET$, we introduce a trigger that
requires two central PF jets with $\pt>30\GeV$ and
$\MET>80\GeV$, for early 2012 data. This trigger recovers events at
lower $\MET$. The efficiency curve, shown in
Fig.~\ref{fig:ZnunuHbbTrigEff2012} (center) reaches a plateau of
95\% at $\MET{\approx}150\GeV$.

For late 2012 running, jets due to pileup caused an increase in
trigger rates, and a more complicated trigger,
requiring at least two central PF jets with  $\pt>60 (25)\GeV$
for the leading (subleading) jet, was introduced. At least
one calorimeter dijet pair with $|\sum_i\vec{p}_{\mathrm{T}_i}|>100\GeV$ is
required. The minimum $\Delta\phi$ between the $\MET$ and the closest
calorimeter jet with $\pt>40\GeV$ is required to be greater than 0.5.
Finally, we require PF $\MET>100\GeV$. The obtained turn-on curve
for this trigger is shown in Fig.~\ref{fig:ZnunuHbbTrigEff2012} (right).
The  trigger achieves 90\% efficiency at
$\MET\approx170\GeV$, with roughly 80\% efficiency for $\MET$ in the
range of 130--170\GeV.

To accept events with even lower $\MET$ (down to 100\GeV) we exploit
triggers with a b-tag online requirement (Section~\ref{sec:BTag}): two jets
with $\pt > 20\,(30)\GeV$ and $\MET>80\GeV$ for early (late) data.
These triggers by themselves achieve an efficiency of roughly 50\% at
$\MET= 100\GeV$ and 60\% efficiency for \MET between 100 and $130\GeV$ when
requiring at least one PF jet with a high value of the b-tagging
discriminator (tight CSV $>0.898$) offline.
The trigger strategy for the full 2012 period  used the
combination of all the aforementioned triggers to collect events
with $\MET>100\GeV$.

Rather than measuring the efficiency curves directly in data and
applying them to the simulation, the efficiencies of the simulated
triggers are parametrized and corrected as a function of $\MET$ and
the CVS  b tagging discriminator to match the efficiencies measured
in data (described below). This approach takes into account the
non-negligible correlations among the various trigger paths. It also
characterizes the online b tagging efficiency and its dependence on
jet \pt and $\eta$, as the geometry and trigger algorithm are
simulated in a way that are as close as possible to the actual trigger
environment.
Studies show that the data and simulation agree to within less than
5\%, except for the b tag trigger, where the agreement is approximately
10--20\%.

\begin{figure}[tbp]
  \centering
    \includegraphics[width=0.49\textwidth]{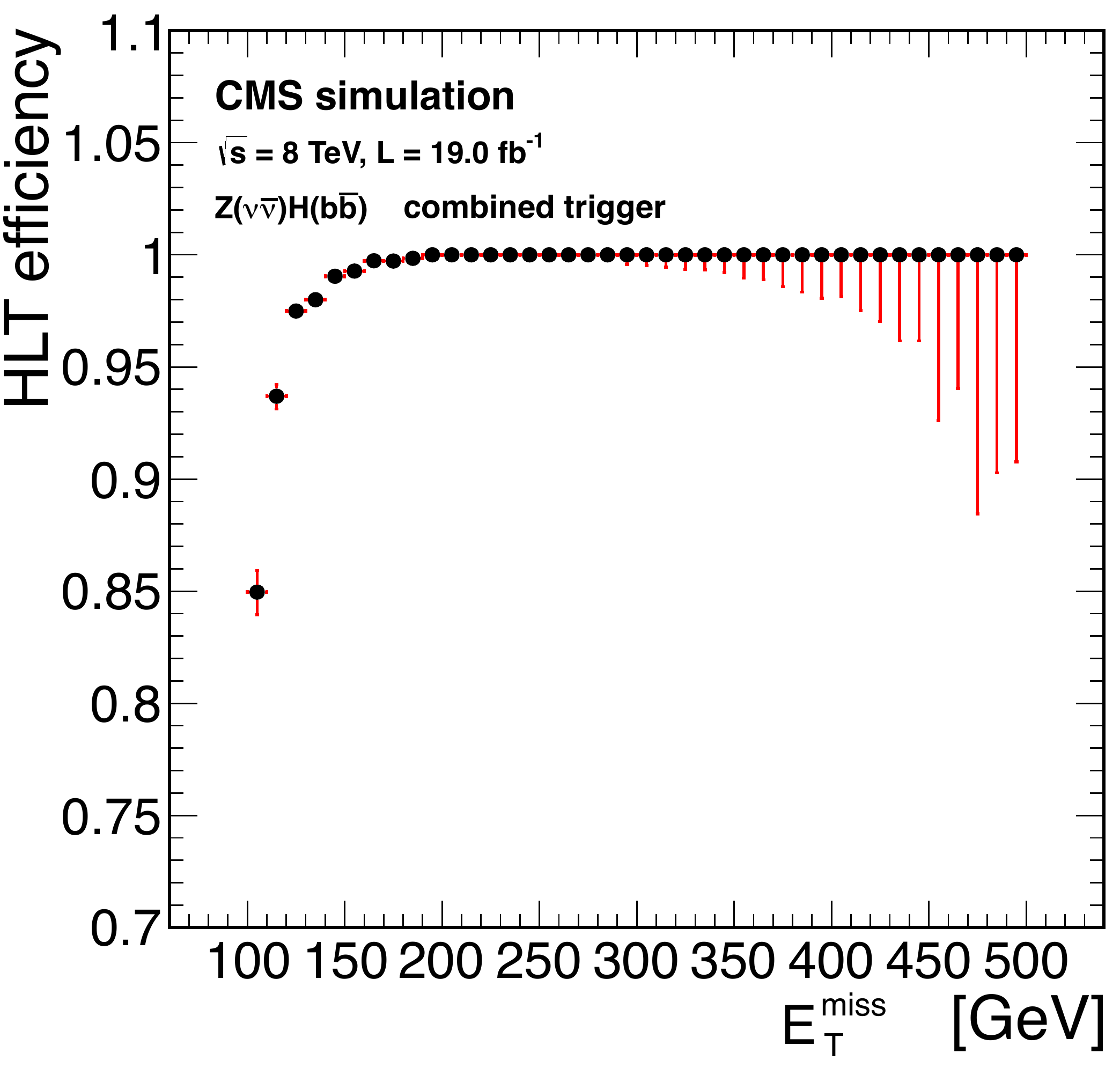}
    \caption{Efficiency as function of
      $\MET$ for the $\Z(\PGn\PAGn)\PH(\bbbar)$ signal events. An efficiency greater than 99\% is obtained for
      $\MET>170\GeV$.
    }
    \label{fig:ZnunuHbbTrigEff2012OR}
\end{figure}
In Fig.~\ref{fig:ZnunuHbbTrigEff2012OR}, we show the total trigger
efficiency as a function of $\MET$ for signal
$\Z(\PGn\PAGn)\PH(\bbbar)$ events. The cumulative efficiency is
99\% for $\MET> 170\GeV$, 98\% for events with $130<\MET<170\GeV$, and
88\% for events with $100<\MET<130\GeV$.

The total systematic uncertainty in the trigger efficiency is of the
order of a few percent in the high-\pt  ($\MET$ $>
170\GeV$), and not more than 7\% in the intermediate-\pt ($130 < \MET <
170\GeV$), and 10\% in the
low-\pt regions ($100 < \MET < 130\GeV$) search regions.

\subsection{Top  quark triggers}
\label{sec:toppag}

Measurement of the properties of the top quark are among the most
important standard model measurements in CMS. The LHC is a top factory, and the large number of top quark pairs created allows detailed studies
of its properties. One of the most fundamental measurements is the top
quark pair production cross section. The most accurate measurement of this cross section can
be made in the so-called `lepton + jets' decay mode, where one of the
W bosons from the top quark decays to a lepton and a neutrino, and
the other W decays hadronically, leading to a final state with a
well-isolated lepton, large missing transverse energy, and four hadronic jets (two
of which are b jets)~\cite{cms_ljets2010,Chatrchyan:2013faa}. In
Run~1, \ttbar production studies used several trigger paths for the
semileptonic top quark decay channels, to ensure that \ttbar signal events
were recorded as efficiently as possible. To maximize the acceptance of the transverse energy (momentum) requirement applied to
leptons, measurements used trigger paths requiring one online
reconstructed lepton ($\Pe$ or $\mu$) as well as at least 3 online
reconstructed jets.
\begin{table}[tbp]
\topcaption{Unscaled cross-triggers used for the \ttbar (lepton plus jets)
  cross section measurement in 2012. All leptons use tight or very tight
  identification, and lepton and calorimeter isolation
  requirements. All jets are PF jets and restricted to the
  central region. At L1, single electrons or muons are required with
  the denoted thresholds and the L1 muons are required to be central
  ($\abs{\eta}<2.1$). When two  thresholds are listed at L1, they include
  a lower (possibly prescaled) threshold and a higher unscaled threshold.
}
\centering
\begin{tabular}{|ccccc|cc|}
  \hline
  \multicolumn{5}{|c|}{HLT} & \multicolumn{2}{c|}{L1} \\ \hline
  $\Pe/\mu$ & Threshold & $n_\text{jet}$ & Jet  & Jet threshold & L1 Seed & Threshold\\
            &  (\GeVns{})   &               & corrections &        &         &  (\GeVns{}) \\
  \hline
  \multirow{4}{*}{$\Pe $} & 25 & 3 & & 30 & EG & 20, 22 \\
         & 25 & 3 & pileup subtracted & 30 & EG & 20, 22 \\
         & 25 & 3 & pileup subtracted & $30,30,20$ & EG & 20, 22 \\
         & 25 & 3 & pileup subtracted & $45,35,25$ & EG & 20, 22 \\
  \hline
  \multirow{6}{*}{$\mu$}  & 20 & 3 & & 30 & MU & 14, 16 \\
         & 20 & 3 & pileup subtracted & 30 & MU & 16 \\
         & 17 & 3 & pileup subtracted & 30 & MU & 14 \\
         & 17 & 3 & pileup subtracted & $30,30,20$ & MU & 14 \\
         & 17 & 3 & pileup subtracted & $45,35,25$ & MU & 14 \\
  \hline
\end{tabular}
\label{tab:TopLepJetTrigger}
\end{table}

Table~\ref{tab:TopLepJetTrigger} summarizes the main paths used for
the triggers deployed to accommodate the high instantaneous luminosity
and pileup of the 2012 run. All leptons triggers had tight or very
tight lepton identification and calo\-ri\-meter isolation
requirements, comparable to those used offline. Jets in PF jet
triggers were restricted to the central region. At L1, single
electrons or muons are required with the denoted thresholds. The L1
muons are central ($\abs{\eta}<2.1$).  Charged-hadron
subtraction~\cite{CMS-PAS-JME-14-001} (labeled `pileup subtracted' in
the Table) was implemented for pileup mitigation. Additionally, the
introduction of jet energy calibrations online in the second half of
2012 resulted in higher \ET thresholds in the three-jet paths;
however, the data from that period were not used in the cross section
measurements due to systematic uncertainties associated with the large
pileup.

Simulated events are used to estimate the top quark acceptance, and were corrected
for the trigger efficiency measured in data. To estimate the trigger
efficiency, simulated Drell--Yan and \ttbar samples were used to
compare with data collected with single lepton triggers. The overall
efficiency for the lepton plus jets paths is parametrized as a
product of two independent efficiencies for the leptonic and hadronic
legs of the trigger, $\epsilon_{\rm lep}~\times~\epsilon_{\rm had}$. A
cleaning requirement based on the \DR\xspace distance between the leptons and jets
motivates this approach.

The leptonic leg efficiency is measured using a tag-and-probe method
with $\Z/\Pgg^*$ events, as described in Sections~\ref{sec:egammaHLT}
(\Pe) and \ref{sec:muHLT} ($\mu$).

While the lepton trigger was not changed during the 2012
data-taking period, the jet trigger changed as shown in
Table~\ref{tab:TopLepJetTrigger}. Similar to the measurement for the
lepton leg, the efficiency of the jet leg of the associated cross-trigger is measured in an unbiased data sample. The
reference sample is required to pass a single lepton trigger, to assure
a data set independent of the hadronic trigger which fulfills the lepton
leg of the cross-trigger simultaneously.

\begin{figure}[tbp]
  \centering
    \includegraphics[width=0.45\textwidth]{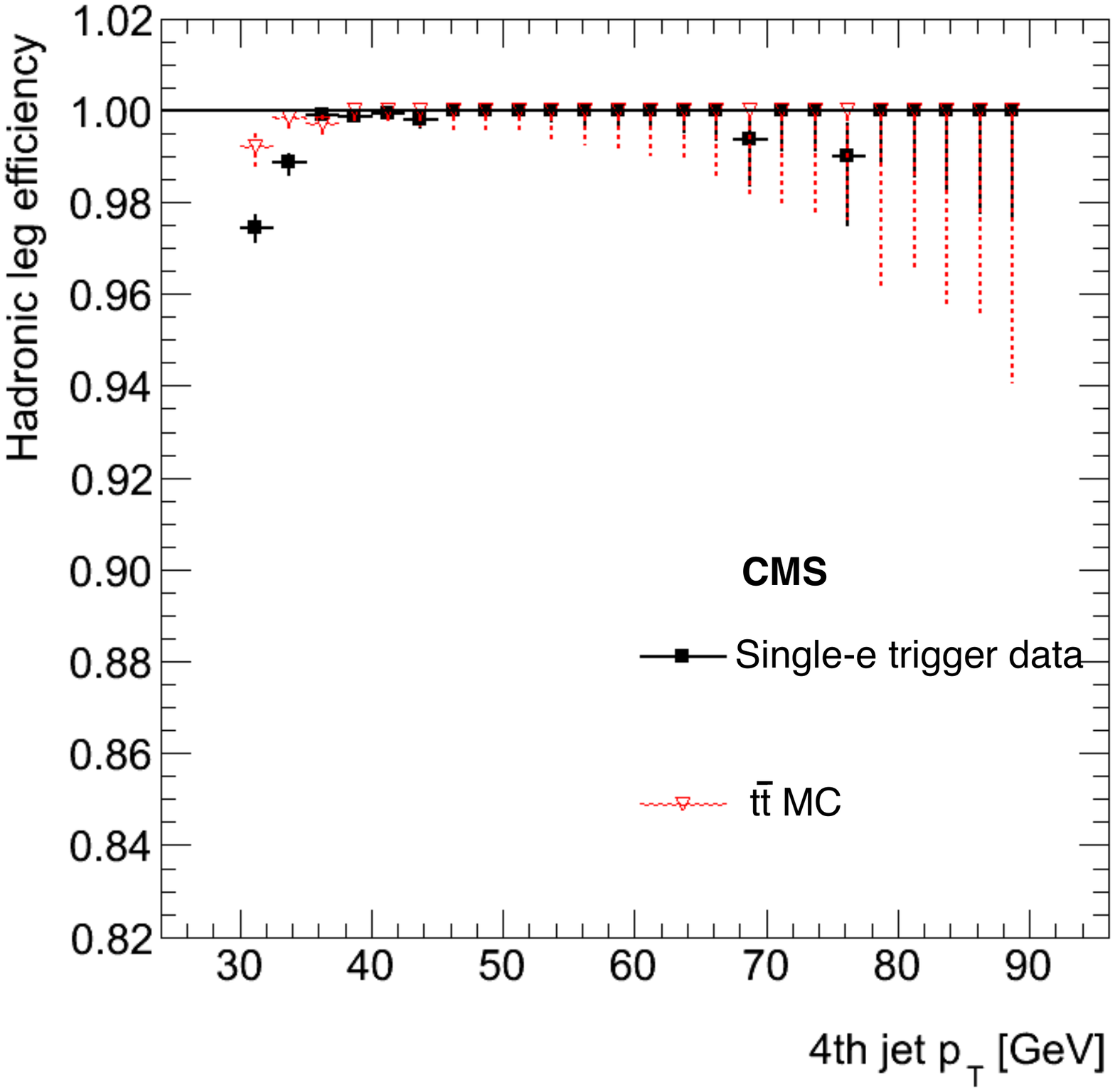}
    \includegraphics[width=0.45\textwidth]{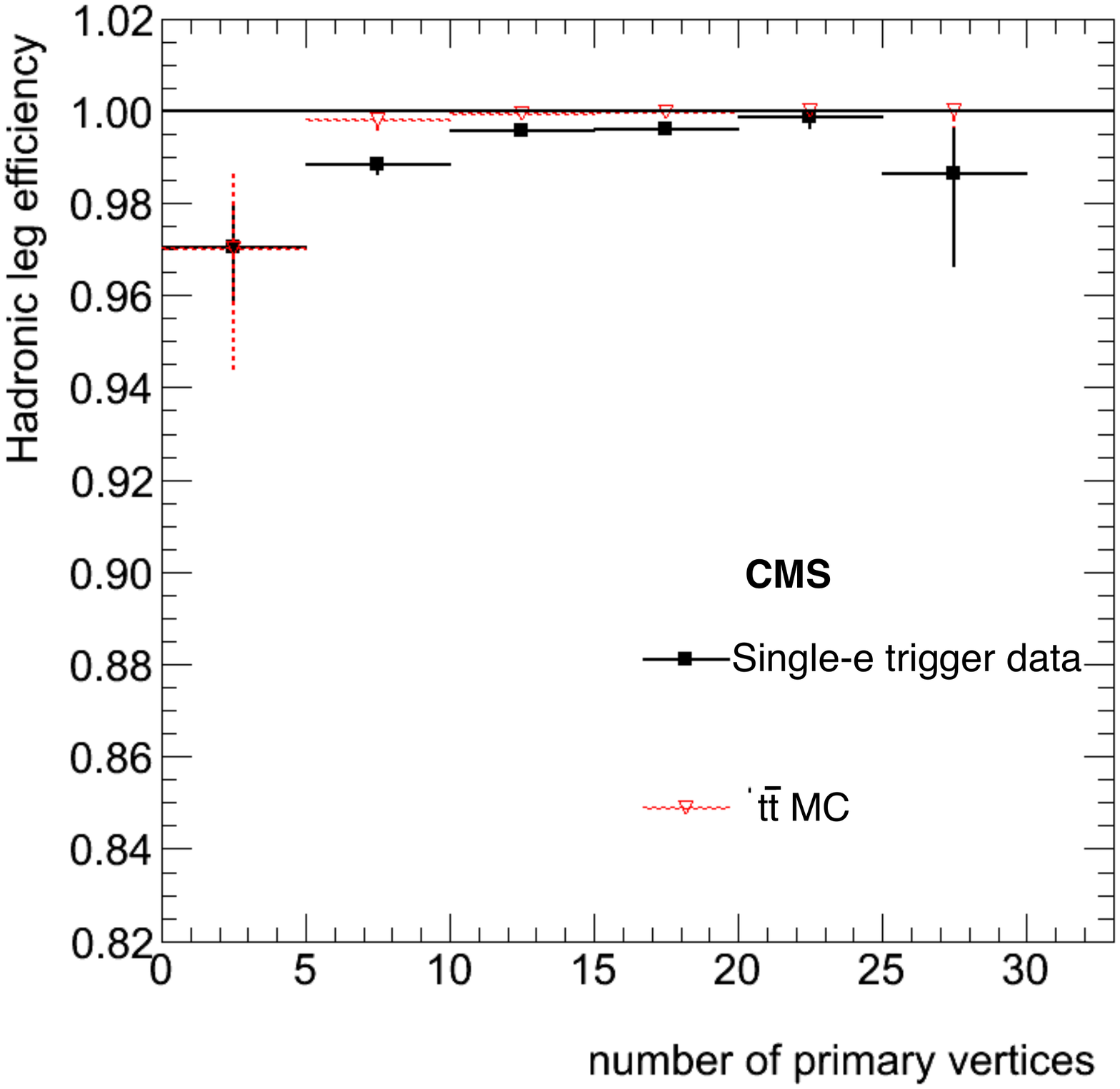}
    \caption{Top quark triggers: Efficiency of the hadronic leg for the
      electron plus jets paths in 2012 as a function of the \pt of the 4th jet
      (left) and of the number of reconstructed vertices
      (right).}
    \label{figure:topEff}
\end{figure}
As an example, Fig.~\ref{figure:topEff} shows the efficiency turn-on
curve of the hadronic leg (transverse momentum of the 4th jet) for the
electron plus jets paths in 2012, and its dependence with respect to
the number of reconstructed vertices, both for a selection based on the combination of
the PF jets without and with charged-hadron
subtraction. The offline selection of the transverse
momentum requirements on the offline jets was devised to assure a
plateau behavior of the scale factors, meaning no variation of the
scale factor with respect to the MC sample or jet energy
calibrations. From the variation of the scale factors it was concluded
that a systematic uncertainty of 2\% (1.5\%) in electron (muon)
scale factors covered the variations around their value of 0.995
(0.987).

\subsection{Triggers for supersymmetry searches}
\label{sec:susypag}

Supersymmetry (SUSY) is one of the most appealing extensions to the
standard model, as it solves the mass hierarchy problem, offers a
path towards grand unification, and can provide candidate dark matter
particles.  During the years 2010--2012, many SUSY searches
were performed with CMS data. Exclusion limits were set in the context
of the mSUGRA model of SUSY breaking and also on the masses of the
particles involved in specific cascade decays (simplified
models~\cite{sms}).

For the allowed parameter space, SUSY signatures~\cite{glennis} are characterized by
the presence and decay of heavy particles. If R-parity is conserved,
stable, invisible particles are expected. Most of the final states
contain significant hadronic activity and \MET. At
CMS, SUSY searches were divided into leptonic, hadronic, and photonic
categories, depending on the event content.

In addition, some supersymmetric models predict the existence of heavy
stable charged particles, \eg, the gluino, top quarks, or $\tau$ sleptons.
Their mass is expected to be of the order of a few hundred\GeV,
therefore their velocity would be significantly smaller than the speed
of light. The signature of heavy stable charged particles would look
like a non-relativistic ionizing particle, with hits in the chambers being
delayed by about one bunch crossing, either in all the layers or
in the outermost one(s), with respect to an ordinary ``prompt''
minimum ionizing particle.

In this section we discuss the CMS trigger performance collecting
events for searches for supersymmetry. Most leptonic searches in CMS
were performed using the same triggers as the Higgs boson leptonic searches
and therefore are not documented here. For hadronic and photonic
searches, we have selected three representative triggers: the $\alpT$
trigger, the ``Razor'' trigger, and the photon trigger. The $\alpT$ and
photon analyses were performed using a data sample corresponding to
an integrated luminosity of 4~\fbinv, while the Razor
analysis used an integrated luminosity of 20~\fbinv, all collected at CMS during
2012 at a center-of-mass energy of $8\TeV$.

\subsubsection{Triggers for all-hadronic events with
  \texorpdfstring{$\alpT$}{alphaT}}

We present a typical example of a purely hadronic search, where events
with leptons are vetoed and events with a high jet multiplicity, large
\MET, and large \HT are selected~\cite{cmsalphaT}. Multijet events are
the most important background in this region of the phase space. To
suppress these events, the analysis uses a kinematical variable called
$\alpT$. For events with exactly two jets, $\alpT$ is defined as the
transverse energy of the subleading jet divided by the transverse mass
of the dijet system. For events with two or more jets, two pseudo-jets
are created combining jet components and selecting the configuration
that minimizes the energy between the two.  The value of
$\alpha_\mathrm{T}$ is equal to 0.5 in balanced multijet events and less
than 0.5 in multijet events with jet energy mismeasurement. For SUSY
signal events with genuine \MET, $\alpha_\mathrm{T}$ tends to values
$> 0.5$, thus providing a good discrimination between signal and
background.
To estimate the remaining significant backgrounds (\PW+jets,
top quark pair, single top quark , and $\Z \to \PGn\PAGn$), data control
regions are used.

A cross-trigger  based on the quantities \HT and $\alpT$
is used to record the candidate event sample. A
prescaled \HT trigger, labeled henceforth as \HT, is used with
various thresholds to record events for the control region. The \HT
thresholds of the \HT and $\HT$--$\alpT$ cross-triggers are chosen to
match where possible, and are 250, 300, 350, 400, and
450\GeV. The $\alpT$ thresholds of the $\HT$-$\alpT$ trigger are tuned
according to the threshold on the \HT leg in order to suppress
QCD multijet events (whilst simultaneously satisfying other criteria,
such as sensitivity to trigger rates).

\begin{figure}[tbp]
  \centering
      \includegraphics[width=0.48\textwidth]{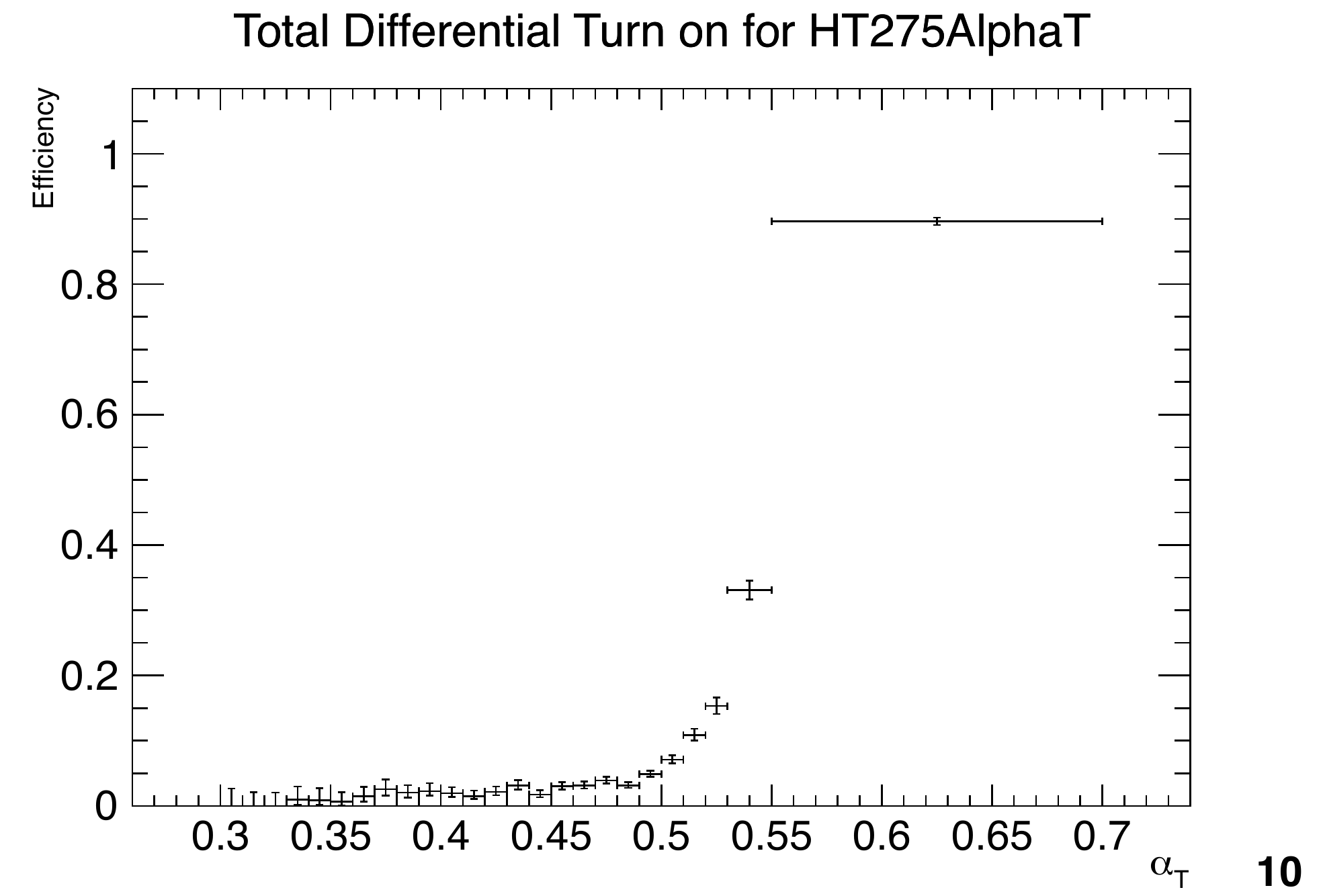}
      \includegraphics[width=0.48\textwidth]{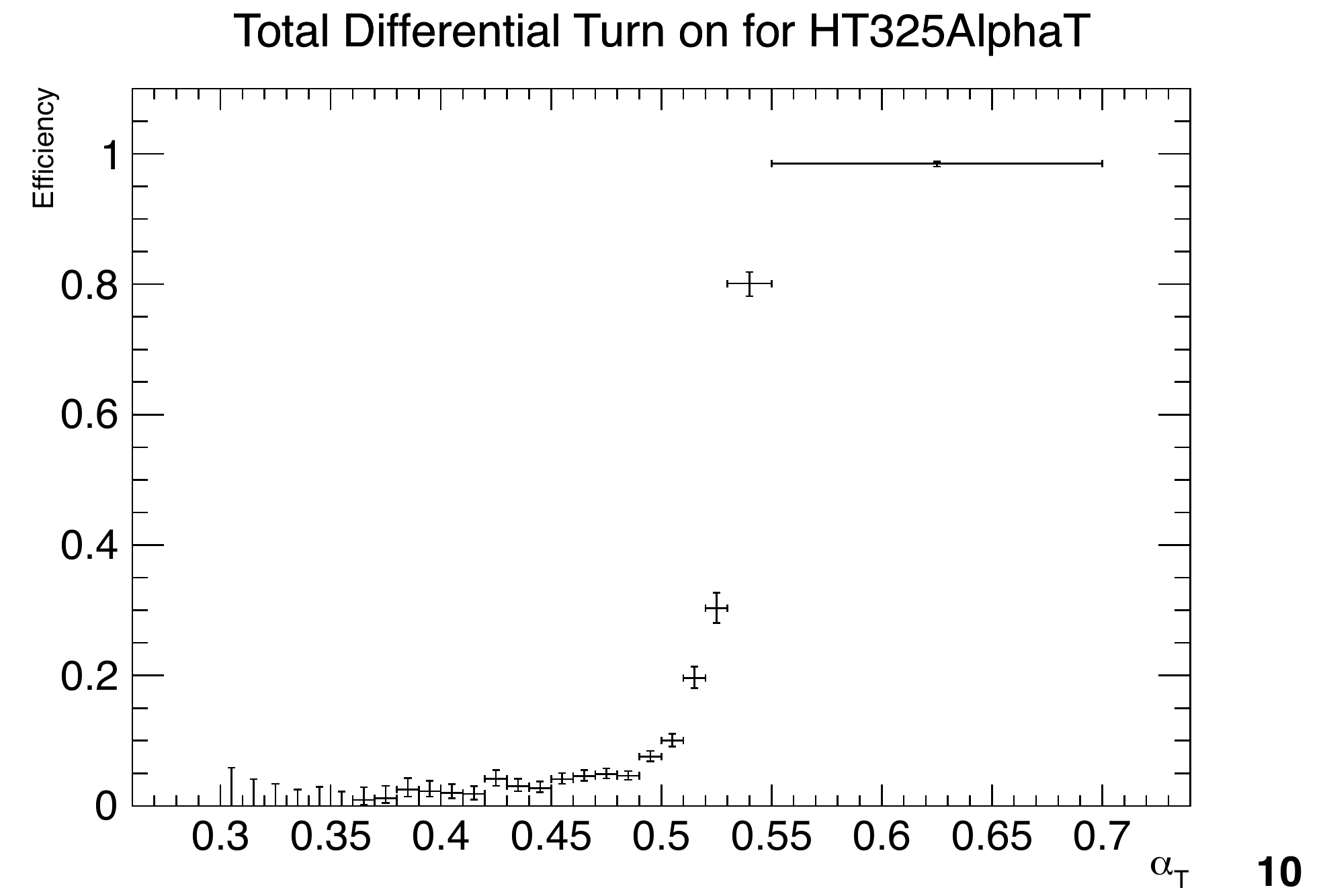}
      \includegraphics[width=0.48\textwidth]{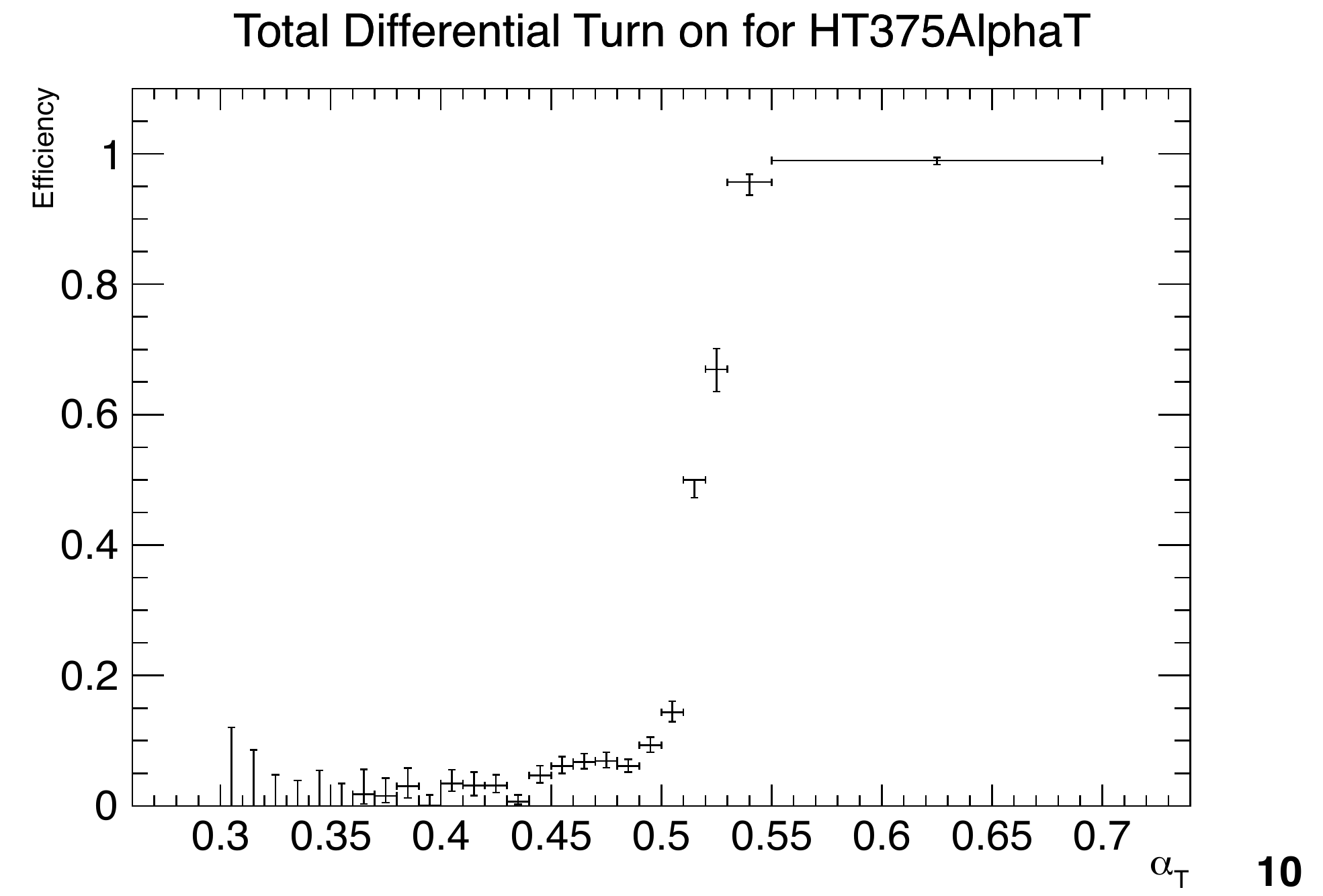}
      \includegraphics[width=0.48\textwidth]{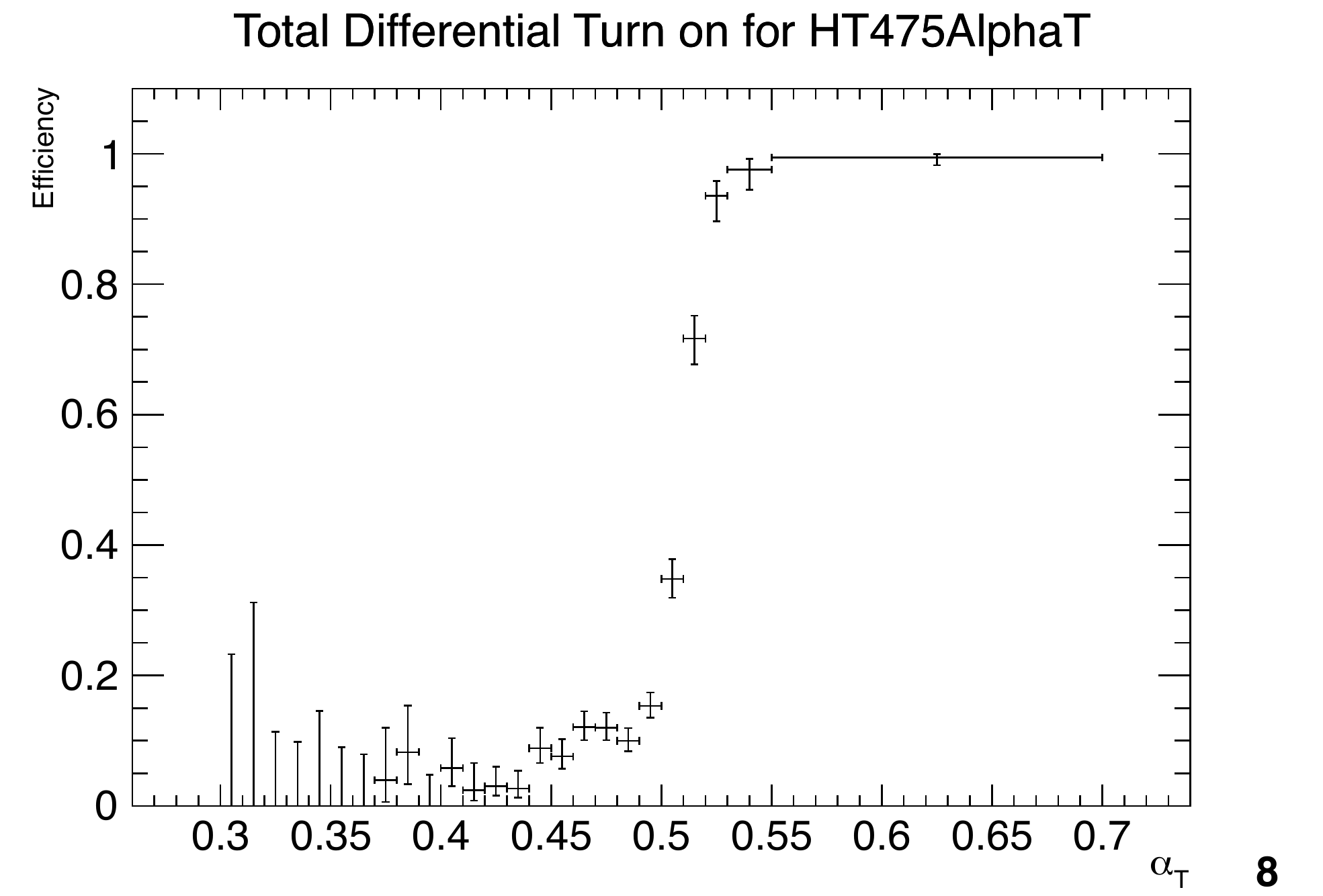}
    \caption{Efficiency turn-on curves for the $\alpha_T$ triggers
      used to collect events for four different \HT regions:
      $275<\HT<325\GeV$ (upper left),
      $325<\HT<375\GeV$ (upper right), $375<\HT<475\GeV$ (lower left), and $\HT>475\GeV$ (lower right).}
    \label{fig:alphaT}
\end{figure}

To ensure that the \HT leg of the $\HT$-$\alpT$ cross-trigger and the \HT prescaled trigger
are efficient for the final event selection, the lower bounds of the offline
\HT bins are offset by 25\GeV with respect to the online thresholds. Figure~\ref{fig:alphaT} shows
the turn-on curves of the \HT and $\alpT$ legs of the trigger with respect to the offline selection.

Efficiencies for the $\HT$-$\alpT$ triggers were calculated using an
orthogonal data set based on single muons, by requiring a matching to an isolated
single-muon trigger. Exactly one isolated muon that is well
separated from all jets is required to ``tag'' the event. This muon is
not considered in the calculations of \HT, \MET-like quantities,
and $\alpT$,
thereby miscalculating genuine \MET by ignoring the muon. The
assumption for the \HT triggers is that their efficiency is not
sensitive to whether there is genuine \MET in the event or not. The
results (efficiencies with respect to offline selection) are shown in
Table~\ref{tab:alphaT}.

\begin{table}[tbp]
\topcaption{Measured efficiencies of the \HT and \HT-\alpT triggers, as a
  function of \alpT and \HT, as measured with respect to the offline selection used in the $\alpha_\mathrm{T}$ analysis.}
\label{tab:alphaT}
\centering
\begin{tabular}{|ccc|} \hline\
$\alpT$ lower threshold  &  \HT range (\GeVns{})  & Efficiency(\%)  \\
\hline
$0.55       $ & 275--325   & $89.6^{+0.5}_{-0.6}$ \\
$0.55       $ & 325--375   & $98.5^{+0.3}_{-0.5}$ \\
$0.55       $ & 375--475   & $99.0^{+0.5}_{-0.6}$ \\
$0.55       $ & 475--$\infty$  & $99.4^{+0.5}_{-1.2}$ \\
\hline
\end{tabular}
\end{table}

\subsubsection{Triggers for inclusive search with Razor variables}

The Razor variables $R^2$ and $M_R$ were introduced in CMS to
complement other variables that can be used to  probe SUSY production at the
LHC~\cite{RazorPRD,RazorPRL}.
The analyses are designed to kinematically discriminate the pair production of heavy particles from
SM backgrounds, without making strong assumptions about the
$\MET$
spectrum or details of
the decay chains of these particles.  The baseline selection requires two or more reconstructed
objects, which can be calorimetric jets, isolated electrons or isolated muons.
The Razor kinematic construction
exploits the transverse momentum imbalance of SUSY events more
efficiently than the traditional $\MET$-based variables, retaining events
with as low as $\MET \approx50\GeV$ while reducing the background from
QCD multijet events to a negligible level. Details of the definition
of $R^2$ and $M_R$ can be found in the above references.

The use of $\MET$ and $\HT$ triggers alone would not be practical
for a Razor-based search, resulting in a nontrivial dependence of the
trigger efficiency on $R$ and $M_R$. Instead, a set of dedicated triggers was
developed, both for the fully hadronic and the leptonic final
states considered in the analysis.

The Razor triggers are based on the events with two central jets with
$\pt>64\GeV$, selected at  L1. The calorimetric towers in the
event are clustered using the  anti-\kt algorithm with a distance
parameter of $0.5$. The two highest $\pt$ jets are required to have
$\pt>65\GeV$, which is  fully efficient for PF
jets with $\pt>80\GeV$. If an event has more than seven jets with
$\pt>40\GeV$, it is accepted by the trigger. Otherwise, we consider
all the possible ways to divide the reconstructed jets in two
groups. We then form a \emph{mega-jet} summing the four-momenta of the
jets in one group. The mega-jet pair with the smallest sum of
invariant masses is used to compute the values of $R$ and $M_R$. A
selection on $R$ and $M_R$ is applied to define a leptonic
Razor trigger. A looser version of this selection is used for the
lepton Razor triggers, in association with one isolated muon or
electron with $\pt>12\GeV$. Electrons are selected with a loose calorimeter
identification requirement and a very loose isolation requirement.
The kinematic selection includes cuts on both on $R$ and $M_R$: $R^2>0.09$ and
  $M_R>150\GeV$ (inclusive trigger); $R^2>0.04$ and $M_R>200\GeV$
  (leptonic triggers).
A ``parked'' version (as described in Section~\ref{sec:HLTDAQ}) of the inclusive Razor trigger was
also implemented, requiring $R^2>0.04$.

\begin{figure}[tbp]
\centering
\includegraphics[width=0.4\textwidth]{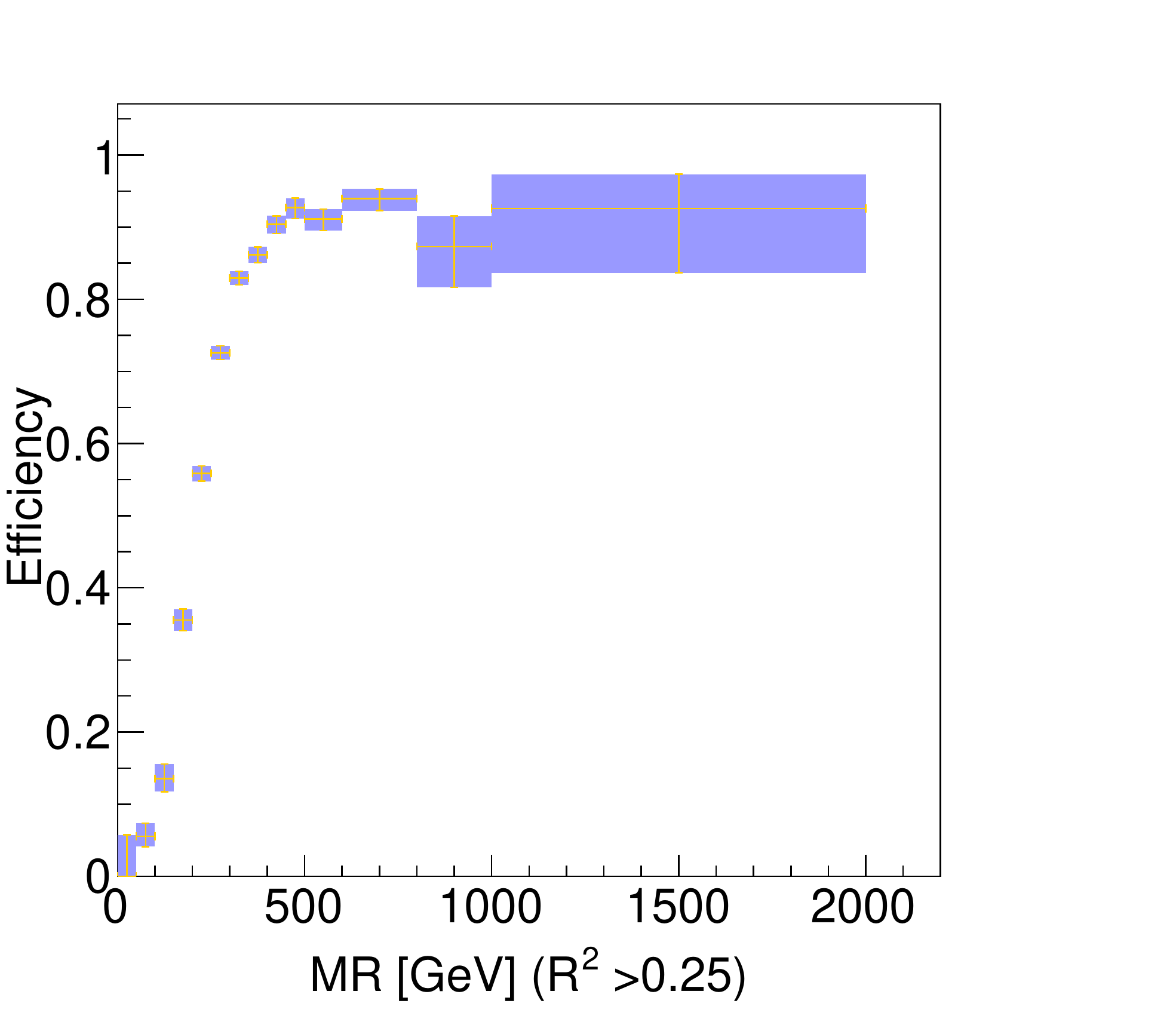}
\includegraphics[width=0.4\textwidth]{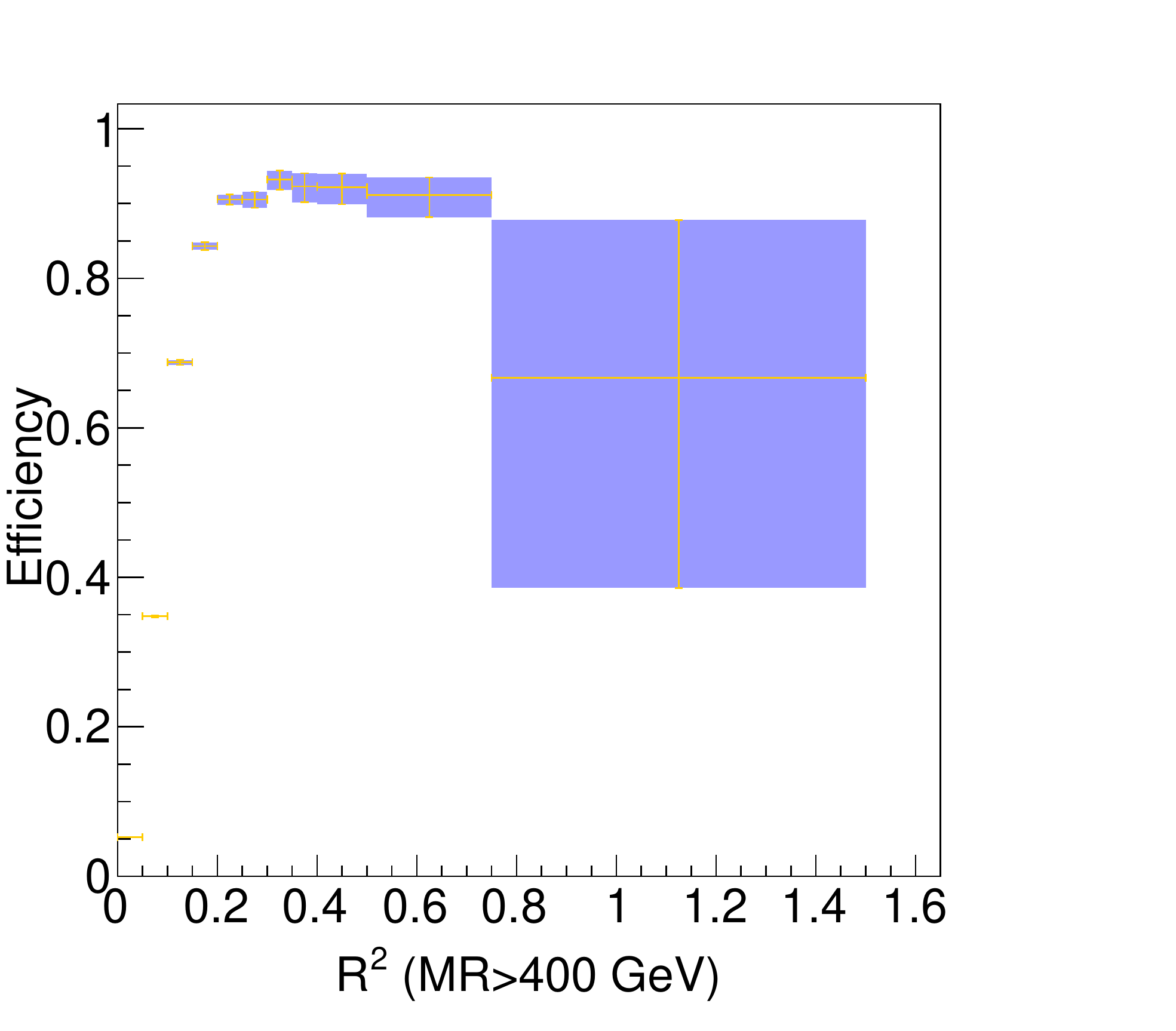}
\caption{\label{fig:razorturnon} Turn-on curve for $M_R$ (left) and $R^2$
  (right) for the inclusive Razor trigger, after requiring $R^2>0.25$
  (left) and $M_R>400\GeV$ (right). Events passing the
  single-electron triggers are selected to define the denominator of
  the efficiency, together with the dijet requirement. The
  requirement of satisfying the Razor trigger defines the
  numerator.}
\end{figure}

Events selected by the single-electron (single-muon) triggers are used
to measure the efficiency of the inclusive and electron (muon) Razor
paths. The baseline sample for the efficiency measurement is defined
requiring two jets of $\pt>80\GeV$, passing the reference trigger, and
not rejected by the event cleanup requirements (designed to remove the
noisy calorimeter events from the offline analysis). The numerator of
the efficiency is defined from this sample, with the requirement that the
relevant Razor trigger condition is satisfied. Figure~\ref{fig:razorturnon} shows
the efficiency versus $M_R$ and $R^2$ for the inclusive Razor trigger,
also requiring $M_R>400\GeV$ ($R^2>0.25$) in order for the $R^2$ ($M_R$)
plot to match the selection applied in the analysis. The efficiency is
found to be flat within the statistical precision, limiting the
precision on the tail or $R^2$ after the applied $M_R$ requirement. The analysis uses $(95 \pm 5)\%$ as an estimate of the efficiency.

\subsubsection{Triggers for photons and missing energy}

We present the triggers used in  a search for supersymmetry in events with at
least one isolated photon, jets, and \MET. Dominant standard model
background processes are direct photon production and QCD
multijet events where
a jet is misreconstructed as a photon. Multijet events have small
intrinsic \MET, but the finite resolution of the jet energy measurement
together with the large cross section leads to a significant
contribution in the tail of the \MET. Other backgrounds arise from
electroweak electron production, \eg, $\PW\to \Pe\nu$, where
an electron is misreconstructed as a photon. Additional contributions
are expected from initial- or final-state photon radiation in various
QCD and electroweak processes. Single-photon trigger
thresholds are too high for the efficient selection of
many SUSY benchmark points, so that for this analysis a cross-trigger
based on a single photon and \HT is
used. The main backgrounds are modeled using data control samples.

To trigger on the signal as well as to collect the control samples used for
estimation of the QCD multijet and electroweak backgrounds, a cross-trigger is used, requiring at least one photon with $\pt > 70\GeV$ and
$\HT>400\GeV$.
The control region is defined by events containing at least one
isolated photon with $\pt > 80\GeV$ and $\abs{\eta}<1.4$, two or more jets
with $\pt > 30\GeV$ and $\abs{\eta}<2.6$, and $\HT>450\GeV$. The signal
region includes an additional $\MET>100\GeV$ requirement.

The trigger efficiency was measured in data for the photon and \HT
legs, using a single-photon baseline trigger, which requires a single
photon with $\pt > 50\GeV$ and is expected to be fully efficient in
the kinematic region of interest. As the statistical
power of the data sample is limited by the large prescale of the
baseline trigger (prescale of 900), a cross-check is performed using a
less prescaled single photon trigger with $\pt > 75\GeV$ (prescale of
150). In this case, it is not possible to observe the  \pt turn-on of the
photon leg efficiency, as the baseline selection is more restrictive than the online
selection used by the analysis; however this is a valid check
of the \HT leg.
\begin{figure}[tbp]
\centering
\includegraphics[width=0.48\textwidth]{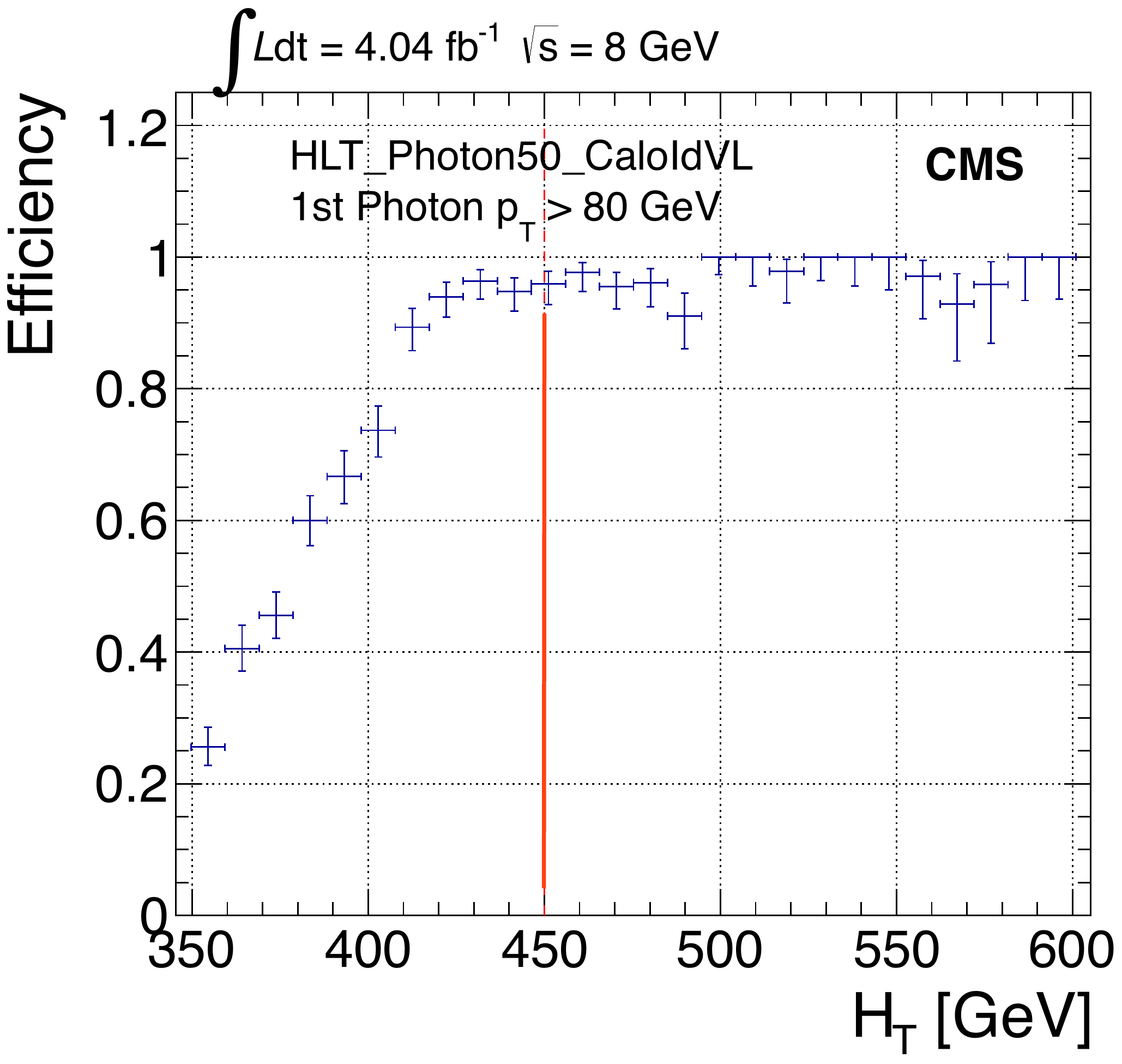}
\includegraphics[width=0.48\textwidth]{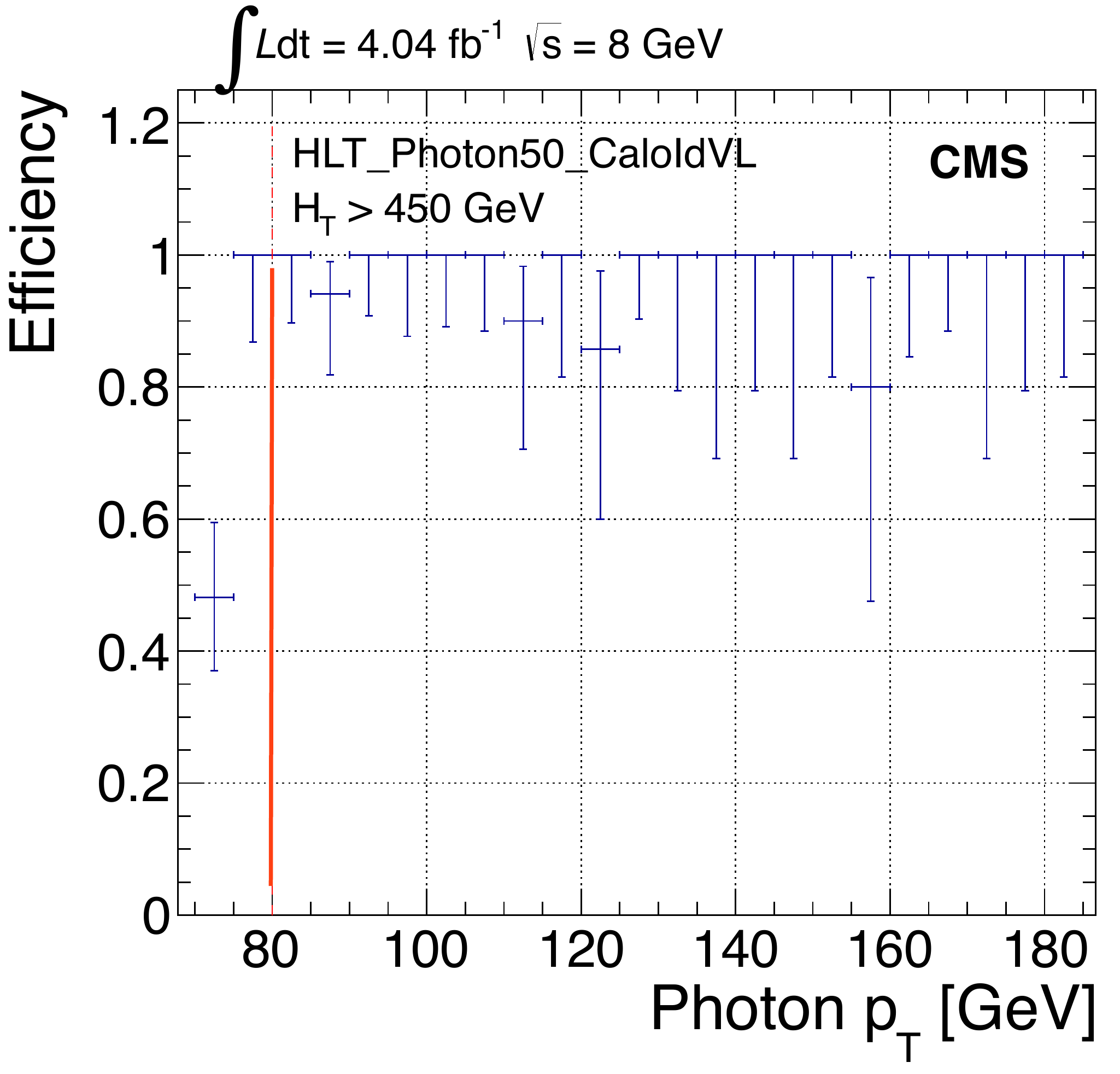}
\includegraphics[width=0.48\textwidth]{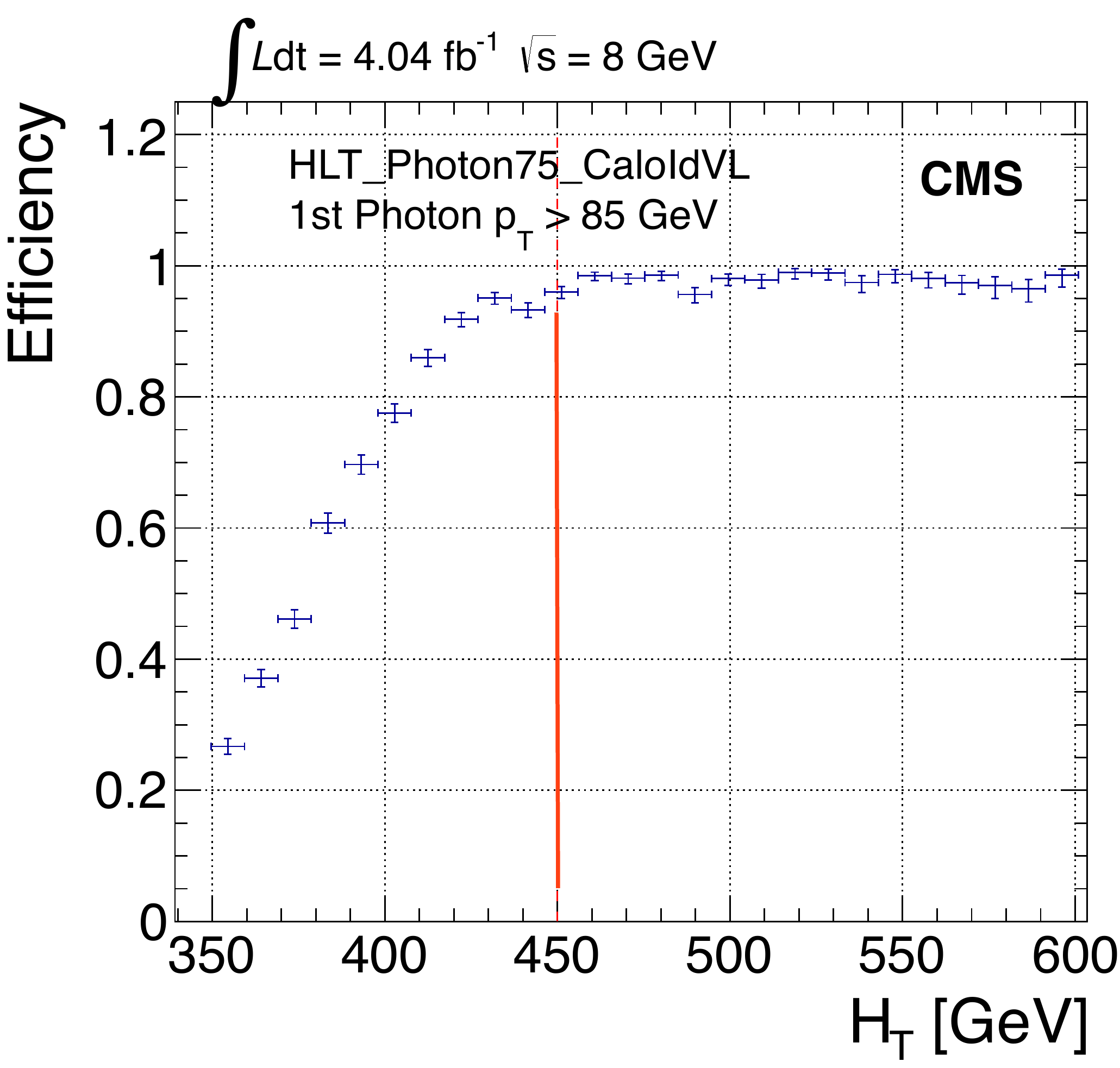}
\includegraphics[width=0.48\textwidth]{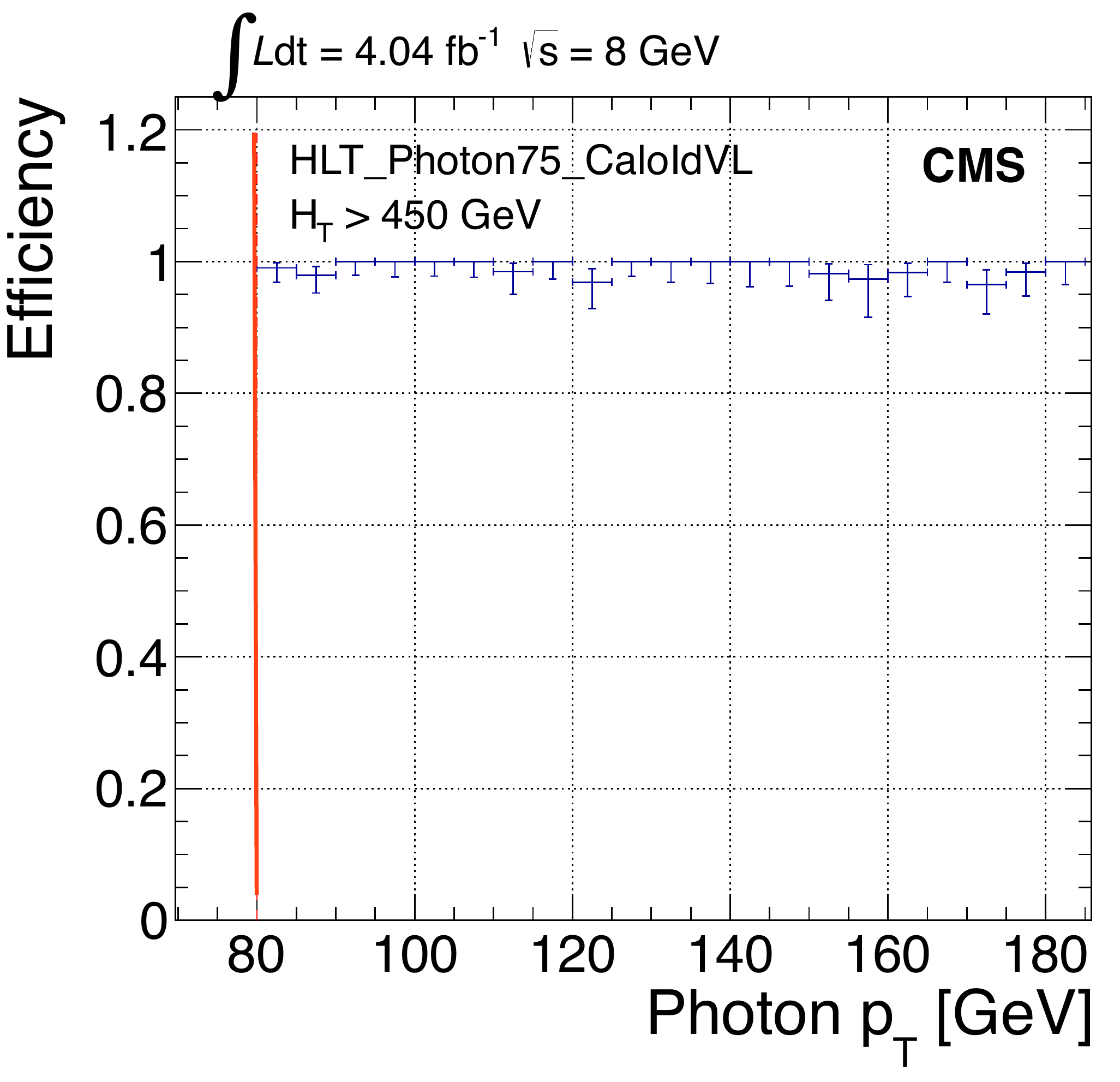}
\caption{Supersymmetry search in the $\gamma$ + \MET channel: trigger
  efficiency of the \HT leg (left column), and the photon leg (right column), using as
  a reference the single-photon trigger with $\pt > 50\GeV$ (top row) and
  $\pt > 75\GeV$ (bottom row). The red lines indicate offline
  requirements.}\label{fig:PhotonHT}
\end{figure}
Figure~\ref{fig:PhotonHT} shows the turn-on curve for the
\HT and photon \pt legs, both single-photon triggers. Only in the \HT
leg for the single-photon trigger with the $\pt > 75\GeV$
requirement, a higher threshold in the photon $\pt > 85\GeV$ is
used to avoid regions with inefficiencies due to the cross-trigger. After applying the offline analysis requirements on the
photon momentum of $\pt> 80\GeV$ and on $\HT > 450\GeV$, indicated
in the figure, the trigger
is fully efficient within an uncertainty of 4\%. The uncertainty is
due to the low statistical power of the data set.

\subsubsection{Triggers for heavy stable charged particles}
\label{sec:HSCP}
\begin{figure}[tbph]
	\centering
          \includegraphics[width=\textwidth]{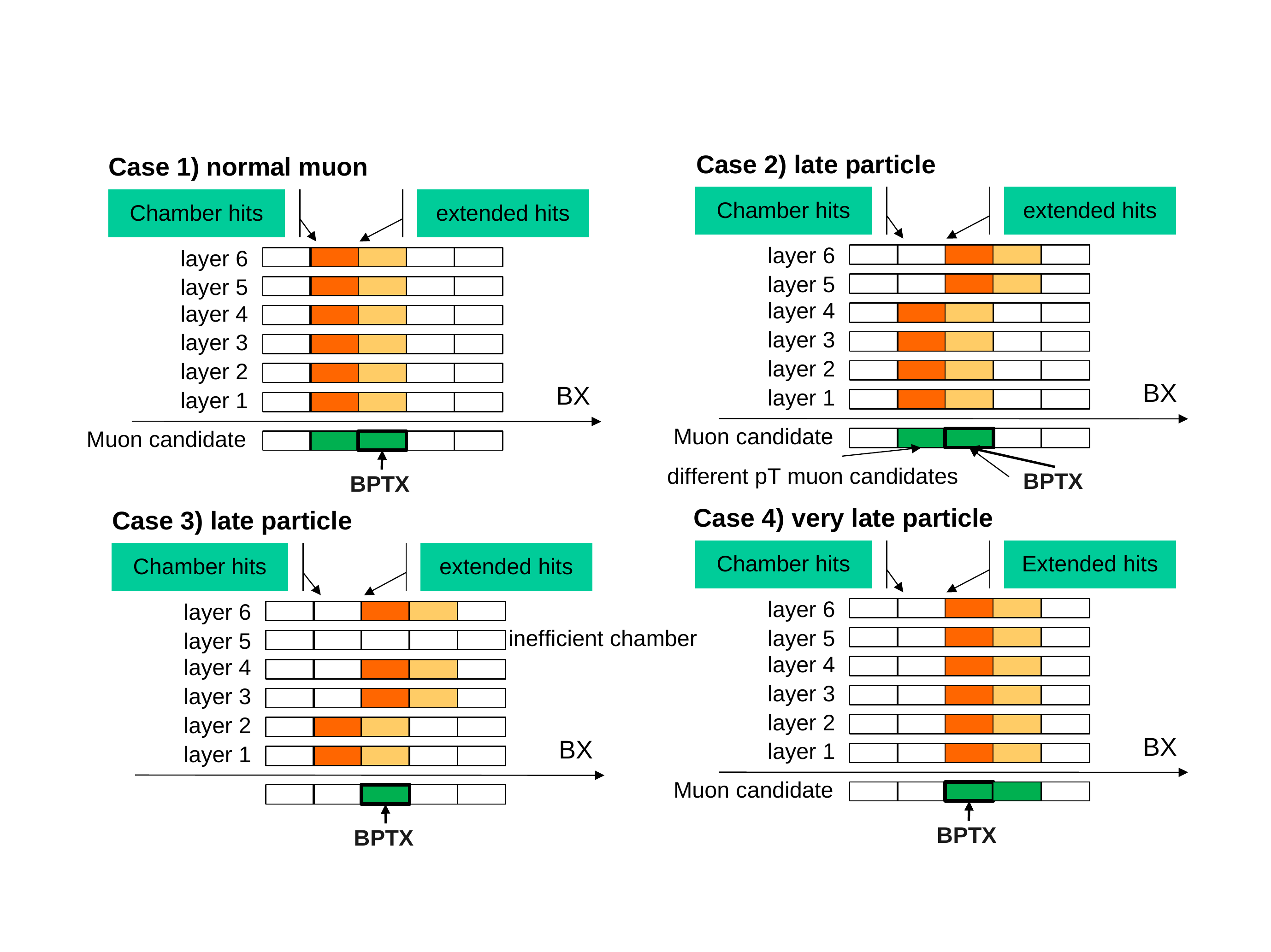}
	\caption{The principle of operation of the RPC HSCP trigger
          for an ordinary muon (case 1), and a slow minimum ionizing
          particle, which produces hits across two consecutive bunch crossings
          (cases 2, 3) or in the next BX (case 4). Hits that would be
          seen in the standard PAC configuration are effectively those
          shown in pale orange; additionally observed hits in the HSCP
          configuration are those shown in dark orange. In case 1 the
          output of both configurations is identical, in case 2 the
          HSCP configuration uses the full detector information, in
          case 3 only the HSCP configuration can issue a trigger, and in
          case 4 the HSCP configuration brings back the event to the
          correct BX.}
	\label{fig:rpc-hscp}
\end{figure}

The CMS experiment has a specific RPC muon trigger configuration to increase the
efficiency for triggering on heavy stable charged particles (HSCP)
using the excellent time resolution of detected muon
candidates. Double-gap RPCs operating in avalanche mode have an
intrinsic time resolution of around 2\unit{ns}. This, folded with the
uncertainty coming from the time propagation along the strip, which
contributes about 2\unit{ns}, and the additional jitter that comes from
small channel-by-channel differences in the electronics and cable
lengths, again of the order of 1--2\unit{ns}, give an overall time resolution
of about 3\unit{ns}--much lower than the 25\unit{ns} timing window of the RPC
data acquisition system.

If hits are not in coincidence within one BX, the RPC PAC algorithm is
likely to fail because the minimum plane requirements would not be
met, or if the algorithm does succeed, a lower quality value and
possibly a different \pt will be assigned to the trigger particle. In addition, if the muon trigger is one BX
late with respect to the trigger clock cycle, the pixel hits will not
be recorded and the muon chamber calibration constant will be
suboptimal, resulting in a poor offline reconstruction of late
``muon-like'' candidates. The functionality to extend the RPC hits to
two (or more) consecutive BXs, plus the excellent intrinsic timing
capabilities of the RPCs, allow the construction a dedicated physics
trigger for such ``late muons''. In the PAC logic the RPC hits are
extended in time to 2 BXs, hence the plane requirements are met for at
least one BX and triggers can be issued. On the GMT input, the RPC
candidates are advanced by one BX with respect to DT and CSC candidates,
so hits of a ``late muon'' generate a trigger in the proper BX.
Ordinary ``prompt'' muons will produce two trigger candidates: one in
the proper BX and one in the previous BX. Misreconstructed candidates can, however, be
suppressed at the GT level by a veto operated on the basis of BPTX coincidences
(Section~\ref{sec:bptx}). Figure~\ref{fig:rpc-hscp} shows the
principle of
operation of the RPC-based HSCP trigger. Studies with simulated data
indicate that the HSCP trigger configuration significantly increases
the CMS capability to detect a slow HSCP, for example, for an 800\GeV
long-lived gluino, the overall improvement in trigger efficiency ranges from 0.24
to 0.32. The gain is the largest within the range
$200 < \pt < 600\GeV$ and for gluino velocities $0.4 < \beta <
0.7$.
The HSCP trigger configuration was the main RPC operation mode during
data-taking in most of the 2011 and the entire 2012 run.

\subsection{Exotic new physics scenarios}

Models of physics beyond the standard model that are not
supersymmetric are called `exotic" in CMS. In this section we describe
three exotic physics scenarios and the triggers used in
searches for these signals.

\subsubsection{Triggers for dijet resonance searches}

During the 7\TeV run, the search for heavy resonances decaying to jet
pairs was performed on events triggered by the single-jet
trigger. With increasing peak luminosity, the tighter threshold
applied on the jet $\pt$ became a major problem for the analysis. At
the same time, the analysis was improved by introducing the so-called
\emph{wide jets} to take into account the presence of additional jets
from final-state radiation.
Wide jets are formed around a given set of \emph{seed} jets, taking as
input the other jets in the event. The four-momentum of each seed
jet is summed with the four-momenta of other jets within $\DR <1.1$ of the seed
jet and with $\pt>40$\GeV. A jet close to more than one seed jet is
associated with the closest seed.
With this new approach, a trigger based on $\HT=\sum_\text{jet} |\pt|$
is more efficient. A further improvement in the analysis was
obtained by implementing a dedicated topology-based trigger, applying
a looser version of the analysis reconstruction and selection
requirements at the HLT:
\begin{itemize}
\item Wide jets were built by looking for jets with $\pt>40$\GeV in a
  cone of size $\DR=1.1$ around the two highest $\pt$ jets;
\item Multijet events were removed by requiring that the two wide
  jets fulfill $\Delta \eta<1.5$.
\end{itemize}
During the 8\TeV run, events were kept if the wide jets built around
the two highest $\pt$ jets had an invariant mass larger than
$750$\GeV ({\it Fat750}). While this trigger alone would have
performed similarly to  the \HT trigger already in use, the
combination of the two triggers in a logical OR allowed us to recover
the inefficiency for mass values close to the applied threshold,
making the overall efficiency turn-on curve sharper.
The loosest \HT-based L1 path (L1\_HTT150) was used as a seed for all
triggers.
The trigger efficiency was measured in data, taking the events
triggered by the prescaled $\HT>550\GeV$ trigger as a baseline. These
events were filtered by applying the analysis selection (particularly,
the $\Delta \eta$ requirement on the two wide jets) to define the
denominator of the efficiency curve. The subset of these events also
satisfying the analysis requirements defines the numerator of the efficiency.
\begin{figure}[tb]
\centering
\includegraphics[width=0.7\textwidth]{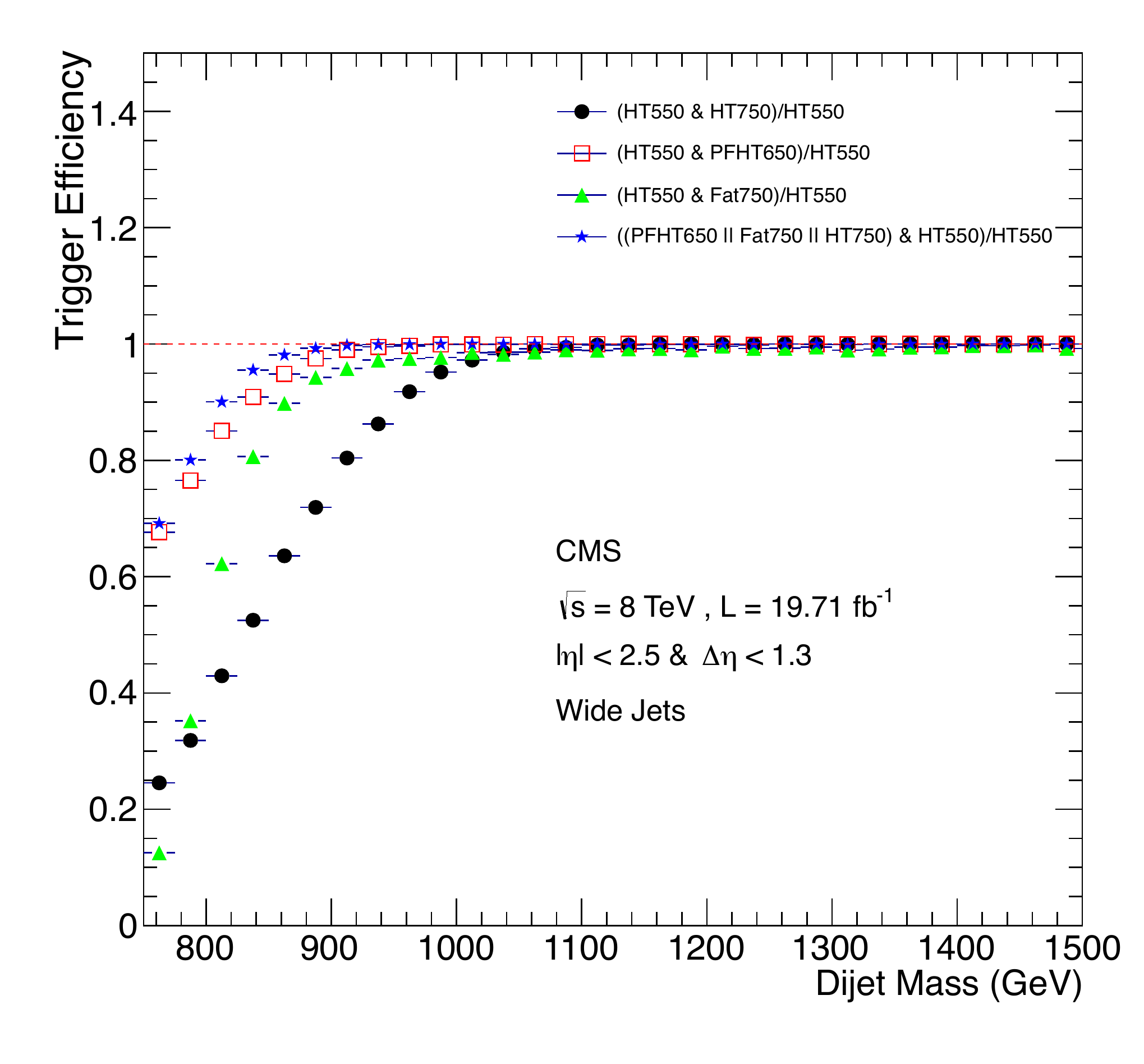}
\caption{ Dijet resonance search triggers.
  The HLT efficiency of $\HT>650\GeV$, $\HT>750\GeV$, and
  \emph{Fat750} triggers individually, and their
  logical OR as a function of the offline dijet mass. The efficiency
  is measured with the data sample collected with a trigger
  path that requires $\HT>550\GeV$.
  The horizontal dashed line marks the trigger efficiency ${\geq}
  99\%$. }
\label{fig:dijet}
\end{figure}
Figure~\ref{fig:dijet} shows the trigger efficiency as a function of the offline
dijet mass for individual triggers and for their logical OR.
While the combination of the \HT and Fat750 triggers already
represents a sizable improvement with respect to the individual
triggers, a further increase in the efficiency was obtained with the
introduction of the PF-based \HT trigger. The combination of the three
triggers made the analysis $\geq 99\%$ efficient for invariant masses
above 890\GeV. As a result of the trigger improvements, the threshold
for the dijet resonance search for the 8\TeV run was 100\GeV lower
than would have been possible if the 7\TeV strategy had been used.
\subsubsection{Triggers for black hole search}
\label{sec:blackholehpa}

If the scale of quantum gravity is as low as a few \TeVns, it is possible
for the LHC to produce microscopic black holes or their quantum
precursors (``string balls") at a significant
rate~\cite{Dimopoulos:2001hw,Giddings:2001bu,Dimopoulos:2001qe}. Black holes decay
democratically, \ie, with identical couplings to all standard
model degrees of freedom.
Roughly 75\% of the black holes decay products are jets. The
average number of particles in the final state varies from roughly two
(in case of quantum black holes) to half a dozen (semiclassical black
holes and string balls). The microscopic black holes are massive
objects, thus at least a few hundred\GeV of visible energy in the
detector is expected.

Since \textit{a priori} we do not know the precise final state, we
trigger on the total jet activity in an event. The common notation of
such triggers is \texttt{HLT\_HT}$x$, \texttt{HLT\_PFHT}$x$, and
\texttt{HLT\_PFNoPUHT}$x$, where $x$ denotes the total energy in
\GeV.
All energies of HLT jets are fully corrected, and in the case of the
\texttt{HLT\_PFNoPUHT}$x$ paths, pileup corrections are also applied
to the HLT PF jets. The pileup subtraction is performed by
first removing all of the jet's charged hadrons not associated to the
primary vertex, then calculating an energy offset based on the jet
energy density distribution to remove the remaining pileup
contribution. More details of the jet reconstruction at L1 and
 HLT are given in Section~\ref{sec:JetMET}.

After the jets are selected at both the L1 and the HLT, an \HT
variable is calculated.
In Ref.~\cite{bh_2012_legacy}, the jet \ET threshold at L1 is 10\GeV and the \HT thresholds are 150,
175, and 200\GeV  (Section~\ref{sec:JetHLT}.) These L1 triggers are
used as seeds to the HLT algorithms. At the HLT, the jet \ET threshold is 40\GeV and the \HT thresholds have a range of
650--750\GeV. The unprescaled HLT paths and their L1 triggers are
summarized in Table~\ref{tab:BHTrigger}. The L1 triggers for some of
the ``total jet activity'' paths were updated in the middle of 2012 to
account for higher instantaneous luminosity of the LHC. For
simplicity, we refer to the data taking periods before (after) that
change as ``early" (``late"). In the previous iterations of the
analysis~\cite{bh_full2010_plb,bh_full2011_jhep}, the \HT thresholds
at the HLT were as low as 100--350\GeV.

\begin{table}[tbp]
\centering
\topcaption{Black Hole trigger: Unprescaled total jet activity HLT paths
  and their respective L1 seeds. The L1 seeds for a number of the HLT paths were
  revised during the data taking to account for higher instantaneous
  luminosity.}
\begin{tabular}{ | l  l  l | }
        \hline
        Path name & \multicolumn{1}{c}{L1 seed} & \multicolumn{1}{c}{Data-taking period}\\
        \hline
	\texttt{HLT\_HT750} & \texttt{L1\_HTT150 OR L1\_HTT175} & Early \\
	\texttt{HLT\_HT750} & \texttt{L1\_HTT150 OR L1\_HTT175 OR L1\_HTT200} & Late \\		
	\hline
	\texttt{HLT\_PFHT650} & \texttt{L1\_HTT150 OR L1\_HTT175} & Early \\
	\texttt{HLT\_PFHT650} & \texttt{L1\_HTT150 OR L1\_HTT175 OR L1\_HTT200} & Late  \\
	\texttt{HLT\_PFHT700} & \texttt{L1\_HTT150 OR L1\_HTT175} & Early  \\
	\texttt{HLT\_PFHT700} & \texttt{L1\_HTT150 OR L1\_HTT175 OR L1\_HTT200} & Late  \\	
	\texttt{HLT\_PFHT750} & \texttt{L1\_HTT150 OR L1\_HTT175} & Early \\
	\texttt{HLT\_PFHT750} & \texttt{L1\_HTT150 OR L1\_HTT175 OR L1\_HTT200} & Late  \\		
	\hline
	\texttt{HLT\_PFNoPUHT650} & \texttt{L1\_HTT150 OR L1\_HTT175} &   \\
	\texttt{HLT\_PFNoPUHT700} & \texttt{L1\_HTT150 OR L1\_HTT175} &   \\		
	\texttt{HLT\_PFNoPUHT750} & \texttt{L1\_HTT150 OR L1\_HTT175} &   \\
	\hline
\end{tabular}
\label{tab:BHTrigger}
\end{table}

\begin{figure}[htbp]
\centering
\includegraphics[width=0.4\textwidth]{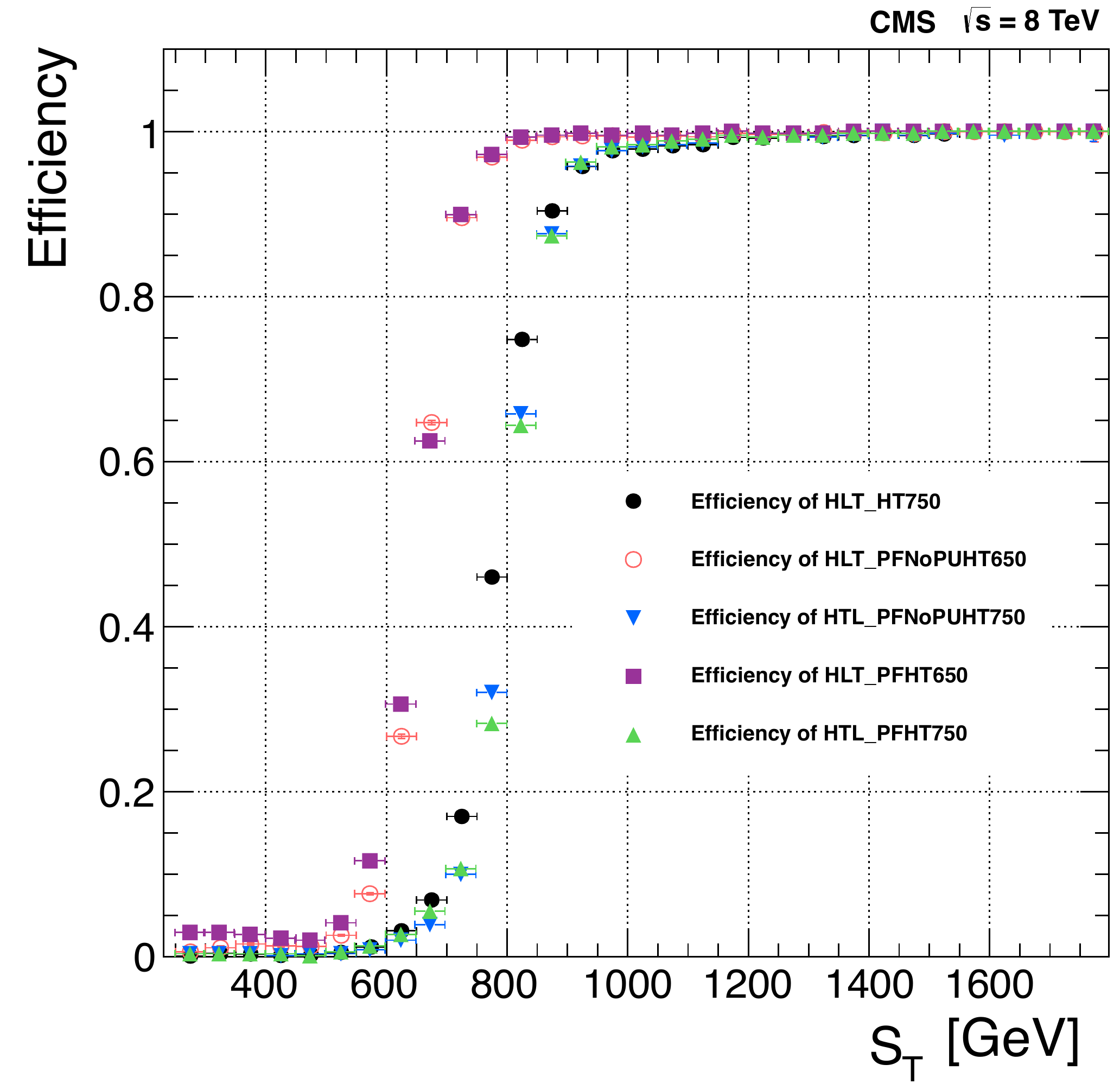}
\includegraphics[width=0.4\textwidth]{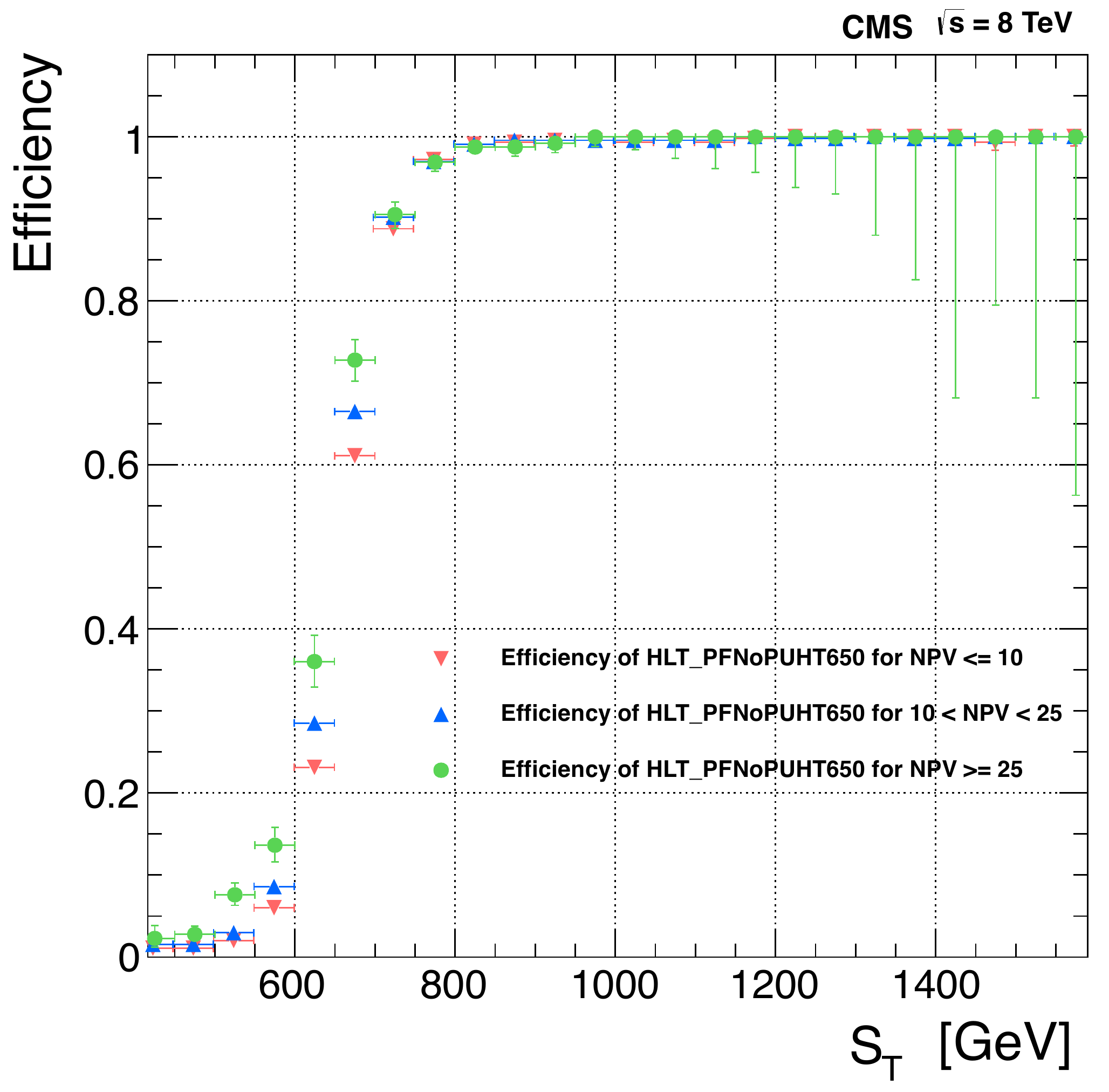}
\caption{Left: Efficiency of unscaled total jet activity HLT paths as a function of \ST. Right: Efficiency of \texttt{HLT\_PFNoPUHT650} as a function of \ST in three bins of number of primary vertices, $N_\mathrm{PV}$: (i) $N_\mathrm{PV} \le 10$, (ii) $10 < N_\mathrm{PV} < 25$, and (iii) $N_\mathrm{PV} \ge 25$. All efficiencies are calculated with respect to a prescaled total activity path with $\HT = 450$\GeV threshold.}
\label{fig:ST}
\end{figure}

As the majority of the final-state objects are jets, we use
jet-enriched collision data to search for black holes. These data are
recorded using a logical OR of the following trigger groups, whose
triggers only differ by a threshold: i) total jet activity
triggers, ii) paths that select high-mass dijet events, iii) triggers
that require presence of significant \MET and a jet with \pt above a few
hundred\GeV. The main offline quantities that describe the black hole
are the multiplicity of the final-state objects, $N$, and a scalar sum of
transverse momenta of all objects (jets, leptons, and photons) and the
\MET reconstructed in the event, $\ST = \sum \pt^{\text{jets}} + \sum
\pt^{\text{leptons}} + \sum \pt^{\text{photons}} + \MET$. We apply a
50\GeV requirement on all final-state objects \pt and \MET, and select events
with a multiplicity greater than one. Note that \MET is not counted
towards the multiplicity. The relative efficiency of unprescaled HLT paths
that are used in the analysis as a function of the $\ST$ is shown in
Fig.~\ref{fig:ST}~(left). The efficiencies are calculated using the
same jet-enriched data set with respect to prescaled total jet
activity path with the \HT threshold of 450\GeV. The paths with the
\HT threshold of 650 (750)\GeV are fully efficient starting from $\ST
= 1000~(1200)$\GeV, respectively, which is significantly below the
low-\ST boundary of 1500\GeV that is used in the search. To
check the pileup dependence of the trigger turn-on, we plot the
efficiency of the selection path \texttt{HLT\_PFNoPUHT650} as a function of
\ST in three bins of the number of reconstructed primary vertices, $N_\mathrm{PV}$: i)
$N_\mathrm{PV} \le 10$, ii) $10 < N_\mathrm{PV} < 25$, and iii) $N_\mathrm{PV}
\ge 25$ (Fig.~\ref{fig:ST}~(right)). Although the trigger turn-on curves
become less sharp when $N_\mathrm{PV}$ increases, this does not affect
the point when the trigger becomes fully efficient. Thus, the pileup
dependence of total jet activity triggers can be neglected in the
black holes analysis.

\subsection{B physics and quarkonia triggers}
\label{sec:bphbpag}

\begin{figure}[tbp]
\centering
\includegraphics[width=0.8\linewidth]{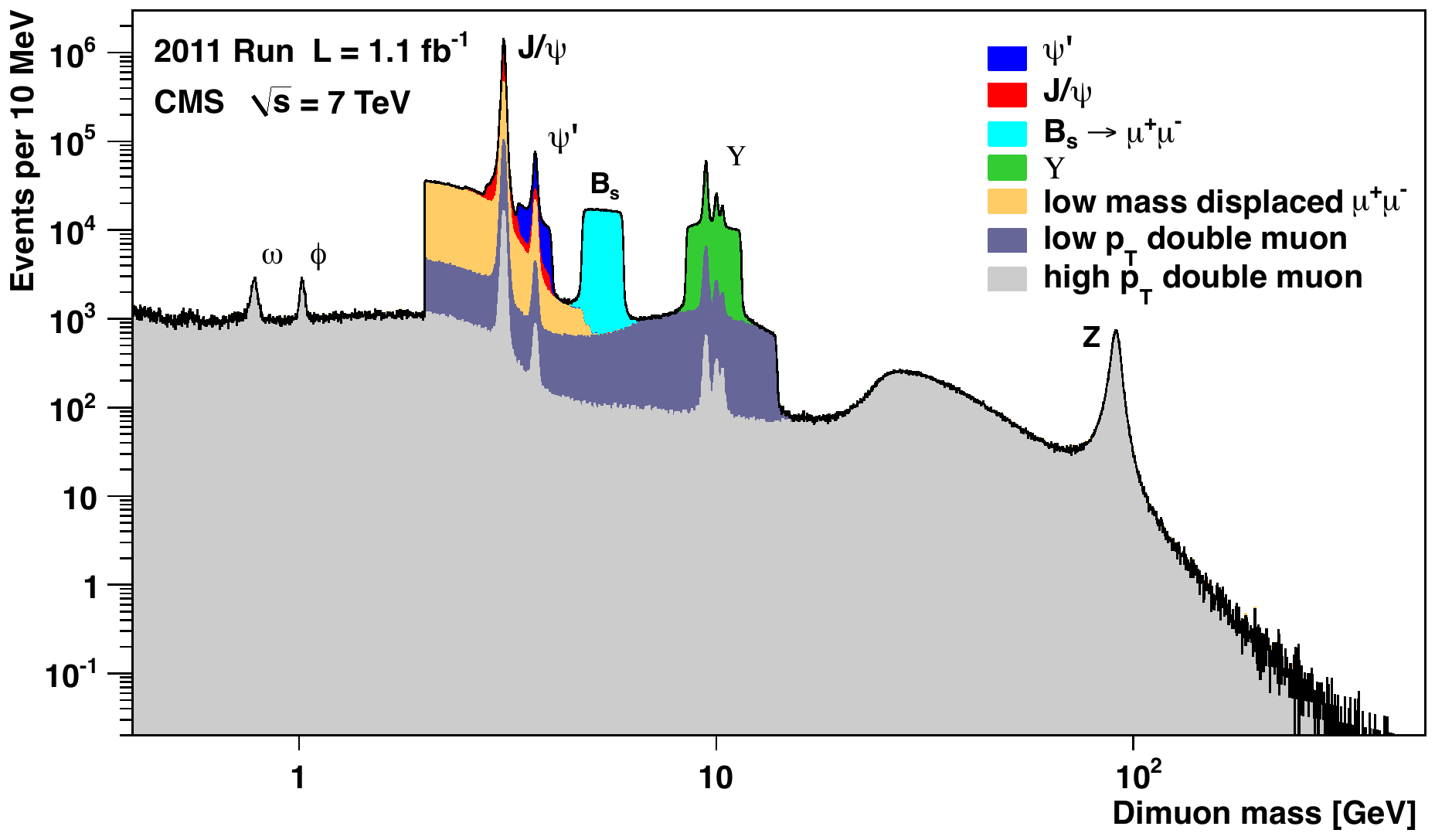}
\caption{\label{fig:BPH:DimuonMass} Dimuon mass distributions
  collected with the inclusive double-muon trigger used during early data taking in 2011. The colored areas
  correspond to triggers requiring dimuons in specific mass windows,
  while the dark gray area represents a trigger only operated during
  the first 0.2\fbinv of the 2011 run. }
\end{figure}
The CMS analyses in the fields of B physics and quarkonium production
are mostly based on data samples collected with double-muon
triggers. In the 2010 run, the LHC instantaneous luminosity was
sufficiently low such that relatively loose triggers could be used.
Essentially all the analyses made in the B physics group were
based on one inclusive trigger, which requires two high quality
muons. The resulting dimuon mass distribution covers dimuon mass
values from threshold all the way to 200\GeV, displaying ``needles''
caused by the dimuon decays of resonances on top of a smooth
underlying continuum.

The significantly higher collision rates of the 2011 LHC run, and the
ceiling of around 25--30\unit{Hz} for the total trigger bandwidth allocated
for B physics, required the development of several specific HLT
paths, each devoted to a more or less exclusive set of physics
analyses. Figure~\ref{fig:BPH:DimuonMass} illustrates the
corresponding dimuon mass distributions, stacked on each other. The
high-rate ``low-\pt double muon'' path was in operation only during
the first few weeks of the run; the others had their rates reduced
through suitable selection requirements on the dimuon mass and on the
single-muon and/or dimuon \pt.

The quarkonia trigger paths (\JPsi, $\psi^\prime$ and $\PgU$) had explicit requirements on the \pt of the dimuon system but not of the
single muons. First, because the analyses are made as a function of the
dimuon \pt and second, because the single-muon \pt requirements induce a
significant restriction of the covered phase space in terms of the
angular decay variable $\cos\theta$, and this is crucial for the measurements of
quarkonium polarization. To further reduce the rate, the two muons
were required to bend away from each other because the ones bending
towards each other have lower
efficiencies.
The dimuon was required to have a central rapidity,
$\abs{y} < 1.25$. This is particularly useful to distinguish the
\PgUb\xspace and \PgUc\xspace resonances, as well as for analyses of
P-wave quarkonia production, which require the measurement of the
photon emitted in the radiative decays (\eg, $\chi_\mathrm{c} \to \JPsi \Pgg$). In fact, to resolve the \Pcgci and \Pcgcii\xspace peaks
(or, even more challenging, the \Pbgci\xspace
and \Pbgcii\xspace
peaks), it is very important to have a high-resolution measurement of
the photon energy, possible through the reconstruction of the
conversions into $\Pep\Pem$ pairs in the barrel section of the silicon
tracker.

In addition to the quarkonia resonances, Fig.~\ref{fig:BPH:DimuonMass}
shows a prominent ``peak'' labeled $\mathrm{B}_\mathrm{s}$, which represents the
data collected to search for the elusive $\mathrm{B}_\mathrm{s} \to \Pgm\Pgm$ and
$\mathrm{B}_\mathrm{d} \to \Pgm\Pgm$ decays. These triggers had no restrictions
on the dimuon rapidity or relative curvature and kept \pt requirements much looser
than those applied in the offline
analysis.
The total rate of the $\mathrm{B}_\mathrm{s}$ trigger paths remained
relatively small, of the order of 5\unit{Hz}, even when the LHC
instantaneous luminosity exceeded $7 \times 10^{33}$\percms, at the
end of the 2012 run.

The other prominent trigger path illustrated in
Fig.~\ref{fig:BPH:DimuonMass}, the ``low-mass displaced
dimuons'', selected events with a pair of opposite-sign muons with a
dimuon vertex pointing back to and displaced from the interaction
point by more than three standard deviations. These events were
collected to study decays of B mesons into final states containing a
pair of muons plus one or more kaon and/or pion, as well as to measure
the $\Lambda_\mathrm{b}$ cross section, lifetime, and polarization. This is the
most challenging trigger path because of its very high rate, which
cannot be reduced through the increase of muon \pt requirements without
a significant loss of signal efficiency.

The main difference between the 2011 and 2012 runs, from the
perspective of B physics, was the availability of the
so-called ``parked data'' (Section~\ref{sec:HLTDAQ}). The resulting
increase in available HLT bandwidth meant that most trigger paths
could have looser requirements in 2012 than in 2011. Additionally, several new
triggers were added, including a like-sign dimuon trigger to study the
``anomalous dimuon charge asymmetry'' observed at the
Tevatron~\cite{Abazov:2010hj}.

Two special calibration triggers were developed to study the single-muon detection efficiencies in an unbiased way. One is a single muon
trigger that requires the presence of an extra track such that the
invariant mass of the muon-track pair is in the \JPsi mass region; the
existence of a \JPsi peak in this event sample ensures that the track
is likely to be a muon that can be used to provide an unbiased
assessment of the muon-related efficiencies (offline reconstruction in
the muon detectors, as well as L1 and L2 trigger efficiencies as described in Section~\ref{sec:muHLT}). The
other is a dimuon trigger for those low-mass dimuons in which the
muons are reconstructed without using any information from the
silicon tracker hits, thereby allowing the study of the offline
tracking and track quality selection efficiencies, as well
as the L3 trigger efficiency (Section~\ref{sec:muHLT}). These efficiency measurements are made
using a tag-and-probe methodology.
\begin{figure}[tbp]
\centering
\resizebox{0.5\linewidth}{!}{%
\includegraphics{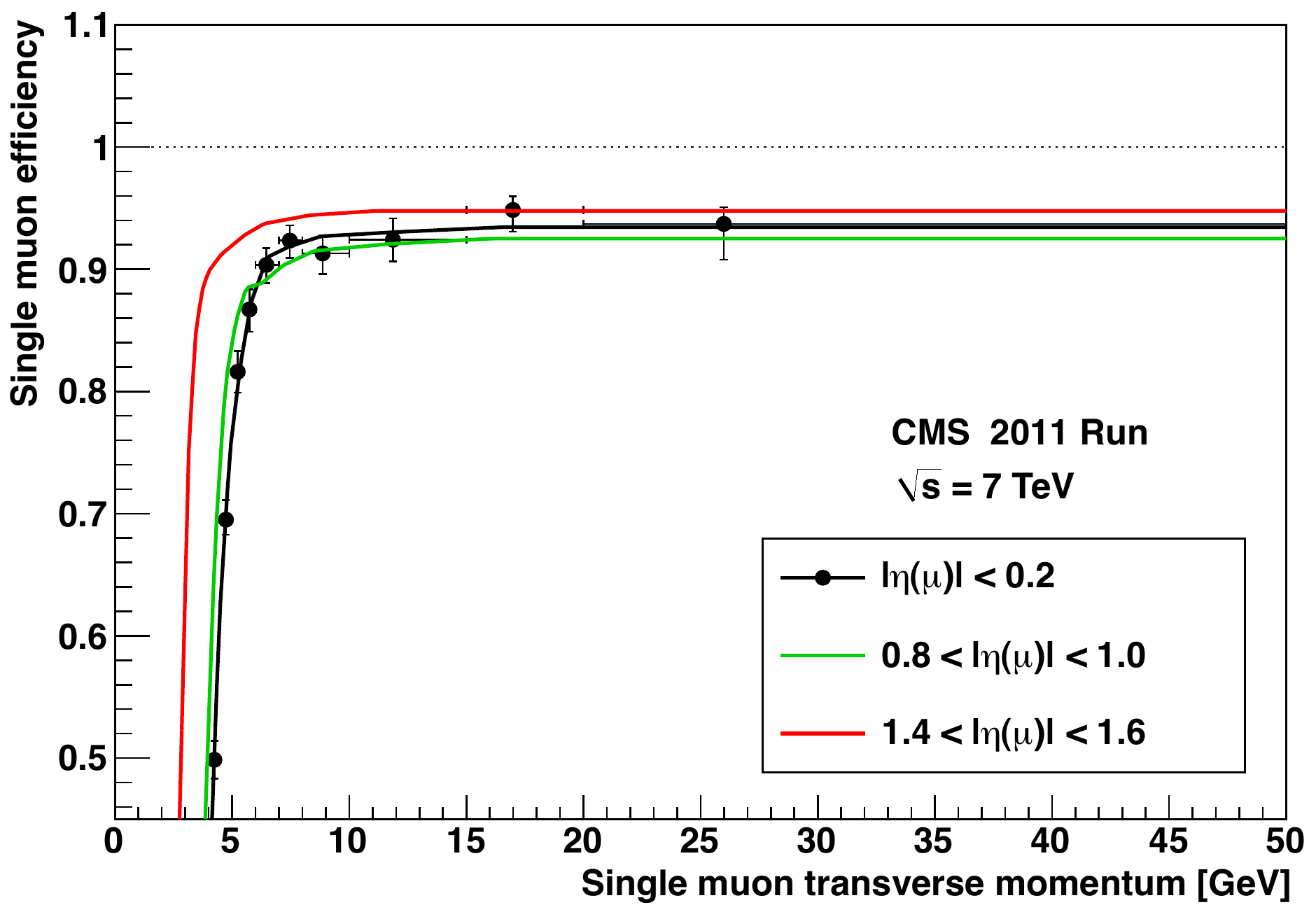}}
\caption{\label{fig:BPH:2011MuonEff} Single-muon detection
  efficiencies (convolving trigger, reconstruction, and selection
  requirements) as a function of \pt, as obtained from the data, using the
  tag-and-probe method. Data points are shown for the pseudorapidity
  range $\abs{\eta} < 0.2$, while the curves (depicting a parametrization
  of the measured efficiencies) correspond to the three ranges
  indicated in the legend.}
\end{figure}
As an illustration, Fig.~\ref{fig:BPH:2011MuonEff} shows the single-muon
detection efficiency as a function of \pt for three muon
pseudorapidity ranges.

The rate of events with single muons is very large and it might happen
that a muon is mistakenly identified as two close-by muons. To prevent
such events from increasing the rate of dimuon triggers, the
trigger logic at L1 and L2 discards dimuon signals if the two muon
trajectories are too close to each other. The drawback is that this
significantly reduces the efficiency of the dimuon trigger for signal
dimuons where the two muons are close to each other, which happens
quite often for low-mass dimuons of high \pt.
\begin{figure}[tbp]
\centering
\includegraphics[width=0.65\textwidth]{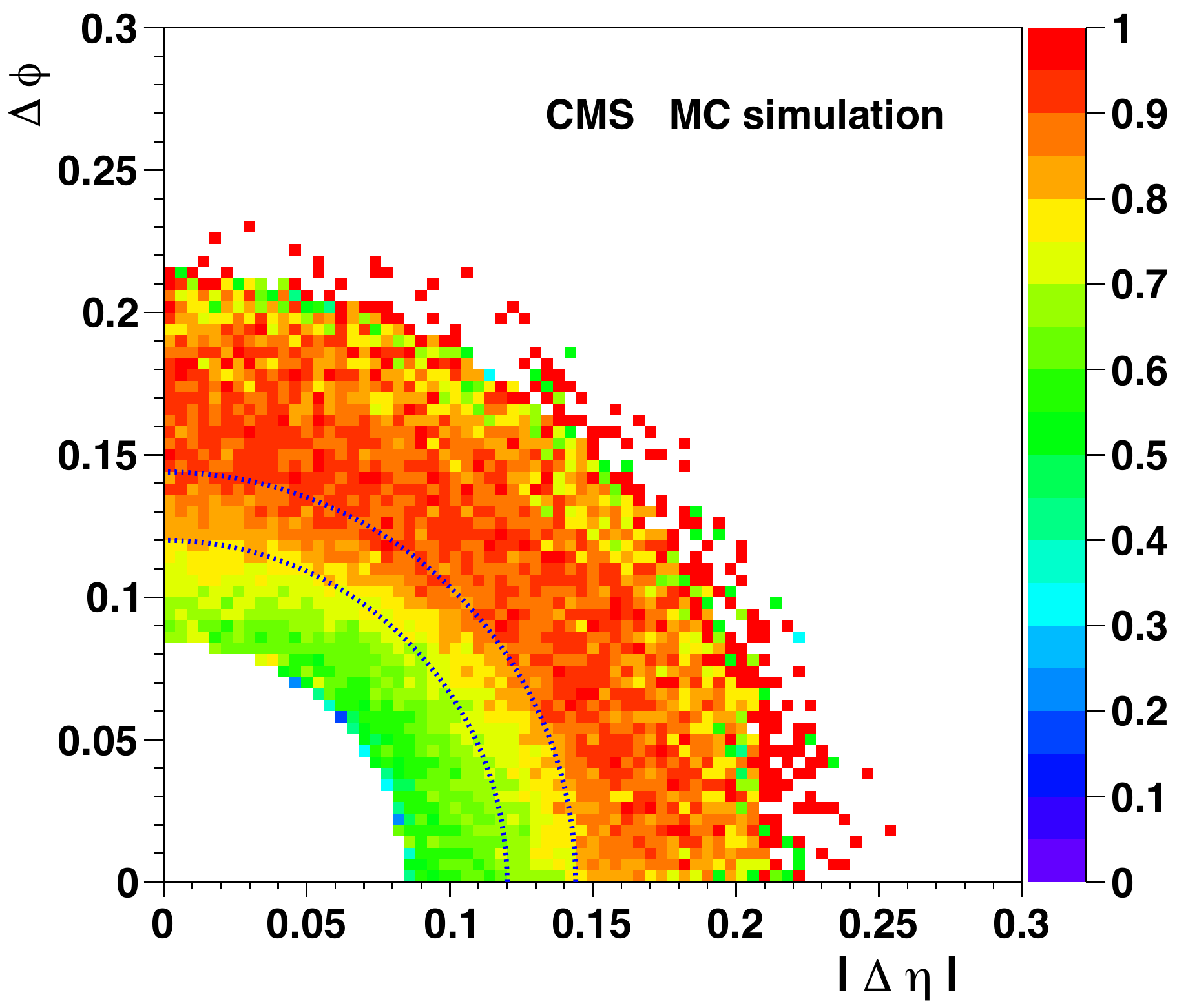}
\caption{Dimuon trigger efficiencies in the $\Delta\phi$ versus
  $\Delta\eta$ plane for \JPsi events generated in the kinematic
  region $\pt > 50$\GeV and $\abs{y}<1.2$, illustrating the efficiency
  drop when the two muons are too close to each other.}
\label{fig:BPH:rho-factor}
\end{figure}
This drop in the dimuon trigger efficiency, shown in
Fig.~\ref{fig:BPH:rho-factor}, is induced through a muon pair
correlation and, hence, is not taken into consideration through the
simple product of the efficiencies of the two single muons. The
corresponding correction can be evaluated by MC simulation
and validated by studying distributions of measured events as a
function of the distance between the two muon tracks. In the 2012
run, a new trigger was developed, in which a high-\pt single muon
selected at L1 and L2 is associated with a tracker muon at L3 before a
dimuon mass range is imposed. In such events, there is only a single
muon required at the L1 and L2 steps, so that the event is not
rejected in case that there is a second muon very close by. This trigger
path is ideally suited to study charmonium production at very high
\pt.

\section{Trigger menus}
\label{sec:triggermenus}
A trigger menu is defined as the sum of all object definitions and
algorithms that define a particular configuration of the CMS trigger
system. The menu consists of definitions of L1 objects and the algorithms
that are used to render the L1 decision, as well as the configuration of
the software modules that are used in the HLT. Sets of prescale columns
for different instantaneous luminosities are also included. By means of
such a prescale set the data-archiving rate of the readout chain could
be adjusted and maximized during a LHC fill as the  instantaneous
luminosity drops along with the current trigger rate.

In this section, we describe the L1 and HLT menus and how they have
evolved in response to the physics goals and significant performance
improvements of the LHC machine during the first run.

\subsection{L1 menus}

From 2010 to 2012, several L1 menus (and corresponding prescale
columns) were developed to meet the experiment's physics goals and to
cope with the evolution of the LHC operational conditions, \ie,
the change of the center-of-mass energy between 2011 and 2012,
the varying number of colliding bunches for LHC fills, and the growth
of luminosity per bunch.
While designing new L1 menus, improved algorithms and thresholds were
utilized to continuously maintain the L1 trigger output rate within
the 100\unit{kHz} bandwidth limit. When the luminosity ramp-up phase
stabilized in 2011 and 2012, the strategy focused on reducing the
number of L1 menus being developed to a few per year, and adapting for
different machine operational conditions by using multiple prescale
columns rather than different L1 menus.
At the end of 2012, during a twelve-hour-long fill, the instantaneous
luminosity delivered by LHC varied significantly spanning from
${\approx}7\times10^{33}\percms$ to
${\approx}2.5\times10^{33}\percms$. The average number of pileup
events per interaction ranged from ${\approx}$30 at the beginning to
${\approx}$12 at the end of the run.

To aid the L1 menu development using data, a special reduced-content
event data format (containing only GCT, GMT and GT readout payloads)
was defined and used to record events in a special data set. These
events were collected on the basis of BPTX and L1 trigger GT decision
only.
Hence, with such recorded zero bias and L1 bias data sets, it was
possible to properly account for rate overlaps of the algorithms
operated in parallel in the GT~(Section~\ref{sec:global_trigger_desc})
while designing new menus.  Additionally, since the event size was
significantly smaller than the standard event sizes~\cite{DAQ-TDR}, it
was possible to collect a much higher trigger rate of these events
than the standard event-data payload, enabling frequent offline
analysis and cross-checks of the L1 trigger decision.

\subsubsection{Menu development}
The L1 menu development for the first LHC run was to a large extent based on data. Data recorded during standard collision runs and from
special LHC setups including high pileup runs. To better understand the features of the LHC machine, different magnet and collimator settings were used. In addition, some data were taken with very few proton bunches. Large number of protons per bunch lead to significantly more collisions per bunch crossing, resulting in high-pileup events. These events were used to project trigger rates at improved LHC performance. Simulated data samples were also used to evaluate the
impact of the 7\TeV to 8\TeV LHC energy increase in 2012.

For the L1 menu development, as well as for the development of the L1 trigger algorithms,
we followed the following principles and strategy:
\begin{itemize}
\item use single-object triggers as baseline algorithms and adjust
  thresholds to be sensitive to the electroweak physics as well as new
  physics, \eg, heavy particles, multi-object final states,
  events with large missing transverse energy;
\item in case the thresholds of the single-object triggers are too
  high with respect to the given physics goals (or if the acceptance for a
  given signal can be largely increased), use multi-object triggers,
  \eg, two muons or one muon plus two jets;
\item prefer algorithms which are insensitive to changing LHC run
  conditions, \eg, prefer algorithms that are less sensitive to
  pileup events; and
\item the algorithms and thresholds in a new L1 menu developed, \eg,
  for a different instantaneous luminosity, should result, if
  possible, in a similar sharing of rates for the same type of
  triggers: \ie, the muon triggers, $\Pe/\Pgg$ and jet/sum
  triggers should have the same rate at a different instantaneous
  luminosity  compared to the existing L1 menu.
\end{itemize}

\begin{table}[tbp]\centering
\topcaption{Machine operational conditions, target instantaneous luminosity used for
  rate estimation, and approximate overall L1 rate for three sample L1
  menus, representative of the end of the year data-taking conditions
  for 2010, 2011, and 2012.}
\begin{tabular}{ | l  c  c  c  c |}
\hline
Year & $\sqrt{s}$ [\TeVns{}] & Ref. \lumi [\!\percms] & $\langle$pileup$\rangle$ & $\langle$L1 rate$\rangle$ [kHz] \\
\hline
2010 & 7 & $0.15 \times  10^{33}$ & ${\approx}$2.5 & 56.9 \\
2011 & 7 & $3.00 \times  10^{33}$ & ${\approx}$14  & 80.9 \\
2012 & 8 & $5.00 \times  10^{33}$ & ${\approx}$23  & 56.5 \\
\hline
\end{tabular}
\label{tab:l1_trigger_menus}
\end{table}

\begin{figure}[tbp]
\includegraphics[width=0.48\linewidth]{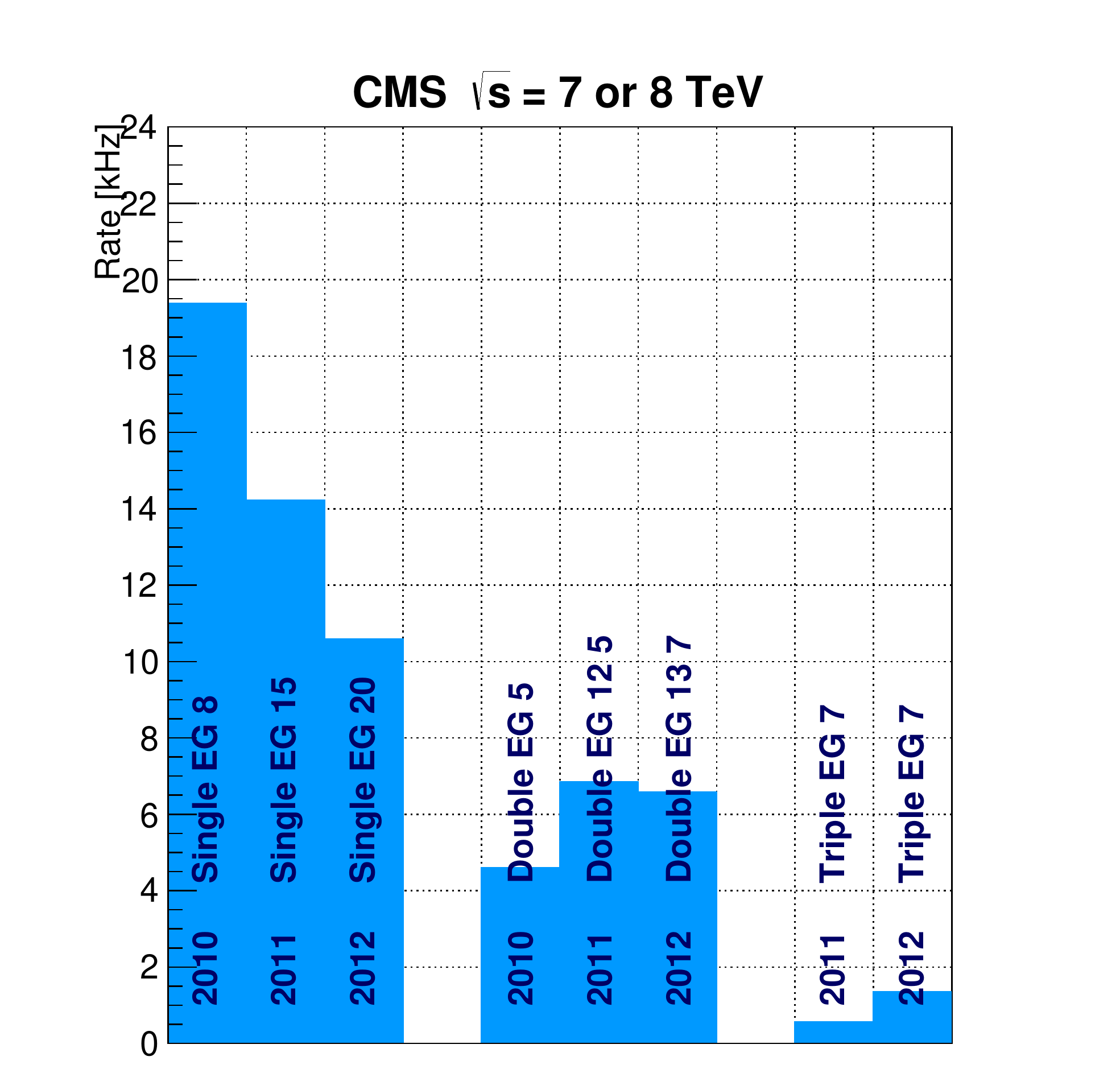}
\includegraphics[width=0.48\linewidth]{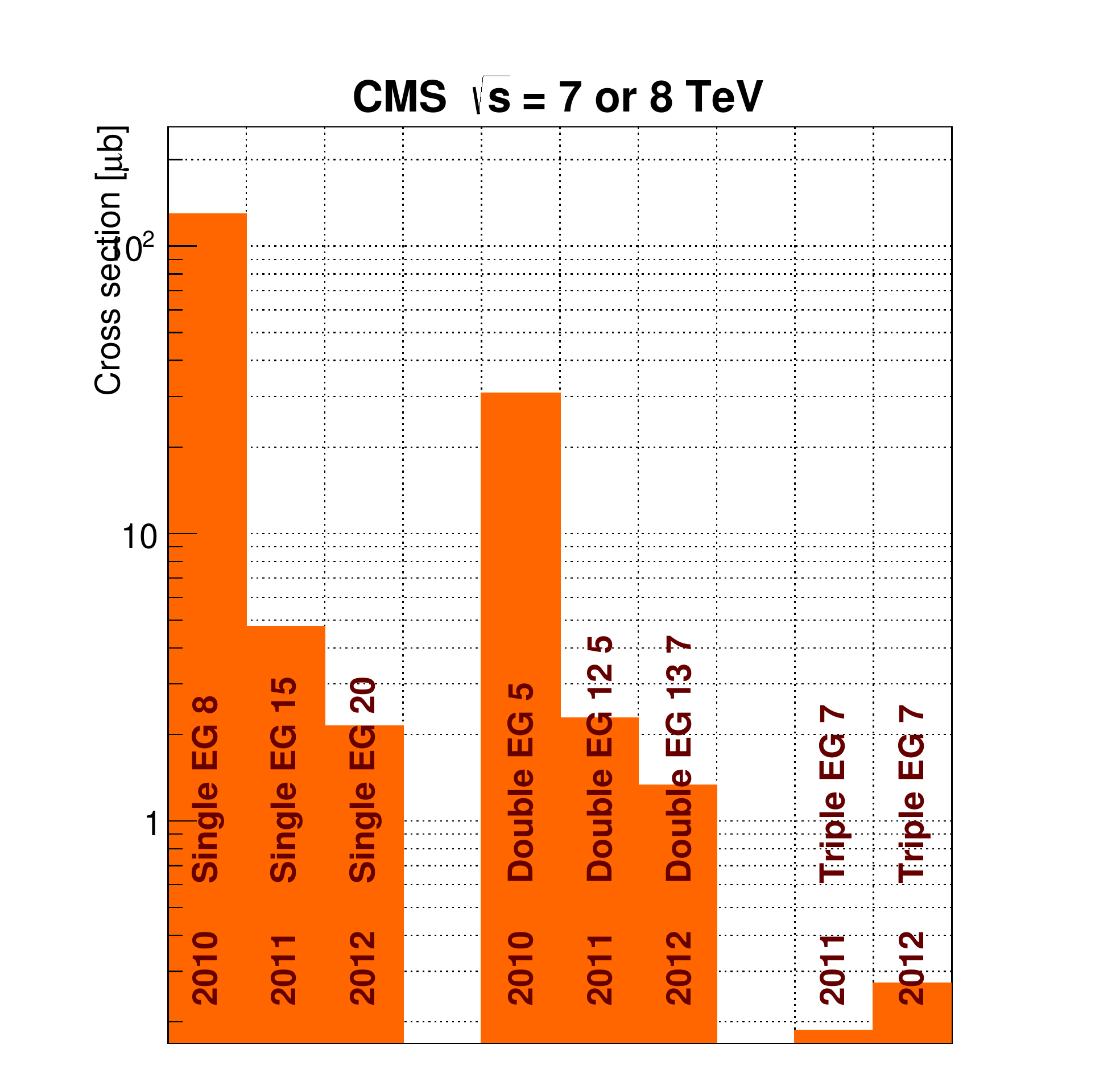}
\caption{Rates (left) and cross sections (right) for a significant
  sample of L1 $\Pe/\Pgg$ triggers
         from 2010, 2011, and 2012 sample menus.}
\label{fig:rate_evolution_eg}
\end{figure}
\begin{figure}[tbp]
\includegraphics[width=0.48\linewidth]{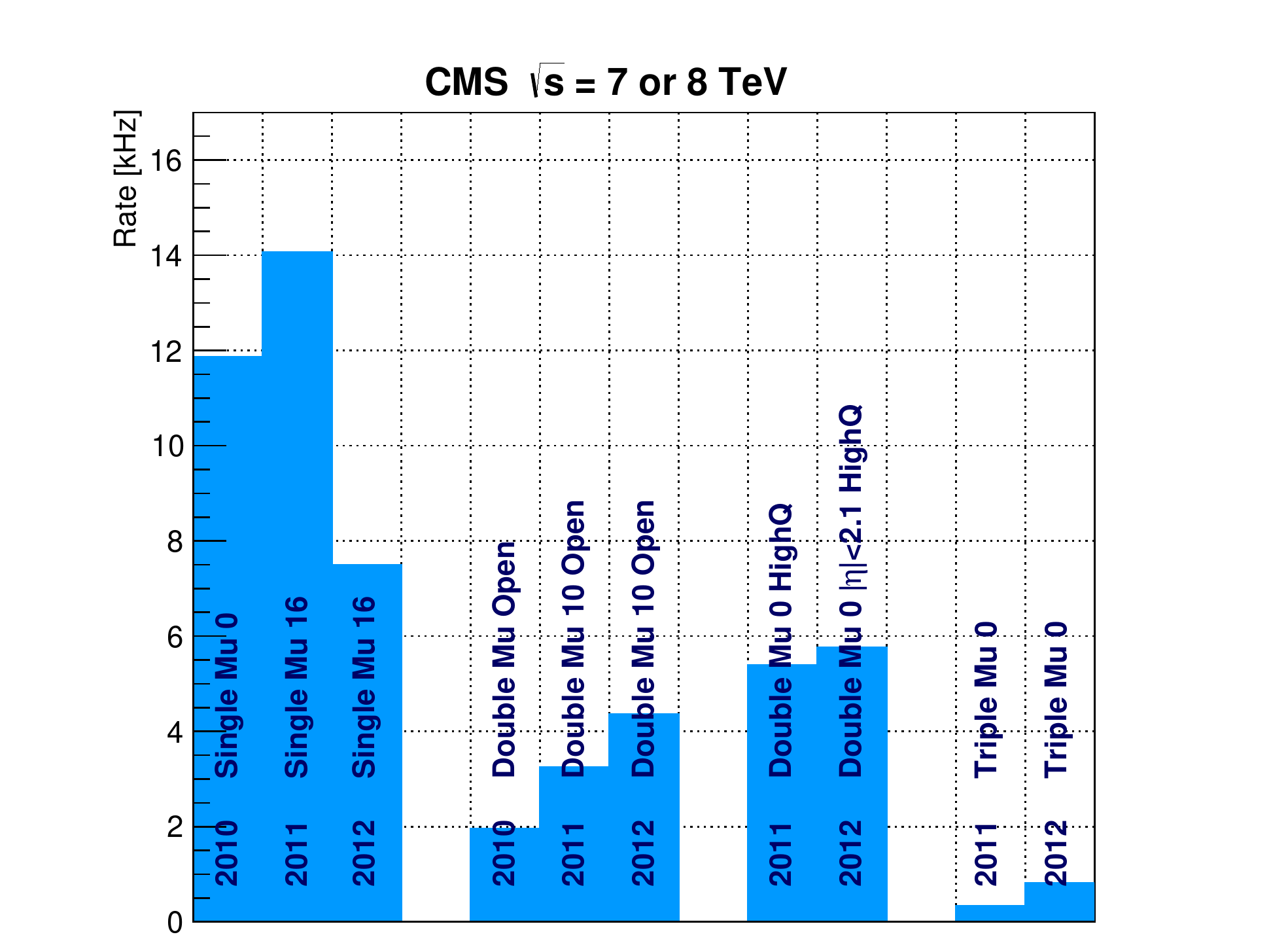}
\includegraphics[width=0.48\linewidth]{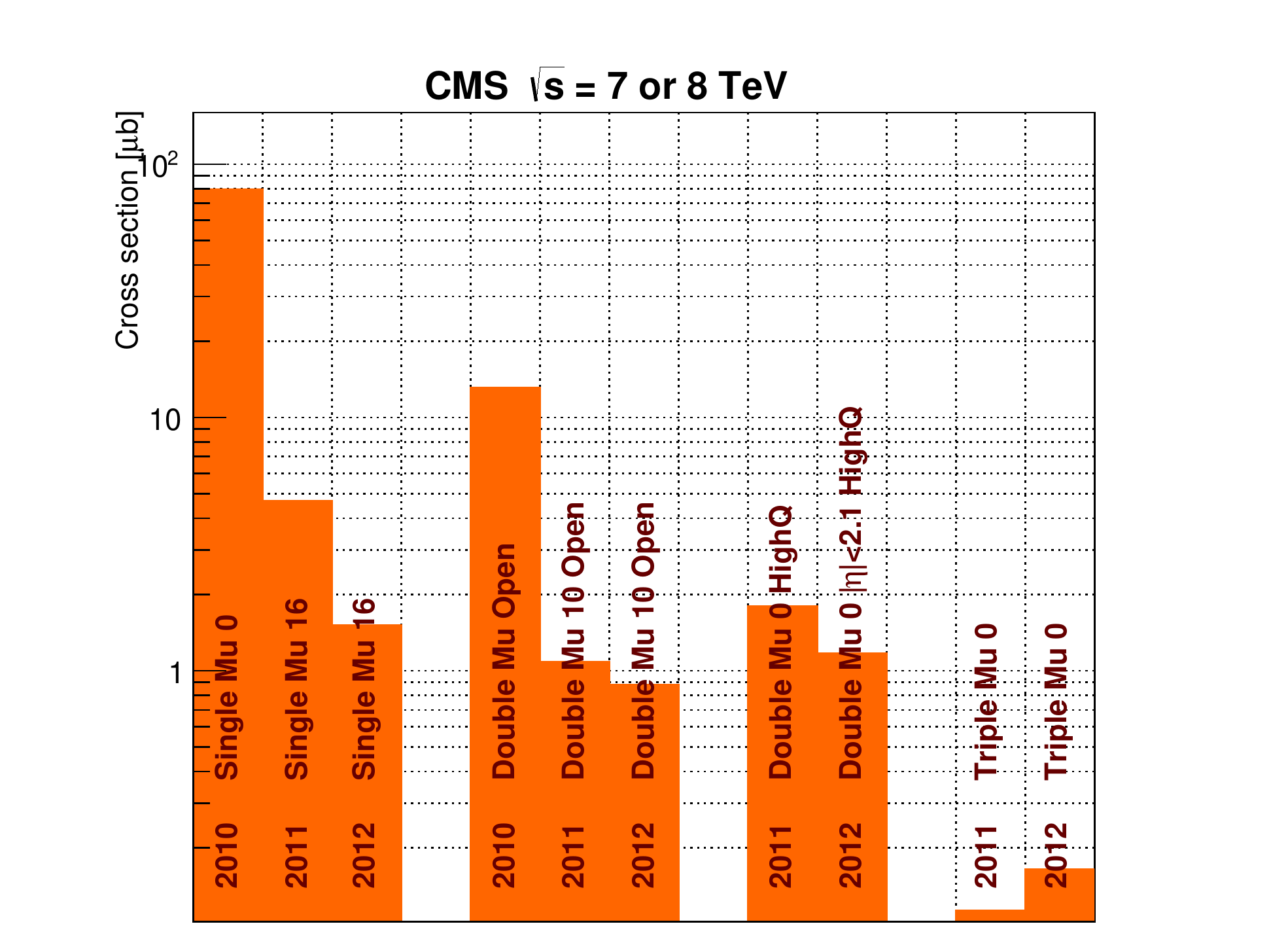}
\caption{Rates (left) and cross sections (right) for a significant sample of L1 muon triggers
         from 2010, 2011, and 2012 sample menus.}
\label{fig:rate_evolution_mu}
\end{figure}
\begin{figure}[tbp]
\centering
\includegraphics[width=0.9\linewidth]{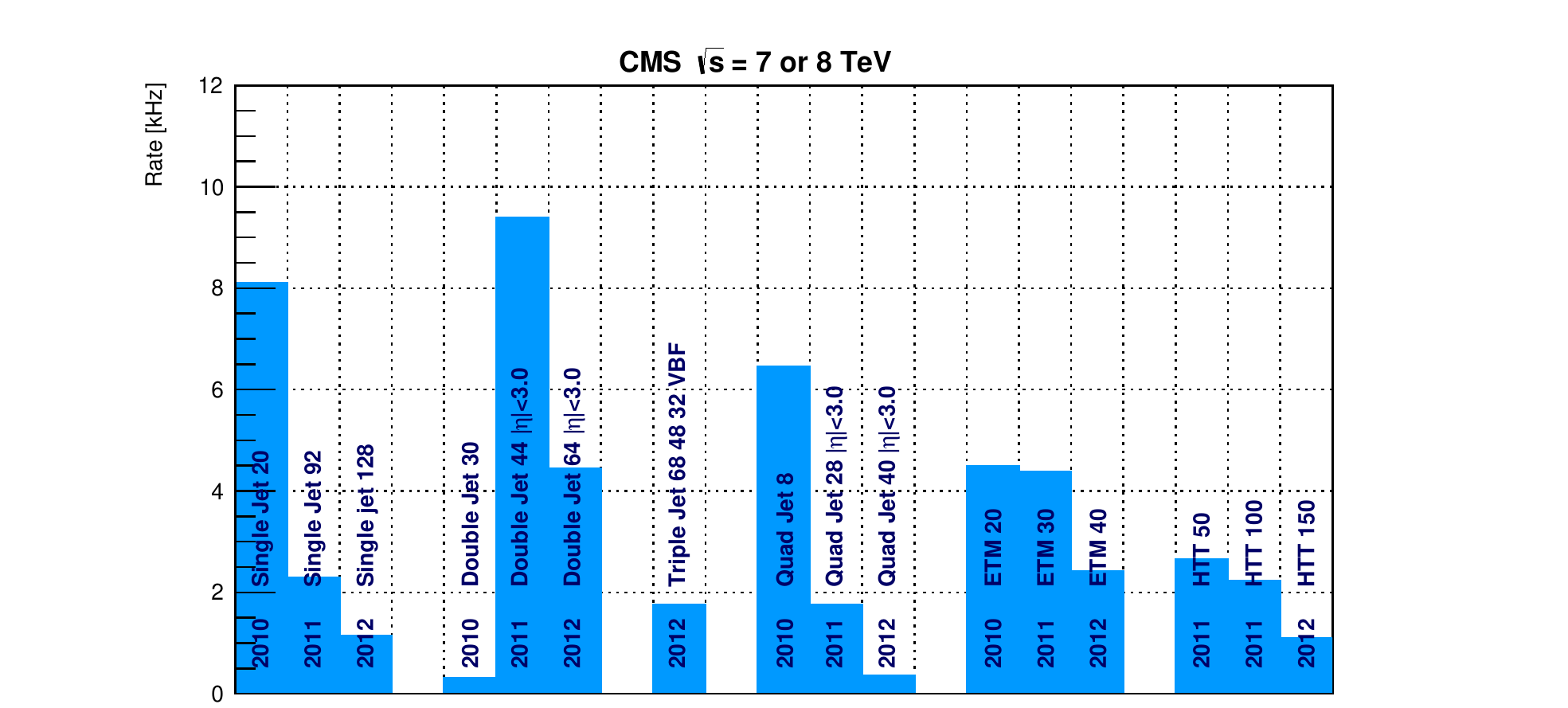}
\includegraphics[width=0.9\linewidth]{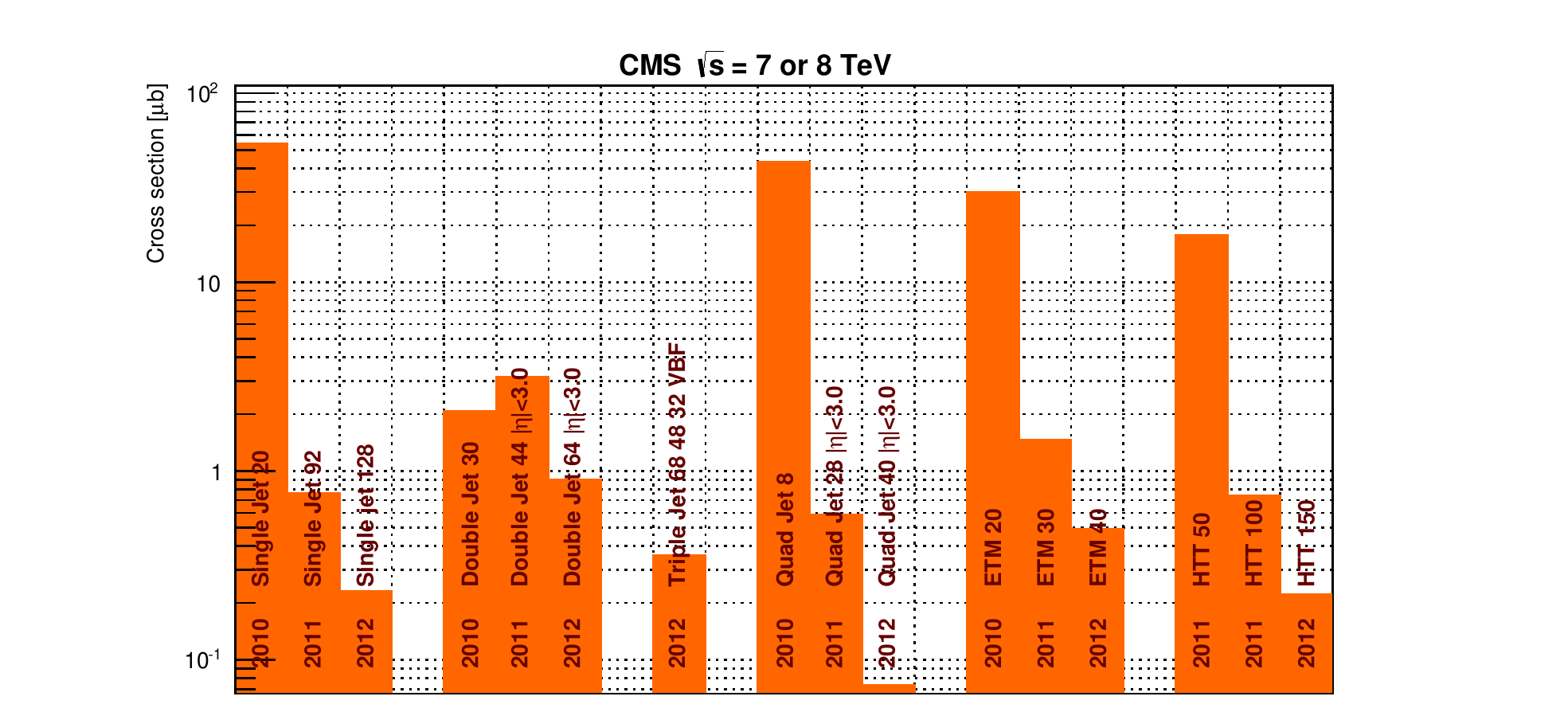}
\caption{Rates (top) and cross sections (bottom) for a significant sample of L1 jet triggers
         from 2010, 2011, and 2012 sample menus.}
\label{fig:rate_evolution_jet_sums}
\end{figure}

Table~\ref{tab:l1_trigger_menus} gives an overview of typical output
rates of the L1 trigger system in 2010, 2011, and 2012, and
Table~\ref{tab:menudetails} shows details for a typical 2012 menu.  The
examples are chosen for LHC run periods where the measured
instantaneous luminosities were close to the ones the different menus
were designed for. The overall L1 trigger output rate was
significantly higher than 50\unit{kHz} and well below the 100\unit{kHz} limit, as
intended. The differences of observed and predicted total trigger
rates largely depended on how the L1 trigger was operated, \ie, if a
prescale column was changed at instantaneous luminosities different
from the desired operating instantaneous luminosity of a specific L1
menu it followed that the total trigger output rate changed
significantly ($\mathcal{O}(10\unit{kHz})$).

The average L1 total trigger output rate varied from year to year due
to adaptations to the changing LHC conditions.
Figures~\ref{fig:rate_evolution_eg}, \ref{fig:rate_evolution_mu}, and
\ref{fig:rate_evolution_jet_sums} show trigger rates and
cross sections of the various lowest threshold, unscaled, single-object,
and multi-object triggers defined for the first LHC run. It was found that for almost
all given triggers in a specific menu, the rates and cross sections had to be similar or lower compared to an earlier used trigger menu. This was required to maintain the overall L1 trigger
output rate of below the 100\unit{kHz} limit taking into account the increasing LHC performance. To achieve this goal for
higher instantaneous luminosities, \ie, later in 2011 and 2012,
multi-object trigger algorithms as well as higher object thresholds
were used.

\begin{table}[tbp]\centering \topcaption{Rates from a significant set of unscaled algorithms
  participating to a typical L1 menu used during 2012 data-taking.
  Rates and cross sections ($\sigma$) are computed for a target luminosity
  of 5$\times 10^{33}\percms$. The overall menu   rate (including
  calibration and monitoring  triggers) is 56.5\unit{kHz}. The corresponding
  average pileup is  approximately 23 interactions per bunch crossing. }
\begin{tabular}{ | l  c  c | }
        \hline
        Seed name & rate @ $5 \times 10^{33}\percms$ & $\sigma$  \\
                  & [kHz]       & [$\mu$b] \\
        \hline
	\texttt{L1\_SingleIsoEG18er} & 7.69 & 1.55  \\
	\texttt{L1\_SingleEG20}      & 10.5 & 2.14  \\
	\texttt{L1\_SingleMu12er}    & 8.11 & 1.64  \\
	\texttt{L1\_SingleMu16}      & 7.49 & 1.51  \\
	\texttt{L1\_SingleJet128}    & 1.15 & 0.232 \\
	\hline
	\texttt{L1\_SingleMu6\_NotBptxOR}    &  0.03 & 0.007 \\
	\texttt{L1\_SingleJetC32\_NotBptxOR} &  0.13 & 0.026 \\
	\hline
	\texttt{L1\_ETM36}  &  4.35 & 0.881 \\
	\texttt{L1\_HTT150} &  1.10 & 0.223 \\
	\texttt{L1\_ETT300} &  0.21 & 0.043 \\
	\hline	
	\texttt{L1\_DoubleEG\_13\_7}    &  6.58 & 1.33  \\
	\texttt{L1\_DoubleMu\_10\_Open} &  4.36 & 0.882 \\
	\texttt{L1\_DoubleMu0er\_HighQ} &  5.77 & 1.16  \\
	\texttt{L1\_DoubleJetC56}       &  7.59 & 1.53  \\
	\texttt{L1\_DoubleTauJet44er}   &  1.88 & 0.381 \\
	\hline
        \texttt{L1\_TripleMu0}                  &  0.81 & 0.165 \\
	\texttt{L1\_TripleEG\_12\_7\_5}         &  2.19 & 0.444 \\
	\texttt{L1\_TripleEG7}                  &  1.35 & 0.273 \\
	\texttt{L1\_TripleJet\_64\_48\_28\_VBF} &  2.28 & 0.462 \\
	\hline
	\texttt{L1\_QuadJetC36} &  0.74 & 0.150 \\	
	\hline
	\texttt{L1\_Mu3p5\_EG12}      &  2.34 & 0.474 \\
	\texttt{L1\_Mu12\_EG7}        &  1.03 & 0.208 \\
	\texttt{L1\_Mu0\_HTT100}      &  0.46 & 0.094 \\
	\texttt{L1\_Mu7er\_ETM20}     &  1.19 & 0.241 \\
	\texttt{L1\_IsoEG12er\_ETM30} &  1.54 & 0.311 \\
	\texttt{L1\_EG22\_ForJet24}   &  2.42 & 0.489 \\
	\hline
	\texttt{L1\_DoubleMu5\_EG5}      &  0.54 & 0.109 \\
	\texttt{L1\_Mu5\_DoubleEG6}      &  0.96 & 0.194 \\
	\texttt{L1\_DoubleEG6\_HTT100}   &  1.32 & 0.266 \\
	\texttt{L1\_DoubleJetC36\_ETM30} &  3.40 & 0.688 \\
	\texttt{L1\_Mu10er\_JetC12\_WdEtaPhi1\_DoubleJetC\_20\_12} &  1.02 & 0.207 \\
	\hline
\end{tabular}
\label{tab:menudetails}
\end{table}

\subsection{HLT menus}

The configuration of all the HLT paths that are run online at one time is called the  HLT menu of CMS. This menu was initially prepared, based on simulated data, before the first data was taken in 2010, and has  continuously evolved since then. This evolution is driven mainly by the changes in the machine conditions, namely $\sqrt{s}$, luminosity, bunch spacing, and pileup conditions.
Moreover, timing improvements in the software-based algorithms and analysis techniques allowed the online algorithms to be brought much closer to the ones adopted offline, leading to better performance, as well as closer correspondence between the online and the offline selections. In addition to the trigger paths designed to preselect the events to be used in the physics analyses, calibration and monitoring paths for the different CMS subdetectors are also necessary and were included in all menus.

The first menus in 2010 consisted of fewer than 60 separate trigger
paths. The low instantaneous luminosity supplied by the LHC at that
time allowed the use of several ``pass-through'' paths, in which the
events accepted by the L1 trigger are accepted also by
the HLT without further requirements and restrictions. In addition to the pass-through paths, single ``physics object" triggers started to
be developed, meant to trigger on inclusive isolated or non-isolated
electrons, photons, muons, and jets. As the instantaneous luminosity
increased, the strategies used to control the trigger rate consisted
of: raising \pt thresholds, adding isolation and quality conditions in
the identification of jets, leptons, and photons, using prescales,
introducing cross-triggers (triggers which require several physics
objects of different types), and defining dedicated $\tau$-like jets.
Moreover, a few other paths were included in the menu to study
possible implementations for future data-taking periods at higher
rate. During 2010, 12 different trigger menus were developed, covering
the wide range of instantaneous luminosity scenarios provided by the
LHC (from $1\times10^{27}$ to $2\times10^{32}$\percms). In
addition, prescale values were designed according to the LHC luminosity.

In 2011, with the LHC still operating at 7\TeV center-of-mass energy,
six different HLT and L1 menus were designed, aimed at instantaneous
luminosities ranging from $5\times10^{32}$ to
$5\times10^{33}$\percms. Tighter selections were therefore needed, and
the refinement of the trigger requirements was achieved by gradually
introducing analysis-like selection criteria at the trigger
level. Besides the usual ``physics object''-oriented paths, the
presence of cross-triggers and more complex trigger paths, based on
algorithms similar to those applied in the offline analyses, became
more and more relevant in the menu. A few refined techniques used
offline could therefore be brought to the HLT, after adapting them to
reduce the CPU time needed, at the expense of very little performance, and without
greatly compromising their final response. Amongst those techniques,
particle flow reconstruction~\cite{CMS-pf1} was used since the beginning to characterize the hadronically decaying
$\tau$ leptons at the HLT. Towards the end of the 2011 data-taking
period, strategies designed to mitigate the effect of pileup were also
included in several trigger paths, with the intent of studying their
performance in view of the 2012 data taking when the pileup effect was
expected to become more relevant. In this respect the so-called
  \FASTJET corrections~\cite{fastjetmanual}, offset corrections which
take into account the average energy density in the event and the area
of each jet in order to correct its energy on a jet-by-jet basis,
proved very successful.

The number of paths deployed in the 2011 menus rose from about 310 at
the beginning of the data taking to approximately 430 towards the end
of the year. A few paths were included specifically to monitor and
calibrate CMS subdetector components. For example, the response of the
electromagnetic calorimeter, which is fundamental for the selection
and analysis of the Higgs boson decaying in two photons, is
continuously monitored by some dedicated paths, in order to provide
updates to the calibrations in a timely manner.

When the 8\TeV run began in 2012, because of the higher instantaneous luminosity achieved by the LHC, pileup effects became much more important and therefore improvements in the design of most of the paths included in the menu were required. Ideally, the acceptance rate of a trigger should be proportional to the instantaneous luminosity, however, due to pileup it may increase non-linearly. This effect, together with the higher LHC luminosity, would give rise to unacceptably high trigger rates.
The rate increases can only be mitigated by either raising the
acceptance thresholds in the path themselves (with the unwanted effect
of reducing the physics reach of the events selected), or by improving
the performance of the selections, with sharper turn-on curves at the
thresholds and less sensitivity to pileup. The main handle used to
achieve this goal, without affecting the CMS physics potential, was
the extension of the implementation of particle flow reconstruction to
most jet- and \MET-based triggers. The replacement of
calorimeter-based jet triggers with PF-based ones was introduced gradually during the year. An additional advantage of using the PF in the trigger selection is that the selection algorithms are mostly the same as those used offline for the final analysis; however, reconstruction algorithms based on PF methods eventually do need more time compared to more ``classical'' one-object-per-detector-type algorithms. One idea used in the HLT to reduce the overall CPU time consumption was to move the PF reconstruction after all other possible selections, which were based on more classical quantities, which are faster to calculate. Among the other technical improvements to the HLT algorithms that allowed the rate to be kept low and the CPU time manageable, the following were particularly relevant: the optimization of lepton identification and isolation; the use of a filter to select leptons coming from the same vertex in several dilepton paths; weekly updates to ECAL transparency corrections, which allows efficiently compensating for the transparency loss in the endcap region affecting the electromagnetic energy scale; and the dedicated $\tau$ lepton reconstruction for the double-$\tau$ and lepton+$\tau$ triggers.

Different menus were used in 2012 for four different LHC peak luminosities, ranging from $5\times10^{33}$ to $8\times10^{33}$\percms. The number of different HLT paths during 2012 was approximately 400 at the beginning, increasing to about 450 by the end of the year.

In addition to the proton-proton triggers, dedicated menus for the heavy ion (lead-lead) collisions in 2010 and 2011 and for the proton-lead collisions in the first months of 2013 were created. The different running conditions and physics requirements led to different menus for the ion-ion and the proton-ion runs. The final heavy ion menus in 2010 and 2011 consisted of 58 and 77 HLT paths, respectively, while the proton-ion menu of 2013 contains about 150 paths.

\section{Trigger system operation and evolution}
\label{sec:operations}

\subsection{Trigger monitoring and operations}
During data taking the angular distributions of objects satisfying the
trigger and the trigger rates were monitored. As these two kinds of
information are produced using two different software tools, they
provide complementary information about the behavior of the trigger system that
are useful in diagnosing problems.

We use the central CMS data quality monitoring (DQM) tools~\cite{DQM} to
monitor the angular distributions. The DQM tools process a small subset of
events selected by the HLT and produce plots of $\eta$ and $\phi$ of the
trigger objects. The distributions are monitored for regions with abnormal
appearances of either too many or too few events.

The rates of each HLT path are  monitored in each node of the HLT computing
cluster, where the CMS data acquisition software records how many times each
trigger path was successful. The path information is summed over all nodes to
give the total rate of each path. The summation occurs at every fixed integrated luminosity,
and the results are written into a database. The HLT group has developed
customized software to extract the rates from the database and compare them to
expectations. The expected rate behavior is defined by fitting the trigger
rates as a function of instantaneous luminosity using  previously recorded data certified as
good. Uncertainties from the fit provide an envelope of expected rate variations
which sets the threshold for displaying warnings in the control room. A
selected set of approximately 20 different representative triggers, out of the 400
of the HLT menu, is used for regular online monitoring. The selected HLT paths
have either a large rate or an important physics signature. For instance, some
of the closely monitored triggers includes single-muon, single-electron, and
diphoton triggers, where rate  variations are identified  with a 5--10\%
sensitivity.

\subsection{Technical performance}
\subsubsection{The L1 trigger deadtime, downtime and reliability}
``Deadtime during active beam" is defined as the percentage of time
during normal data taking (data acquisition system in ``run" mode)
when collisions occur but CMS is not ready to record triggers. There
are several contributions to this deadtime:
\begin{itemize}
\item Partition controller deadtime: It arises when any CMS subsystem
  (such as a subdetector or trigger subsystem) asserts ``not ready"
  because of a transient problem (\eg, ``out of sync", requiring
  a ``resync" command) or because the instantaneous trigger rate is
  too high.
\item Trigger rules deadtime: A set of
  trigger rules of the type ``not more than $m$ triggers within $n$
  bunch crossings" limits the instantaneous trigger rate.
\item Calibration deadtime: At a rate of 100\unit{Hz}, calibration
  triggers are sent (required mainly by the electromagnetic
  calorimeter) and a small part of the orbit is blocked for this
  purpose.
\end{itemize}
Usually, deadtime was kept to approximately 1\%, only a small fraction
of which was due to trigger subsystems.

``Downtime" is the percentage of time when data acquisition system
cannot be put into run mode during active beams because of a
malfunctioning subsystem.  During regular running, the downtime due to
trigger subsystems was well fewer than one percent. In most cases, the
trigger downtime caused by hardware or software crashes, which could
be fixed by restarting the electronics subsystem or the software
process, respectively. To take care of the rare cases where an
electronic module is faulty, spare modules for all systems are kept in
the electronics cavern. For the GT, a fully equipped spare crate is
kept running and ready to take over at any time in case of a hardware
fault.  Empirically, we observe that the L1 trigger system contributes only a
small fraction to the total experiment downtime.

\subsubsection{The HLT resources and optimization}

As described in Section~\ref{sec:HLTDAQ}, the HLT runs on an EVF consisting of three different types of machines. Two complementary methods are used to monitor the usage of this farm by the HLT menu. The first method directly measures the time taken by the HLT selection and reconstruction steps for each event  during data taking. The second method rapidly samples every CPU in the farm to determine its state, and the time per event is calculated based on the frequency of finding the CPU in a non-idle state. The two methods give consistent results. Using the second method, the total busy fraction of the EVF can also be determined.

\begin{figure}[tbp]
  \centering
    \includegraphics[width=0.6\textwidth]{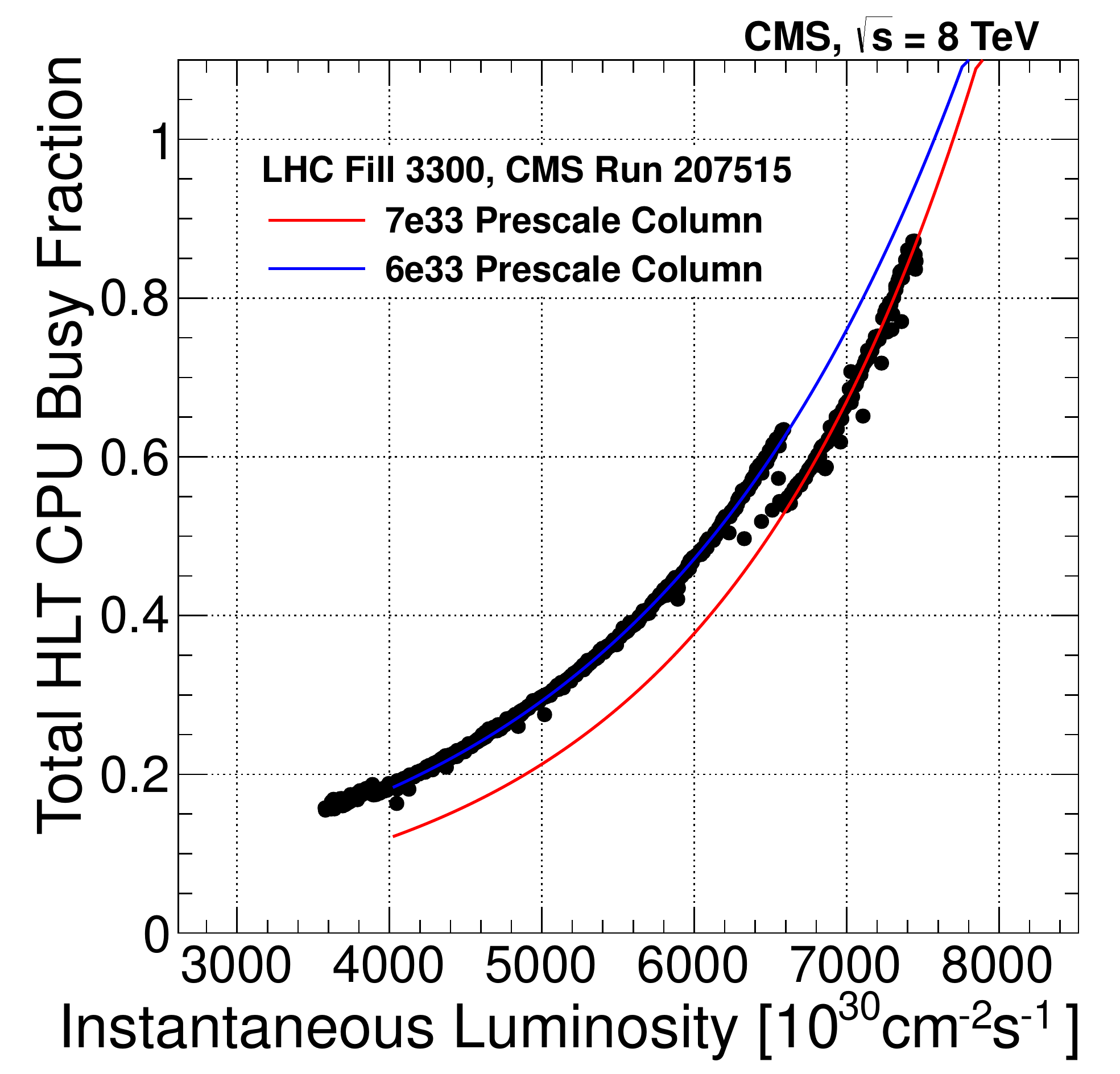}
  \caption{The average CPU busy fraction as a function of instantaneous luminosity for one LHC fill. Luminosity sections with data-taking deadtime $>$40\% are removed.}
  \label{fig:HLTCPUFraction}
\end{figure}

\begin{figure}[tbp]
  \centering
    \includegraphics[width=0.6\textwidth]{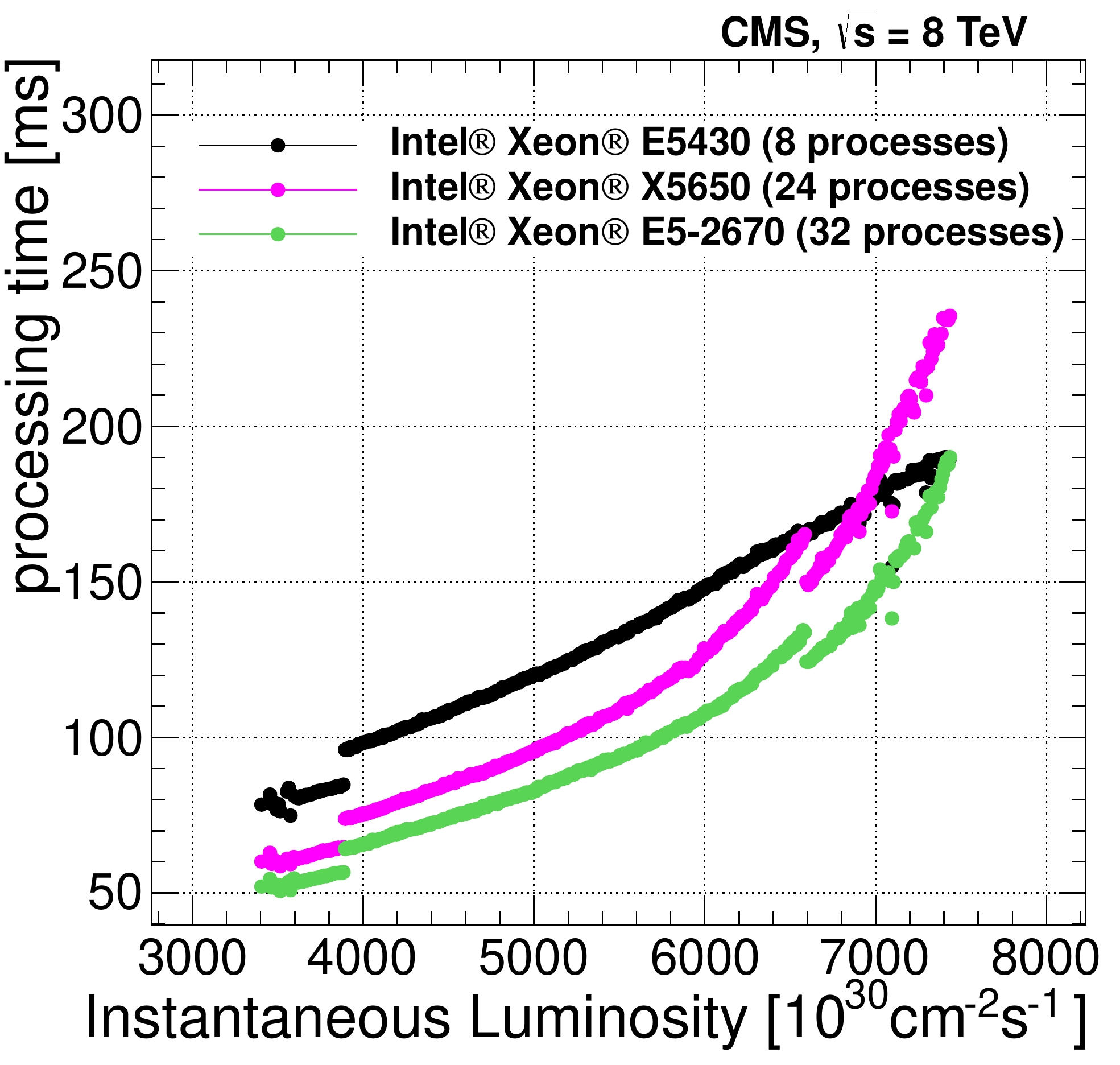}
  \caption{The HLT processing time per event as a function of instantaneous luminosity
           for  the three different machine types used in the filter farm.}
  \label{fig:HLTMachineProcessingTimes}
\end{figure}

To estimate the CPU usage of an HLT menu at a future (higher) instantaneous luminosity value, the average busy fraction over the course of an LHC fill is measured, and a fit is performed as a function of instantaneous luminosity, as shown in Fig.~\ref{fig:HLTCPUFraction}.
An exponential function is found to give a good description of the data over a wide range of instantaneous luminosity and allows extrapolation to higher luminosities. In addition, we also measure the time per event for each type of machine used in the filter farm as shown in Fig.~\ref{fig:HLTMachineProcessingTimes}.
The time per event is observed to be approximately linear as a function of luminosity on the Intel Xeon E5430 CPUs. The other two types of CPUs employ Intel's hyper-threading to run twice as many concurrent processes as there are physical cores by using parts of the CPU that would otherwise be idle. As a result, the time per event for these hyper-threaded CPUs increases faster than linearly as the CPU is saturated with increasing luminosity and input rate. Using this information, it is possible to calculate the maximum time per event of the HLT menu for a given L1 input rate, and also the instantaneous luminosity at which this limit would be reached. The figure of merit used is the time per event for an Intel Xeon E5430 CPU. The filter farm configuration used during 2012 data taking was able to sustain an average processing time per event of approximately 200~ms for an L1 input rate of 100\unit{kHz}.

\begin{figure}[tbp]
  \centering
    \includegraphics[width=0.6\textwidth]{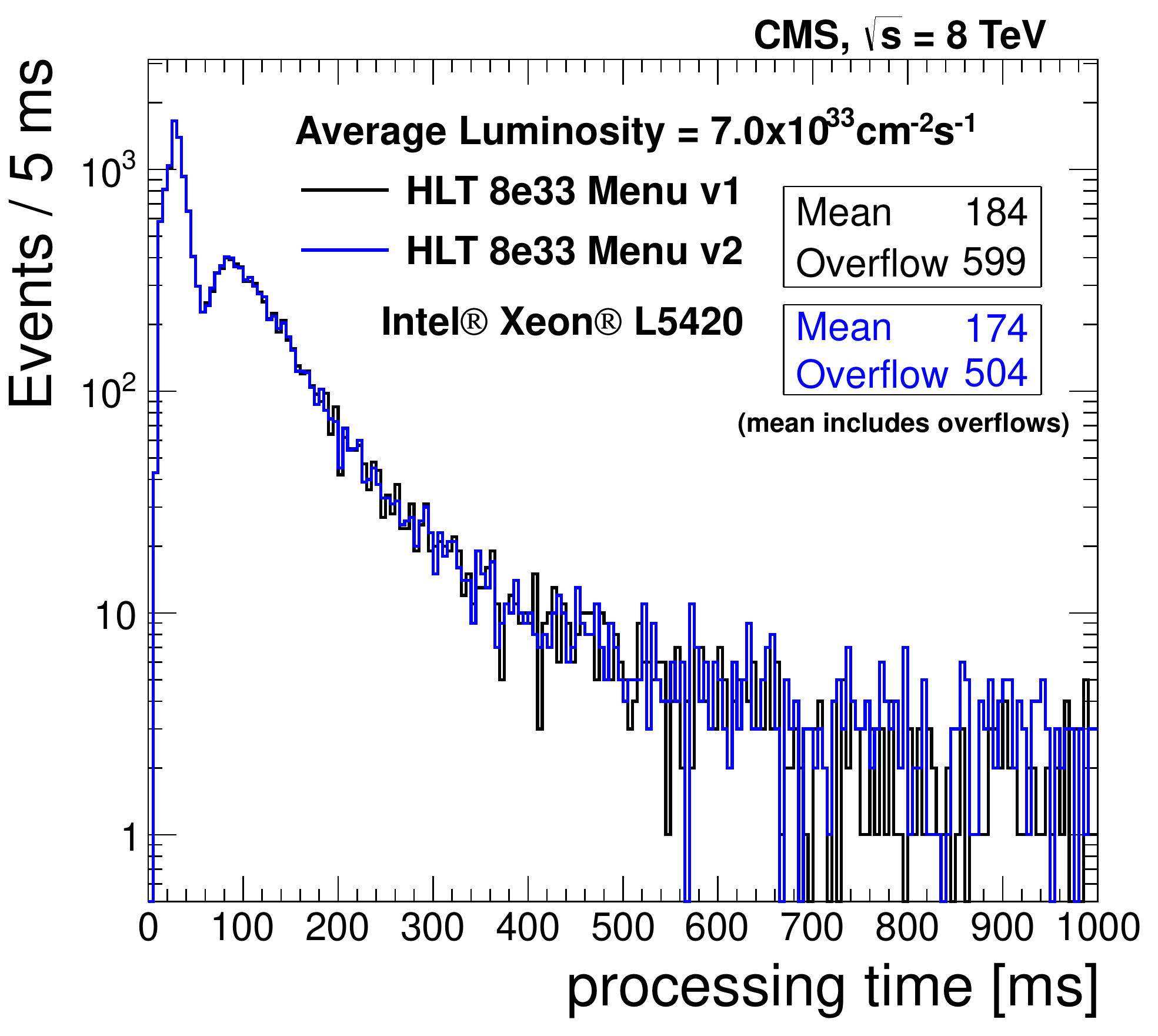}
  \caption{Comparison of the time per event measured for two different HLT menus using a validation
           machine outside of the event filter farm.}
  \label{fig:HLTOfflineComparison}
\end{figure}

In addition to the online monitoring of the HLT menus, each menu is validated in an offline environment before being used for online data taking. Each new version of the menu is compared to a previous version on a single machine to ensure that the CPU consumption does not exceed expectations. The menus are tested by running the HLT once with each menu over the same sample of previously collected events. The measurement is done using a machine with similar core architecture to the Intel Xeon E5430 CPU, and is performed using the direct timing measurement described above. New instantaneous luminosity and L1 input rate limits can then be determined by using the relative performance of the new menu and the measured performance of the older menu. An example of an offline comparison of the times per event for two different HLT menus is shown in Fig.~\ref{fig:HLTOfflineComparison}.  When testing a new menu, the time per event for each HLT path is also checked to determine which paths are the most CPU intensive. The algorithms for CPU-intensive paths are then optimized to ensure that the total processing time does not exceed the limitations of the system.

\subsubsection{The HLT operations}
Following offline validation, HLT menus are validated in an online environment using the HLT online (``HiLTOn'') test stand. In order to be as close as possible to the online environment, the HiLTOn is operated using the same run control interface as the CMS DAQ system. The HiLTOn hardware consists of 30 Dell PE 1950 machines with dual quad-core 2.66~GHz CPUs and 16~GB of RAM. The test stand system is subdivided into three groups of ten machines. The first group is always kept identical to the PE 1950 machines used for data taking operations and thus can be used to validate menus on the current online software environment. The second group of machines may additionally be used to validate the performance of software updates, and HLT menus that depend on the updates, in an online environment. Finally, the third subdivision of the HiLTOn is used to evaluate changes made to the HiLTOn itself. Two machines of each group are dedicated to the building of new online software releases. A third machine is always used to collect the output of the HLT and save events to disk. The remaining seven machines in each group are able to process events via seven instances of the HLT per machine, although only four machines are used for typical menu validation.

The HLT validation is designed to maximize the performance and stability of HLT algorithms. As every event satisfying L1 trigger requirements is examined by the HLT, and several HLT decisions are based on analysis-quality physics objects reconstructed using information from all CMS subdetectors, HLT reliability is critical to the success of the experiment. On rare occasions, typically below a few events per month during data taking, one or more HLT algorithms will experience a processing error while examining a collision event. These events are stored locally for later analysis and are used to improve the reliability of the HLT software.

Roughly 0.5\% of the downtime during collision data taking operations from 2009 until 2012 (including all proton-proton and heavy ion collision operations at any center-of-mass energy) was due to problems with the HLT; 95\% of this downtime was due to a single incident when corrupt detector input resulted in the HLT failure for every incoming collision event. Prior and subsequent data taking using the same HLT menu resulted in no loss of data. Ignoring this incident, the HLT was responsible for a negligible loss of collision data.

\section{Summary}
\label{sec:summary}

The CMS trigger system consists of two levels: an L1 custom
hardware trigger, and an HLT system with custom
\textsc{c++} software routines running on commodity CPUs.

The L1 trigger takes input from the calorimeters and the muon system
to select the events of physics interest. To do this, it uses
identified leptons, photons, and jet candidates, as well as
event-level information such as missing transverse energy.  Trigger
primitives are generated on the front-ends of the subdetectors and then
processed in several steps before a final decision is rendered in the
global trigger.

The L1 calorimeter trigger comprises two stages, a regional
calorimeter trigger (RCT) and a global calorimeter trigger (GCT).  The
RCT processes the regional information in parallel and sends as output
$\Pe/\Pgg$ candidates and regional \ET sums.  The GCT sorts the
$\Pe/\Pgg$ candidates further, finds jets (classified as central,
forward, and tau) using the \ET sums, and calculates global quantities
such as \MET.

Each of the three muon detector systems in CMS participates in the L1
muon trigger to ensure good coverage and redundancy. For the DT and
CSC systems, the front-end trigger electronics identifies track
segments from the hit information registered in multiple detector
planes.  Track finders apply pattern recognition algorithms which
identify muon candidates, and measure their momenta from the amount
of their bending in the magnetic field of the return yoke between measurement
locations.  The RPC hits are directly sent from the
front-end electronics to pattern-comparator logic boards which
identify muon candidates.  The global muon trigger merges muon
candidates, applies quality criteria, and sends the muon candidates to
the global trigger.

The global trigger implements the menu of selection requirements
applied to all objects. A maximum of 128 separate selections can be
implemented simultaneously. Overall, the L1 decision is rendered within
$4\mus$ after the collision. At most 100\unit{kHz} of events are sent
to the HLT for processing.

The HLT is implemented in software, and further refines the purity of
the physics objects. Events are selected for offline storage at an
average rate of 400\unit{Hz}. The HLT event selection is performed in a
similar way to that used in the offline processing. For each event,
objects such as leptons, photons, and jets are reconstructed and
identification criteria are applied in order to select only those
events which are of possible interest for data analysis. The HLT
hardware consists of a processor farm using commodity PCs running
Scientific Linux. The subunits are called builder and filter units. In
the builder units, event fragments are assembled to complete
events. Filter units then unpack the raw data and perform event
reconstruction and trigger filtering.  Both the L1 triggers and HLT
include prescaling of events passing defined selection criteria.

The performance of the CMS trigger system has been evaluated in two
stages. First, the performance of the L1 and HLT systems has been
evaluated for individual trigger objects such as electrons, muons,
photons, or jets, using tag-and-probe techniques. Most of the
measurements considered come from the 2012 CMS data set, where data
have been collected at $\sqrt{s}=8$\TeV.  Performance has been
evaluated in terms of efficiency with respect to offline quantities
and to the appropriate trigger rate. Both L1 and HLT performance have
been studied, showing the high selection efficiency of the CMS
trigger system.  Second, the performance of the trigger system has
been demonstrated by considering key examples across different physics
analyses. In CMS, the HLT decisions often are derived from complex
correlated combinations of single objects such as electrons, muons, or
$\tau$ leptons. The broad range of capabilities of the trigger system
has been shown through examples in Higgs boson, top-quark, and
B physics, as well as in searches for new physics.

The trigger system has been instrumental in the successful collection
of data for physics analyses in Run~1 of the CMS experiment at the
LHC. Efficiencies were measured in data and compared to simulation,
and shown to be high and well-understood. Many physics signals were
collected with high efficiency and flexibility under rapidly-changing
conditions, enabling a diverse and rich physics program, which has led
to hundreds of publications based on the Run~1 data samples.

\begin{acknowledgments}
\hyphenation{Bundes-ministerium Forschungs-gemeinschaft Forschungs-zentren}
\hyphenation{Rachada-pisek}
We congratulate our colleagues in the CERN accelerator departments for the excellent performance of the LHC and thank the technical and administrative staffs at CERN and at other CMS institutes for their contributions to the success of the CMS effort. In addition, we gratefully acknowledge the computing centers and personnel of the Worldwide LHC Computing Grid for delivering so effectively the computing infrastructure essential to our analyses. Finally, we acknowledge the enduring support for the construction and operation of the LHC and the CMS detector provided by the following funding agencies: the Austrian Federal Ministry of Science, Research and Economy and the Austrian Science Fund; the Belgian Fonds de la Recherche Scientifique, and Fonds voor Wetenschappelijk Onderzoek; the Brazilian Funding Agencies (CNPq, CAPES, FAPERJ, and FAPESP); the Bulgarian Ministry of Education and Science; CERN; the Chinese Academy of Sciences, Ministry of Science and Technology, and National Natural Science Foundation of China; the Colombian Funding Agency (COLCIENCIAS); the Croatian Ministry of Science, Education and Sport, and the Croatian Science Foundation; the Research Promotion Foundation, Cyprus; the Secretariat for Higher Education, Science, Technology and Innovation, Ecuador; the Ministry of Education and Research, Estonian Research Council via IUT23-4 and IUT23-6 and European Regional Development Fund, Estonia; the Academy of Finland, Finnish Ministry of Education and Culture, and Helsinki Institute of Physics; the Institut National de Physique Nucl\'eaire et de Physique des Particules~/~CNRS, and Commissariat \`a l'\'Energie Atomique et aux \'Energies Alternatives~/~CEA, France; the Bundesministerium f\"ur Bildung und Forschung, Deutsche Forschungsgemeinschaft, and Helmholtz-Gemeinschaft Deutscher Forschungszentren, Germany; the General Secretariat for Research and Technology, Greece; the National Scientific Research Foundation, and National Innovation Office, Hungary; the Department of Atomic Energy and the Department of Science and Technology, India; the Institute for Studies in Theoretical Physics and Mathematics, Iran; the Science Foundation, Ireland; the Istituto Nazionale di Fisica Nucleare, Italy; the Ministry of Science, ICT and Future Planning, and National Research Foundation (NRF), Republic of Korea; the Lithuanian Academy of Sciences; the Ministry of Education, and University of Malaya (Malaysia); the Mexican Funding Agencies (BUAP, CINVESTAV, CONACYT, LNS, SEP, and UASLP-FAI); the Ministry of Business, Innovation and Employment, New Zealand; the Pakistan Atomic Energy Commission; the Ministry of Science and Higher Education and the National Science Centre, Poland; the Funda\c{c}\~ao para a Ci\^encia e a Tecnologia, Portugal; JINR, Dubna; the Ministry of Education and Science of the Russian Federation, the Federal Agency of Atomic Energy of the Russian Federation, Russian Academy of Sciences, and the Russian Foundation for Basic Research; the Ministry of Education, Science and Technological Development of Serbia; the Secretar\'{\i}a de Estado de Investigaci\'on, Desarrollo e Innovaci\'on and Programa Consolider-Ingenio 2010, Spain; the Swiss Funding Agencies (ETH Board, ETH Zurich, PSI, SNF, UniZH, Canton Zurich, and SER); the Ministry of Science and Technology, Taipei; the Thailand Center of Excellence in Physics, the Institute for the Promotion of Teaching Science and Technology of Thailand, Special Task Force for Activating Research and the National Science and Technology Development Agency of Thailand; the Scientific and Technical Research Council of Turkey, and Turkish Atomic Energy Authority; the National Academy of Sciences of Ukraine, and State Fund for Fundamental Researches, Ukraine; the Science and Technology Facilities Council, UK; the US Department of Energy, and the US National Science Foundation.

Individuals have received support from the Marie-Curie program and the European Research Council and EPLANET (European Union); the Leventis Foundation; the A. P. Sloan Foundation; the Alexander von Humboldt Foundation; the Belgian Federal Science Policy Office; the Fonds pour la Formation \`a la Recherche dans l'Industrie et dans l'Agriculture (FRIA-Belgium); the Agentschap voor Innovatie door Wetenschap en Technologie (IWT-Belgium); the Ministry of Education, Youth and Sports (MEYS) of the Czech Republic; the Council of Science and Industrial Research, India; the HOMING PLUS program of the Foundation for Polish Science, cofinanced from European Union, Regional Development Fund, the Mobility Plus program of the Ministry of Science and Higher Education, the National Science Center (Poland), contracts Harmonia 2014/14/M/ST2/00428, Opus 2013/11/B/ST2/04202, 2014/13/B/ST2/02543 and 2014/15/B/ST2/03998, Sonata-bis 2012/07/E/ST2/01406; the Thalis and Aristeia programs cofinanced by EU-ESF and the Greek NSRF; the National Priorities Research Program by Qatar National Research Fund; the Programa Clar\'in-COFUND del Principado de Asturias; the Rachadapisek Sompot Fund for Postdoctoral Fellowship, Chulalongkorn University and the Chulalongkorn Academic into Its 2nd Century Project Advancement Project (Thailand); and the Welch Foundation, contract C-1845.
\end{acknowledgments}

\bibliography{auto_generated}

\cleardoublepage \appendix\section{The CMS Collaboration \label{app:collab}}\begin{sloppypar}\hyphenpenalty=5000\widowpenalty=500\clubpenalty=5000\textbf{Yerevan Physics Institute,  Yerevan,  Armenia}\\*[0pt]
V.~Khachatryan, A.M.~Sirunyan, A.~Tumasyan
\vskip\cmsinstskip
\textbf{Institut f\"{u}r Hochenergiephysik der OeAW,  Wien,  Austria}\\*[0pt]
W.~Adam, E.~Asilar, T.~Bergauer, J.~Brandstetter, E.~Brondolin, M.~Dragicevic, J.~Er\"{o}, M.~Flechl, M.~Friedl, R.~Fr\"{u}hwirth\cmsAuthorMark{1}, V.M.~Ghete, C.~Hartl, N.~H\"{o}rmann, J.~Hrubec, M.~Jeitler\cmsAuthorMark{1}, V.~Kn\"{u}nz, A.~K\"{o}nig, M.~Krammer\cmsAuthorMark{1}, I.~Kr\"{a}tschmer, D.~Liko, T.~Matsushita, I.~Mikulec, D.~Rabady\cmsAuthorMark{2}, B.~Rahbaran, H.~Rohringer, J.~Schieck\cmsAuthorMark{1}, R.~Sch\"{o}fbeck, J.~Strauss, W.~Treberer-Treberspurg, W.~Waltenberger, C.-E.~Wulz\cmsAuthorMark{1}
% ...
\vskip\cmsinstskip
% ...
1:~~Also at Vienna University of Technology, Vienna, Austria\\
2:~~Also at CERN, European Organization for Nuclear Research, Geneva, Switzerland\\
% ...

\end{sloppypar}
\end{document}